\documentclass[aps,prb,reprint, floatfix]{revtex4-1}

\usepackage{amsmath, amssymb}
\usepackage{bm}
\usepackage{graphicx}
\usepackage{color}
\usepackage{dcolumn}
\usepackage{bm}
\usepackage[hidelinks]{hyperref}
\hypersetup{
  colorlinks   = true, 
  urlcolor     = blue, 
  linkcolor    = blue, 
  citecolor   = red 
}
\usepackage{multirow}

\allowdisplaybreaks[1]

\begin{document}

\title{
 Dynamic Jahn-Teller Phenomena in Heavy Transition Metal Compounds
}

\author{Naoya Iwahara}
\email[]{naoya.iwahara@gmail.com}
\affiliation{Graduate School of Engineering, Chiba University, 1-33 Yayoi-cho, Inage-ku, Chiba-shi, Chiba 263-8522, Japan}

\begin{abstract}
 This paper reviews recent experimental and theoretical developments of the dynamic Jahn-Teller effect-driven phenomena in heavy transition metal based spin-orbit Mott insulators. 
 In cubic $4d/5d$ transition metal compounds, the spin, orbital, and lattice degrees of freedom can form quantum entanglement on metal sites and induce unconventional quantum phenomena. 
 Fingerprints of orbital-lattice entanglement called the dynamic Jahn-Teller effect appear in spectroscopic data such as resonant inelastic x-ray scattering spectra. 
 In cubic $5d^1$ compounds with the fcc structure, the dynamic Jahn-Teller states on metal sites behave cooperatively and exhibit rich ordered phases.
\end{abstract}

\maketitle

\section{Introduction}
\label{Sec:intro}
Heavy transition metal compounds have attracted significant attention for their exotic quantum magnetism, such as Kitaev spin-liquid in $d^5$ compounds, excitonic magnetism in the compounds with nonmagnetic $d^4$ ions, and multipolar ordering in $d^1$ and $d^2$ compounds \cite{Witczak-Krempa2014, Rau2016, Takagi2019, Motome2020, Takayama2021, Trebst2022, Chen2024}.
The critical ingredient for the emergence of these phenomena has been the strong spin-orbit entangled states on $4d$ or $5d$ metal centers in octahedral environments.
Based on the idea, diverse phenomena in the lattices of the spin-orbit entangled $4d/5d$ octahedra have been studied \cite{Jackeli2009, Chen2010, Chen2011, Khaliullin2013, Ishizuka2014, Nasu2014, Natori2016, Romhanyi2017, Yamada2018, Liu2019, Svoboda2021, deCarvalho2023, Kubo2023, Pourovskii2023, Paddinson2024}.

\begin{figure}[bt]
\begin{tabular}{lllll}
(a) &~& (b) \\
\includegraphics[height=0.35\linewidth, bb = 0 0 315 266]{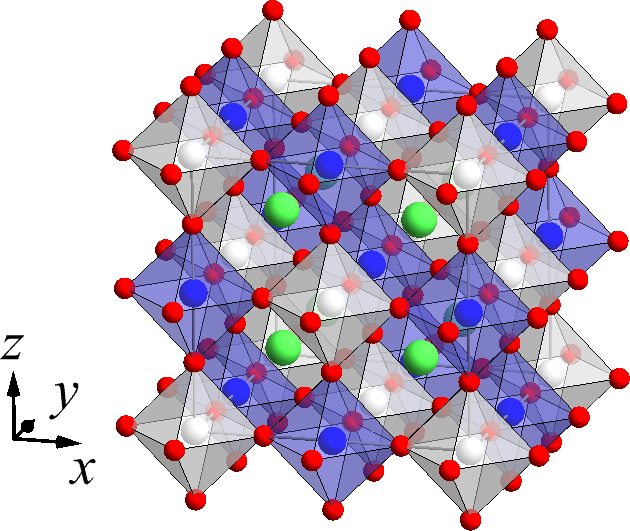}
& &
\includegraphics[height=0.35\linewidth, bb = 0 0 337 285]{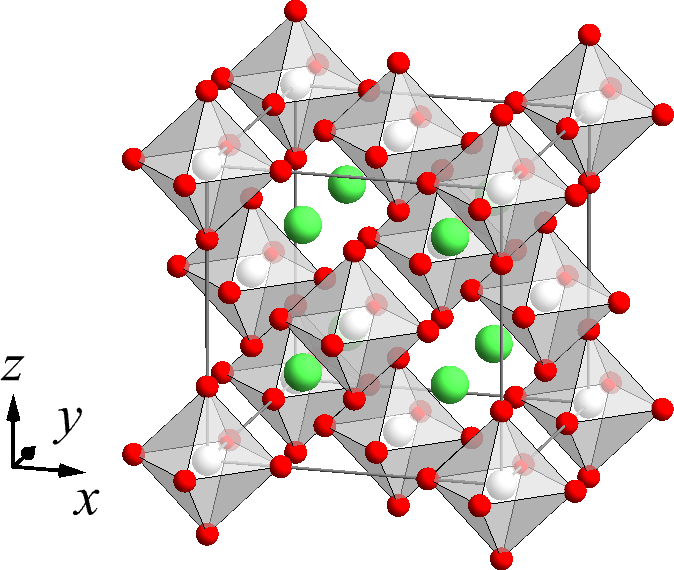}
\end{tabular}
\caption{
  (Color online)
  (a) Double perovskites $A_2BB'X_6$ and (b) antifluorite $A_2B'X_6$. 
  $A$ and $B$ ions are diamagnetic ions, $B'$ the magnetic ions, and $X$ the ligand atoms. 
}
\label{Fig:DP}
\end{figure}

In the transition metal compounds, the electron-phonon (or vibronic) coupling could also contribute to the emergence of unconventional phenomena; however, the insight into the vibronic coupling in spin-orbit Mott insulators remains limited. 
For example, in cubic $d^1$ (Ba$_2$YMoO$_6$, Ba$_2$MgReO$_6$) and $d^2$ (Ba$_2$YReO$_6$) double perovskites [Fig. \ref{Fig:DP}(a)], the electronic states of metal sites are orbitally degenerate and Jahn-Teller (JT) active, whereas the neutron diffraction measurements detected no JT deformation down to low temperature, which was called ``violation of the JT theorem'' \cite{Aharen2010a, Aharen2010b, Marjerrison2016a}. 
Later, various spectroscopic measurements of $d^1$ double perovskites and antifluorites [Fig. \ref{Fig:DP}(b)] detected tiny lattice deformations.
Nuclear magnetic resonance (NMR) measurements of Ba$_2$NaOsO$_6$ revealed that the ``breaking of the local point group symmetry'' phase starts to develop above the magnetic transition and continues to grow by lowering temperature  
\cite{Lu2017, Liu2018}. 
X-ray diffraction and resonant elastic x-ray scattering (REXS) measurements of Cs$_2$TaCl$_6$ \cite{Ishikawa2019, Tehrani2023} and Ba$_2$MgReO$_6$ \cite{Hirai2020, Soh2023} detected the development of the quadrupolar orderings accompanied by structural deformations.
The magnetic and quadrupolar phase in Cs$_2$TaCl$_6$ and the quadrupolar orderings in Ba$_2$MgReO$_6$ differ from the theoretical predictions based on the spin-orbit entangled states. 
Although quadrupolar orderings with small deformations develop in a couple of $5d^1$ double perovskites, the other members, such as Ba$_2$ZnReO$_6$ do not exhibit apparent deformations \cite{Barbosa2022}.

Theoretical works predict that the tiny structural deformations or their absence in a family of heavy $d^1$ and $d^2$ compounds could originate from the development of the dynamic JT effect. 
If the dynamic JT effect \cite{Moffitt1957a, Moffitt1957b, Longuet-Higgins1958} develops, it smears out the structural anisotropy of the crystals. 
{\it Ab initio} calculations supported the possibility \cite{Xu2016, Iwahara2018}. 
Recent resonant inelastic x-ray scattering (RIXS) measurements of several $5d^1$ Re and Os compounds exhibit unusual fine structures and their weak temperature dependence across the quadrupolar transition \cite{Frontini2024, Iwahara2024, Agrestini2024, Zivkovic2024}.
These features are the fingerprints of the dynamic JT effect \cite{Frontini2024, Iwahara2024}.

The dynamic JT effect could emerge in other cubic compounds with $d^4$ and $d^5$ ions. 
The structures of $d^4$ (K$_2$RuCl$_6$) and $d^5$ (K$_2$IrCl$_6$, K$_2$IrBr$_6$, Ba$_2$CeIrO$_6$) are cubic according to the x-ray or neutron diffraction, whereas the RIXS spectra, Raman spectra, and optical conductivity of these compounds show low-symmetric ligand-field-like splittings in their spectra \cite{Takahashi2021, Reigiplessis2020, Khan2021, Raveli2019, Lee2022, Warzanowski2024}. 
These works proposed that either the random local deformation or the dynamic JT effect causes the splitting, while analysis based on the dynamic JT effect has been missing until recently. 
We will explain the spectra in detail below.

Therefore, the family of cubic $4d/5d$ double perovskites and antifluorites is the playground to investigate the emergent phenomena driven by the dynamic JT effect. 
The JT effect in various metal compounds has been an important topic to describe the electronic properties when the symmetry lowering of the structure (the static JT effect \cite{Opik1957}) develops \cite{Kanamori1960, Englman1972, Gehring1975, Kugel1982, Kaplan1995, Tokura2000, Bersuker2006, Khomskii2021}. 
Contrary, the dynamic JT effect \cite{Moffitt1957a, Moffitt1957b, Longuet-Higgins1958} with pronounced quantum effect of the lattice degrees of freedom has been an important topic in the fields of molecules and impurities in crystals. 
The dynamic JT effect modulates the energy spectra \cite{Longuet-Higgins1958, Bersuker1962, OBrien1964, Caner1966, Muramatsu1978, Pooler1978, Pooler1980, Auerbach1994, Koizumi1994, Koizumi1995, OBrien1996, Sato2005, Iwahara2012, Requist2016, Requist2017} and 
gives rise to unusual spectroscopic features and physical properties \cite{Child1961, Ham1965, Toyozawa1966, Cho1968, Ham1968, Fukuda1969, Kahn1972, Gunnarsson1995, Worner2006, Fu2009, Iwahara2010, Abtew2011, Saha2015, Kayanuma2017, Zinchenko2023, Nandipati2023, Lyakhov2023, Magoni2023, Sarychev2024, Silkinis2024, Yanagisawa2024, Abragam1970, Bersuker1989, Bersuker2006}.
Nevertheless, the impact of the dynamic JT effect on the cooperative phenomena in crystals is unexplored.

This review discusses the recent developments on the dynamic JT effect and the related phenomena in cubic $4d$ and $5d$ compounds. 
Section \ref{Sec:JT} describes the dynamic JT effect on the heavy transition metal sites with an octahedral environment. 
This section introduces basic knowledge of the JT effect, such as vibronic coupling and the static and dynamic JT effect, and then, describes the experimental evidence of the development of the dynamic JT effect in heavy transition metal compounds. 
Section \ref{Sec:CDJT} focuses on the cooperative phenomena driven by the dynamic JT effect mainly in cubic $5d^1$ double perovskites.

\section{Dynamic Jahn-Teller effect in cubic site}
\label{Sec:JT}
In spin-orbit assisted Mott insulating phases \cite{Kim2008, Kim2009, Gangopadhyay2015, Gangopadhyay2016} of cubic heavy transition metal compounds, the low-energy phenomena originate from the competition of the spin, orbital, and lattice degrees of freedom on metal sites. 
The electronic degrees of freedom usually dominate the nature of the local quantum states, while the lattice vibrational degrees of freedom can be unquenched in high-symmetric environments. 
Under this situation, the quantum entanglement of the electronic (orbital) and lattice degrees of freedom, which is called the dynamic JT effect \cite{Moffitt1957a, Moffitt1957b, Longuet-Higgins1958, Englman1972, Bersuker1989, Grosso2014}, can develop. 
In crystals, the intersite interactions between the orbital-lattice entangled sites give rise to cooperative phenomena. 
In this section, keeping heavy transition metal double perovskites and antifluorite compounds in mind (Fig. \ref{Fig:DP}), we explain the dynamic JT effect on metal sites with octahedral ligand field. 
Although we sometimes use the irreducible representations of the octahedral group \cite{Sugano1970, Tanabe1954I, Koster1963, Inui1990}, detailed knowledge of group theory is not required. 
For general information on the vibronic coupling and JT effect, see Refs. \onlinecite{Sugano1970, Englman1972, Bersuker1989, Grosso2014}.

\begin{figure*}[tb]
\begin{tabular}{lcllc}
(a) & &~~& (b) \\
& \includegraphics[height=0.21\linewidth, bb = 0 0 465 216]{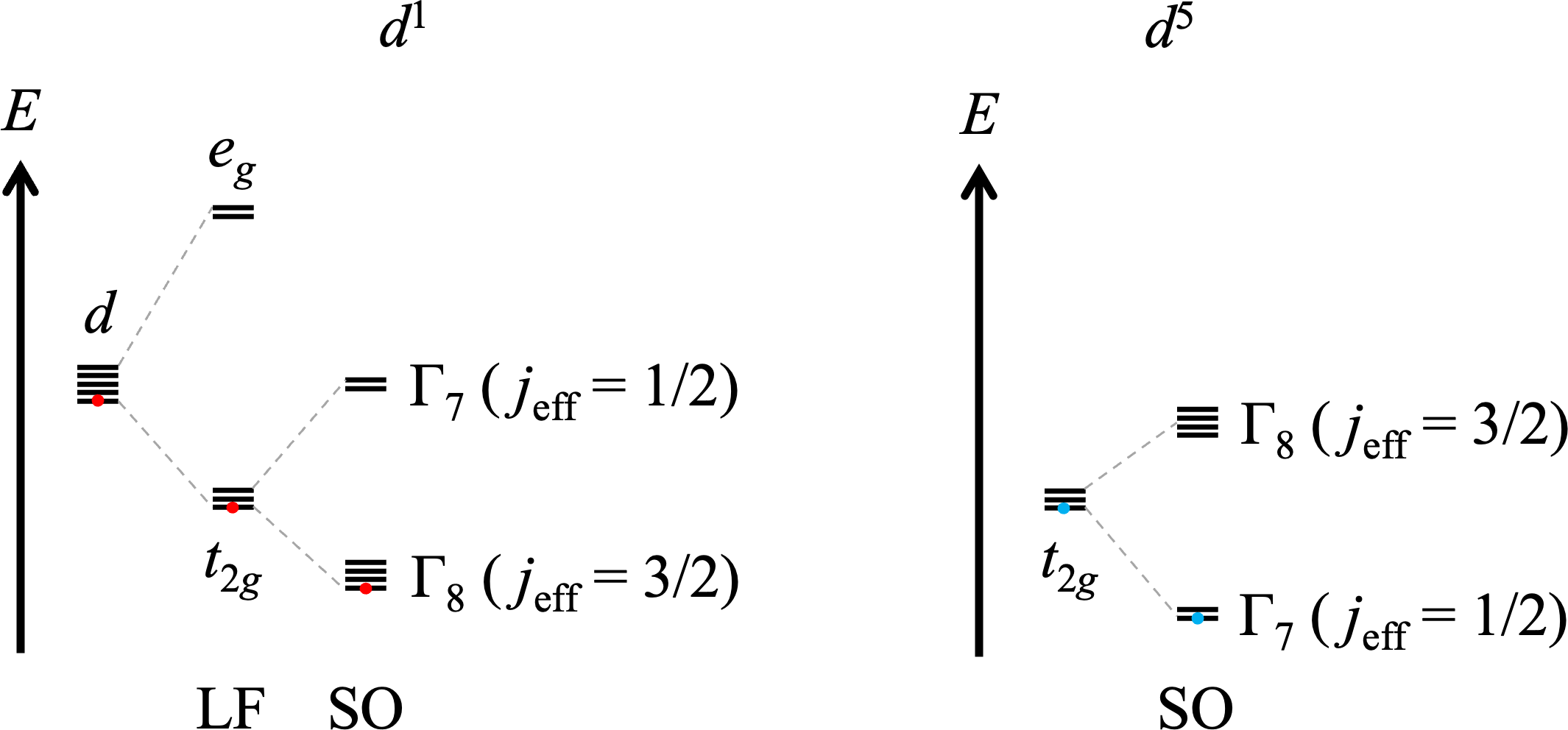} 
& &
& \includegraphics[height=0.21\linewidth, bb = 0 0 718 342]{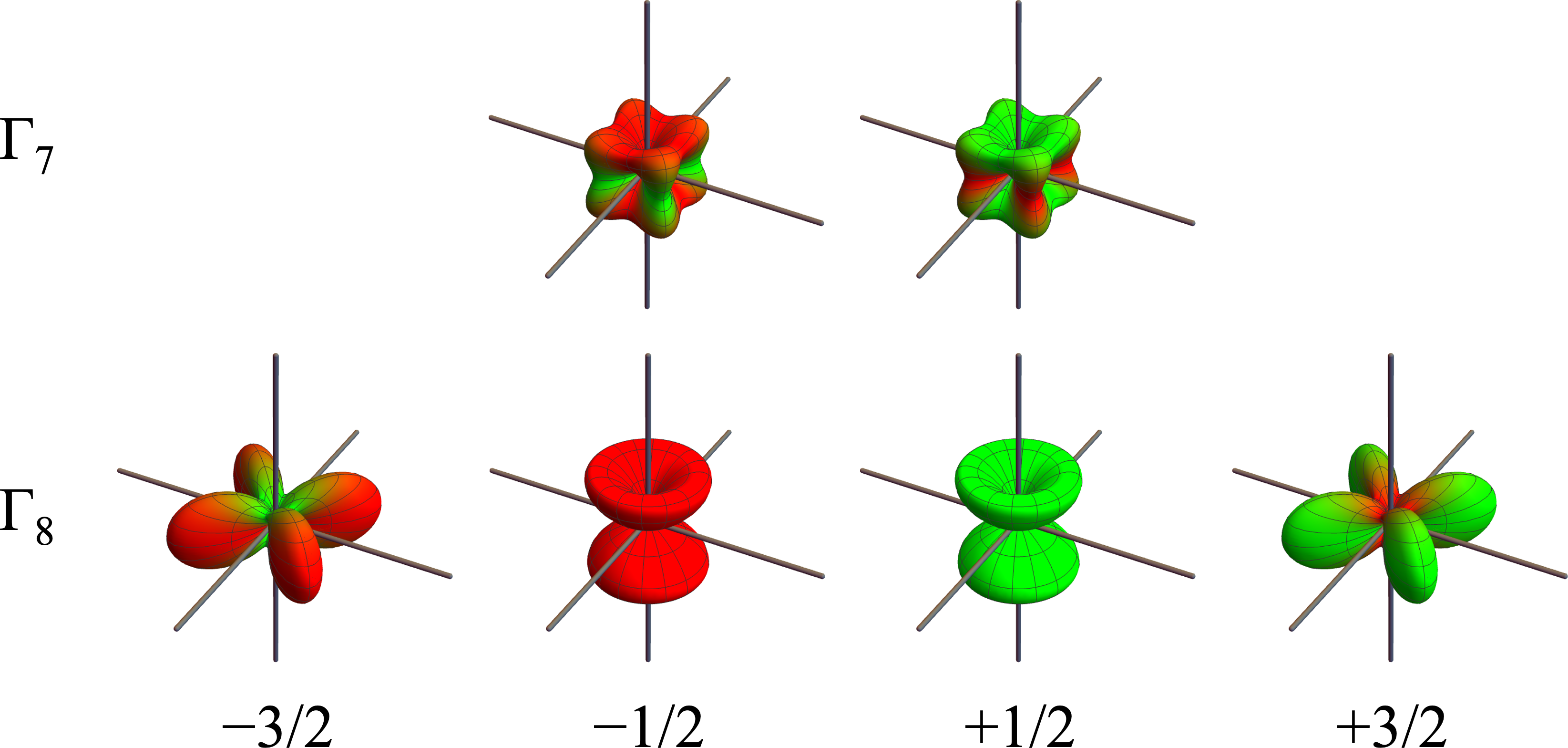}
\end{tabular}
\caption{
  (Color online)
  (a) The electronic energy levels and (b) the electron (hole) distributions of the spin-orbit coupled states of $d^1$ the $d^5$ octahedra. 
We use the hole picture for the $d^5$ ion. 
The red and green areas in the spatial distributions indicate the spin-up and spin-down distributions, respectively.  
}
\label{Fig:electronic_d1}
\end{figure*}

\begin{figure*}[tb]
\begin{tabular}{lcllc}
(a) & &~~& (b) \\
& \includegraphics[height=0.23\linewidth, bb = 0 0 498 235]{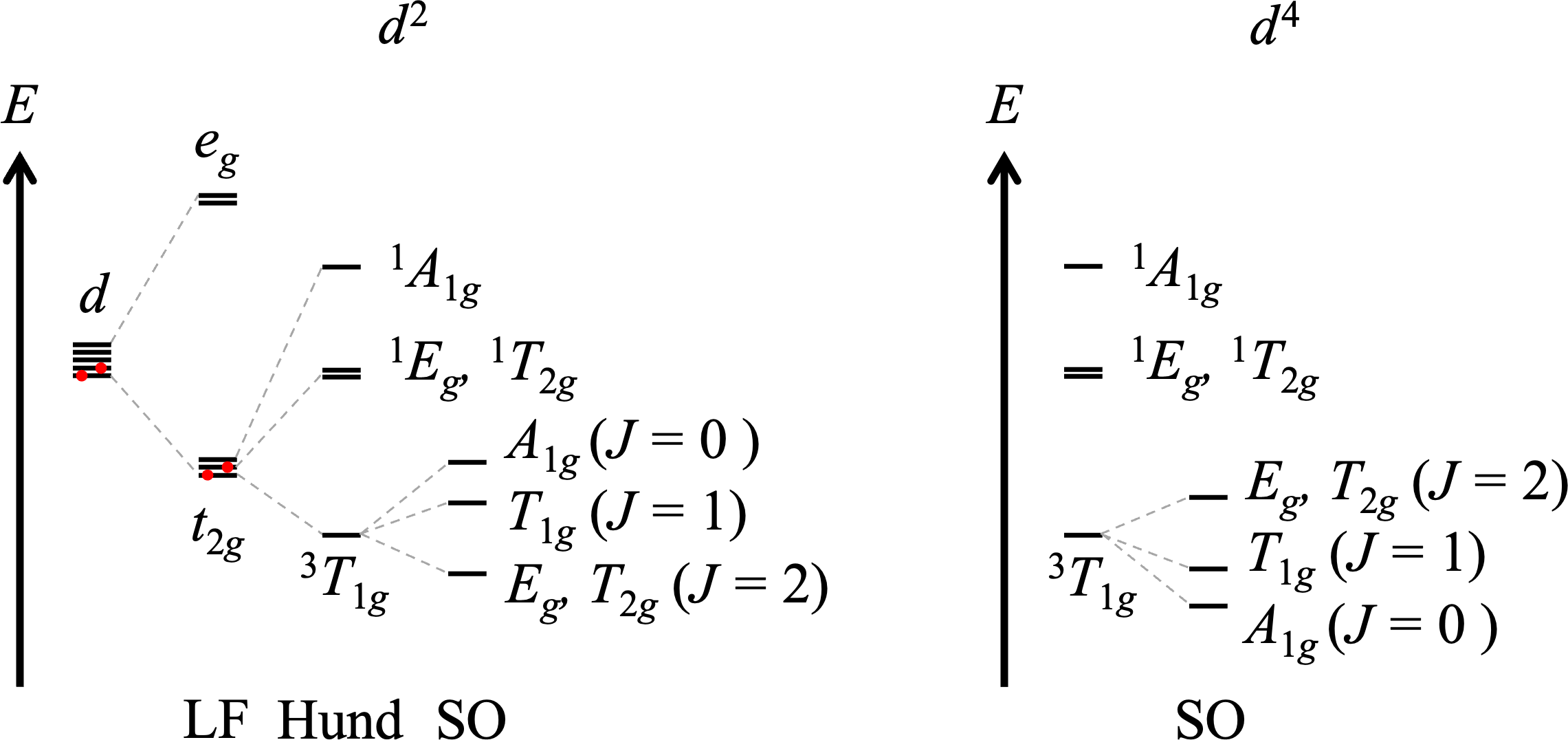} 
& &
& \includegraphics[height=0.23\linewidth, bb = 0 0 552 357]{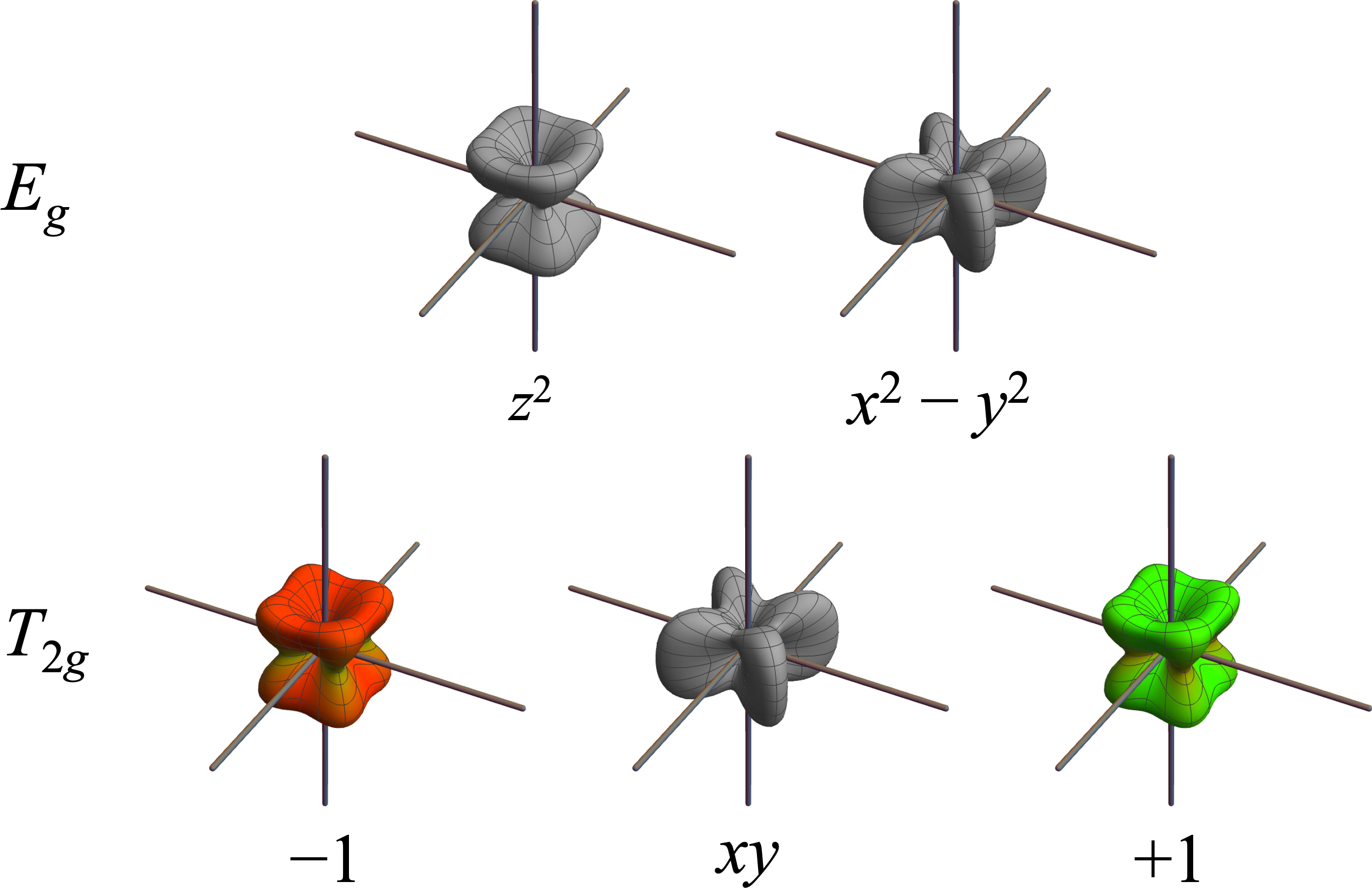}
\end{tabular}
\caption{
  (Color online)
  (a) The electronic energy levels and (b) the electron (hole) distributions of the spin-orbit coupled states of the $d^2$ and $d^4$ octahedra. 
We use the hole picture for the $d^4$ system. 
The red and green areas indicate the spin-up and spin-down distributions, respectively, and the gray area shows the cancellation of the spin-up and spin-down components.
}
\label{Fig:electronic_d2}
\end{figure*}

\subsection{Spin-orbit multiplet states}
\label{Sec:SO}
In spin-orbit Mott insulators, the coexisting spin, orbital, and lattice degrees of freedom on metal sites characterize the local quantum states. 
These physical degrees of freedom correlate via the ligand field, Hund's, spin-orbit, and vibronic (on-site electron-phonon) couplings. 
Here, we review the nature of the low-energy electronic structure of $4d$ and $5d$ metal ions in an octahedral environment. 
For details about the electronic structure of the transition metal ions in octahedral environments, see the book by Sugano, Tanabe, and Kamimura \cite{Sugano1970}.

The ligand-field splits the $d$ orbital levels into doubly degenerate $e_g$ and triply degenerate $t_{2g}$ levels \cite{Sugano1970} [Figs. \ref{Fig:electronic_d1} and \ref{Fig:electronic_d2}].
We can qualitatively understand the ligand-field splitting based on the Coulomb repulsion between the $d$ orbitals and the negatively charged ligand atoms. 
Note, however, that the ligand field comes from various effects, such as the covalency between the $d$ and ligand $p$ orbitals \cite{Sugano1963, Sugano1970}.
The Coulomb repulsion between the ligand atoms and the $e_g$ orbitals delocalized over the metal-ligand bond is stronger than the repulsion between the ligand atoms and the $t_{2g}$ orbitals residing between the bonds
\footnote{
In the octahedral environment, the degeneracy of the $t_{2g}$ orbitals ($yz$, $zx$, and $xy$) is intuitively clear because their physical equivalence in the octahedral environment is apparent. 
In contrast, the degeneracy of the $e_g$ orbitals is unclear.  
We can understand the degeneracy of the $2z^2-x^2-y^2$ ($z^2$ for simplicity) and $x^2-y^2$ orbitals from the fact that their linear combinations generate physically equivalent pairs, $x^2$ and $y^2-z^2$ orbitals and $y^2$ and $z^2-x^2$ orbitals. 
}.
The ligand field does not fully quench the orbital angular momentum of the $d$ orbitals within the $t_{2g}$ orbitals, and the $t_{2g}$ orbitals possess the effective $\tilde{l}=1$ angular momentum.
Although the effective $\tilde{l}=1$ angular momentum resembles the $l_p=1$ orbital angular momentum for the atomic $p$ orbitals, the former behaves as the latter with the opposite sign, $\tilde{\bm{l}} = - \hat{\bm{l}}_p$, under the commutations.

The low-energy electronic states of the embedded $d^1$ ions are the $t_{2g}^1$ configuration type. 
In the multielectronic $d$ ions, the interplay of the ligand field and Hund's rule coupling determines the electronic structures \cite{Tanabe1954II}. 
The ligand field stabilizes the low-spin $t_{2g}^N$ configurations, and the latter does high-spin $t_{2g}^{N-1}e_g^1$ configurations. 
In the $4d$ and $5d$ sites, the ligand field splitting is one order of magnitude larger than the stabilization by the Hund's coupling, and the low-spin $t_{2g}^N$ configurations become more stale than the $t_{2g}^{N-1}e_g^1$ configurations. 
Among the $t_{2g}^N$ configurations, the Hund's coupling between the $t_{2g}$ electrons stabilizes the high-spin term states.

The Tanabe-Sugano diagrams provide the nature of the low-energy term states of embedded multielectronic $d$ ions in octahedral environments \cite{Tanabe1954I, Tanabe1954II, Sugano1970}.
For sufficiently strong ligand field in the $d^5$ systems, $t_{2g}^5$ electron configurations ($t_{2g}^1$ hole states) become the ground state [See Fig. 9 in Ref. \onlinecite{Tanabe1954II}]. 
Thus, the ground states of the $d^1$ and $d^5$ metal ions are the ${}^2T_{2g}$ term states with effective $\tilde{l}=1$ orbital angular momentum and spin $1/2$ [Fig. \ref{Fig:electronic_d1}(a)]. 
In the $d^2$ and $d^4$ metal ions, the ${}^3T_{1g}$ term states become the most stable [Fig. \ref{Fig:electronic_d2}(a) and Figs. 3 and 7 in Ref. \onlinecite{Tanabe1954II}] with the effective angular momentum of $\tilde{L}=1$ and spin 1.
Finally, in the $d^3$ metal ions, the ground states are high-spin ${}^4A_{2g}$ [Fig. 5 in Ref. \onlinecite{Tanabe1954II}]. 
This term states are orbitally nondegenerate, and the JT effect does not appear. Thus, we do not discuss the $d^3$ systems further 
\footnote{ 
In the orbitally nondegenerate high-spin states, a weak JT effect can arise via the hybridization between the nondegenerate states and excited orbitally degenerate states \cite{Warren1984, Halliday1988, Iwahara2017, Streltsov2020, Streltsov2022, Warzanowski2024b}.
}.

Due to the unquenched orbital angular momentum of the $d^N$ states, the spin-orbit entanglement develops \cite{Sugano1970}.
In the $d^1$ and $d^5$ octahedra, the spin-orbit coupling splits the ground ${}^2T_{2g}$ term states into $j_\text{eff}=1/2$ doublet ($\Gamma_7$) and $j_\text{eff}=3/2$ quartet ($\Gamma_8$) multiplet states [Fig. \ref{Fig:electronic_d1}(a)]. 
In the $d^1$ systems, the $\Gamma_8$ multiplet states become the ground states, and, in the $d^5$ systems, the $\Gamma_7$ states do due to the inversion of the sign of the spin-orbit coupling (single electron operators) \cite{Sugano1970}.
Note also that the order of the $j_\text{eff}$ energy levels is opposite to that for the $2p$ orbitals \cite{LandauQM} due to the difference between the $\tilde{l}=1$ and $l_p=1$.

Figure \ref{Fig:electronic_d1}(b) shows the spatial distributions of the $\Gamma_7$ and $\Gamma_8$ electron (hole) densities in the $d^1$ ($d^5$) octahedra.
We can classify the $\Gamma_8$ states into two pairs of Kramers doublets with different spatial distributions, which implies the possibility of decoupling the $\Gamma_8$ states into the spatial and magnetic degrees of freedom. 
We specify the spatial states and the magnetic states by using the pseudo-orbital angular momenta $\tilde{\tau}_z = + 1/2$ $(z^2)$, $-1/2$ $(x^2-y^2)$ and pseudo-spin angular momentum $\tilde{s}_z = \pm 1/2$, respectively \cite{Romhanyi2017, Natori2016}. 
The pseudo-spin states differ from the pure spin states because the former contains information on the orbital and spin degrees of freedom. 
We define the $\tilde{\tau}=1/2$ pseudo-orbital angular momenum operators $\tilde{\tau}_{x/z}$ and $\tilde{s}=1/2$ pseudo-spin angular momentum operators $\tilde{s}_{\gamma}$ ($\gamma = x, y, z$) to have the same form as the spin operators while acting on the pseudo-orbital and pseudo-spin states, respectively. 
The symmetry guarantees the decoupling of the $\Gamma_8$ states into the pseudo orbital and pseudo spin parts: $\Gamma_8 = E \otimes \Gamma_6$, where $E$ stands for the spatial part, and $\Gamma_6$ does the magnetic part \cite{Koster1963}.

In the $d^2$ and $d^4$ octahedra, the spin-orbit coupling splits the $^3T_{1g}$ term states into the effective $J=0, 1, 2$ multiplet states [Fig. \ref{Fig:electronic_d2} (a)]. 
The spin-orbit multiplet states of the $d^2$ ($d^4$) ions are the $J=$ 2, 1, 0 (0, 1, 2) types in the increasing order of energy. 
Furthermore, the five-fold degenerate $J=2$ multiplet states split into $E_g$ doublet and $T_{2g}$ triplet states in an octahedral environment [Fig. \ref{Fig:electronic_d2}(a)]. 
The splitting of the $J=2$ states could arise from the nonspherical Coulomb interaction for the $t_{2g}$ electrons and the spin-orbit hybridization between the $t_{2g}^2$ and $t_{2g}^1e_g^1$ term states \cite{Voleti2020, Takayama2021}.
The spin-orbit coupling stabilizes the $E_g$ states, which occurs in various $5d^2$ Os double perovskites \cite{Maharaj2020} and a W antifluorite \cite{Pradhan2024}. 
In Sec. \ref{Sec:DJT}, we show that the vibronic (electron-phonon) coupling also stabilizes the $E_g$ (vibronic) states. 
As Fig. \ref{Fig:electronic_d2}(b) shows, the $E_g$ states are nonmagnetic. 
The magnetic $T_{2g}$ states are the ground state in $5d^2$ Re compounds \cite{Thompson2014, Marjerrison2016b, Nilsen2021}, while the stabilization mechanism of the $T_{2g}$ states is under debate.

\begin{figure}[tb]
\includegraphics[width=\linewidth, bb = 0 0 557 372]{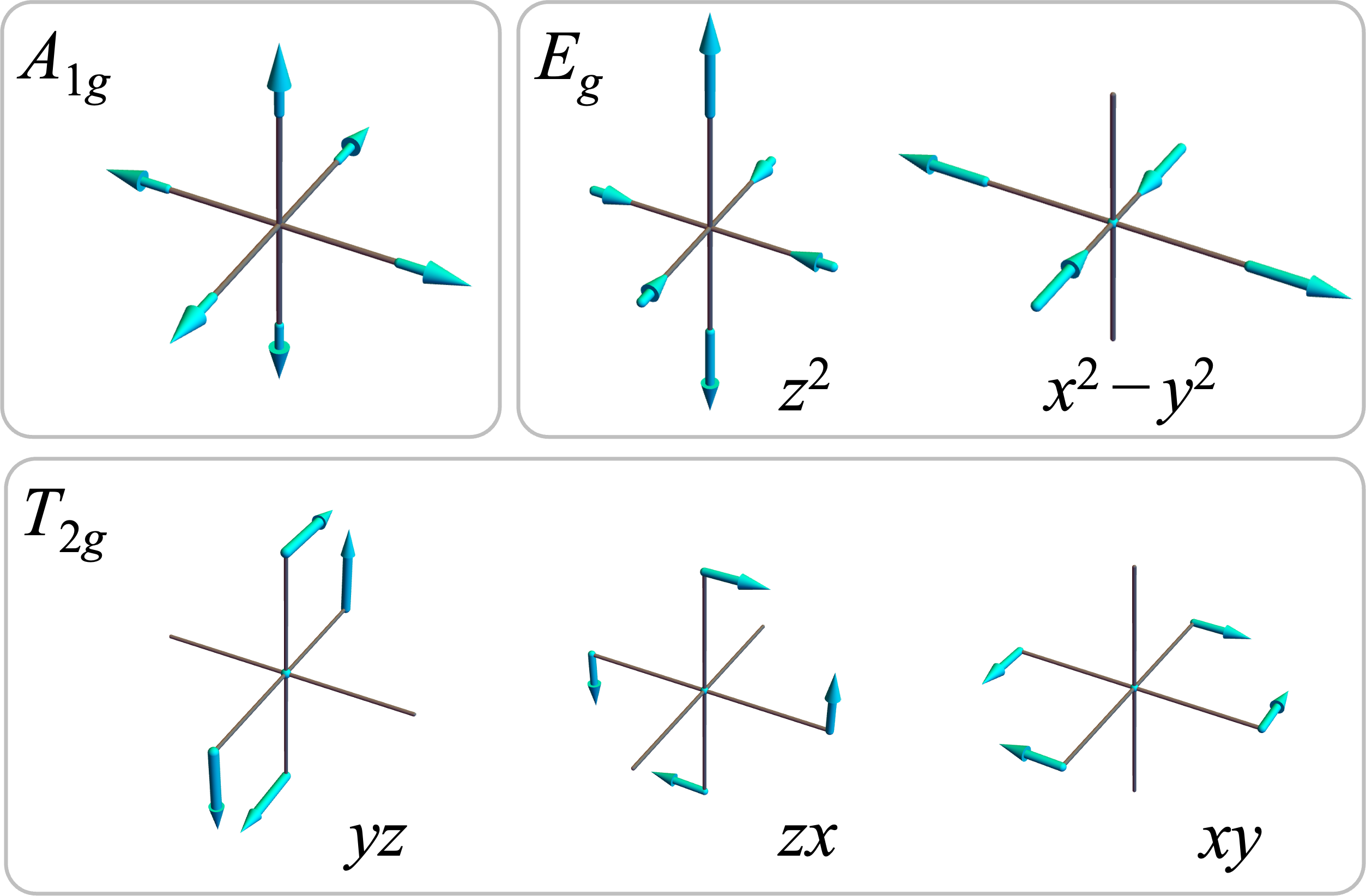}
\caption{
  (Color online)
  The vibronically active modes. Among them, the JT active modes are the $E_g$ and $T_{2g}$ modes.}
\label{Fig:mode}
\end{figure}

\begin{figure}[tb]
\begin{tabular}{lc}
(a) \\
& \includegraphics[width=0.85\linewidth, bb = 0 0 339 230]{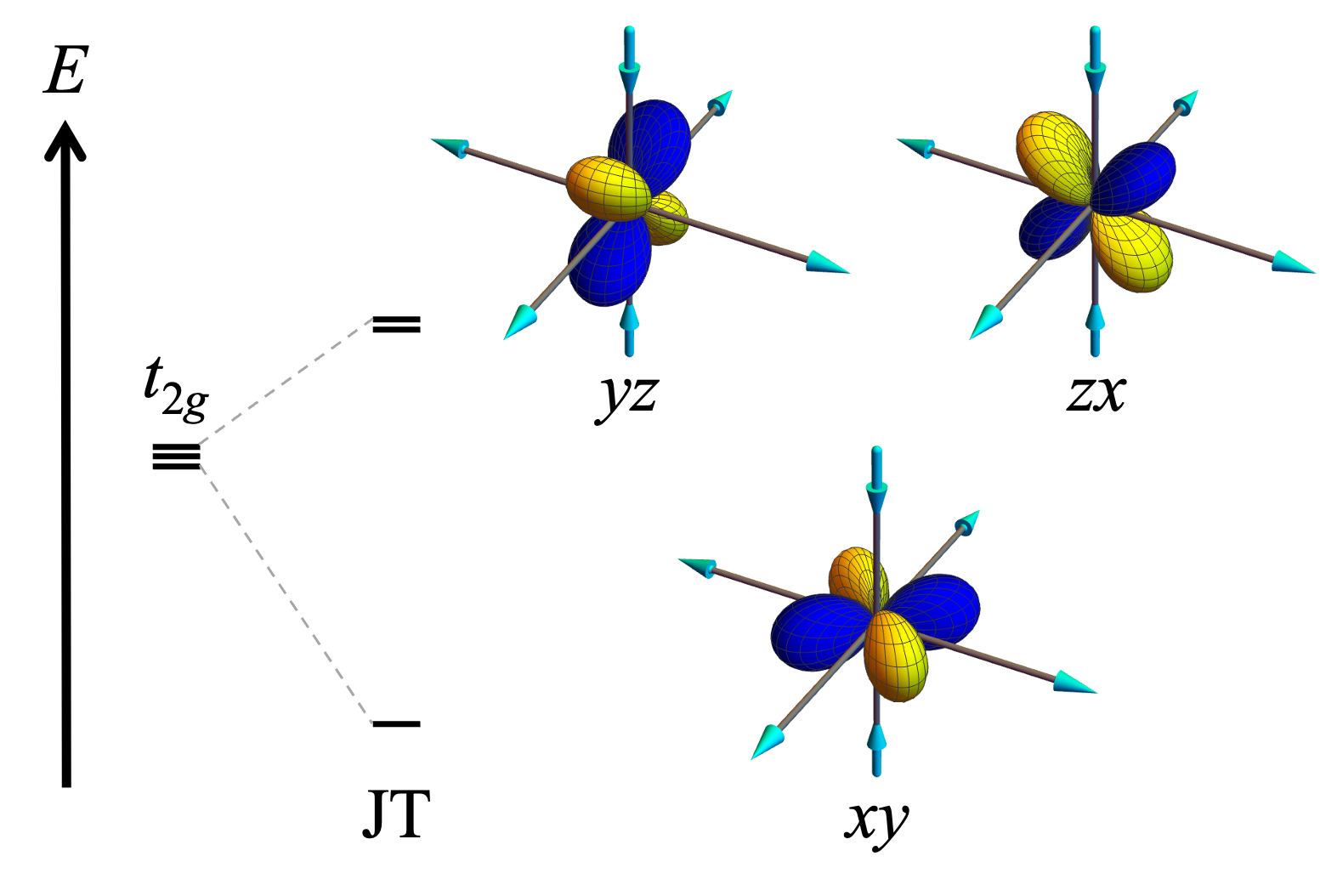}
\\
(b) \\
& \includegraphics[width=0.85\linewidth, bb = 0 0 336 230]{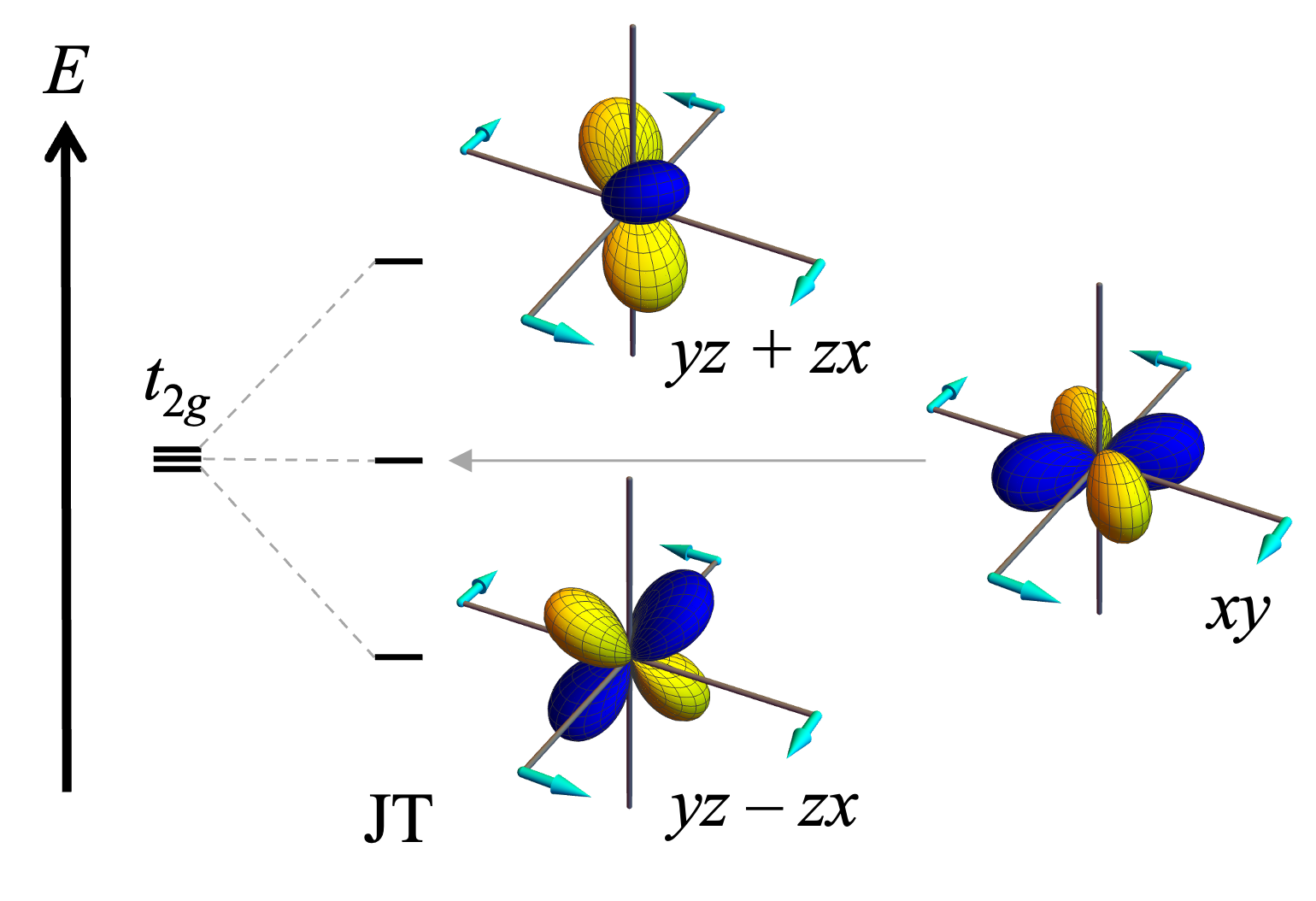}
\end{tabular}
\caption{
  (Color online)
  The Jahn-Teller splittings of the $t_{2g}$ orbitals under (a) the $E_g z^2$ and (b) the $T_{2g} xy$ deformations.}
\label{Fig:vibronic}
\end{figure}

\subsection{Vibronic coupling}
\label{Sec:vibronic}
The $t_{2g}$ orbitals couple to the nuclear vibrations (vibronic coupling), which causes the JT effect \cite{Jahn1937, Jahn1938, LandauQM}. 
The change in the Coulomb interaction under the lattice deformations is the vibronic coupling. 
Below, we explain the vibronic coupling combining the point charge model with lattice deformations. 
See for a standard derivation of the vibronic coupling Appendix \ref{A:vibronic}.

Let us consider how the deformations modulate the ligand field and the $t_{2g}$ orbital levels. 
We first consider tetragonal deformations along the $z$ axis [Fig. \ref{Fig:mode}]. 
With the $E_g z^2$ compression, the ligand atoms in the $xy$ plane leave from the central ion and those on the $z$ axis approach. 
Under this situation, the Coulomb repulsion between the $xy$ orbital and the ligand atoms decreases, and that between the $yz$ or $zx$ orbital and the ligands increases, resulting in the splitting of the $t_{2g}$ orbital levels [Fig. \ref{Fig:vibronic}(a)].
For the $E_g z^2$ elongation, the $t_{2g}$ energy levels take the inverted order: The $xy$ orbital is unstable, and the others are stable. 
Due to the octahedral symmetry of the system, the $x^2$ and $y^2$ deformations give rise to physically equivalent results as the $z^2$ deformation.

Similarly, the $T_{2g}$ deformations splits the $t_{2g}$ orbital levels. 
To understand the response of the $t_{2g}$ orbital levels to the $T_{2g} xy$ deformation [Fig. \ref{Fig:mode}], we use the linear combinations of the $yz$ and $zx$ orbitals: $|yz\rangle - |zx\rangle$ and $|yz\rangle + |zx\rangle$ [See Fig. \ref{Fig:vibronic}(b)].
Under the $T_{2g} xy$ deformation, the Coulomb repulsion between the $xy$ orbital level and the ligand atoms does not linearly change, while that between the $yz-zx$ ($zx+xy$) and the ligands decreases (increases) [Fig. \ref{Fig:vibronic}(b)]. 
We can understand the response of the $t_{2g}$ orbitals to the $yz$ and $zx$ deformations [Fig. \ref{Fig:mode}], in the same manner.

Now, we write down the vibronic interaction model for the $t_{2g}$ orbitals.
To describe the magnitude of the $E_g$ and $T_{2g}$ deformations, we use the mass-weighted normal coordinates $q_{\Gamma\gamma}$ \cite{Inui1990, Wilson1980}. 
Positive $q$ stands for the shift of the ligand atoms along the arrows in Fig. \ref{Fig:mode}.
From the discussions above on the responses of the $t_{2g}$ orbitals to the $E_g$ and the $T_{2g}$ deformations, the vibronic interactions that are linear to the $q$'s are
\begin{align}
 \hat{V}_\text{JT} &= 
 v_E \left( q_{x^2} |yz\rangle \langle yz| + 
            q_{y^2} |zx\rangle \langle zx| + 
            q_{z^2} |xy\rangle \langle xy| 
     \right)
 \nonumber\\
 &+ v_T \big[
    q_{yz} ( |zx\rangle \langle xy| + |xy\rangle \langle zx| )
  + q_{zx} ( |xy\rangle \langle yz| 
  \nonumber\\
 &+ |yz\rangle \langle xy| )
  + q_{xy} ( |yz\rangle \langle zx| + |zx\rangle \langle yz| )
 \big].
 \label{Eq:VJT}
\end{align}
Here $v_E$ and $v_T$ are the linear vibronic coupling parameters for the $E_g$ and $T_{2g}$ modes, 
$|\gamma\rangle$ $(\gamma = yz, zx, xy)$ are the $t_{2g}$ orbital states, 
$q_{z^2}$ and $q_{x^2-y^2}$ the normal coordinates for the $E_g$ modes, 
$q_{yz}$, $q_{zx}$, and $q_{xy}$ the normal coordinates for the $T_{2g}$ modes, 
and $q_{x^2} = -\frac{1}{2} q_{z^2} + \frac{\sqrt{3}}{2} q_{x^2-y^2}$ and $q_{y^2} = -\frac{1}{2} q_{z^2} - \frac{\sqrt{3}}{2} q_{x^2-y^2}$.

Among the JT active modes, the $E_g$ vibrations are strongly coupled to the $t_{2g}$ orbitals compared with the $T_{2g}$ modes because the $E_g$ deformations change the metal-ligand bond lengths. 
Hereafter, we consider only the coupling to the $E_g$ modes for simplicity. 
We also do not consider the nonlinear vibronic couplings (See for details Ref. \onlinecite{Bersuker1989}).

\begin{figure}[tb]
\begin{tabular}{lc}
(a) \\
& \includegraphics[width=0.8\linewidth, bb = 0 0 273 295]{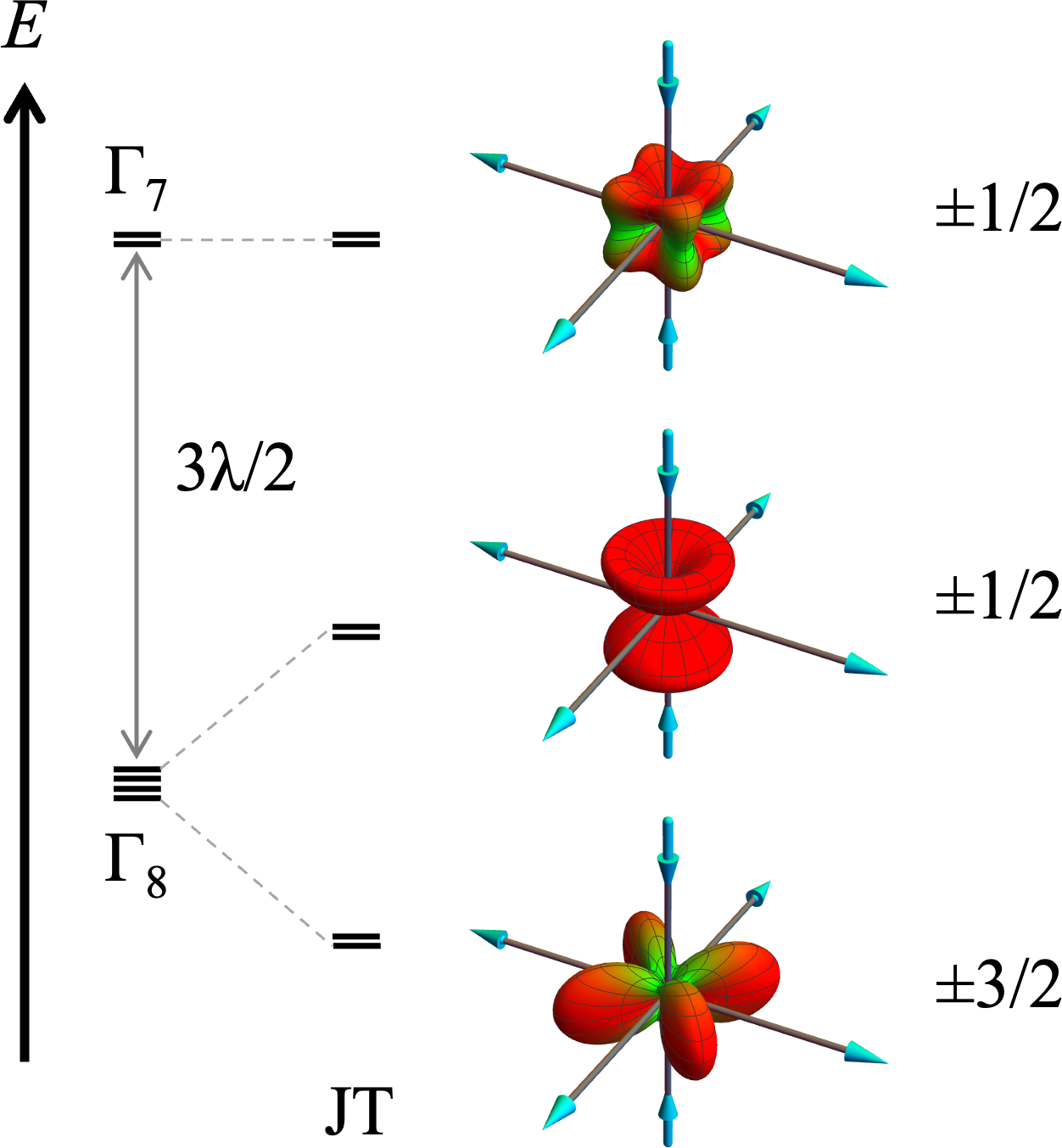}
\\
(b) \\
& \includegraphics[width=0.8\linewidth, bb = 0 0 297 302]{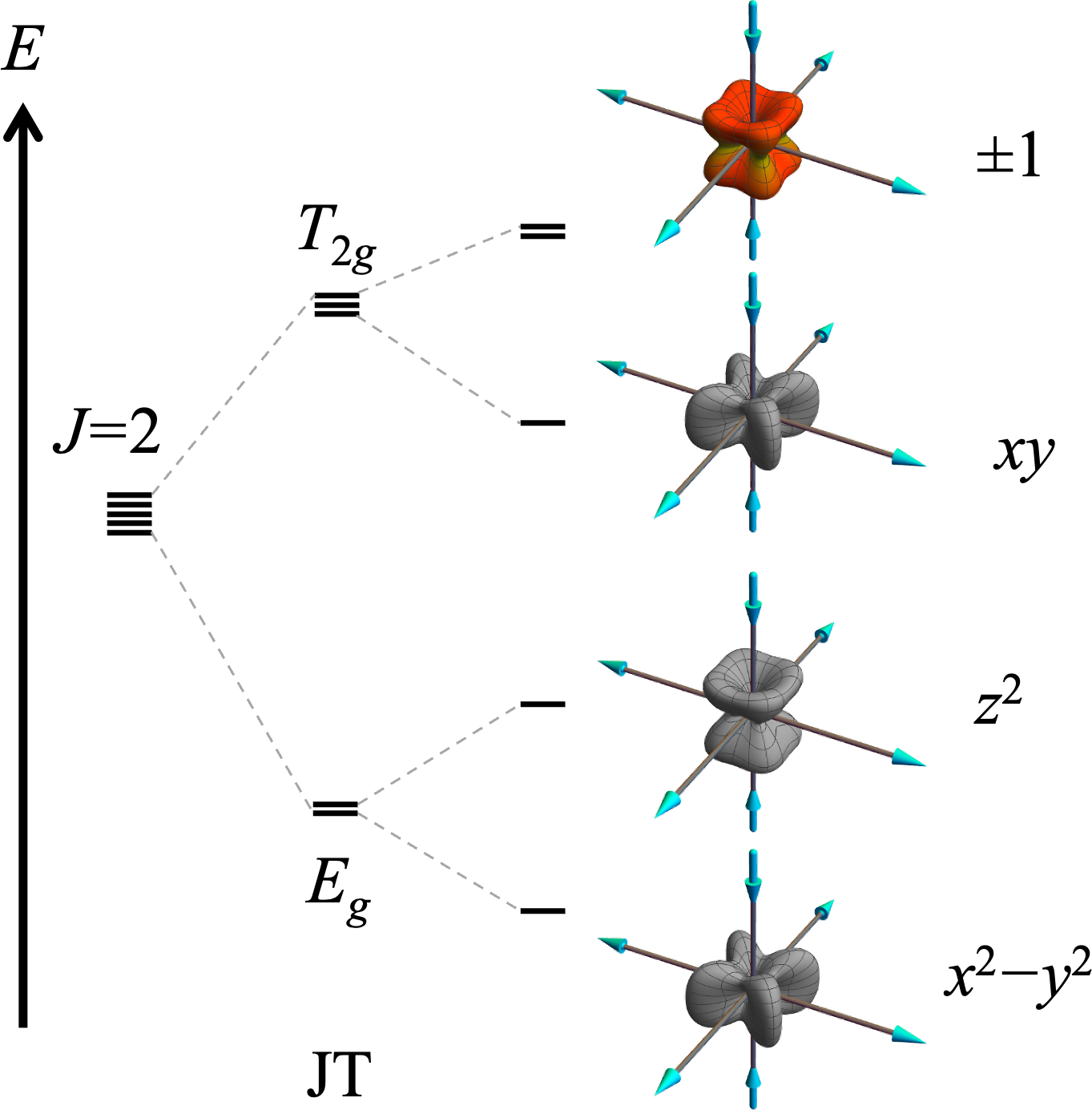}
\end{tabular}
\caption{
  (Color online)
  Vibronic coupling between the ground spin-orbit multiplet states and the $E_gz^2$ deformations in (a) the $d^1$ and (b) the $d^2$ octahedra.}
\label{Fig:JT_SO}
\end{figure}

\begin{figure*}[tb]
\begin{tabular}{lllll}
 (a) &~& (b) &~& (c) \\
 \includegraphics[width=0.33\linewidth, bb = 0 0 463 381]{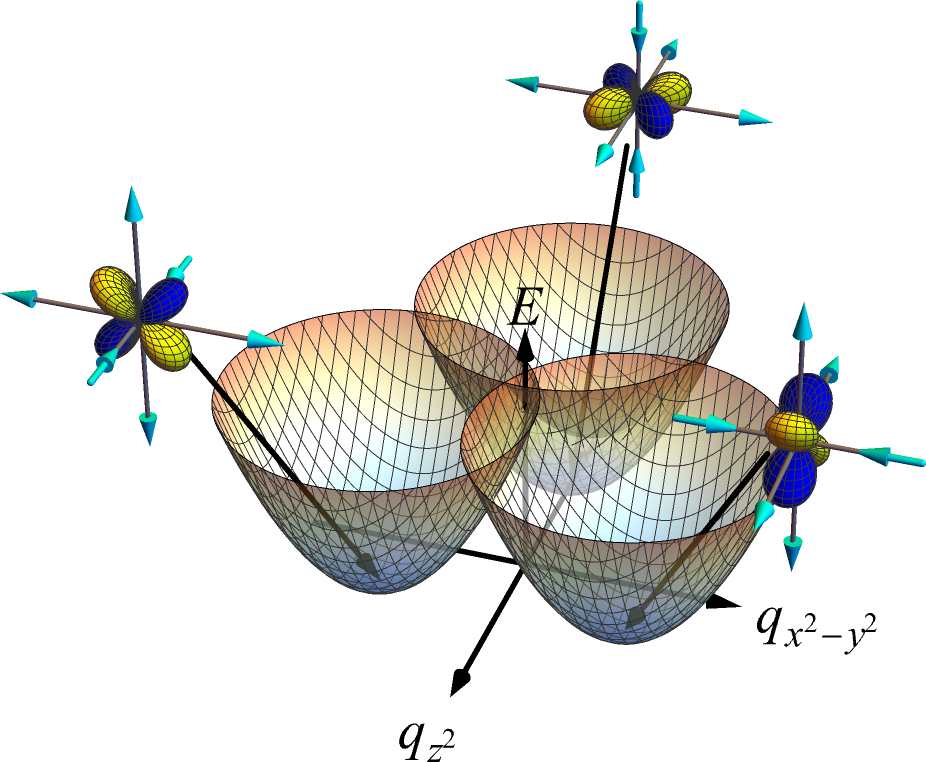}
 & &
 \includegraphics[width=0.33\linewidth, bb = 0 0 466 373]{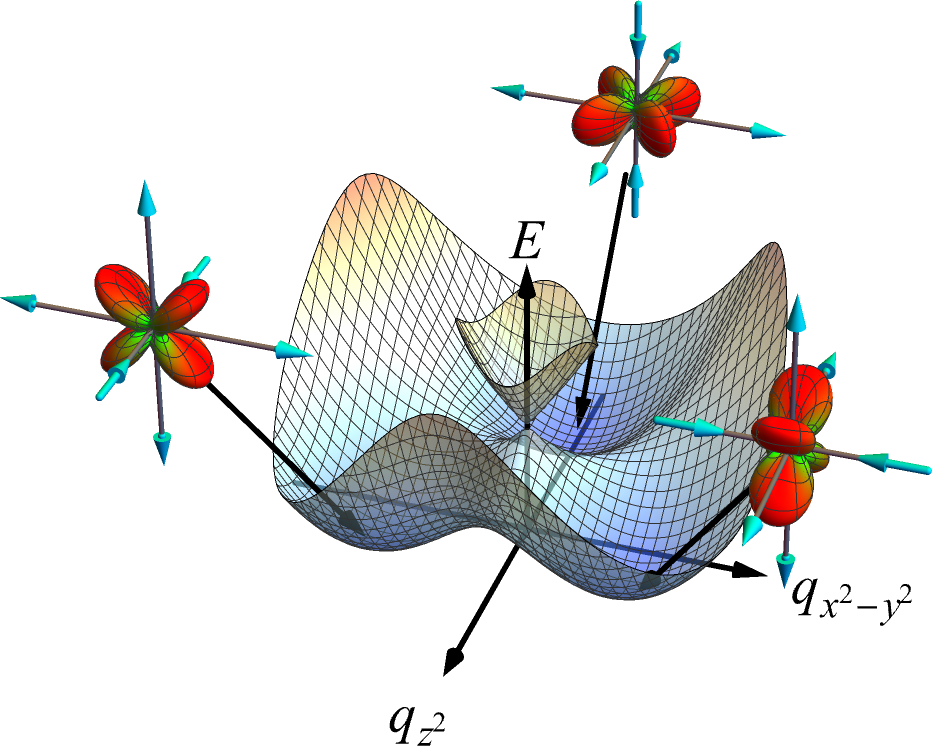}
 & &
 \includegraphics[width=0.26\linewidth, bb = 0 0 365 291]{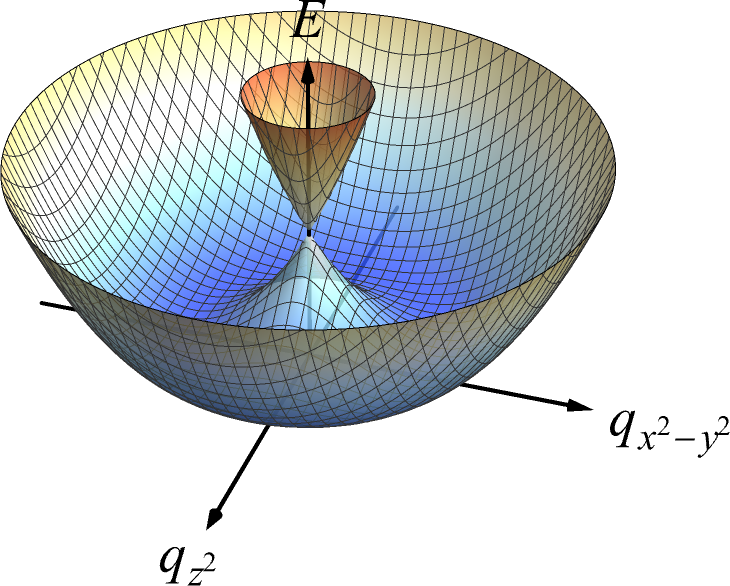}
\end{tabular}
\caption{
  (Color online)
  The APES of $d^1$ octahedron (a) without, (b) with moderate, and (c) with strong spin-orbit coupling. 
The JT deformation stabilizes the system by the energy gap $E_\text{JT}$ between the conical intersection at the origin and the minimum of the potential.
}
\label{Fig:APES}
\end{figure*}

The vibronic and spin-orbit couplings tend to quench each other \cite{Opik1957, Ham1965}, while both effects can persist in a cubic environment regardless of the strength of the interactions. 
In general, the spin-orbit coupling stabilizes the spin-orbital hybridized states, while the vibronic coupling does one of the degenerate orbital states, resulting in their competition.
Both effects can persist in cubic environments: The spatially anisotropic electron distributions [Figs. \ref{Fig:electronic_d1}(b) and \ref{Fig:electronic_d2}(b)] respond to the JT deformations in different manners. 
Figure \ref{Fig:JT_SO} shows how the highly degenerate spin-orbit coupled states in the $4d/5d$ ions respond to the tetragonal compressions due to the Coulomb repulsion between the $d$ electronic states and the negatively charged ligand atoms.

Let us write the vibronic coupling within the ground $\Gamma_8$ multiple states of $d^1$ octahedron, assuming the limit of strong spin-orbit coupling for simplicity. 
Using the pseudo-orbital states, the vibronic coupling to the $E_g$ deformations is 
\begin{align}
 \hat{V}_\text{JT} &= -v_E 
 \left(q_{z^2} \tilde{\tau}_z + q_{x^2-y^2} \tilde{\tau}_x \right).
 \label{Eq:VJT_SO_d1}
\end{align}
Here, $\tilde{\tau}_{z/x}$ are the pseudo-orbital operators described in Sec. \ref{Sec:SO}.
Compared with the orbital splitting by the $E_g$ deformations without spin-orbit coupling (\ref{Eq:VJT}), the splitting within the $\Gamma_8$ multiplet states becomes half by the spin-orbit coupling, and hence, non-negligible.
Moreover, the mathematical structures of Eqs. (\ref{Eq:VJT}) and (\ref{Eq:VJT_SO_d1}) are different, which results in qualitatively different JT effects, as described below. 
The $\Gamma_8$ multiplet states can couple to the $T_{2g}$ modes too \cite{Moffitt1957a}, while we ignore this contribution as mentioned above. 
In the $\Gamma_8$ multiplet states of $d^5$ octahedron, the sign of the vibronic coupling parameter in Eq. (\ref{Eq:VJT_SO_d1}) changes.

In the $d^2$ systems, the $E_g$ and $T_{2g}$ spin-orbit coupled states from the $J=2$ multiplet states interact with the $E_g$ deformations. 
Within the $E_g$ multiplet states, the vibronic coupling has the same form as Eq. (\ref{Eq:VJT_SO_d1}) with $\tilde{\tau}$'s defined using the $E_g$ states. 
Within the $T_{2g}$ multiplet states, the vibronic coupling has the same form as Eq. (\ref{Eq:VJT}), again, with the replacement of $|\gamma\rangle$ orbitals and $|T_{2g}\gamma\rangle$ multiplet states ($\gamma = yz, zx, xy$). 
We ignore the vibronic coupling between the $T_{2g}$ spin-orbit coupled states and the $T_{2g}$ deformations. 
The vibronic coupling to the $E_g$ modes is zero between the $E_g$ and the $T_{2g}$ electronic states. 
In the $d^4$ systems, the excited $J=2$ and $J=1$ states couple to the $E_g$ vibrations in the same manner as the $d^2$ systems with the opposite sign of the vibronic coupling parameters, while the nondegenerate ground $J=0$ state does not couple to the $E_g$ modes (see, e.g., Ref. \onlinecite{Iwahara2023Ru}).

\subsection{Static Jahn-Teller effect}
\label{Sec:SJT}
The vibronic coupling makes the high-symmetric system with degenerate orbital states unstable against symmetry lowering (JT theorem) \cite{Jahn1937, Jahn1938, LandauQM}. 
When the symmetry lowering of the nuclear configuration occurs due to the JT theorem, the phenomenon is called the static JT effect. 
In discussing the static JT effect, the mass-weighted normal coordinates are external variables, and the conjugate momenta are zero.

To visualize the instability of the orbitally degenerate system at a high-symmetric structure, we draw the adiabatic potential energy surface (APES). 
The model Hamiltonian for the octahedral system consists of the electronic Hamiltonian, the harmonic potential ($U_0 = \sum_\gamma \omega^2 q_\gamma^2/2$), and the vibronic coupling (\ref{Eq:VJT}). 
The energy eigenvalues of the model in functions of coordinates are the APESs. 
Figure \ref{Fig:APES}(a) shows the APESs of the $d^1$ octahedron without spin-orbit coupling.
The APES at the octahedral structure ($q_{z^2} = q_{x^2-y^2} = 0$) is tilted, which indicates the structural instability of the orbitally degenerate system with high-symmetry.
The APESs consist of three equivalent paraboloids with three equivalent minima at $x^2$, $y^2$, and $z^2$ compressed positions. 
Each sheet represents Kramers doublet states in the present case because the time-even vibronic coupling does not lift the degeneracy. 
The system lowers the energy by $E_\text{JT}$ with respect to the $O_h$ structure by taking one of the low-symmetric configurations (the static JT effect) \cite{Opik1957, Liehr1963, Bersuker1989}. 
We call $E_\text{JT}$ JT stabilization energy. 
At the minimum, we can express the system with an orbital-lattice configuration: For example, $xy$ orbital state with $z^2$ compression [See Fig. \ref{Fig:APES}(a)].

Turning on the spin-orbit coupling, the shape of the APES gradually changes. 
The spin-orbit coupling splits the cross sections of the paraboloids in Fig. \ref{Fig:APES}(a).
At the origin, the electronic states split into the $\Gamma_8$ and $\Gamma_7$ multiplet states, and due to the pseudo-orbital degeneracy of the $\Gamma_8$ ground states, the JT instability remains [Fig. \ref{Fig:APES}(b)]. 
The minima of the APES are located at $x^2$, $y^2$, and $z^2$ compressed positions as in the case without spin-orbit coupling [Fig. \ref{Fig:APES}(a)], while the magnitude of the deformation reduces due to the spin-orbit coupling \cite{Opik1957} as indicated by the reduction of the vibronic coupling by the spin-orbit coupling (\ref{Eq:VJT_SO_d1}). 
The minima in the lowest APES appear on a single surface, contrary to the APES without the spin-orbit coupling. 
Increasing the spin-orbit coupling, the energy gap between the minima and the saddle decreases.
In the limit of the strong spin-orbit coupling, the warping in the lowest APES disappears [Fig. \ref{Fig:APES}(c)].
The warping of the APES occurs due to the vibronic coupling between the $\Gamma_7$ and $\Gamma_8$ multiplet states. 
For a sufficiently strong spin-orbit gap, the effect of the off-diagonal vibronic coupling becomes negligible, and the vibronic coupling within the low-energy states reduces to Eq. (\ref{Eq:VJT_SO_d1}).

A peculiarity of the static JT system is its relevance to Berry's phase. 
Suppose the nuclear coordinates vary along a closed path encircling the high-symmetric position ($q_{z^2} = q_{x^2-y^2} = 0$) in the space of the normal coordinates [Fig. \ref{Fig:APES}(b), (c)]
\footnote{Note that this rotation is not the real rotation of the octahedra.}, the electronic state changes its sign after the rotation \cite{Longuet-Higgins1958}. 
The sign change is an example of Berry's phase that appears by the adiabatic rotation around the conical intersection \cite{Berry1984}. 
We can see the sign change by observing the ground adiabatic state of Eq. (\ref{Eq:VJT_SO_d1}).
The ground adiabatic state is $|\Phi_\text{ad}(\theta)\rangle = \cos\frac{\theta}{2} |z^2\rangle + \sin \frac{\theta}{2} |x^2-y^2\rangle$ with $\theta = \text{arctan} (q_{x^2-y^2}/q_{z^2})$ and $q = \sqrt{q_{z^2}^2+q_{x^2-y^2}^2} > 0$.
The adiabatic state changes its sign for $\theta \rightarrow \theta + 2\pi$.

The APESs of the excited $\Gamma_8$ multiplet states of the $d^5$ system are similar to those of the ground state of the $d^1$ system. 
The positions of the minima and the saddle points for the single-hole system are opposite to those of the single-electron system.

In the $d^2$ system, the $E_g$ and the $T_{2g}$ multiplet states from the $J=2$ spin-orbit multiplet form two sets of the APESs. 
These sets of APESs are independent of each other due to the absence of the vibronic coupling between the two electronic states (Sec. \ref{Sec:vibronic}). 
Due to the equivalence of the vibronic models for the $\Gamma_8$ multiplet states of the $d^1$ octahedra and the $E_g$ states of the $d^2$ octahedra, the APESs become the last type (c) in Fig. \ref{Fig:APES}.
The vibronic coupling between the $T_{2g}$ states and the $E_g$ vibrations form the APES consisting of three paraboloids as Fig. \ref{Fig:APES}(a).

See for further information on the interplay of the static JT effect and spin-orbit couplings, e.g., Refs. \onlinecite{Opik1957, Warren1984, Streltsov2020, Streltsov2022}.

\begin{figure}[htb]
\includegraphics[width=\linewidth, bb = 0 0 2378 1012]{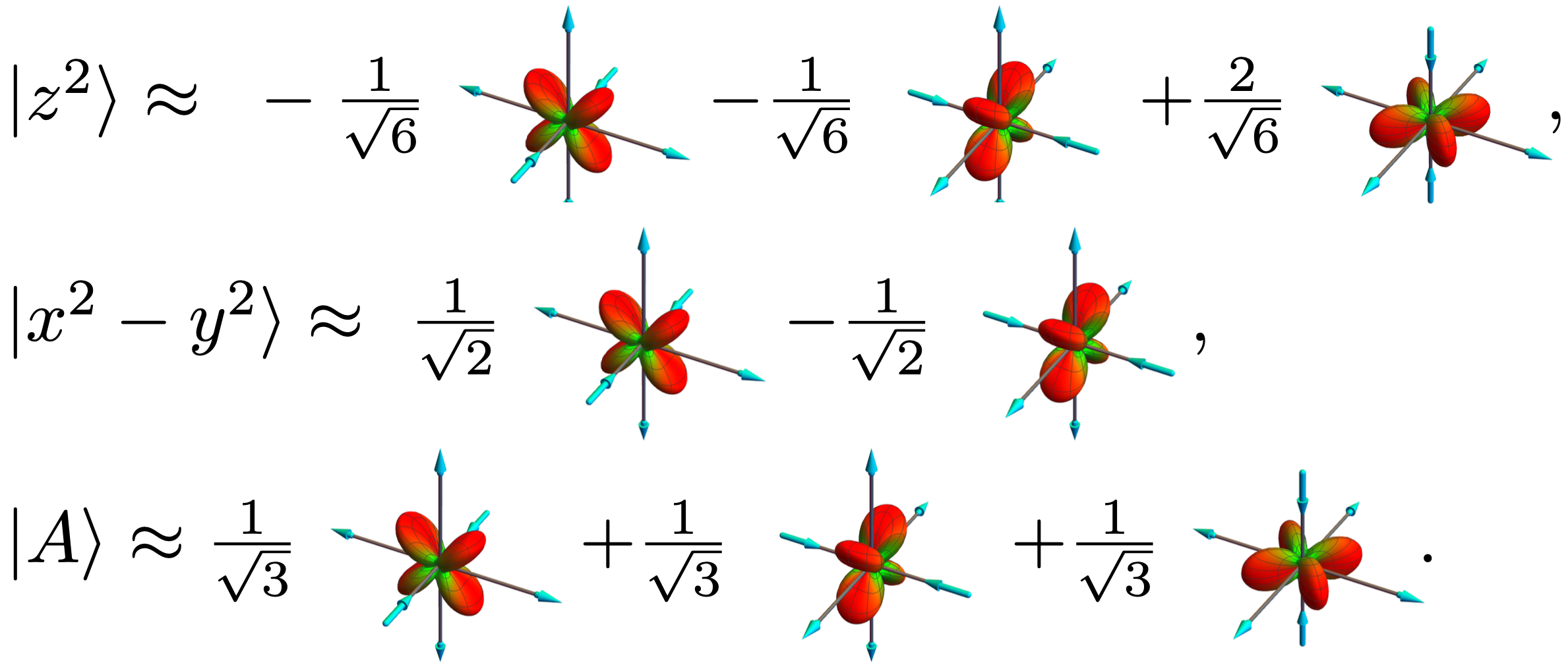}
\caption{
  (Color online)
  Low-energy vibronic states of $d^1$ system within the tunneling splitting approach. 
$|z^2\rangle$ and $|x^2-y^2\rangle$ are the vibronic doublet and $|A\rangle$ the vibronic singlet.}
\label{Fig:vibronic_wf}
\end{figure}

\begin{figure}[!tbh]
\begin{tabular}{lll}
(a) \\
& \includegraphics[width=0.90\linewidth, bb = 0 0 450 419]{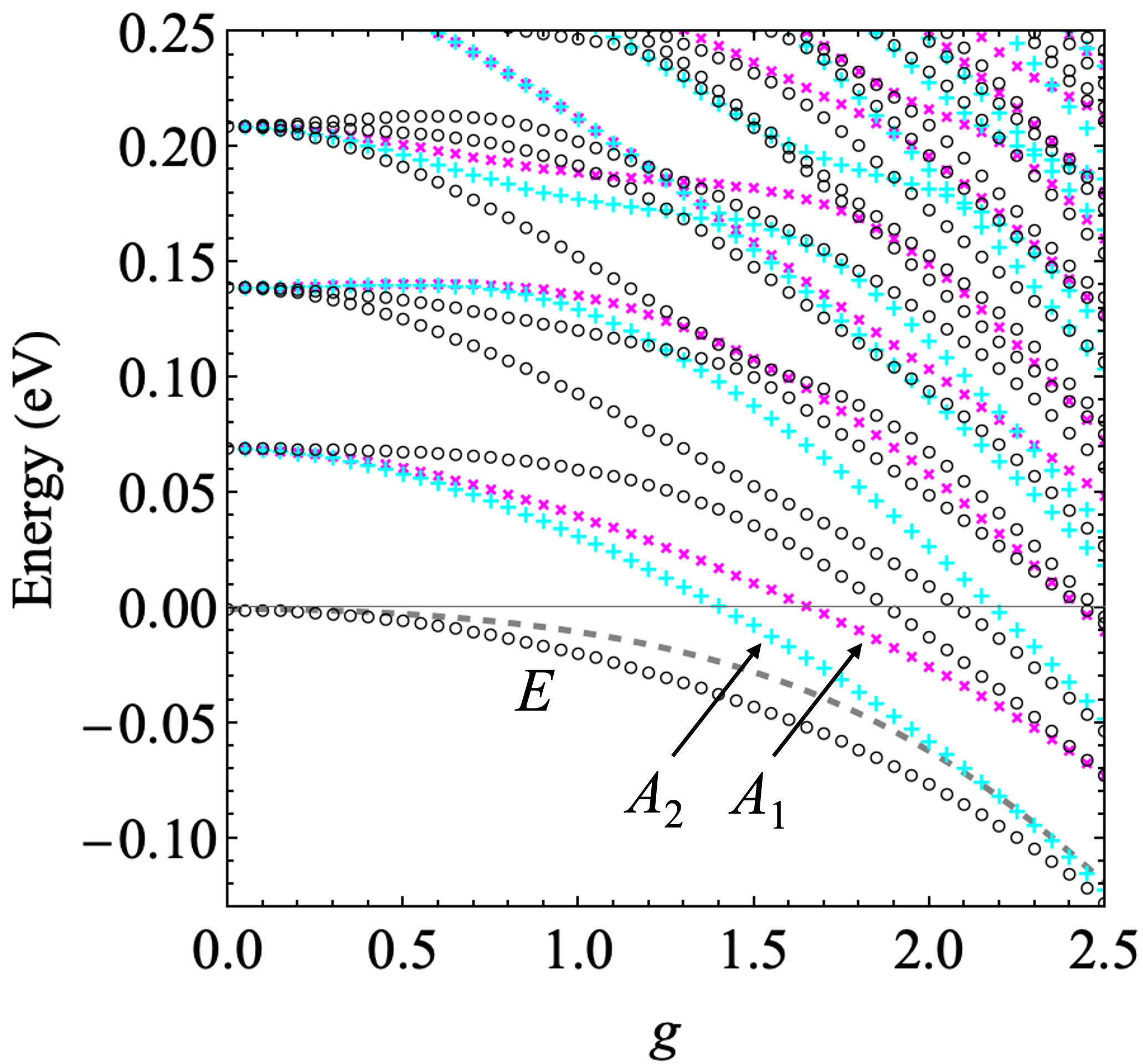}
\\
(b) \\
& \includegraphics[width=0.90\linewidth, bb = 0 0 450 397]{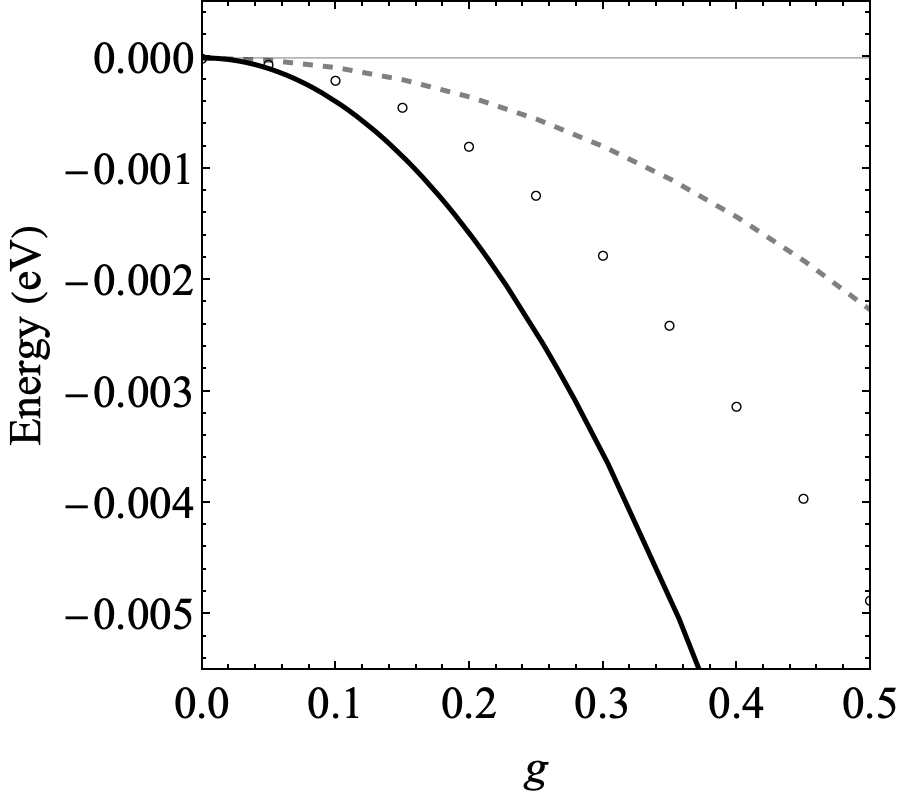}
\end{tabular}
\caption{
  (Color online)
  (a) Numerical vibronic levels and the static JT energy of $d^1$ octahedron in function of the dimensionless vibronic coupling parameter $g = v_E/\sqrt{\hslash \omega^3}$ (Appendix \ref{A:vibronic}). 
The cyan cross ($\times$), the magenta plus ($+$), and the black circle ($\circ$) indicate the doubly degenrate $\Gamma_6$ ($A_1$), doubly degenerate $\Gamma_7$ ($A_2$), and four-fold degenerate $\Gamma_8$ ($E$) vibronic states, and the gray dashed line the static JT energy. 
The irreducible representations in the brackets are for the time-even part without pseudo-spin. 
We used the interaction parameters close to the experimental data of $5d^1$ Re compounds: Frequency $\omega = 70$ meV \cite{Pasztorova2023}, and spin-orbit coupling parameter $\lambda = $ 0.32 eV \cite{Frontini2024}. 
We set the ground energy at $g=0$ as the origin of the energy. 
(b) Comparison between the lowest numerical vibronic level and an approximate solution within the second-order perturbation theory (the black line). 
}
\label{Fig:energy_d1}
\end{figure}

\begin{figure}[htpb]
 \includegraphics[width=0.9\linewidth, bb = 0 0 450 419]{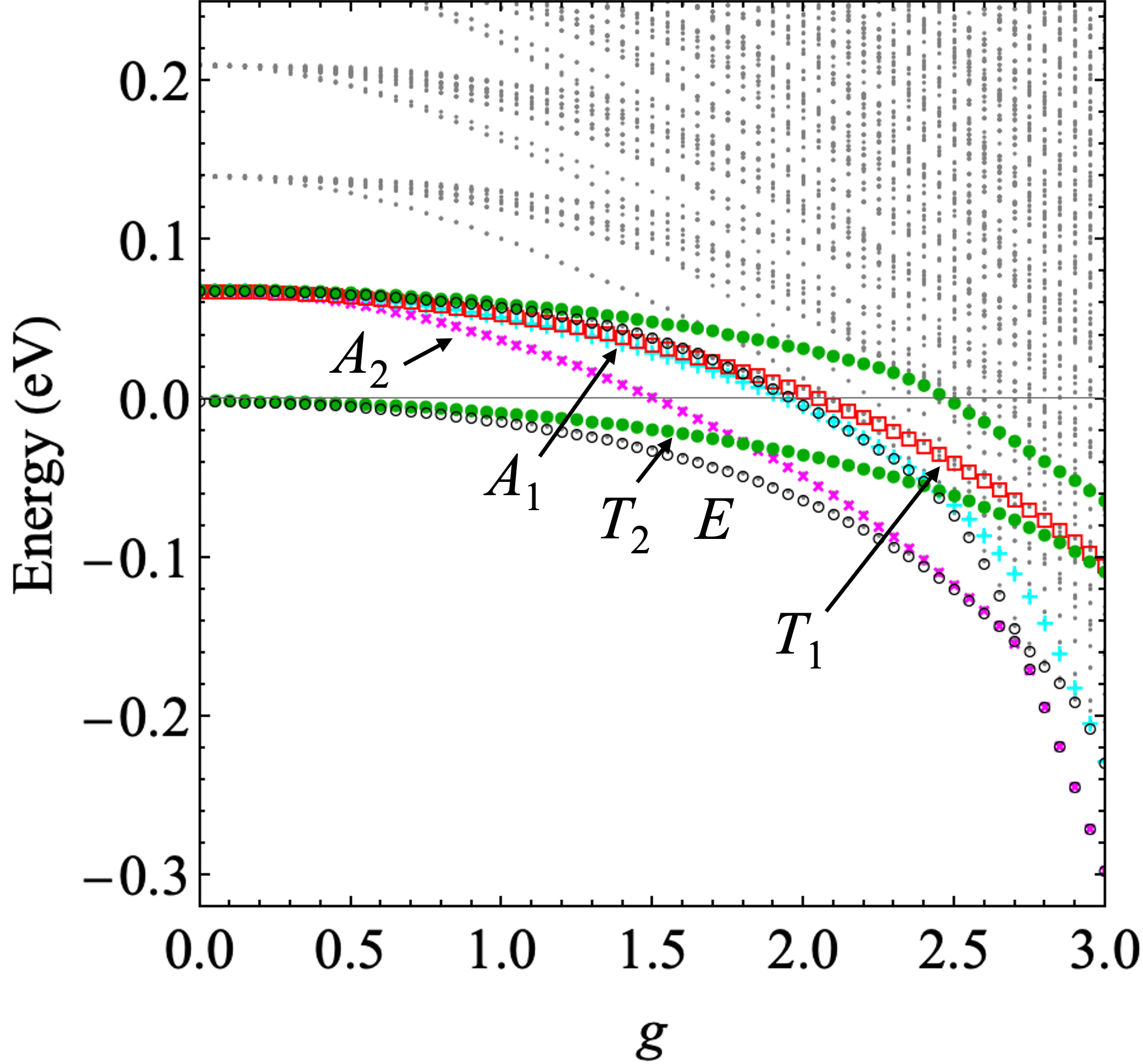}
\caption{
  (Color online)
  Low-energy vibronic levels of $d^2$ octahedron in function of $g$. 
Frequency $\omega = 70$ meV \cite{Pasztorova2023}, $\lambda = 0.4$ eV, and Hund's parameter $J_\text{H} = 0.25$ eV \cite{Yuan2017}. 
The cyan cross ($\times$), the magenta plus ($+$), the black circle ($\circ$), the red square ($\square$), and the green bullet ($\bullet$) are the $A_1$, $A_2$, $E$, $T_1$, and $T_2$ vibronic states, respectively.
$A_1$, $A_2$ are nondegenerate, $E$ doubly degenerate, and $T_1$ and $T_2$ triply degenerate. 
The crossing of the $E$ and the $A_2$ vibronic states occur at $g \approx 2.7$ with the present parameters.
} 
\label{Fig:energy_d2}
\end{figure}

\subsection{Dynamic Jahn-Teller effect}
\label{Sec:DJT}
In this section, we explain the dynamic JT effect \cite{Moffitt1957a, Moffitt1957b, Longuet-Higgins1958}. 
The dynamic JT effect is the phenomenon that emerges due to the quantum effect of the lattice degrees of freedom on the JT potential energy surface.
Contrary to the static JT effect where the normal coordinates $q$ are classical, and the conjugate momenta $p$ are zero, both $q$ and $p$ are quantum dynamical variables in the dynamic JT effect. 
The model Hamiltonian of the system consists of the electronic term, harmonic oscillator Hamiltonian [$\hat{H}_0 = \sum_\gamma (\hat{p}^2_\gamma + \omega^2 \hat{q}_\gamma^2)/2$], and the vibronic coupling (\ref{Eq:VJT}). 
We discuss the dynamic JT effect based on three different approaches: tunneling splitting \cite{Bersuker1962}, pseudo rotation \cite{Longuet-Higgins1958}, and numerical diagonalization \cite{Longuet-Higgins1958, Muramatsu1978}.

Let us consider the lattice dynamics in the warped APES of $d^1$ octahedron [Fig. \ref{Fig:APES}(b)].
When the energy barriers between the minima are high enough to localize the system at one of the minima, the ground states of the system are, in a good approximation, the direct product states of the electronic and the ground harmonic vibrational states in the minima. 
This description is valid when the vibronic coupling $v_E$ is strong, and we assume this condition to be satisfied.
Lowering the height of the energy barriers between the minima, the localized wave functions in different minima weakly overlap, giving nonzero transfer parameter $t$ ($>0$) between different minima \cite{Bersuker1962, Bersuker1989}. 
The transfer between the minima splits the three localized orbital-lattice states into the ground vibronic doublet ($-t$) and excited vibronic singlet ($2t$), which is called tunneling splitting 
\footnote{We do not consider the pseudo-spin sector because it is irrelevant to the vibronic coupling.}.
The energy eigenstates of the dynamic JT model (vibronic states) are linear combinations of the local orbital-lattice states as shown in Fig. \ref{Fig:vibronic_wf}, indicating the formation of the quantum entanglement of the orbital and lattice degrees of freedom. 
The delocalization of the ground wave function over the APES stabilizes the system by $t$ compared with the localized orbital-lattice states.

The tunneling splitting vibronic states [Fig. \ref{Fig:vibronic_wf}] indicate that the system is simultaneously deformed and structurally isotropic. 
Each orbital-lattice configuration expresses structural anisotropy due to the deformation. 
At the same time, the density matrix for the ground state contains all the orbital-lattice configurations with equal weight, wiping out the structural anisotropies. 
Consequently, we cannot detect the deformation of the dynamic JT systems with, e.g., x-ray or neutron diffractions.

By lowering the energy barriers further, the minima of the APES form a trough [Fig. \ref{Fig:APES}(c)] and the vibronic states become pseudo rotational type. 
The structure of the APES with continuous minima suggests the presence of the rotational invariance in the two-dimensional space of the normal coordinates around the origin. 
Hence, the system has simultaneous eigenstates of the Hamiltonian and the angular momentum $j$ for the pseudo-orbital and lattice degrees of freedom.
The low-energy lattice dynamics is a combination of radial harmonic oscillation and pseudo rotation. 
The corresponding vibronic states are approximately $|\Phi_\text{ad}(\theta)\rangle \chi_0(q) e^{ij\theta}/\sqrt{2\pi}$ with energy levels of $E_{j} \approx -E_\text{JT} + \hslash \omega [-1/2 + j^2/(2(g/2)^2) ]$. 
Here, $\chi$ is the zero-point energy state for the radial oscillation, $e^{ij\theta}/\sqrt{2\pi}$ the pseudo rotational wave function, $E_\text{JT}$ the static JT stabilization energy, and $g = v_E/\sqrt{\hslash \omega^3}$ is the dimensionless vibronic coupling (Appendix \ref{A:vibronic}).
We set the ground state energy of the model Hamiltonian with $v_E = 0$ to the origin of the energy. 
Under the $2\pi$ rotation around the high-symmetric position, the wave function of the entire system must be single-valued, while the adiabatic electronic state $|\Phi_\text{ad}(\theta)\rangle$ changes the sign \cite{Longuet-Higgins1958} as demonstrated above (Sec. \ref{Sec:SJT}).
To compensate for the sign change in the adiabatic electronic state, the lattice part in the wave function also has to change the sign.
The latter condition makes the angular momentum $j$ half-integer ($j = \pm 1/2, \pm 3/2, ...$) and the pseudo rotational vibronic levels doubly degenerate \cite{Note4}. 
The two-fold degeneracy characterized by the half-integer $j$ is the consequence of the presence of Berry's phase. 
The second term in the vibronic level ($E_j$) indicates the stabilization due to the delocalization of the lattice wave function over the trough.

With a finite warping of the trough, the ground state remains degenerate, while the first excited states split into two singlets \cite{OBrien1964}. 
The ground vibronic doublet and the first excited vibronic singlet in the warped APES are consistent with the prediction within the tunneling splitting.

We show accurate vibronic levels obtained by numerical diagonalization of the dynamic JT Hamiltonian for $d^1$ octahedron. 
Figure \ref{Fig:energy_d1}(a) shows the low-energy vibronic energy spectra in functions of the dimensionless vibronic coupling parameter, $g = v_E/\sqrt{\hslash \omega^3}$ (see Appendix \ref{A:vibronic}). 
We conducted the simulations with finite spin-orbit coupling, so the orbital-lattice dynamics are on a warped APES [Fig. \ref{Fig:APES}(b)].
The ground state is doublet (quartet including the Kramers degeneracy), and the first excited state is a singlet. 
The qualitative features agree with the analytical solutions described above. 
The calculated spectra show that the distribution of the energy levels significantly differs from that of the harmonic oscillator. 
Figure \ref{Fig:energy_d1}(a) shows the static JT stabilization energy with the gray dashed line. 
As expected from the analyses above, the ground vibronic level stabilizes more than the static JT system.
Figure \ref{Fig:energy_d1}(b) compares the numerical ground level and the ground level within the second-order perturbation theory. 
The latter can describe the dynamic JT effect with very weak vibronic coupling ($g < 0.05$).

As we have seen, the ground vibronic states are doubly degenerate as the tunneling splitting and pseudo-orbital states are. 
The degeneracy comes from the presence of Berry's phase and does not depend on the presence/absence of weak warping of the APES \cite{Ham1987} 
\footnote{The warping is much smaller than the JT stabilization energy.}. 
The Berry's phase also determines the nature of the low-lying vibronic states:
In the increasing order of the energy, the vibronic states are $E$, $A_{2(1)}$, $A_{1(2)}$, $E$, ...
The coincidence of the degeneracies (symmetries) of the ground pseudo-orbital and the ground vibronic states breaks down for very strong nonlinear vibronic coupling \cite{Koizumi1999} or the presence of the electronic multiplet structures \cite{OBrien1996}.
As an example of the latter case, see the vibronic spectra of $d^2$ octahedron below.
See for further information on Berry's phase in dynamic JT systems, e.g., Refs. \onlinecite{OBrien1993, Sakurai1994, Bohm2003, Grosso2014}.

With the knowledge of the vibronic states, let us discuss a theoretical tool to describe the physical phenomena within the low-lying vibronic states. 
For the description of the local physical properties of the dynamic JT systems, we project the relevant physical quantities $\hat{O}$ into the space of the low-energy vibronic states, $\hat{O} \rightarrow \hat{\mathcal{O}} = \hat{\mathcal{P}} \hat{O} \hat{\mathcal{P}}$, where $\hat{\mathcal{P}}$ is the projection operator into the low-energy vibronic states.  
Due to the hybridizations of the orbital-lattice configurations, the physical quantities of the pseudo-orbital origin tend to be reduced by the vibronic dynamics \cite{Child1961, Ham1968}. 
Contrary to the case of the electronic operators, the dynamic JT effect enhances the lattice-related quantities.
With the vibronic states of tunneling splitting type (Fig. \ref{Fig:vibronic}), we can explicitly evaluate the pseudo-orbital and the normal coordinates of the $d^1$ systems: We can treat both $\hat{\tau}_\gamma$ and $q_\gamma$ ($\gamma = z^2, x^2-y^2$) with the vibronic quadrupole moments $\hat{\mathcal{T}}_\gamma$ \cite{Iwahara2023}.
The vibronic quadrupole moments enable a unified treatment of the orbital and lattice degrees of freedom in the crystals with the dynamic JT effect on sites.

Now, we move to the dynamic JT effect in the $d^2$ systems.
The ground $J=2$ electronic states comprise of $E_g$ and $T_{2g}$ multiplets, and both of them couple to the $E_g$ vibrations as Eq. (\ref{Eq:VJT_SO_d1}) and (\ref{Eq:VJT}), respectively: Needless to say, the electronic operators in these model Hamiltonians act on the $d^2$-$E_g$ and $d^2$-$T_{2g}$ multiplet states. 
The vibronic states from the $E_g$ multiple states resemble those of the $d^1$ system. 
The ground states of the $T_{2g}$ electronic states coupled to the $E_g$ vibrations are, in a good approximation, harmonic oscillations in one of the minima [Fig. \ref{Fig:APES}(a)]: In this case, the stabilization by the delocalization over the minima is weak \cite{Caner1966}. 
Therefore, the $E_g$ vibronic states tend to be more stabilized than the $T_{2g}$ vibronic states due to the stronger quantum effect, giving rise to the splitting between the nonmagnetic $E_g$ and the magnetic $T_{2g}$ states. 
This splitting can contribute to the $E_g$-$T_{2g}$ energy gap in the $5d^2$ double perovskites.

Figure \ref{Fig:energy_d2} shows the numerical vibronic levels of the $d^2$ system. 
The ground $E_g$ vibronic level is lower than the $T_{2g}$ vibronic level. 
For realistic $g$ ($\approx 1-1.5$) for the $5d$ compounds, the splitting is about 5-13 meV in the present calculations, which is not negligible compared with the experimental $E_g$-$T_{2g}$ splitting of 10-17 meV in various $5d^2$ Os double perovskites \cite{Maharaj2020}.  
With the present model, however, we cannot invert the $E_g$ and the $T_{2g}$ states as observed in the $5d^2$ Re compounds \cite{Thompson2014, Marjerrison2016b, Nilsen2021}.

Figure \ref{Fig:energy_d2} also indicates that the ground state can change for large linear vibronic couplings. 
Within the present choice of the interaction parameters, the vibronic doublet ($E_g$) is the most stable up to $g = g_c \approx 2.7$, while the nondegenerate $A_2$ vibronic state becomes the ground state for $g > g_c$.
This change in the ground vibronic states resembles the case with multiplet splitting in Ref. \onlinecite{OBrien1996}.

The dynamic JT effect is also relevant to the $d^5$ and $d^4$ systems.
In the $d^5$ octahedron, the dynamic JT effect can develop in the excited $\Gamma_8$ quartet states.
In the $d^4$ octahedron, the dynamic JT can arise in the excited $J=1$ and $J=2$ states and also the excited ${}^1E_g$ and ${}^1T_{2g}$ term states above the ground ${}^3T_{2g}$ term states [See Fig. 7 in Ref. \onlinecite{Tanabe1954II}].

\begin{figure}[htpb]
\begin{tabular}{lll}
(a) &~& (b) \\
 \raisebox{15mm}{
\includegraphics[width=0.25\linewidth, bb = 0 0 114 160]{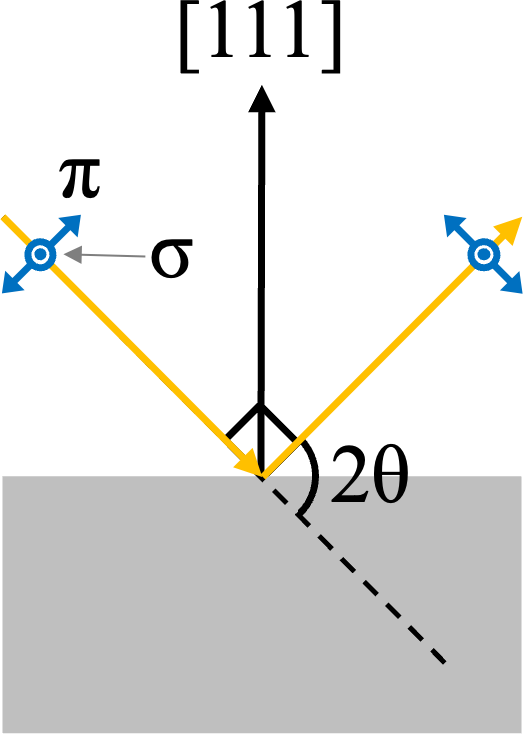}
 }
&&
\includegraphics[width=0.50\linewidth, bb = 0 0 211 251]{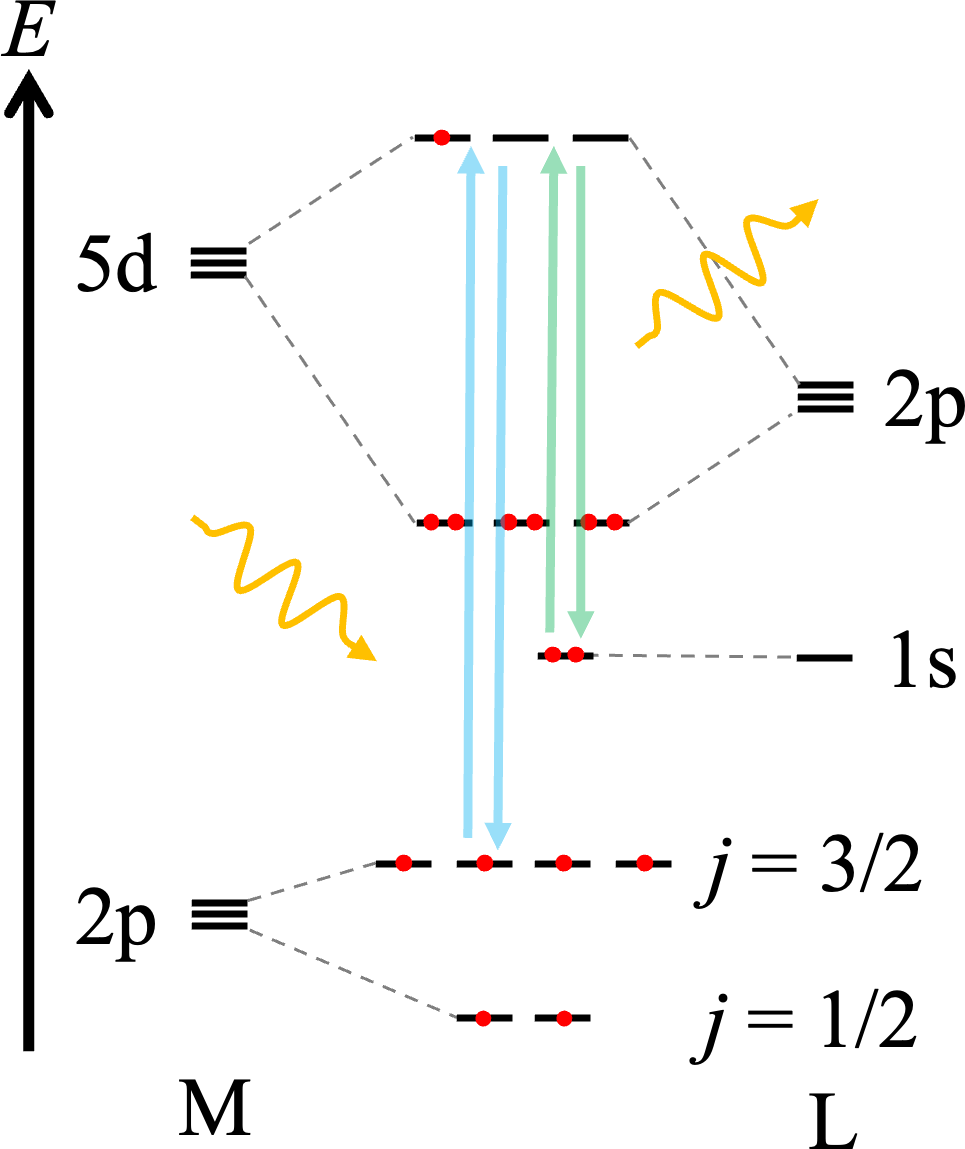}
\end{tabular}
\caption{
  (Color online)
  (a) The scattering geometry for the simulations of RIXS in Figs. \ref{Fig:d1_RIXS} and \ref{Fig:d2_RIXS} and (b) the $L_3$ and $K$-edge RIXS processes.}
\label{Fig:RIXS}
\end{figure}

\subsection{Fingerprints of the dynamic JT effect in spectroscopic data}
\label{Sec:RIXS}
The fingerprints of the dynamic JT effect on metal sites can appear in various spectroscopic data. 
The RIXS spectroscopy is an indispensable tool for studying the local dynamic JT effect on metal sites because the RIXS can capture arbitrary type of elementary excitations. 
Recently reported RIXS spectra of several cubic $5d$ compounds show the fingerprints of the dynamic JT effect on metal sites.
Below, we review the vibronic effect in the RIXS spectra on $5d$ metal sites.
Many of the experimental RIXS spectra we discuss do not show band dispersion and provide only information on the local energy spectra.

The RIXS is a photon-in and photon-out process, and the energy losses of the emitted photons directly determine the distribution of the energy levels \cite{RIXS, RIXS2}. 
Figure \ref{Fig:RIXS}(b) is a schematic picture of the RIXS processes. 
For the studies of the $4d$ and $5d$ transition metal compounds, metal $L_3$ and ligand $K$-edge RIXS are frequently used. 
The $L_3$-edge RIXS is a process of $(2p_{3/2})^4d^N \rightarrow (2p_{3/2})^3d^{N+1} \rightarrow (2p_{3/2})^4d^N$, and the $K$-edge RIXS corresponds to $(1s)^2d^N \rightarrow (1s)^1d^{N+1} \rightarrow (1s)^2d^N$, where the $2p_{3/2}$ and $d$ are the metal orbitals, and the $1s$ is the ligand $1s$ orbital. 
In dynamic JT systems, the initial and the final states must be vibronic, and such RIXS spectra show fine structures absent within the electronic models.

\begin{figure}[htpb]
\begin{tabular}{lc}
(a) \\
& \includegraphics[width=0.9\linewidth, bb = 0 0 1717 1096]{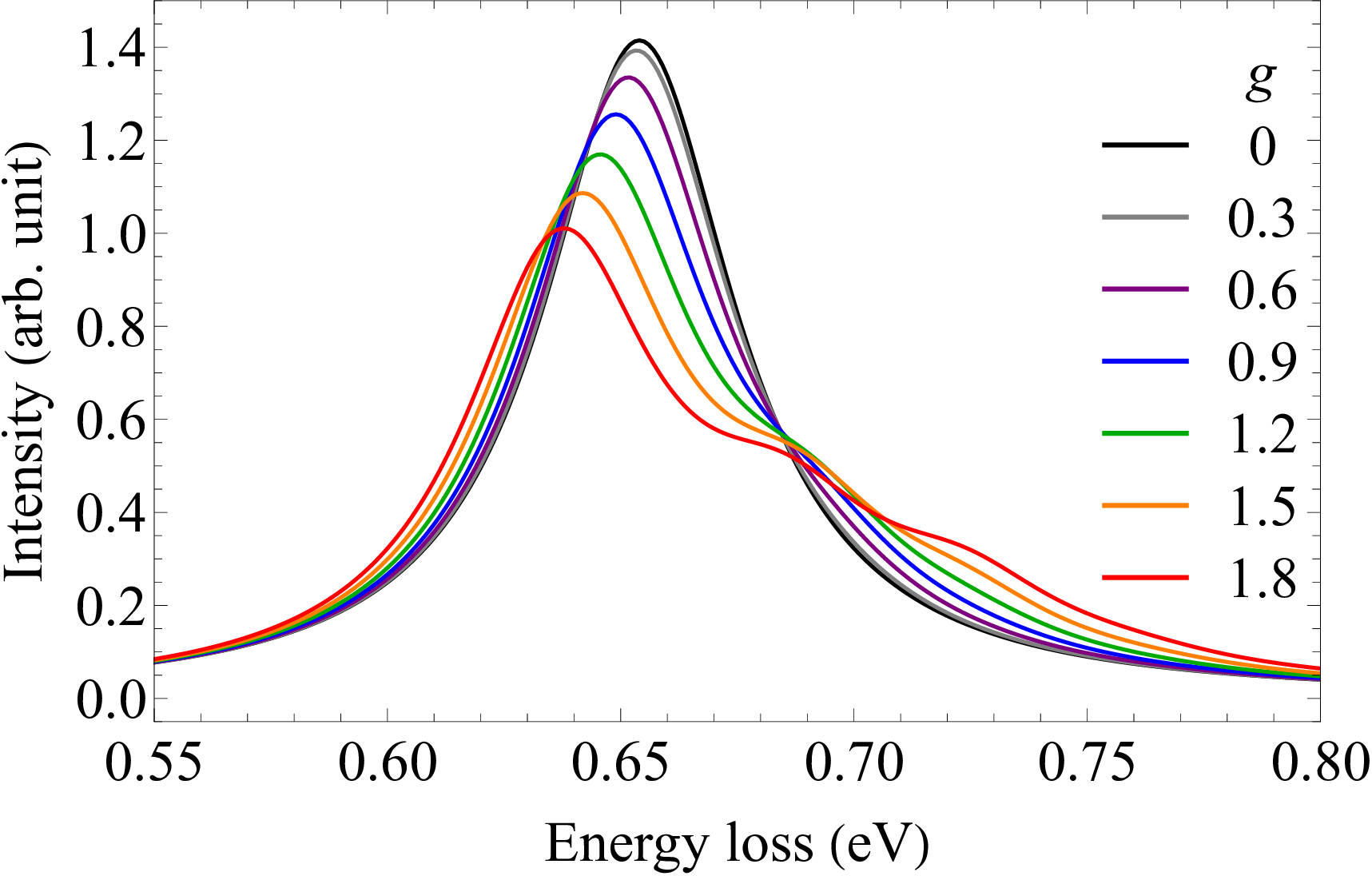}
\\
(b) \\
& \includegraphics[width=0.9\linewidth, bb = 0 0 1717 1096]{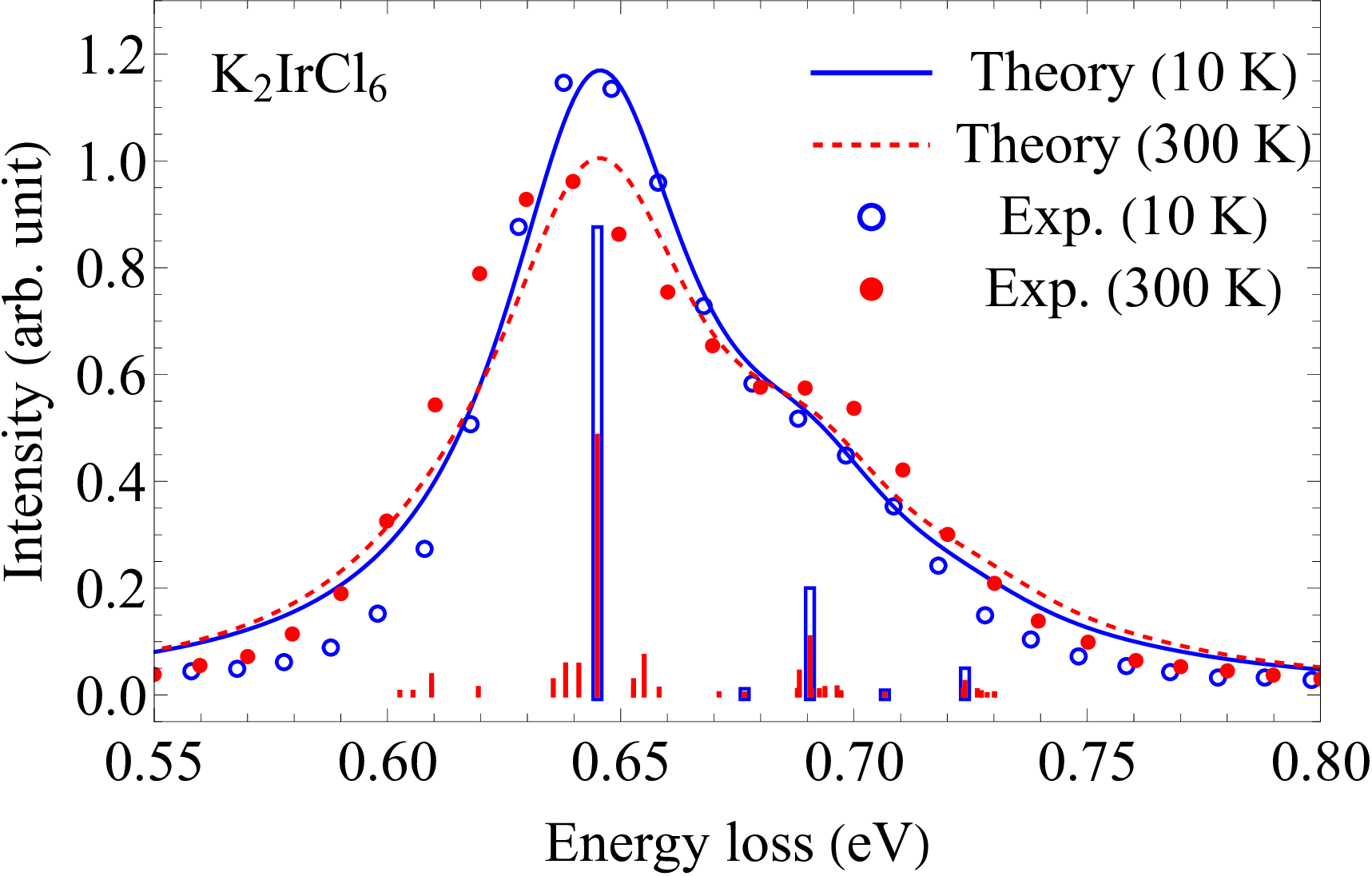}
\end{tabular}
\caption{
  (Color online)
  (a) The evolution of the $L_3$-edge RIXS spectra in function of the dimensionless vibronic coupling $g$. 
(b) Ir $L_3$-edge RIXS spectra of K$_2$IrCl$_6$. We took the experimental spectra from Ref. \onlinecite{Reigiplessis2020}. 
For the simulation, we took $\omega = 37.1$ meV from Raman scattering spectra\cite{Lee2022}.
The figures are taken from Ref. \onlinecite{Iwahara2023Ir} (c) 2023 The American Physical Society.}
\label{Fig:Ir}
\end{figure}

At a glance, the RIXS spectra of some $4d/5d$ octahedra look contradictory to the other experimental data. 
For example, in the case of cubic K$_2$IrCl$_6$ \cite{Khan2019} [Fig. \ref{Fig:DP}(b)], the transition from the ground $j_\text{eff}=1/2$ doublet ($\Gamma_7$) to the spin-orbit $j_\text{eff}=3/2$ quartet ($\Gamma_8$) on Ir sites [Fig. \ref{Fig:electronic_d1}(a)] has two peaks \cite{Reigiplessis2020, Khan2021, Warzanowski2024} [The blue and red circles in Fig. \ref{Fig:Ir}(b)].
Although such shoulder peaks often appear due to low-symmetric ligand field, symmetry lowering is inconsistent with the structural data down to 0.3 K \cite{Reigiplessis2020} \footnote{Recent high-resolution x-ray diffraction data suggest the development of tiny deformation below $T_N \approx 3.2$ \cite{Wang2024}.} 
and magnetic resonance data \cite{Bhaskaran2021}.

The dynamic JT effect changes the shapes of the RIXS spectra without introducing static local deformations. 
Figure \ref{Fig:Ir}(a) shows the $L_3$-edge RIXS spectra of the $5d^5$ dynamic JT system around the $j_\text{eff}=3/2$ states \cite{Iwahara2023Ir}.
For the simulation, we employed the Kramers-Heisenberg formula \cite{Sakurai1967, Blume1985, RIXS} within the fast collision approximation \cite{Luo1993, vanVeenendaal2006}, which is valid for the $5d$ elements \cite{Clancy2012}. 
At $g = 0$, the RIXS spectrum is symmetric and has only one peak. 
Increasing $g$, new shoulder peaks corresponding to the transition from $j_\text{eff}=1/2$ doublet to the excited vibronic states around the $j_\text{eff}=3/2$ quartet grow, and the spectrum becomes asymmetric.
As Fig. \ref{Fig:energy_d1} indicates, the energy gaps between the vibronic levels are not equally separated, and the peaks in the RIXS spectra are neither, which is distinct from the harmonic peaks within the Franck-Condon mechanism for the non-JT systems \cite{Sugano1970, Toyozawa2003, Grosso2014}.

Figure \ref{Fig:Ir}(b) shows a good agreement between the theoretical and experimental RIXS spectra of K$_2$IrCl$_6$ \cite{Iwahara2023Ir}. 
The solid lines indicate the simulated spectra with $g = 1.2$ at low- and room temperatures (10 and 300 K). 
The theoretical data reproduce both the fine structures and the temperature evolution.
The interpretation based on the dynamic JT effect is also consistent with the structural and magnetic data.
Besides, we could qualitatively reproduce the RIXS spectra of isostructural K$_2$IrBr$_6$ by only considering the effect of the change in mass of ligand atoms (Cl $\rightarrow$ Br). 
The dynamic JT effect can arise in other cubic Ir compounds such as Ba$_2$CeIrO$_6$ \cite{Raveli2019} and also $\alpha$-RuCl$_3$ \cite{Suzuki2021}. 
In the former, the RIXS spectra of the Ir$^{4+}$ sites also show a similar splitting pattern to K$_2$IrCl$_6$, which implies the development of the dynamic JT effect in the excited spin-orbit state of the compounds.

Similar splitting arises in the Raman peaks for the $\Gamma_8$ states \cite{Lee2022}, and vibronic progression appears in optical conductivity \cite{Warzanowski2024} of K$_2$IrCl$_6$. 
These features could result from the dynamic JT effect, while analysis considering the dynamic JT effect is missing. 
Further analysis is called for.

The vibronic effect also appears in the excited states of $d^4$ compounds. 
The Ru $L_3$-edge RIXS spectra of $4d^4$ Ru sites in K$_2$RuCl$_6$ antifluorite [Fig. \ref{Fig:DP}(b)] shows broad or split peaks around the excited spin-orbit multiplet/term states \cite{Takahashi2021}.
Assuming the dynamic JT effect in the excited states, we could reproduce the fine structures and the temperature dependence of the RIXS spectra \cite{Iwahara2023Ru}.

\begin{figure*}[htpb]
\begin{tabular}{lllll}
(a) &~& (b) &~& (c) \\ 
\includegraphics[height=0.25\linewidth]{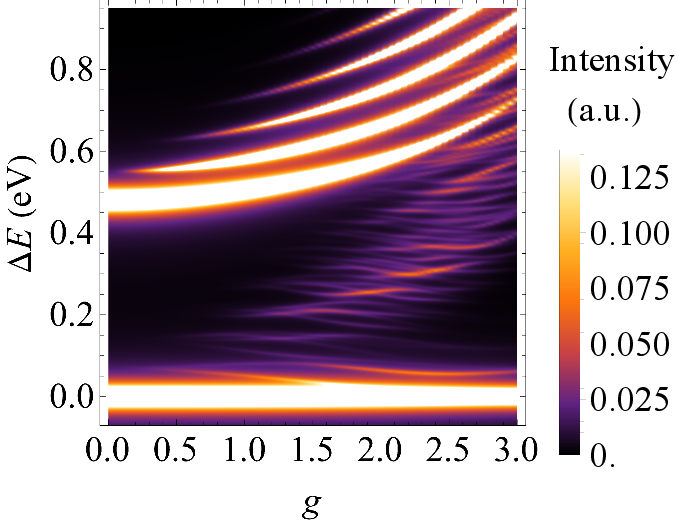}
& &
\includegraphics[height=0.25\linewidth]{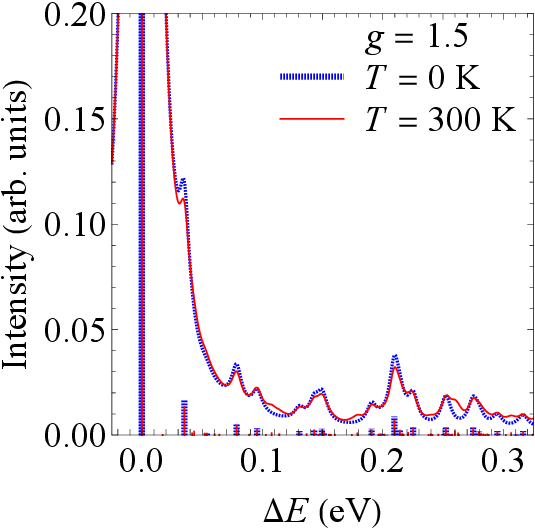}
& &
\includegraphics[height=0.25\linewidth]{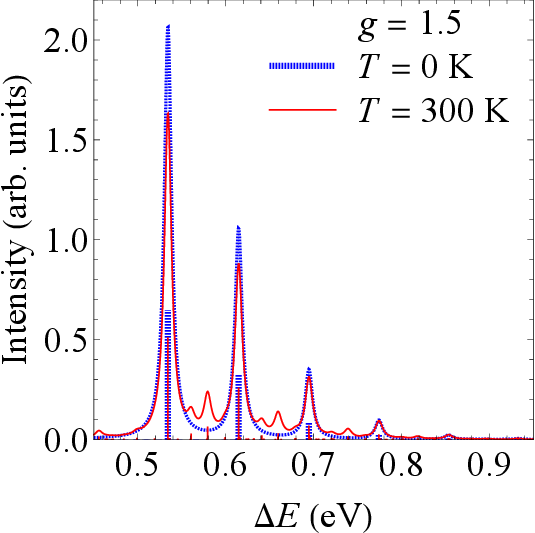}
\\
(d) &~& (e) &~& (f) \\ 
\includegraphics[height=0.25\linewidth]{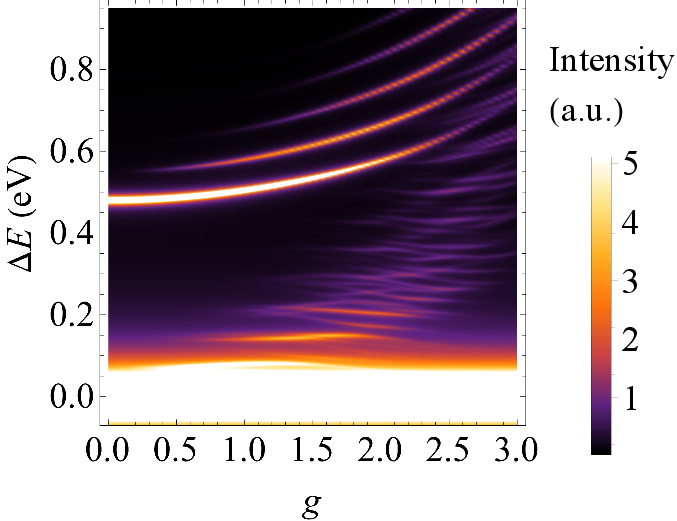}
& &
\includegraphics[height=0.25\linewidth]{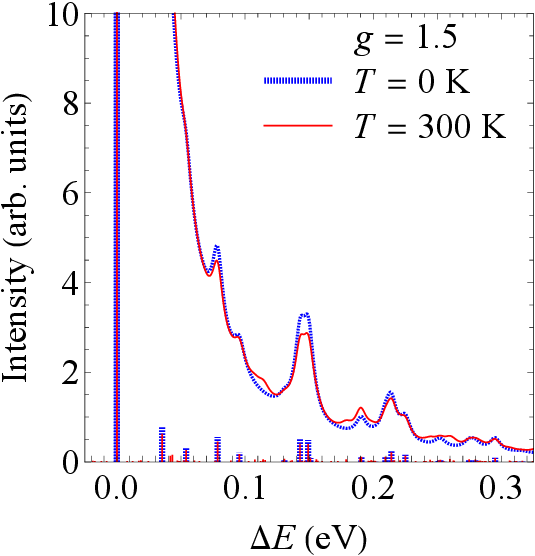}
& &
\includegraphics[height=0.25\linewidth]{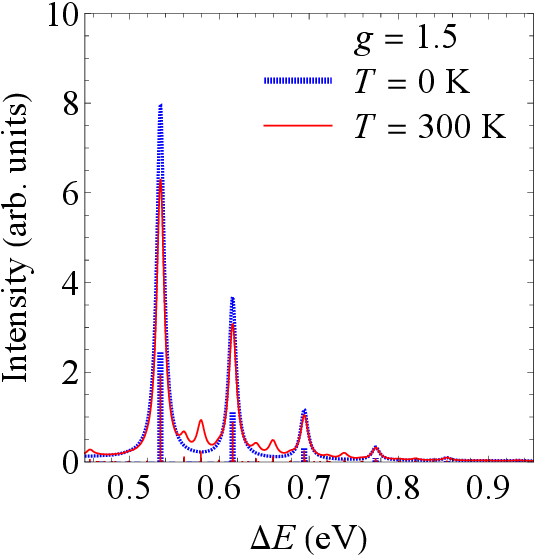}
\end{tabular}
\caption{
  (Color online)
  The simulated RIXS spectra of $d^1$ octahedron. (a), (b) The metal $L_3$-edge RIXS spectra, and (c), (d) the ligand $K$-edge RIXS spectra. 
The incident light is $\pi$ and $\sigma$ polarized for the $L_3$- and $K$-edge RIXS spectra, respectively, and $2\theta = \pi/2$.
In both spectra, 
$\omega = 70$ meV \cite{Pasztorova2023}, $\lambda = 0.32$ eV \cite{Frontini2024}, and the linewidth $\Gamma = 10$ meV.
}
\label{Fig:d1_RIXS}
\end{figure*}

\begin{figure*}[htpb]
\begin{tabular}{lll}
(a) &~& (b) \\
\includegraphics[height=0.25\linewidth]{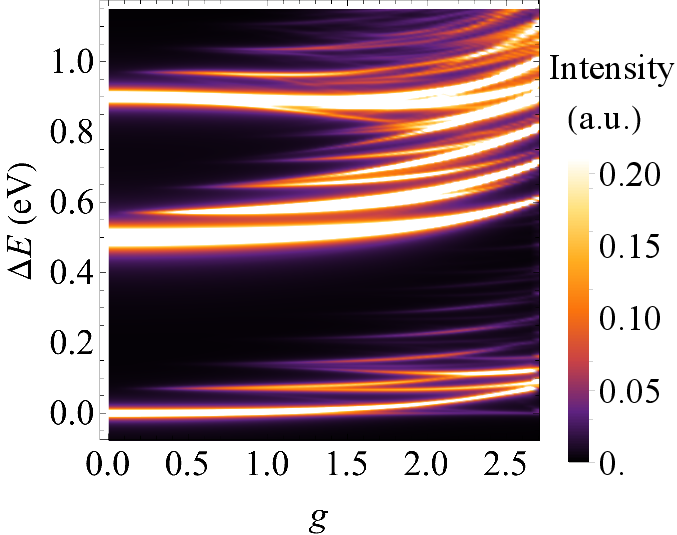}
& &
\includegraphics[height=0.25\linewidth]{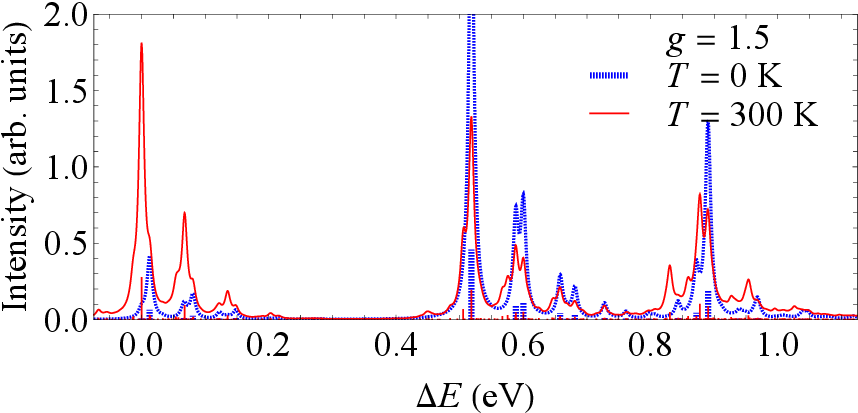}
\\
(c) &~& (d) \\
\includegraphics[height=0.25\linewidth]{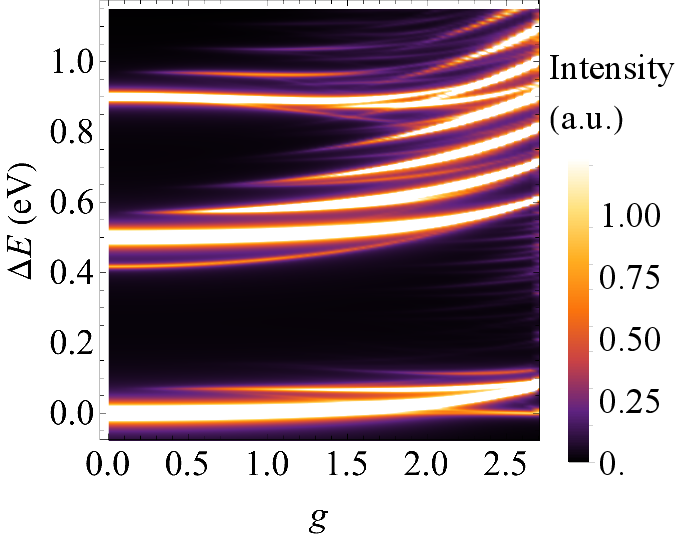}
& &
\includegraphics[height=0.25\linewidth]{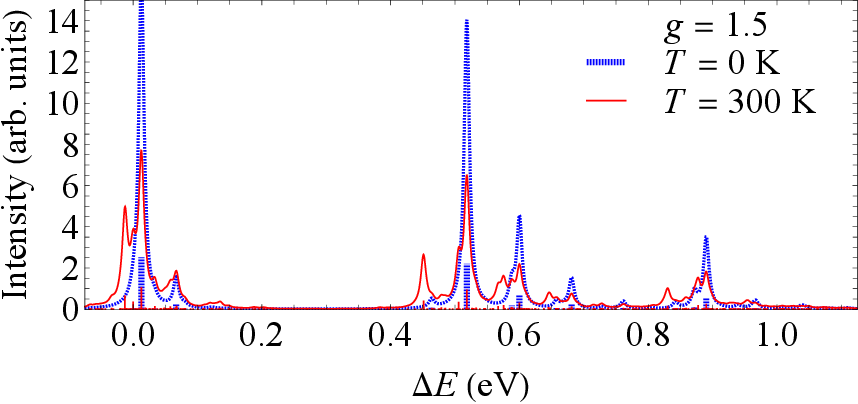}
\end{tabular}
\caption{
  (Color online)
  The simulated RIXS spectra of $d^2$ octahedron. (a), (b) The metal $L_3$-edge RIXS spectra, and (c), (d) the ligand $K$-edge RIXS spectra. 
In both cases, the incident light is $\pi$ polarized and $2\theta = \pi/2$.
$\omega = 70$ meV \cite{Pasztorova2023}, $J_H = 0.25$ eV \cite{Yuan2017}, $\lambda = 0.40$ eV \cite{Yuan2017}, and the linewidth $\Gamma = 10$ meV.}
\label{Fig:d2_RIXS}
\end{figure*}

The RIXS measurements can detect the dynamic JT effect in the ground spin-orbit multiplet states of cubic $d^1$/$d^2$ compounds.
Figures \ref{Fig:d1_RIXS} and \ref{Fig:d2_RIXS} show the simulated $L_3$ and $K$-edge RIXS spectra for the $d^1$ and $d^2$ systems, respectively, with the scattering geometry shown in Fig. \ref{Fig:RIXS}(a).
For the $K$-edge RIXS calculations, we utilized the method for the $L_3$-edge RIXS spectra to simulate the $K$-edge RIXS spectra.
The calculated spectra show many transitions between the vibronic states: The $L_3$ and $K$-edge RIXS intensities are different.

\begin{figure*}[!tbh]
\begin{tabular}{lll}
(a) &~& (b) \\
\includegraphics[height=0.29\linewidth, bb = 0 0 2852 1662]{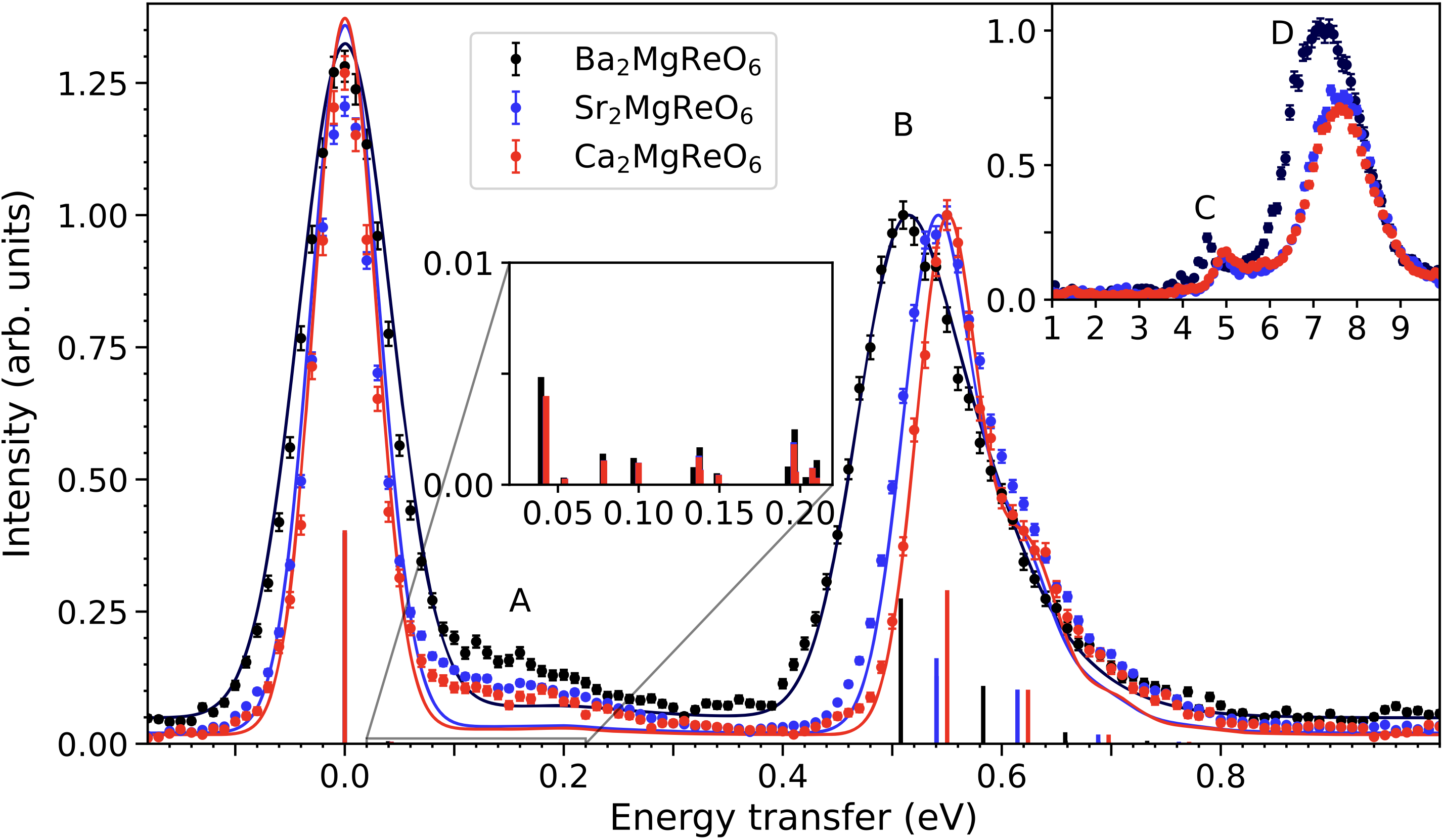}
&&
\includegraphics[height=0.29\linewidth, bb = 0 0 1562 1714]{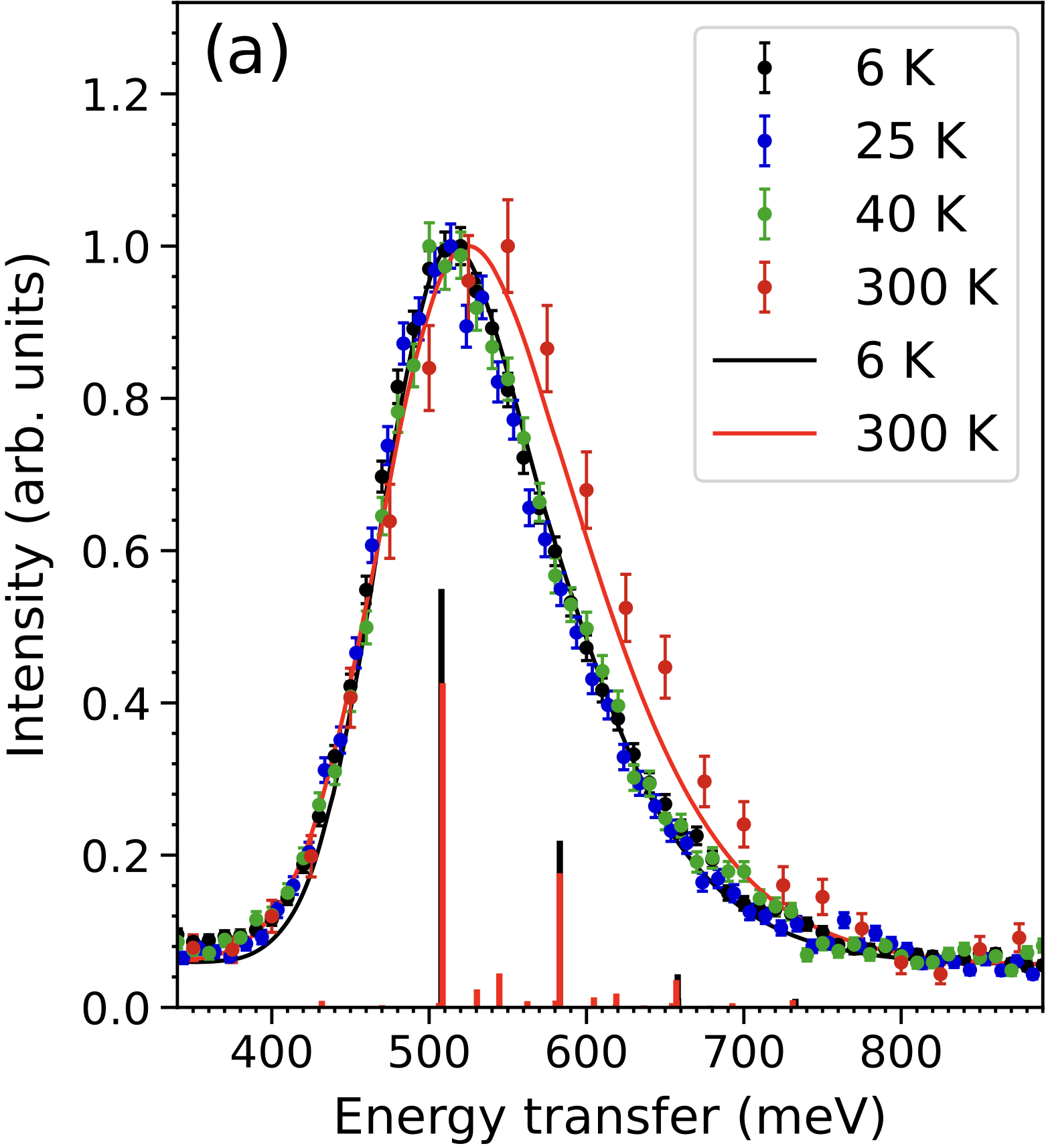}
\end{tabular}
\caption{
  (Color online)
(a) The experimental and simulated Re $L_3$-edge RIXS spectra of Ba$_2$MgReO$_6$, Ca$_2$MgReO$_6$, and Sr$_2$MgReO$_6$. 
The symbols indicate the experimental data, and the lines are the simulated data. 
(b) The temperature dependence of the RIXS spectra of Ba$_2$MgReO$_6$. 
The figures are taken from Ref. \onlinecite{Frontini2024} (c) 2024 The American Physical Society.}
\label{Fig:Re}
\end{figure*}

Indeed, recent $L_3$-edge RIXS spectra of the $5d^1$ rhenium (Ba$_2$MgReO$_6$ \cite{Frontini2024, Zivkovic2024}, Ba$_2$CaReO$_6$ \cite{Iwahara2024}) and $5d^1$ osmium (Ba$_2$NaOsO$_6$ \cite{Agrestini2024}) oxides show asymmetric $\Gamma_7$ peaks. 
In the $L_3$-edge RIXS spectra, the peak for the spin-orbit excitation around $0.5$ eV does not split, while it becomes asymmetric. 
Figure \ref{Fig:Re} shows the Re $L_3$-edge RIXS spectra of Ba$_2$MgReO$_6$ \cite{Frontini2024}. 
With the dynamic JT effect, we could reproduce the asymmetry and the temperature evolution (slight broadening) of the peak. 
The situation is similar to the vibrational spectra of the non-JT systems within Franck-Condon approximation \cite{Sugano1970, Toyozawa2003, Grosso2014}.  
In the present case, the transitions are from the ground dynamic JT states [Fig. \ref{Fig:energy_d1}] to the non-JT $j_\text{eff}=1/2$ multiple states with vibrational excitations.

The $L_3$-edge RIXS spectra, however, do not show the fine structures from the low-energy vibronic levels: 
They appear in the O $K$-edge RIXS spectra. 
As Fig. \ref{Fig:d1_RIXS}(d) shows, the intensities of the low-energy vibronic states are stronger in the $K$-edge spectra than in the $L_3$-edge spectra. 
Such low-energy peaks were observed in the O $K$-edge RIXS spectra of Ba$_2$CaReO$_6$ \cite{Iwahara2024},  
Ba$_2$NaOsO$_6$ \cite{Agrestini2024}, and Ba$_2$MgReO$_6$ \cite{Zivkovic2024}.
The low-energy experimental RIXS spectra of Ba$_2$CaReO$_6$ and the simulation fully considering the dynamic JT effect show a good agreement, indicating the development of the dynamic JT effect in the family of cubic $5d^1$ double perovskites.

Another important feature is that the RIXS spectra do not change across the quadrupolar transition \cite{Frontini2024} [Fig. \ref{Fig:Re}(b)]. 
If the quadrupolar transition corresponds to the cooperative JT deformations, the asymmetric peak should become symmetric above the transition temperature. 
The lack of the change suggests the presence of a unified mechanism for the fine structure of the RIXS both below and above the transition temperature.
The dynamic JT effect naturally explains both the fine structure and the temperature dependence of the RIXS spectra.

The $5d^2$ Re $L_2$ and $L_3$-edge RIXS measurements have been reported for Ba$_2$YReO$_6$ \cite{Yuan2017, Paramekanti2018}. 
The $L_3$-edge RIXS spectra show two peaks at about 0.5 eV, which could be relevant to vibronic effects [Fig. \ref{Fig:d2_RIXS}(b)], whereas the peaks are too broad to unambiguously conclude the presence/absence of the dynamic JT effect in the compound. 
Recent O $K$-edge RIXS spectra on Ba$_2$CaOsO$_6$ show vibronic progression \cite{Fujimori}, which is consistent with the simulated data [Fig. \ref{Fig:d2_RIXS}(d)].

\section{Cooperative phenomena}
\label{Sec:CDJT}

\subsection{Multipolar interaction}
\label{Sec:multipolar}
The vibronic states on metal sites correlate via intersite interactions and can give rise to collective phenomena.
Since the local vibronic states contain information on the spin, orbital, and lattice degrees of freedom, they respond to the magnetic, electric, and elastic interactions.
To incorporate the dynamic JT effect into intersite interactions, we project the interactions into the space of the low-energy vibronic states as we have discussed for the single-site physical properties in Sec. \ref{Sec:DJT}. 
In the present case, we replace the pseudo-orbital operators $\tilde{\tau}_\gamma$ and normal coordinates $q_\gamma$ with the vibronic quadrupole moments $\hat{\mathcal{T}}_\gamma$. 
The derivation of the effective model follows the general procedure for correlated insulators such as spin-orbit coupled systems (see, e.g., Refs. \onlinecite{Moriya1960, Jackeli2009, Chen2010, Khaliullin2021}). 
A similar approach has been employed for alkali-doped fullerides \cite{Chibotaru2005, Iwahara2013} and a $3d$ transition metal oxide \cite{Nasu2013, Nasu2014e}.

Let us derive the spin-orbital exchange and the elastic interaction models between the vibronic states on metal sites. 
We obtain the spin-orbital exchange interaction by projecting the superexchange exchange interaction \cite{Anderson1959} between the $t_{2g}$ orbitals into the ground $\Gamma_8$ multiple on sites. 
Considering the most significant electron transfer between the pair of $t_{2g}$ orbitals with $\sigma$ type overlap, the exchange interaction reduces to a simple Kugel-Khomskii-like form: The model contains the isotropic antiferromagnetic term, symmetric term, and pseudo-orbital quadrupolar interaction term \cite{Romhanyi2017}. 
By replacing the pseudo-orbital moments with the vibronic quadrupole moments, we obtain a pseudo-spin-vibronic exchange model \cite{Iwahara2023}:
\begin{align}
 \hat{\mathcal{H}}^{ij}_\text{ex} &=
 \tilde{\mathcal{J}}^{ij} \tilde{\bm{s}}^{i} \cdot \tilde{\bm{s}}^{j}
 +
 \tilde{\bm{s}}^{i} \tilde{\mathcal{K}}^{ij} \tilde{\bm{s}}^{j}
 +
 \tilde{\mathcal{Q}}^{ij},
 \label{Eq:Hex}
\end{align}
where $\mathcal{J}$, $\mathcal{K}$, and $\mathcal{Q}$ are the isotropic, symmetric exchange, and quadrupolar operators, which are functions of the vibronic quadrupole moments, $\hat{\mathcal{T}}$.

We construct the elastic interaction between the nearest neighbor using the Slater-Koster method \cite{Slater1954}.
The interaction between the nearest neighbor sites in the $xy$ plane takes the form of 
$(q_{z^2}^i, q_{x^2-y^2}^i) \bm{D}^{ij}_0(\theta) (q_{z^2}^j, q_{x^2-y^2}^j)^T$ with $\bm{D}^{ij}_0(\theta) = d_0(\cos \theta \bm{\sigma}_0 + \sin \theta \bm{\sigma}_z)$.
Replacing the normal coordinates with the vibronic quadrupole moments, $q_{z^2/x^2-y^2} \rightarrow \hat{\mathcal{T}}_{z/x}$, we obtain vibronic quadrupolar interaction:
\begin{align}
 \hat{\mathcal{H}}^{ij}_{\text{vib}} &= 
 (\hat{\mathcal{T}}_{z}^i, \hat{\mathcal{T}}_{x}^j)
 \bm{D}^{ij}(\theta)
 (\hat{\mathcal{T}}_{z}^j, \hat{\mathcal{T}}_{x}^j)^T,
 \label{Eq:Hvib}
\end{align}
where $\bm{D}^{ij} = g^2\bm{D}^{ij}_0$. 
This Hamiltonian has mathematically the same form as the electronic quadrupolar model in Refs. \onlinecite{Tsunetsugu2021, Hattori2023}. 
By varying the type of the elastic coupling via $\theta$ in $\bm{D}$, the nature of the quadrupolar interaction changes much compared with $\hat{Q}$ in Eq. (\ref{Eq:Hex}).

In the $d^2$ system, the situation resembles the $d^1$ system. 
The vibronic states on metal sites should interact via the spin-orbital exchange and elastic couplings. 
The elastic interaction reduces to the electronic quadrupolar interaction when the vibronic coupling is weak \cite{Sugihara1959}.
The elastic interaction simply enhances the quadrupolar interaction \cite{Khaliullin2021}, while the nature can vary in the case with strong vibronic coupling.

\begin{figure*}[!tbh]
 \begin{tabular}{lll}
 (a) &~~~& (b) \\
 \includegraphics[bb = 0 0 782 382, height=0.27\linewidth]{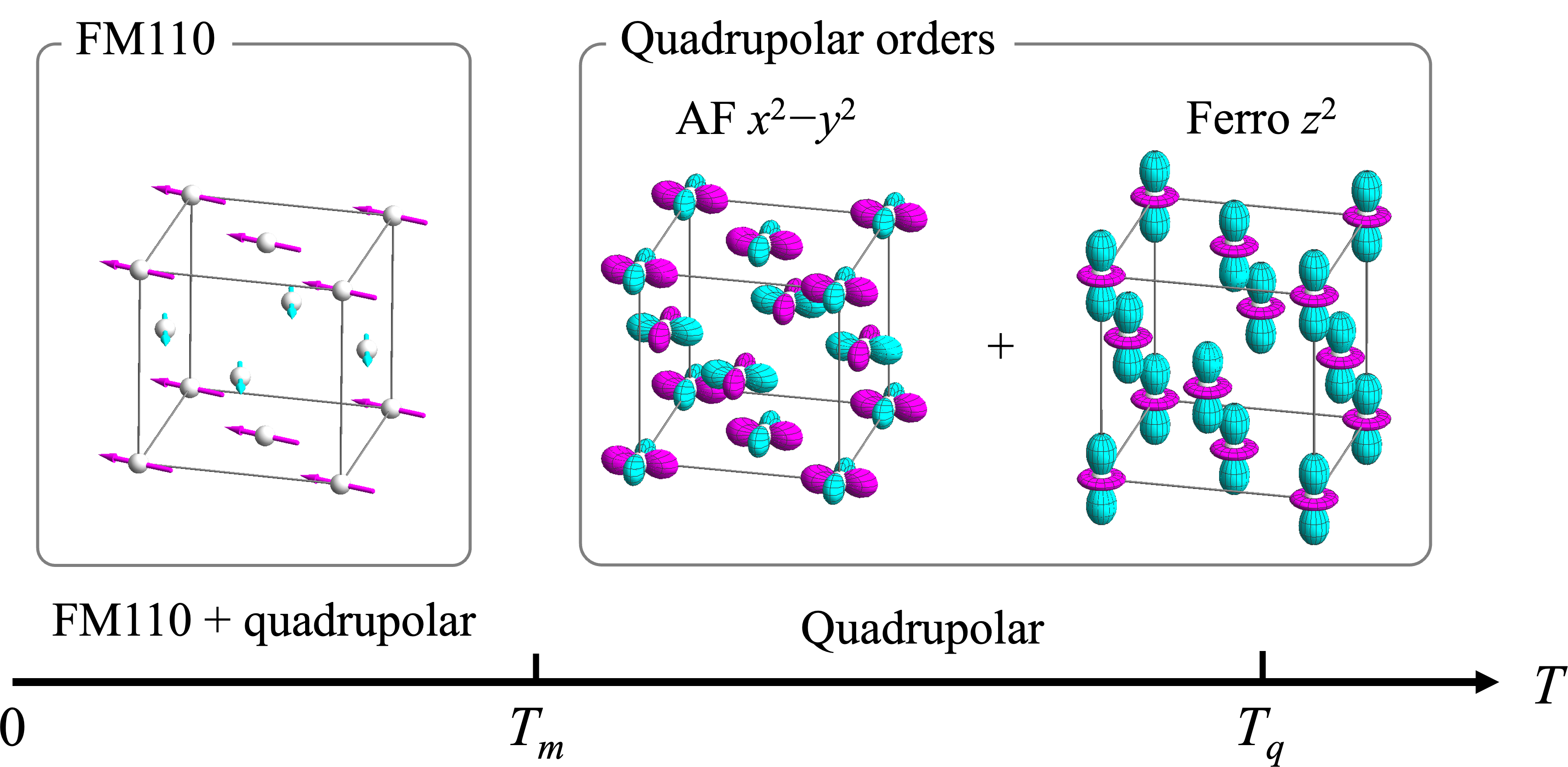}
 &&
 \includegraphics[bb = 0 0 569 382, height=0.27\linewidth]{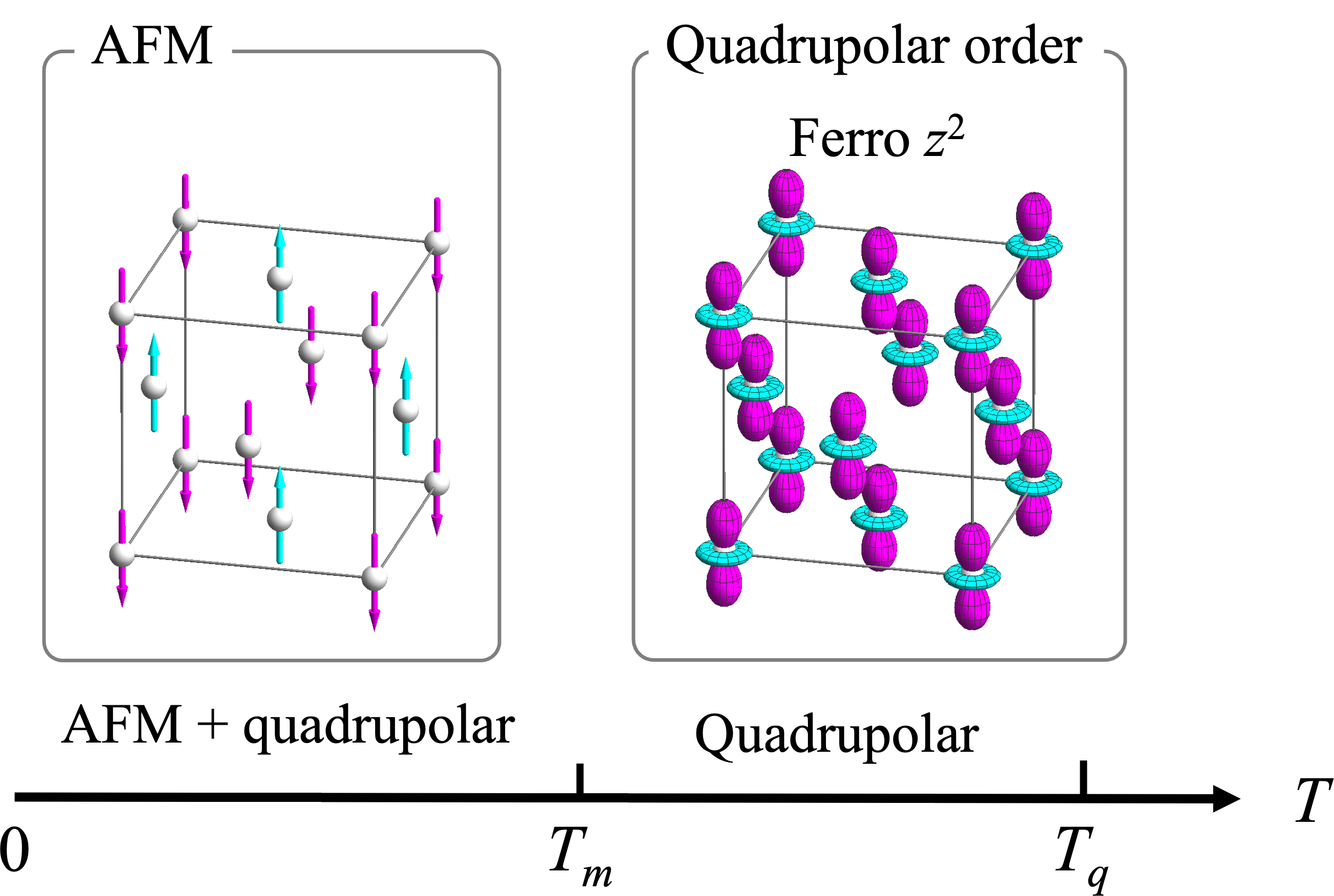}
 \end{tabular}
 \caption{
  (Color online)
 (a) The FM110 and (b) AFM-FQ orderings. 
 The magenta arrows indicate the magnetic dipole moments, and the lobes express the vibronic quadrupole moments (the magenta indicates positive, and the cyan does negative). 
 }
 \label{Fig:orderings}
\end{figure*}

\subsection{Vibronic order}
\label{Sec:vibronic_order}
The dynamic JT states on metal sites show cooperative behaviors through intersite interactions. 
Here, we mainly discuss the ordered phases in cubic $5d^1$ double perovskites.

The $5d^1$ double perovskites exhibit various ordered phases at low temperature. 
Among them, two ordered phases are well established. 
The first ordered phase appears in Ba$_2$NaOsO$_6$ \cite{Erickson2007} and in Ba$_2$MgReO$_6$ \cite{Marjerrison2016a}.  
In the phase, weak net magnetic moment aligns along the [110] crystallographic direction. 
Microscopic theory based on the $j_\text{eff}=3/2$ ($\Gamma_8$) spin-orbit states on metal sites revealed that the weak ferromagnetism (FM) arises as a consequence of the canted ferro- or antiferromagnetic (AFM) orderings \cite{Chen2010}.
In each $xy$-plane, the magnetic moments and $x^2-y^2$ quadruple moments align in a ferroic way, and the orderings of the neighboring $xy$-planes are reflected each other with respect to the plane spanned by $z$ and $[110]$ axes [Fig. \ref{Fig:orderings}(a)].
Besides the antiferro $x^2-y^2$ quadrupolar orderings along the $z$ axis, ferro $z^2$ quadrupolar ordering develops at 0 K. 
This phase is called FM110 or canted AFM phase. 
Within the spin-orbit theory, by raising the temperature, magnetic transition occurs, and above the temperature, only the antiferro $x^2-y^2$ quadrupolar ordering persists. 
The theoretical predictions on the orderings and phase transitions look to be consistent with the NMR  \cite{Lu2017, Liu2018} and thermodynamic \cite{Willa2019} data of Ba$_2$NaOsO$_6$.

\begin{figure}[!tbh]
 \begin{tabular}{lc}
 (a) \\
& \includegraphics[bb = 0 0 1496 1296, height=0.7\linewidth]{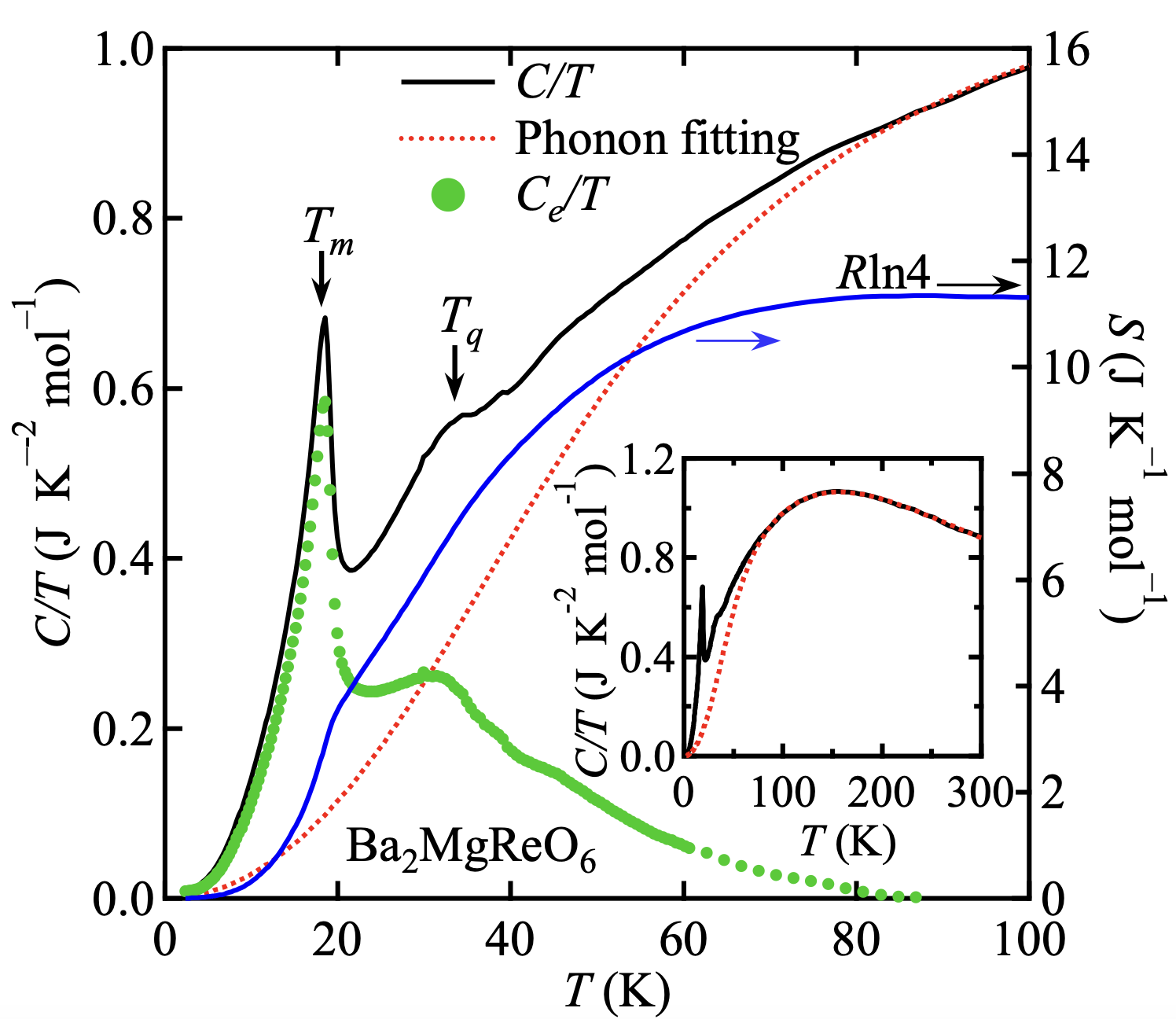} \\
 (b) \\
&  \includegraphics[bb = 0 0 1134 1244, height=0.7\linewidth]{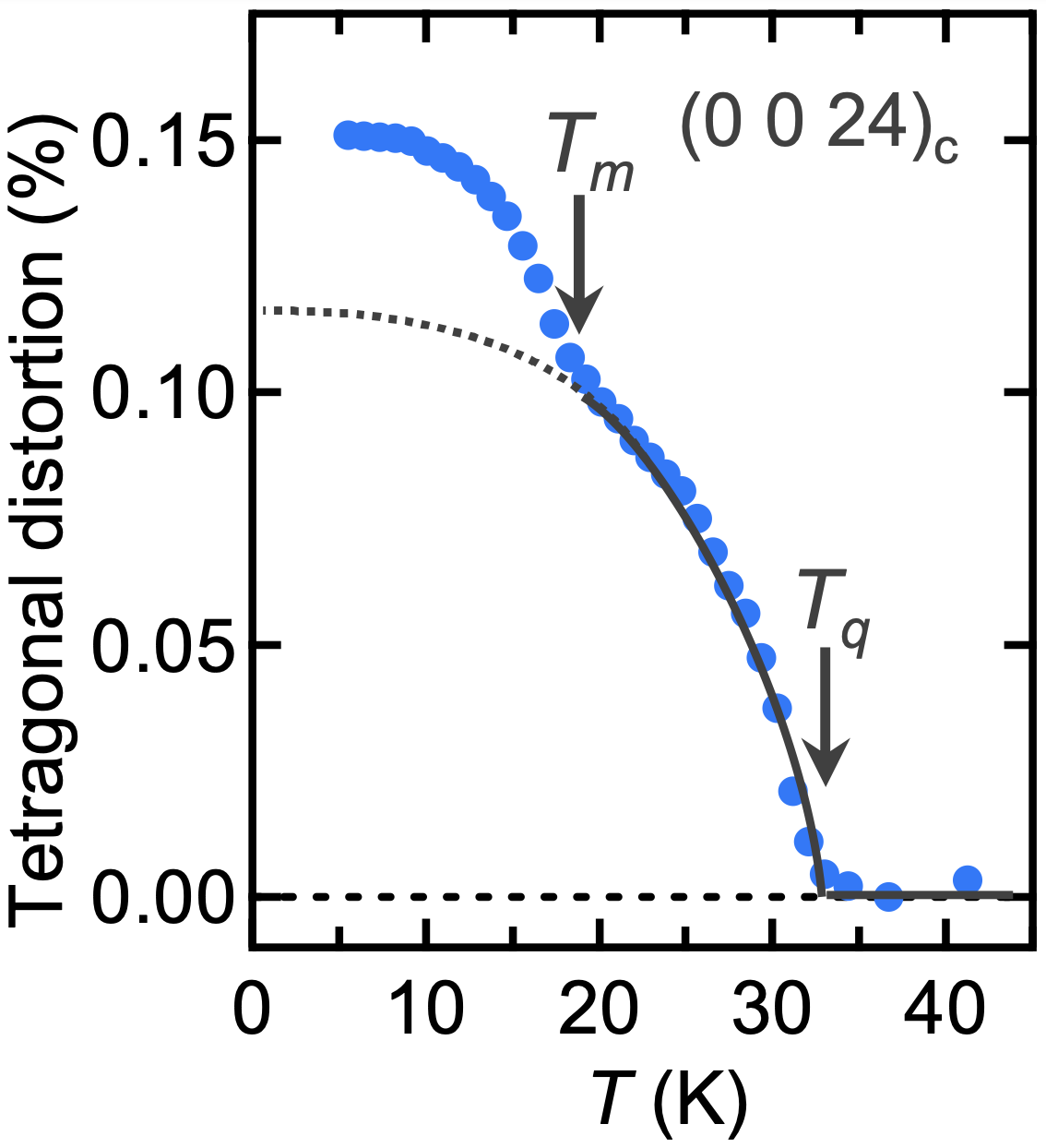}
 \end{tabular}
 \caption{
  (Color online)
   The thermodynamic properties \cite{Hirai2019} and an order parameter \cite{Hirai2020} of single crystalline Ba$_2$MgReO$_6$. 
 (a) The specific heat $C_p$ and entropy $S$. (b) The $z^2$ quadrupolar moments (the tetragonal deformations).
 The figures (a), (b) are, respectively, taken from Ref. \onlinecite{Hirai2019} (c) (2019) The Physical Society of Japan
 and Ref. \onlinecite{Hirai2020} (c) (2020) The American Physical Society.
 }
 \label{Fig:exp}
\end{figure}

The FM110 phase appears in other compounds such as Ba$_2$MgReO$_6$ at low-temperature \cite{Marjerrison2016a, Hirai2019}, while the quadrupolar ordered phase above the magnetic transition [Fig. \ref{Fig:exp}(a)] differs from the theoretical predictions based on the spin-orbit model \cite{Hirai2020}. 
High-resolution x-ray diffraction data showed that two types of quadrupole orderings coexist in the high-temperature quadrupolar phase \cite{Hirai2020} [Fig. \ref{Fig:exp}(b)]. 
Recent REXS measurements support the coexistence of the two quadrupole orderings between the magnetic and quadrupolar transitions \cite{Soh2023}.

\begin{figure}[!tbh]
\centering
 \begin{tabular}{lll}
 (a) &~& (b) \\
 \includegraphics[bb = 0 0 720 613, height=0.42\linewidth]{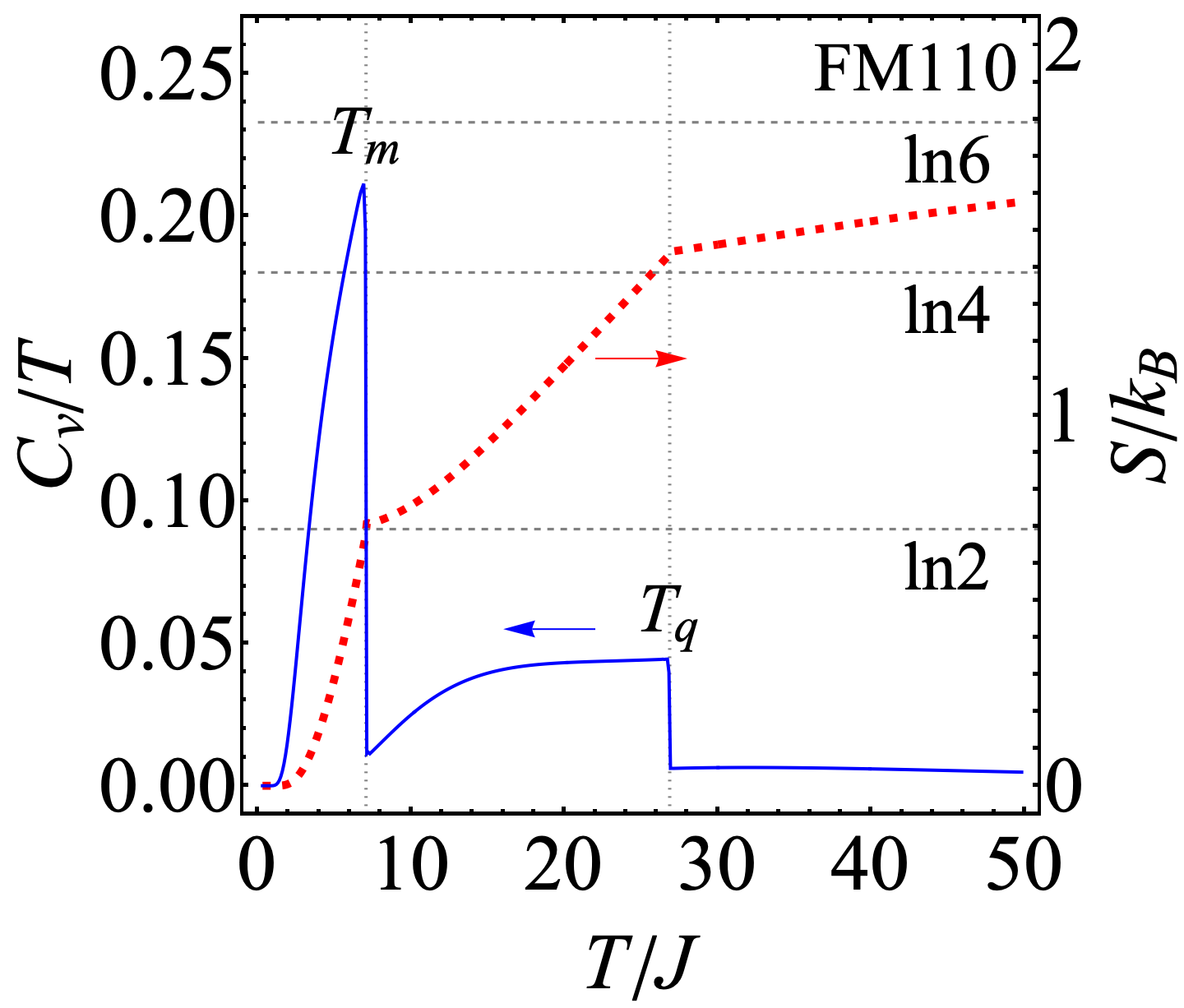}
 & &
 \includegraphics[bb = 0 0 720 659, height=0.42\linewidth]{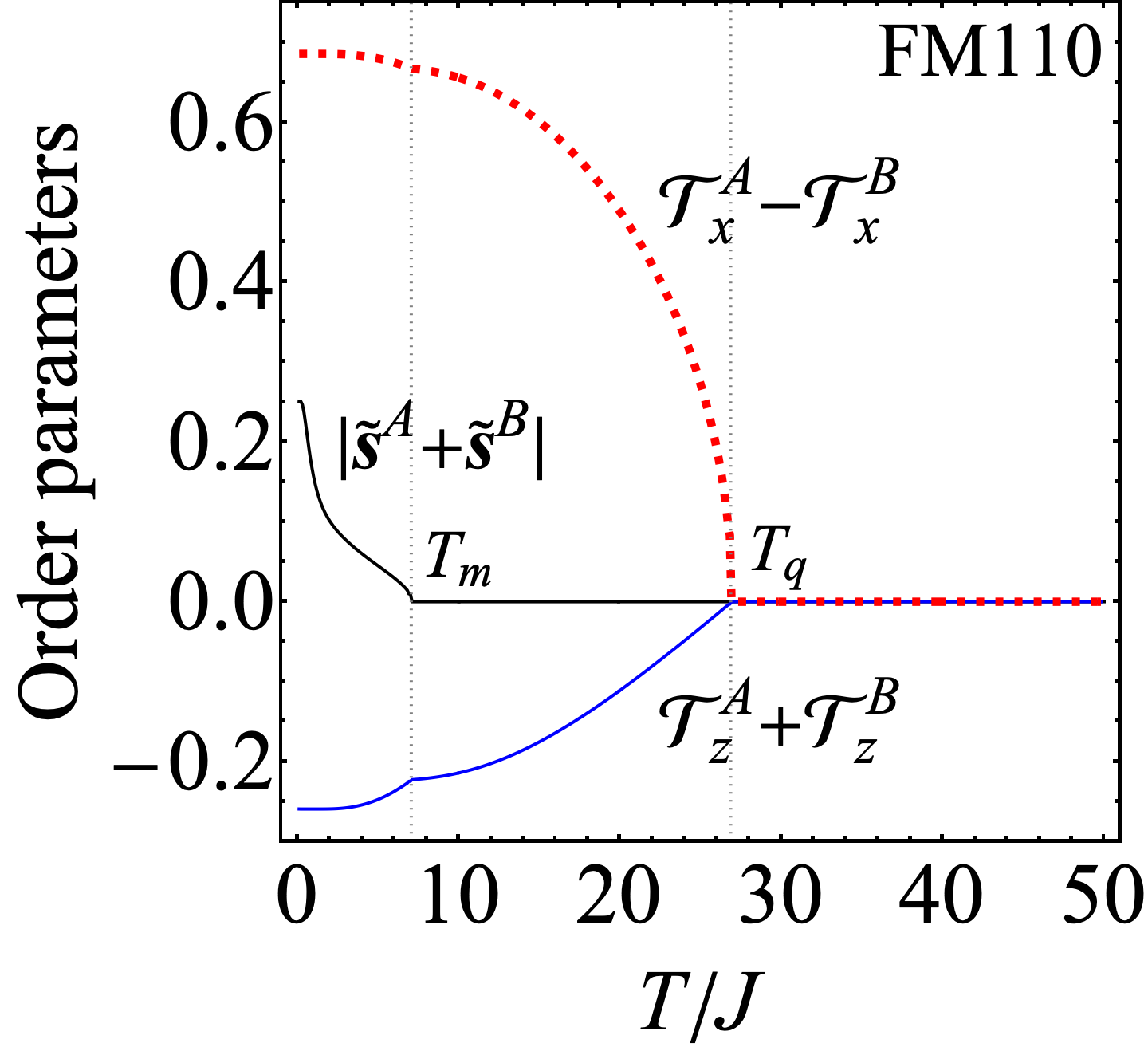}
 \\
 (c) &~& (d) \\
 \includegraphics[bb = 0 0 720 613, height=0.42\linewidth]{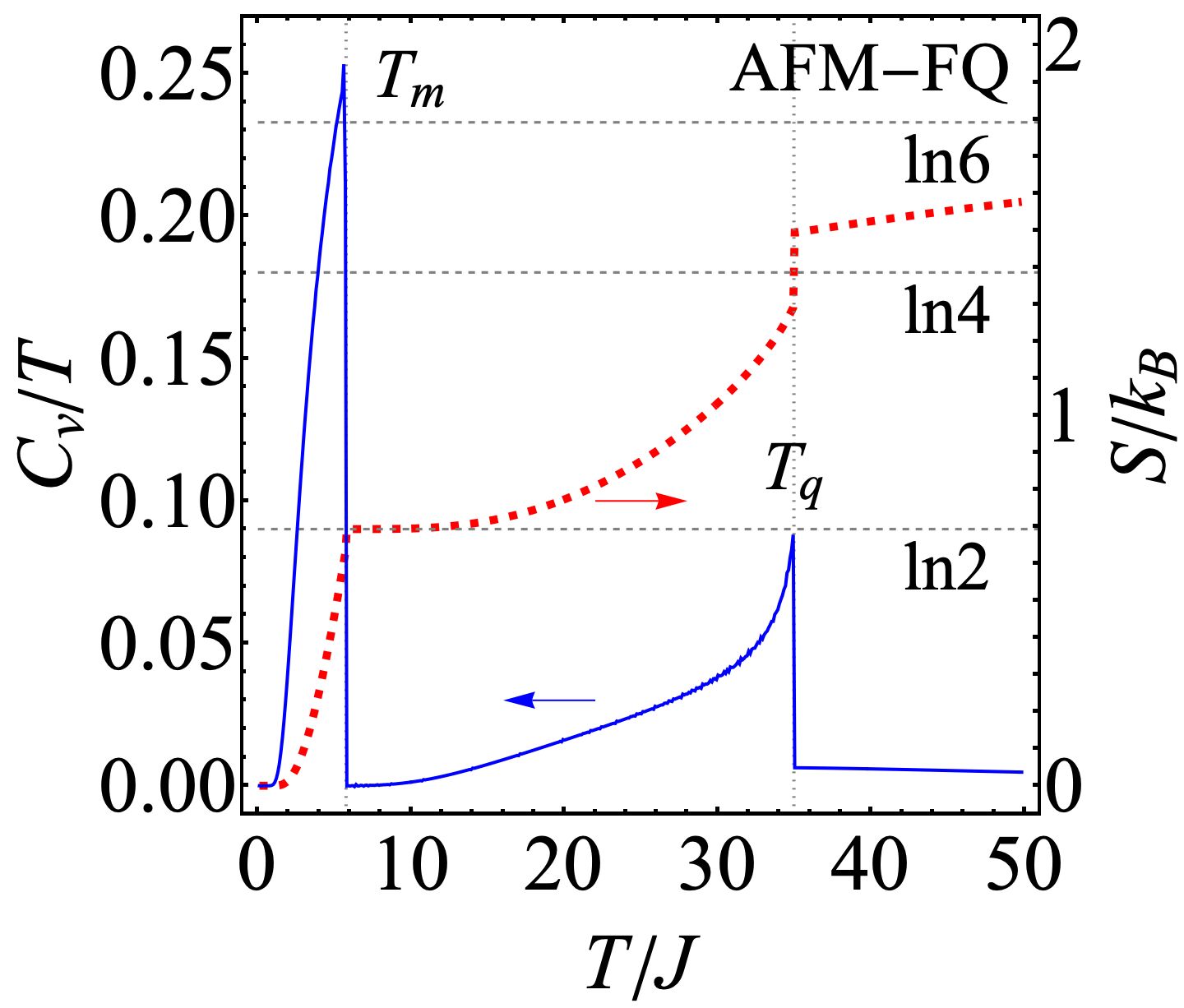}
 & &
 \includegraphics[bb = 0 0 720 716, height=0.42\linewidth]{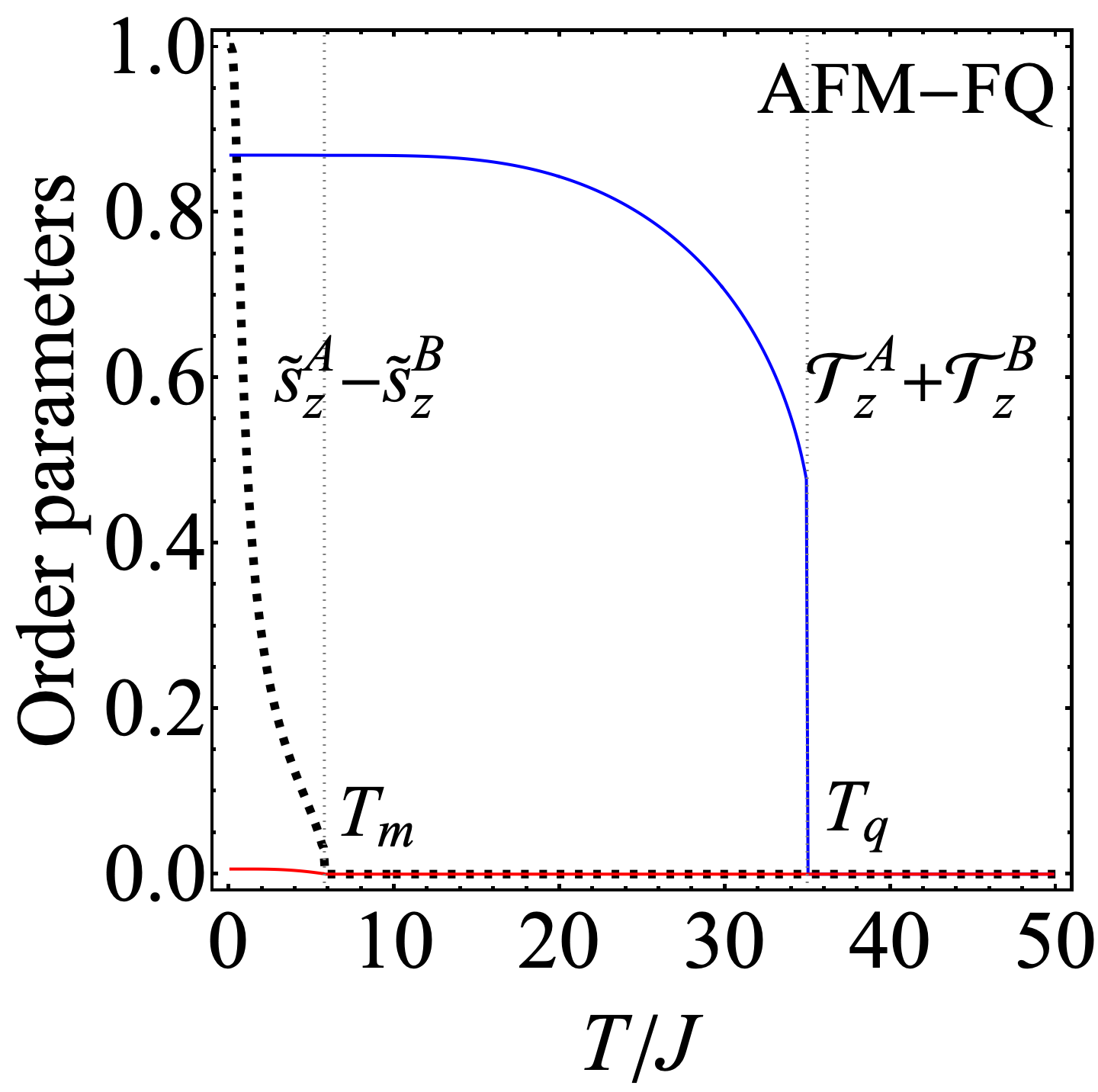}
 \end{tabular}
 \caption{
  (Color online)
  Temperature evolution of the order parameters and thermodynamic quantities in the FM110 and AFM-FQ phases.  
  The blue solid and red dotted lines in (a) and (c) indicate $C_v$ and $S$, respectively. 
  In (b) and (d), the red and blue lines indicate vibronic quadrupole moments, and the black lines pseudo-spins. 
  The FM110 and AFM-FQ phases have two sublattices, $A$ and $B$ (see Fig. \ref{Fig:orderings}). 
  $\theta = 12\pi/25$ for FM110 phase and $\pi$ for the AFM-FQ phase. 
  The figures are taken from Ref. \onlinecite{Iwahara2023} (c) The American Physical Society.
  }
 \label{Fig:OP}
\end{figure}

The second one is the AFM phase with ferro $z^2$ quadrupole (FQ) order found in Cs$_2$TaCl$_6$.  
Lowering the temperature from the room temperature, Cs$_2$TaCl$_6$ shows the FQ order with tetragonal compression along [001] axis, and the quadrupole order remains across the N\'eel transition  \cite{Ishikawa2019, Tehrani2023}. 
Such a phase, however, has not been predicted by the theories based on the spin-orbit entangled states.

To reveal the mechanism behind the ordered phases of $5d^1$ double perovskites, we calculated the mean-field phase diagram of the effective vibronic multipolar model described in Sec. \ref{Sec:multipolar} \cite{Iwahara2023}. 
The model Hamiltonian consists of the local dynamic JT Hamiltonian, the intersite spin-orbital exchange interaction (\ref{Eq:Hex}), and the intersite elastic coupling (\ref{Eq:Hvib}). 
The $T=0$ mean-field theory predicts various ordered phases, including the FM110 and AFM-FQ phases [Fig. \ref{Fig:orderings}]. 
The temperature dependence of the specific heat and magnetic entropy [Fig. \ref{Fig:OP}(a)] are in line with the experimental data [Fig. \ref{Fig:exp}(a)].
The two quadrupolar orderings in the FM110 phase remain above the magnetic transitions [Fig. \ref{Fig:OP} (b)], which is consistent with the quadrupolar ordered phase of Ba$_2$MgReO$_6$ [Fig. \ref{Fig:exp}(b)]. 
The ferro $z^2$ ordering arises due to the hybridization of the vibronic states (Fig. \ref{Fig:vibronic_wf}) by the intersite quadrupolar interaction (\ref{Eq:Hvib}).

Similarly, the vibronic theory explains the AFM-FQ phase found in Cs$_2$TaCl$_6$ \cite{Iwahara2023}. 
The simulated AFM-FQ phase determines the magnetic ordering in Cs$_2$TaCl$_6$ [Fig. \ref{Fig:orderings}(b)]. 
The temperature dependence of the thermodynamic quantities [Fig. \ref{Fig:OP}(c)] is similar to the experimental data \cite{Ishikawa2019, Tehrani2023}.
Above the N\'eel transition, the AFM-FQ phase turns into the FQ phase [Fig. \ref{Fig:OP}(d)], which is consistent with the experimental data of Cs$_2$TaCl$_6$. 
The collective phenomena of the ordering of the vibronic states enable the unified understanding of the two puzzling ordered phases in the $5d^1$ double perovskites.

The vibronic mechanism quantitatively explains the ordered phases in $5d^1$ double perovskites.
In the analysis above, we supposed that the elastic interaction is about 1 meV, supported by recent {\it ab initio} calculations \cite{Huang2024}.

However, many issues about the physical properties of the family of heavy $d^1$ double perovskites and antifluorites remain unsolved.
The nature of the orderings has not been clarified in many $5d^1$ compounds.
Ba$_2$MgReO$_6$ and Cs$_2$TaCl$_6$ show clear magnetic and quadrupolar orderings, while similar compounds do not necessarily show quadrupolar orderings. 
For example, the nature of the quadrupolar orderings of Ba$_2$ZnReO$_6$ and Ba$_2$NaOsO$_6$ 
and the nature of the AFM phase of Ba$_2$LiOsO$_6$ \cite{Stitzer2002, Steele2011, Barbosa2022} are still unclear because these compounds show no explicit quadrupolar orderings, unlike Ba$_2$MgReO$_6$ and Cs$_2$TaCl$_6$.
Recent NMR data of Ba$_2$LiOsO$_6$ show an anomaly in the $T_2$ relaxation time above the N\'eel temperature \cite{Cong2022}. 
Although the anomaly could be related to the quadrupolar transition, the origin is unanswered.

A complete understanding of the nature of the ordered phases will lead to resolve the mystery of small magnetic entropy of Ba$_2$NaOsO$_6$ and the debate on the magnetic entropy of Ba$_2$MgReO$_6$. 
Although the ground states are four-fold degenerated in the $5d^1$ double perovskites, the magnetic entropy per site reaches only half of $k_B \ln 4$ per site even above the quadrupolar transition in Ba$_2$NaOsO$_6$ \cite{Erickson2007}.
The situation of the magnetic entropy of Ba$_2$MgReO$_6$ is controversial:
The experimentally estimated magnetic entropy is either $k_B \ln 4$ \cite{Hirai2019} [Fig. \ref{Fig:exp}(a)] or $k_B \ln 2$ \cite{Marjerrison2016a, Pasztorova2023b, Zivkovic2024}, and our mean-field theory predicts that the magnetic entropy reaches $k_B \ln 4$ at $T_q$ [Fig. \ref{Fig:OP} (a)].

Exploration of the relation between the dynamic JT effect and the physical phenomena awaits in various situations. 
Non-JT deformations of octahedra occur in some $5d^1$ double perovskites, such as Ba$_2$CaReO$_6$ \cite{Yamaura2006, Ishikawa2021b}, Ba$_2$CdReO$_6$ \cite{Hirai2021, Barbosa2022}, Sr$_2$MgReO$_6$ \cite{Bramnik2003, Gao2020}, Ca$_2$MgReO$_6$ \cite{Bramnik2003}, and various Ta hexahalides \cite{Ishikawa2019, Ishikawa2021}. 
The dynamic JT effect could persist and modulate the ordered phases in such noncubic compounds.  
The RIXS data of Sr$_2$MgReO$_6$, Ca$_2$MgReO$_6$, and Ba$_2$CaReO$_6$ suggest the persistence of the vibronic effect \cite{Frontini2024, Iwahara2024} [Fig. \ref{Fig:Re}(a)].
The dynamic JT effect could emerge and influence the disordered phases in cubic $4d^1$ Ba$_2$YMoO$_6$ and Ba$_2$LuMoO$_6$ \cite{Cussen2006, Aharen2010a, deVries2010, Carlo2011, deVries2013, Coomer2013, Mustonen2022}, and Ba$_2$YWO$_6$ \cite{Lee2021}. 
The evolution of the magneto-vibronic orders under external pressure \cite{Arima2022, Mosca2024}, magnetic field \cite{Ishikawa2021b}, and by electron doping \cite{Kesavan2020, Cong2023, Voleti2023, Celiberti2024} should be investigated further.

Another issue is the relation between theoretical results based on the static JT deformation and those considering the dynamic JT effect.
The RIXS measurements and theoretical analyses established the presence of the dynamic JT effect in the family of cubic $5d^1$ double perovskites\cite{Frontini2024, Iwahara2024}, and we reproduce all the reported ordered phases by fully considering the dynamic JT effect \cite{Iwahara2023}, suggesting the presence of the dynamic JT effect in the family of $5d^1$ double perovskites. 
Recent {\it ab initio} study perturbatively included the vibronic effect and predicted pressure induced phases \cite{Mosca2024}.
At the same time, various theories based on the static JT distortions have reproduced some features of magnetic-quadrupole ordered phases \cite{Chen2010, Ishizuka2014, Cong2020, Mosca2021, Weng2021, Tehrani2021, Merkel2023}.
The relation between the dynamic and static JT-based approaches is an open problem.

The nature of the ordered phases of the family of $5d^2$ double perovskites \cite{Aharen2010b, Thompson2014, Marjerrison2016b, Maharaj2020, Nilsen2021, Cong2023, Morgan2023} is under debate. 
For example, in the $5d^2$ osmates, the orderings could be of octupolar \cite{Maharaj2020, Paramekanti2020} or quadrupolar \cite{Khaliullin2021} type. 
The origin of the difference between the nonmagnetic Os and magnetic Re compounds is also unclear. 
The majority of the theoretical analyses has been based on the pure electronic model \cite{Chen2011, Maharaj2020, Paramekanti2020} or with static JT deformations \cite{Pourovskii2021}, while the dynamic JT effect can develop on $d^2$ metal sites as discussed in Sec. \ref{Sec:RIXS} and Ref. \onlinecite{Khaliullin2021}. 
Although the former treatment assumes weak vibronic coupling, the coupling could be far from the weak coupling regime as in the $5d^1$ compounds. 
To fully reveal the nature of the ordered phases in the $5d^2$ double perovskites, quantitative determination of all the interaction parameters and their concomitant treatment will be indispensable.

\section{Conclusion}
In transition metal compounds with $4d$ or $5d$ transition metal ions on the fcc lattice, the dynamic Jahn-Teller effect can develop and give nontrivial influence on the quantum phenomena. 
Recent RIXS measurements and theoretical analysis have revealed the presence of the dynamic JT effect on metal sites in various cubic $d^1$-$d^5$ double perovskites. 
Besides, the cooperative dynamic JT effect explains several multipolar ordered phases in $5d^1$ double perovskites that the conventional electronic theories did not.
These studies on the cooperative dynamic JT effect in spin-orbit Mott insulators suggest the emergence of rich quantum phenomena driven by orbital-lattice entanglement in a wide range of correlated materials.

\appendix

\section{Vibronic coupling operator}
\label{A:vibronic}
Here, we give a general formalism of the vibronic coupling \cite{Englman1972, Azumi1977, Bersuker1989, Sato2009, Grosso2014}. 
In the description below, we employ the crude adiabatic approximation among various adiabatic approximations \cite{Azumi1977, Sato2009}. 
Within the approximation, we set a reference structure (octahedral structure in this work) and use the energy eigenstates of the electronic Hamiltonian for the reference structure as the electronic basis. 

As mentioned in Sec. \ref{Sec:vibronic}, the vibronic coupling indicates the dependence of the Coulomb interactions (nuclei-electron and nuclei-nuclei) on the nuclear deformations. 
Writing the Coulomb interactions for the reference structure and arbitrary deformed structure as $\hat{U}(0)$ and $\hat{U}(\bm{q})$, respectively, their difference $\Delta \hat{U}(\bm{q}) = \hat{U}(\bm{q}) - \hat{U}(0)$ contain all the vibronic couplings. 
We define the linear and nonlinear vibronic coupling and elastic potential by carrying out the Taylor expansion of $\Delta \hat{U}(\bm{q})$ around the reference structure $\bm{q}=\bm{0}$ (Herzberg-Teller expansion), 
\begin{align}
 \Delta \hat{U}(\bm{q}) &= \sum_\alpha \frac{\partial \hat{U}(0)}{\partial q_\alpha} q_\alpha + \sum_{\alpha \alpha'} \frac{1}{2!} \frac{\partial^2 \hat{U}(0)}{\partial q_{\alpha} \partial q_{\alpha'}} q_\alpha q_{\alpha'} + \cdots,
 \label{Eq:U}
\end{align}
where $\bm{q}$ stand for all the mass-weighted normal coordinates \cite{Inui1990, Wilson1980}.
The first-order term of $q$'s is the linear vibronic coupling.
By projecting the linear vibronic coupling into the $t_{2g}$ orbital states at the reference structure, and employing the Wigner-Eckart theorem \cite{Inui1990, Koster1963, Sugano1970, Bersuker1989}, we obtain Eq. (\ref{Eq:VJT}).
With the selection rule for the matrix elements of the vibronic couplings, $\langle t_{2g}\gamma_1|\partial U(0)/\partial q_{\Gamma\gamma}|t_{2g}\gamma_2 \rangle$, we can identify the vibronically active modes \cite{Jahn1937, Jahn1938}: $[t_{2g}^2] = A_g \oplus E_g \oplus T_{2g}$, and hence, $A_g$, $E_g$, and $T_{2g}$ modes are vibronic active modes [Fig. \ref{Fig:mode}].
The second-order term of $q$'s in Eq. (\ref{Eq:U}) contains the quadratic vibronic and elastic couplings. 
The quadratic vibronic coupling is traceless over the electronic states, and the elastic coupling is the remaining part in the second-order term. 

Let us summarize the dimensions of the quantities in the vibronic coupling within the atomic units. 
The dimension of the mass-weighted normal coordinate $q$ is $\sqrt{m_e}a_0$, where $m_e$ is the mass of an electron, and $a_0$ is the Bohr radius. 
The dimensions of the linear vibronic coupling, 
$v=\langle \partial U(0)/\partial q \rangle$, 
and the dynamical matrix, $\omega^2 = \langle \partial^2 U(0)/\partial q^2 \rangle$, are, respectively, $E_h/(\sqrt{m_e} a_0)$ and $E_h/(m_e a_0^2)$, where $E_h$ is the Hartree energy.  
The dimension of the quadratic vibronic coupling corresponds to that of the dynamical matrix.  

We used the dimensionless quantities in the main text. 
Using $q = \sqrt{\hslash/(2\omega)}(b^\dagger + b)$ and $p = i\sqrt{\hslash \omega/2} (b^\dagger - b)$, we introduce the dimensionless normal coordinates and the conjugate momenta as $\bar{q} = (1/\sqrt{2})(b^\dagger + b)$ and $\bar{p} = (i/\sqrt{2})(b^\dagger - b)$, where $b^\dagger$ and $b$ are the creation and annihilation operators of the lattice vibrations. 
Similarly, we define the dimensionless linear vibronic coupling constant by $g = v/\sqrt{\hslash \omega^3}$ with which $vq = \hslash \omega g \bar{q}$.

\begin{acknowledgments}
I thank L. F. Chibotaru, V. Vieru, Z. Huang, D. Hirai, F. I. Frontini, Y.-J. Kim, J.-R. Soh, H. Suzuki, S. Agrestini, D. Fiore Mosca, L. V. Pourovskii, R. Cong, T. Muroi, A. Fujimori, J. Okamoto, J. Kim, and I. \v{Z}ivkovi\'c for fruitful discussions and the referee for valuable comments.
This work was partly supported by the Iketani Science and Technology Foundation, Grant-in-Aid for Scientific Research (Grant No. 22K03507) from the Japan Society for the Promotion of Science, and Chiba University Open Recruitment for International Exchange Program.
\end{acknowledgments}

\bibliography{ref}

\begin{thebibliography}{200}%
\makeatletter
\providecommand \@ifxundefined [1]{%
 \@ifx{#1\undefined}
}%
\providecommand \@ifnum [1]{%
 \ifnum #1\expandafter \@firstoftwo
 \else \expandafter \@secondoftwo
 \fi
}%
\providecommand \@ifx [1]{%
 \ifx #1\expandafter \@firstoftwo
 \else \expandafter \@secondoftwo
 \fi
}%
\providecommand \natexlab [1]{#1}%
\providecommand \enquote  [1]{``#1''}%
\providecommand \bibnamefont  [1]{#1}%
\providecommand \bibfnamefont [1]{#1}%
\providecommand \citenamefont [1]{#1}%
\providecommand \href@noop [0]{\@secondoftwo}%
\providecommand \href [0]{\begingroup \@sanitize@url \@href}%
\providecommand \@href[1]{\@@startlink{#1}\@@href}%
\providecommand \@@href[1]{\endgroup#1\@@endlink}%
\providecommand \@sanitize@url [0]{\catcode `\\12\catcode `\$12\catcode
  `\&12\catcode `\#12\catcode `\^12\catcode `\_12\catcode `\%12\relax}%
\providecommand \@@startlink[1]{}%
\providecommand \@@endlink[0]{}%
\providecommand \url  [0]{\begingroup\@sanitize@url \@url }%
\providecommand \@url [1]{\endgroup\@href {#1}{\urlprefix }}%
\providecommand \urlprefix  [0]{URL }%
\providecommand \Eprint [0]{\href }%
\providecommand \doibase [0]{http://dx.doi.org/}%
\providecommand \selectlanguage [0]{\@gobble}%
\providecommand \bibinfo  [0]{\@secondoftwo}%
\providecommand \bibfield  [0]{\@secondoftwo}%
\providecommand \translation [1]{[#1]}%
\providecommand \BibitemOpen [0]{}%
\providecommand \bibitemStop [0]{}%
\providecommand \bibitemNoStop [0]{.\EOS\space}%
\providecommand \EOS [0]{\spacefactor3000\relax}%
\providecommand \BibitemShut  [1]{\csname bibitem#1\endcsname}%
\let\auto@bib@innerbib\@empty
\bibitem [{\citenamefont {Witczak-Krempa}\ \emph {et~al.}(2014)\citenamefont
  {Witczak-Krempa}, \citenamefont {Chen}, \citenamefont {Kim},\ and\
  \citenamefont {Balents}}]{Witczak-Krempa2014}%
  \BibitemOpen
  \bibfield  {author} {\bibinfo {author} {\bibfnamefont {W.}~\bibnamefont
  {Witczak-Krempa}}, \bibinfo {author} {\bibfnamefont {G.}~\bibnamefont
  {Chen}}, \bibinfo {author} {\bibfnamefont {Y.~B.}\ \bibnamefont {Kim}}, \
  and\ \bibinfo {author} {\bibfnamefont {L.}~\bibnamefont {Balents}},\ }\href
  {\doibase 10.1146/annurev-conmatphys-020911-125138} {\bibfield  {journal}
  {\bibinfo  {journal} {Annual Review of Condensed Matter Physics}\ }\textbf
  {\bibinfo {volume} {5}},\ \bibinfo {pages} {57} (\bibinfo {year}
  {2014})}\BibitemShut {NoStop}%
\bibitem [{\citenamefont {Rau}\ \emph {et~al.}(2016)\citenamefont {Rau},
  \citenamefont {Lee},\ and\ \citenamefont {Kee}}]{Rau2016}%
  \BibitemOpen
  \bibfield  {author} {\bibinfo {author} {\bibfnamefont {J.~G.}\ \bibnamefont
  {Rau}}, \bibinfo {author} {\bibfnamefont {E.~K.-H.}\ \bibnamefont {Lee}}, \
  and\ \bibinfo {author} {\bibfnamefont {H.-Y.}\ \bibnamefont {Kee}},\ }\href
  {\doibase 10.1146/annurev-conmatphys-031115-011319} {\bibfield  {journal}
  {\bibinfo  {journal} {Annu. Rev. Condens. Matter Phys.}\ }\textbf {\bibinfo
  {volume} {7}},\ \bibinfo {pages} {195} (\bibinfo {year} {2016})}\BibitemShut
  {NoStop}%
\bibitem [{\citenamefont {Takagi}\ \emph {et~al.}(2019)\citenamefont {Takagi},
  \citenamefont {Takayama}, \citenamefont {Jackeli}, \citenamefont
  {Khaliullin},\ and\ \citenamefont {Nagler}}]{Takagi2019}%
  \BibitemOpen
  \bibfield  {author} {\bibinfo {author} {\bibfnamefont {H.}~\bibnamefont
  {Takagi}}, \bibinfo {author} {\bibfnamefont {T.}~\bibnamefont {Takayama}},
  \bibinfo {author} {\bibfnamefont {G.}~\bibnamefont {Jackeli}}, \bibinfo
  {author} {\bibfnamefont {G.}~\bibnamefont {Khaliullin}}, \ and\ \bibinfo
  {author} {\bibfnamefont {S.~E.}\ \bibnamefont {Nagler}},\ }\href {\doibase
  10.1038/s42254-019-0038-2} {\bibfield  {journal} {\bibinfo  {journal} {Nat.
  Rev. Phys.}\ }\textbf {\bibinfo {volume} {1}},\ \bibinfo {pages} {264}
  (\bibinfo {year} {2019})}\BibitemShut {NoStop}%
\bibitem [{\citenamefont {Motome}\ and\ \citenamefont
  {Nasu}(2020)}]{Motome2020}%
  \BibitemOpen
  \bibfield  {author} {\bibinfo {author} {\bibfnamefont {Y.}~\bibnamefont
  {Motome}}\ and\ \bibinfo {author} {\bibfnamefont {J.}~\bibnamefont {Nasu}},\
  }\href {\doibase 10.7566/JPSJ.89.012002} {\bibfield  {journal} {\bibinfo
  {journal} {Journal of the Physical Society of Japan}\ }\textbf {\bibinfo
  {volume} {89}},\ \bibinfo {pages} {012002} (\bibinfo {year}
  {2020})}\BibitemShut {NoStop}%
\bibitem [{\citenamefont {Takayama}\ \emph {et~al.}(2021)\citenamefont
  {Takayama}, \citenamefont {Chaloupka}, \citenamefont {Smerald}, \citenamefont
  {Khaliullin},\ and\ \citenamefont {Takagi}}]{Takayama2021}%
  \BibitemOpen
  \bibfield  {author} {\bibinfo {author} {\bibfnamefont {T.}~\bibnamefont
  {Takayama}}, \bibinfo {author} {\bibfnamefont {J.}~\bibnamefont {Chaloupka}},
  \bibinfo {author} {\bibfnamefont {A.}~\bibnamefont {Smerald}}, \bibinfo
  {author} {\bibfnamefont {G.}~\bibnamefont {Khaliullin}}, \ and\ \bibinfo
  {author} {\bibfnamefont {H.}~\bibnamefont {Takagi}},\ }\href {\doibase
  10.7566/JPSJ.90.062001} {\bibfield  {journal} {\bibinfo  {journal} {Journal
  of the Physical Society of Japan}\ }\textbf {\bibinfo {volume} {90}},\
  \bibinfo {pages} {062001} (\bibinfo {year} {2021})}\BibitemShut {NoStop}%
\bibitem [{\citenamefont {Trebst}\ and\ \citenamefont
  {Hickey}(2022)}]{Trebst2022}%
  \BibitemOpen
  \bibfield  {author} {\bibinfo {author} {\bibfnamefont {S.}~\bibnamefont
  {Trebst}}\ and\ \bibinfo {author} {\bibfnamefont {C.}~\bibnamefont
  {Hickey}},\ }\href {\doibase https://doi.org/10.1016/j.physrep.2021.11.003}
  {\bibfield  {journal} {\bibinfo  {journal} {Phys. Rep.}\ }\textbf {\bibinfo
  {volume} {950}},\ \bibinfo {pages} {1} (\bibinfo {year} {2022})}\BibitemShut
  {NoStop}%
\bibitem [{\citenamefont {Chen}\ and\ \citenamefont {Wu}(2024)}]{Chen2024}%
  \BibitemOpen
  \bibfield  {author} {\bibinfo {author} {\bibfnamefont {G.~V.}\ \bibnamefont
  {Chen}}\ and\ \bibinfo {author} {\bibfnamefont {C.}~\bibnamefont {Wu}},\
  }\href {\doibase 10.1038/s41535-023-00614-2} {\bibfield  {journal} {\bibinfo
  {journal} {npj Quantum Mater.}\ }\textbf {\bibinfo {volume} {9}},\ \bibinfo
  {pages} {1} (\bibinfo {year} {2024})}\BibitemShut {NoStop}%
\bibitem [{\citenamefont {Jackeli}\ and\ \citenamefont
  {Khaliullin}(2009)}]{Jackeli2009}%
  \BibitemOpen
  \bibfield  {author} {\bibinfo {author} {\bibfnamefont {G.}~\bibnamefont
  {Jackeli}}\ and\ \bibinfo {author} {\bibfnamefont {G.}~\bibnamefont
  {Khaliullin}},\ }\href {\doibase 10.1103/PhysRevLett.102.017205} {\bibfield
  {journal} {\bibinfo  {journal} {Phys. Rev. Lett.}\ }\textbf {\bibinfo
  {volume} {102}},\ \bibinfo {pages} {017205} (\bibinfo {year}
  {2009})}\BibitemShut {NoStop}%
\bibitem [{\citenamefont {Chen}\ \emph {et~al.}(2010)\citenamefont {Chen},
  \citenamefont {Pereira},\ and\ \citenamefont {Balents}}]{Chen2010}%
  \BibitemOpen
  \bibfield  {author} {\bibinfo {author} {\bibfnamefont {G.}~\bibnamefont
  {Chen}}, \bibinfo {author} {\bibfnamefont {R.}~\bibnamefont {Pereira}}, \
  and\ \bibinfo {author} {\bibfnamefont {L.}~\bibnamefont {Balents}},\ }\href
  {\doibase 10.1103/PhysRevB.82.174440} {\bibfield  {journal} {\bibinfo
  {journal} {Phys. Rev. B}\ }\textbf {\bibinfo {volume} {82}},\ \bibinfo
  {pages} {174440} (\bibinfo {year} {2010})}\BibitemShut {NoStop}%
\bibitem [{\citenamefont {Chen}\ and\ \citenamefont
  {Balents}(2011)}]{Chen2011}%
  \BibitemOpen
  \bibfield  {author} {\bibinfo {author} {\bibfnamefont {G.}~\bibnamefont
  {Chen}}\ and\ \bibinfo {author} {\bibfnamefont {L.}~\bibnamefont {Balents}},\
  }\href {\doibase 10.1103/PhysRevB.84.094420} {\bibfield  {journal} {\bibinfo
  {journal} {Phys. Rev. B}\ }\textbf {\bibinfo {volume} {84}},\ \bibinfo
  {pages} {094420} (\bibinfo {year} {2011})}\BibitemShut {NoStop}%
\bibitem [{\citenamefont {Khaliullin}(2013)}]{Khaliullin2013}%
  \BibitemOpen
  \bibfield  {author} {\bibinfo {author} {\bibfnamefont {G.}~\bibnamefont
  {Khaliullin}},\ }\href {\doibase 10.1103/PhysRevLett.111.197201} {\bibfield
  {journal} {\bibinfo  {journal} {Phys. Rev. Lett.}\ }\textbf {\bibinfo
  {volume} {111}},\ \bibinfo {pages} {197201} (\bibinfo {year}
  {2013})}\BibitemShut {NoStop}%
\bibitem [{\citenamefont {Ishizuka}\ and\ \citenamefont
  {Balents}(2014)}]{Ishizuka2014}%
  \BibitemOpen
  \bibfield  {author} {\bibinfo {author} {\bibfnamefont {H.}~\bibnamefont
  {Ishizuka}}\ and\ \bibinfo {author} {\bibfnamefont {L.}~\bibnamefont
  {Balents}},\ }\href {\doibase 10.1103/PhysRevB.90.184422} {\bibfield
  {journal} {\bibinfo  {journal} {Phys. Rev. B}\ }\textbf {\bibinfo {volume}
  {90}},\ \bibinfo {pages} {184422} (\bibinfo {year} {2014})}\BibitemShut
  {NoStop}%
\bibitem [{\citenamefont {Nasu}\ \emph {et~al.}(2014)\citenamefont {Nasu},
  \citenamefont {Udagawa},\ and\ \citenamefont {Motome}}]{Nasu2014}%
  \BibitemOpen
  \bibfield  {author} {\bibinfo {author} {\bibfnamefont {J.}~\bibnamefont
  {Nasu}}, \bibinfo {author} {\bibfnamefont {M.}~\bibnamefont {Udagawa}}, \
  and\ \bibinfo {author} {\bibfnamefont {Y.}~\bibnamefont {Motome}},\ }\href
  {\doibase 10.1103/PhysRevLett.113.197205} {\bibfield  {journal} {\bibinfo
  {journal} {Phys. Rev. Lett.}\ }\textbf {\bibinfo {volume} {113}},\ \bibinfo
  {pages} {197205} (\bibinfo {year} {2014})}\BibitemShut {NoStop}%
\bibitem [{\citenamefont {Natori}\ \emph {et~al.}(2016)\citenamefont {Natori},
  \citenamefont {Andrade}, \citenamefont {Miranda},\ and\ \citenamefont
  {Pereira}}]{Natori2016}%
  \BibitemOpen
  \bibfield  {author} {\bibinfo {author} {\bibfnamefont {W.~M.~H.}\
  \bibnamefont {Natori}}, \bibinfo {author} {\bibfnamefont {E.~C.}\
  \bibnamefont {Andrade}}, \bibinfo {author} {\bibfnamefont {E.}~\bibnamefont
  {Miranda}}, \ and\ \bibinfo {author} {\bibfnamefont {R.~G.}\ \bibnamefont
  {Pereira}},\ }\href {\doibase 10.1103/PhysRevLett.117.017204} {\bibfield
  {journal} {\bibinfo  {journal} {Phys. Rev. Lett.}\ }\textbf {\bibinfo
  {volume} {117}},\ \bibinfo {pages} {017204} (\bibinfo {year}
  {2016})}\BibitemShut {NoStop}%
\bibitem [{\citenamefont {Romh\'anyi}\ \emph {et~al.}(2017)\citenamefont
  {Romh\'anyi}, \citenamefont {Balents},\ and\ \citenamefont
  {Jackeli}}]{Romhanyi2017}%
  \BibitemOpen
  \bibfield  {author} {\bibinfo {author} {\bibfnamefont {J.}~\bibnamefont
  {Romh\'anyi}}, \bibinfo {author} {\bibfnamefont {L.}~\bibnamefont {Balents}},
  \ and\ \bibinfo {author} {\bibfnamefont {G.}~\bibnamefont {Jackeli}},\ }\href
  {\doibase 10.1103/PhysRevLett.118.217202} {\bibfield  {journal} {\bibinfo
  {journal} {Phys. Rev. Lett.}\ }\textbf {\bibinfo {volume} {118}},\ \bibinfo
  {pages} {217202} (\bibinfo {year} {2017})}\BibitemShut {NoStop}%
\bibitem [{\citenamefont {Yamada}\ \emph {et~al.}(2018)\citenamefont {Yamada},
  \citenamefont {Oshikawa},\ and\ \citenamefont {Jackeli}}]{Yamada2018}%
  \BibitemOpen
  \bibfield  {author} {\bibinfo {author} {\bibfnamefont {M.~G.}\ \bibnamefont
  {Yamada}}, \bibinfo {author} {\bibfnamefont {M.}~\bibnamefont {Oshikawa}}, \
  and\ \bibinfo {author} {\bibfnamefont {G.}~\bibnamefont {Jackeli}},\ }\href
  {\doibase 10.1103/PhysRevLett.121.097201} {\bibfield  {journal} {\bibinfo
  {journal} {Phys. Rev. Lett.}\ }\textbf {\bibinfo {volume} {121}},\ \bibinfo
  {pages} {097201} (\bibinfo {year} {2018})}\BibitemShut {NoStop}%
\bibitem [{\citenamefont {Liu}\ and\ \citenamefont
  {Khaliullin}(2019)}]{Liu2019}%
  \BibitemOpen
  \bibfield  {author} {\bibinfo {author} {\bibfnamefont {H.}~\bibnamefont
  {Liu}}\ and\ \bibinfo {author} {\bibfnamefont {G.}~\bibnamefont
  {Khaliullin}},\ }\href {\doibase 10.1103/PhysRevLett.122.057203} {\bibfield
  {journal} {\bibinfo  {journal} {Phys. Rev. Lett.}\ }\textbf {\bibinfo
  {volume} {122}},\ \bibinfo {pages} {057203} (\bibinfo {year}
  {2019})}\BibitemShut {NoStop}%
\bibitem [{\citenamefont {Svoboda}\ \emph {et~al.}(2021)\citenamefont
  {Svoboda}, \citenamefont {Zhang}, \citenamefont {Randeria},\ and\
  \citenamefont {Trivedi}}]{Svoboda2021}%
  \BibitemOpen
  \bibfield  {author} {\bibinfo {author} {\bibfnamefont {C.}~\bibnamefont
  {Svoboda}}, \bibinfo {author} {\bibfnamefont {W.}~\bibnamefont {Zhang}},
  \bibinfo {author} {\bibfnamefont {M.}~\bibnamefont {Randeria}}, \ and\
  \bibinfo {author} {\bibfnamefont {N.}~\bibnamefont {Trivedi}},\ }\href
  {\doibase 10.1103/PhysRevB.104.024437} {\bibfield  {journal} {\bibinfo
  {journal} {Phys. Rev. B}\ }\textbf {\bibinfo {volume} {104}},\ \bibinfo
  {pages} {024437} (\bibinfo {year} {2021})}\BibitemShut {NoStop}%
\bibitem [{\citenamefont {de~Carvalho}\ \emph {et~al.}(2023)\citenamefont
  {de~Carvalho}, \citenamefont {Freire},\ and\ \citenamefont
  {Pereira}}]{deCarvalho2023}%
  \BibitemOpen
  \bibfield  {author} {\bibinfo {author} {\bibfnamefont {V.~S.}\ \bibnamefont
  {de~Carvalho}}, \bibinfo {author} {\bibfnamefont {H.}~\bibnamefont {Freire}},
  \ and\ \bibinfo {author} {\bibfnamefont {R.~G.}\ \bibnamefont {Pereira}},\
  }\href {\doibase 10.1103/PhysRevB.108.094418} {\bibfield  {journal} {\bibinfo
   {journal} {Phys. Rev. B}\ }\textbf {\bibinfo {volume} {108}},\ \bibinfo
  {pages} {094418} (\bibinfo {year} {2023})}\BibitemShut {NoStop}%
\bibitem [{\citenamefont {Kubo}\ \emph {et~al.}(2023)\citenamefont {Kubo},
  \citenamefont {Ishitobi},\ and\ \citenamefont {Hattori}}]{Kubo2023}%
  \BibitemOpen
  \bibfield  {author} {\bibinfo {author} {\bibfnamefont {H.}~\bibnamefont
  {Kubo}}, \bibinfo {author} {\bibfnamefont {T.}~\bibnamefont {Ishitobi}}, \
  and\ \bibinfo {author} {\bibfnamefont {K.}~\bibnamefont {Hattori}},\ }\href
  {\doibase 10.1103/PhysRevB.107.235134} {\bibfield  {journal} {\bibinfo
  {journal} {Phys. Rev. B}\ }\textbf {\bibinfo {volume} {107}},\ \bibinfo
  {pages} {235134} (\bibinfo {year} {2023})}\BibitemShut {NoStop}%
\bibitem [{\citenamefont {Pourovskii}(2023)}]{Pourovskii2023}%
  \BibitemOpen
  \bibfield  {author} {\bibinfo {author} {\bibfnamefont {L.~V.}\ \bibnamefont
  {Pourovskii}},\ }\href {\doibase 10.1103/PhysRevB.108.054436} {\bibfield
  {journal} {\bibinfo  {journal} {Phys. Rev. B}\ }\textbf {\bibinfo {volume}
  {108}},\ \bibinfo {pages} {054436} (\bibinfo {year} {2023})}\BibitemShut
  {NoStop}%
\bibitem [{\citenamefont {Paddison}\ \emph {et~al.}(2024)\citenamefont
  {Paddison}, \citenamefont {Zhang}, \citenamefont {Yan}, \citenamefont
  {Cliffe}, \citenamefont {McGuire}, \citenamefont {Do}, \citenamefont {Gao},
  \citenamefont {Stone}, \citenamefont {Dahlbom}, \citenamefont {Barros},
  \citenamefont {Batista},\ and\ \citenamefont {Christianson}}]{Paddinson2024}%
  \BibitemOpen
  \bibfield  {author} {\bibinfo {author} {\bibfnamefont {J.~A.~M.}\
  \bibnamefont {Paddison}}, \bibinfo {author} {\bibfnamefont {H.}~\bibnamefont
  {Zhang}}, \bibinfo {author} {\bibfnamefont {J.}~\bibnamefont {Yan}}, \bibinfo
  {author} {\bibfnamefont {M.~J.}\ \bibnamefont {Cliffe}}, \bibinfo {author}
  {\bibfnamefont {M.~A.}\ \bibnamefont {McGuire}}, \bibinfo {author}
  {\bibfnamefont {S.-H.}\ \bibnamefont {Do}}, \bibinfo {author} {\bibfnamefont
  {S.}~\bibnamefont {Gao}}, \bibinfo {author} {\bibfnamefont {M.~B.}\
  \bibnamefont {Stone}}, \bibinfo {author} {\bibfnamefont {D.}~\bibnamefont
  {Dahlbom}}, \bibinfo {author} {\bibfnamefont {K.}~\bibnamefont {Barros}},
  \bibinfo {author} {\bibfnamefont {C.~D.}\ \bibnamefont {Batista}}, \ and\
  \bibinfo {author} {\bibfnamefont {A.~D.}\ \bibnamefont {Christianson}},\
  }\href {\doibase 10.1038/s41535-024-00650-6} {\bibfield  {journal} {\bibinfo
  {journal} {npj Quantum Mater.}\ }\textbf {\bibinfo {volume} {9}},\ \bibinfo
  {pages} {48} (\bibinfo {year} {2024})}\BibitemShut {NoStop}%
\bibitem [{\citenamefont {Aharen}\ \emph
  {et~al.}(2010{\natexlab{a}})\citenamefont {Aharen}, \citenamefont {Greedan},
  \citenamefont {Bridges}, \citenamefont {Aczel}, \citenamefont {Rodriguez},
  \citenamefont {MacDougall}, \citenamefont {Luke}, \citenamefont {Imai},
  \citenamefont {Michaelis}, \citenamefont {Kroeker}, \citenamefont {Zhou},
  \citenamefont {Wiebe},\ and\ \citenamefont {Cranswick}}]{Aharen2010a}%
  \BibitemOpen
  \bibfield  {author} {\bibinfo {author} {\bibfnamefont {T.}~\bibnamefont
  {Aharen}}, \bibinfo {author} {\bibfnamefont {J.~E.}\ \bibnamefont {Greedan}},
  \bibinfo {author} {\bibfnamefont {C.~A.}\ \bibnamefont {Bridges}}, \bibinfo
  {author} {\bibfnamefont {A.~A.}\ \bibnamefont {Aczel}}, \bibinfo {author}
  {\bibfnamefont {J.}~\bibnamefont {Rodriguez}}, \bibinfo {author}
  {\bibfnamefont {G.}~\bibnamefont {MacDougall}}, \bibinfo {author}
  {\bibfnamefont {G.~M.}\ \bibnamefont {Luke}}, \bibinfo {author}
  {\bibfnamefont {T.}~\bibnamefont {Imai}}, \bibinfo {author} {\bibfnamefont
  {V.~K.}\ \bibnamefont {Michaelis}}, \bibinfo {author} {\bibfnamefont
  {S.}~\bibnamefont {Kroeker}}, \bibinfo {author} {\bibfnamefont
  {H.}~\bibnamefont {Zhou}}, \bibinfo {author} {\bibfnamefont {C.~R.}\
  \bibnamefont {Wiebe}}, \ and\ \bibinfo {author} {\bibfnamefont {L.~M.~D.}\
  \bibnamefont {Cranswick}},\ }\href {\doibase 10.1103/PhysRevB.81.224409}
  {\bibfield  {journal} {\bibinfo  {journal} {Phys. Rev. B}\ }\textbf {\bibinfo
  {volume} {81}},\ \bibinfo {pages} {224409} (\bibinfo {year}
  {2010}{\natexlab{a}})}\BibitemShut {NoStop}%
\bibitem [{\citenamefont {Aharen}\ \emph
  {et~al.}(2010{\natexlab{b}})\citenamefont {Aharen}, \citenamefont {Greedan},
  \citenamefont {Bridges}, \citenamefont {Aczel}, \citenamefont {Rodriguez},
  \citenamefont {MacDougall}, \citenamefont {Luke}, \citenamefont {Michaelis},
  \citenamefont {Kroeker}, \citenamefont {Wiebe}, \citenamefont {Zhou},\ and\
  \citenamefont {Cranswick}}]{Aharen2010b}%
  \BibitemOpen
  \bibfield  {author} {\bibinfo {author} {\bibfnamefont {T.}~\bibnamefont
  {Aharen}}, \bibinfo {author} {\bibfnamefont {J.~E.}\ \bibnamefont {Greedan}},
  \bibinfo {author} {\bibfnamefont {C.~A.}\ \bibnamefont {Bridges}}, \bibinfo
  {author} {\bibfnamefont {A.~A.}\ \bibnamefont {Aczel}}, \bibinfo {author}
  {\bibfnamefont {J.}~\bibnamefont {Rodriguez}}, \bibinfo {author}
  {\bibfnamefont {G.}~\bibnamefont {MacDougall}}, \bibinfo {author}
  {\bibfnamefont {G.~M.}\ \bibnamefont {Luke}}, \bibinfo {author}
  {\bibfnamefont {V.~K.}\ \bibnamefont {Michaelis}}, \bibinfo {author}
  {\bibfnamefont {S.}~\bibnamefont {Kroeker}}, \bibinfo {author} {\bibfnamefont
  {C.~R.}\ \bibnamefont {Wiebe}}, \bibinfo {author} {\bibfnamefont
  {H.}~\bibnamefont {Zhou}}, \ and\ \bibinfo {author} {\bibfnamefont
  {L.~M.~D.}\ \bibnamefont {Cranswick}},\ }\href {\doibase
  10.1103/PhysRevB.81.064436} {\bibfield  {journal} {\bibinfo  {journal} {Phys.
  Rev. B}\ }\textbf {\bibinfo {volume} {81}},\ \bibinfo {pages} {064436}
  (\bibinfo {year} {2010}{\natexlab{b}})}\BibitemShut {NoStop}%
\bibitem [{\citenamefont {Marjerrison}\ \emph
  {et~al.}(2016{\natexlab{a}})\citenamefont {Marjerrison}, \citenamefont
  {Thompson}, \citenamefont {Sala}, \citenamefont {Maharaj}, \citenamefont
  {Kermarrec}, \citenamefont {Cai}, \citenamefont {Hallas}, \citenamefont
  {Wilson}, \citenamefont {Munsie}, \citenamefont {Granroth}, \citenamefont
  {Flacau}, \citenamefont {Greedan}, \citenamefont {Gaulin},\ and\
  \citenamefont {Luke}}]{Marjerrison2016a}%
  \BibitemOpen
  \bibfield  {author} {\bibinfo {author} {\bibfnamefont {C.~A.}\ \bibnamefont
  {Marjerrison}}, \bibinfo {author} {\bibfnamefont {C.~M.}\ \bibnamefont
  {Thompson}}, \bibinfo {author} {\bibfnamefont {G.}~\bibnamefont {Sala}},
  \bibinfo {author} {\bibfnamefont {D.~D.}\ \bibnamefont {Maharaj}}, \bibinfo
  {author} {\bibfnamefont {E.}~\bibnamefont {Kermarrec}}, \bibinfo {author}
  {\bibfnamefont {Y.}~\bibnamefont {Cai}}, \bibinfo {author} {\bibfnamefont
  {A.~M.}\ \bibnamefont {Hallas}}, \bibinfo {author} {\bibfnamefont {M.~N.}\
  \bibnamefont {Wilson}}, \bibinfo {author} {\bibfnamefont {T.~J.~S.}\
  \bibnamefont {Munsie}}, \bibinfo {author} {\bibfnamefont {G.~E.}\
  \bibnamefont {Granroth}}, \bibinfo {author} {\bibfnamefont {R.}~\bibnamefont
  {Flacau}}, \bibinfo {author} {\bibfnamefont {J.~E.}\ \bibnamefont {Greedan}},
  \bibinfo {author} {\bibfnamefont {B.~D.}\ \bibnamefont {Gaulin}}, \ and\
  \bibinfo {author} {\bibfnamefont {G.~M.}\ \bibnamefont {Luke}},\ }\href
  {\doibase 10.1021/acs.inorgchem.6b01933} {\bibfield  {journal} {\bibinfo
  {journal} {Inorg. Chem.}\ }\textbf {\bibinfo {volume} {55}},\ \bibinfo
  {pages} {10701} (\bibinfo {year} {2016}{\natexlab{a}})}\BibitemShut {NoStop}%
\bibitem [{\citenamefont {Lu}\ \emph {et~al.}(2017)\citenamefont {Lu},
  \citenamefont {Song}, \citenamefont {Liu}, \citenamefont {Reyes},
  \citenamefont {Kuhns}, \citenamefont {Lee}, \citenamefont {Fisher},\ and\
  \citenamefont {Mitrovi\'{c}}}]{Lu2017}%
  \BibitemOpen
  \bibfield  {author} {\bibinfo {author} {\bibfnamefont {L.}~\bibnamefont
  {Lu}}, \bibinfo {author} {\bibfnamefont {M.}~\bibnamefont {Song}}, \bibinfo
  {author} {\bibfnamefont {W.}~\bibnamefont {Liu}}, \bibinfo {author}
  {\bibfnamefont {A.~P.}\ \bibnamefont {Reyes}}, \bibinfo {author}
  {\bibfnamefont {P.}~\bibnamefont {Kuhns}}, \bibinfo {author} {\bibfnamefont
  {H.~O.}\ \bibnamefont {Lee}}, \bibinfo {author} {\bibfnamefont {I.~R.}\
  \bibnamefont {Fisher}}, \ and\ \bibinfo {author} {\bibfnamefont {V.~F.}\
  \bibnamefont {Mitrovi\'{c}}},\ }\href {\doibase 10.1038/ncomms14407}
  {\bibfield  {journal} {\bibinfo  {journal} {Nat. Commun.}\ }\textbf {\bibinfo
  {volume} {8}},\ \bibinfo {pages} {14407} (\bibinfo {year}
  {2017})}\BibitemShut {NoStop}%
\bibitem [{\citenamefont {Liu}\ \emph {et~al.}(2018)\citenamefont {Liu},
  \citenamefont {Cong}, \citenamefont {Garcia}, \citenamefont {Reyes},
  \citenamefont {Lee}, \citenamefont {Fisher},\ and\ \citenamefont
  {Mitrović}}]{Liu2018}%
  \BibitemOpen
  \bibfield  {author} {\bibinfo {author} {\bibfnamefont {W.}~\bibnamefont
  {Liu}}, \bibinfo {author} {\bibfnamefont {R.}~\bibnamefont {Cong}}, \bibinfo
  {author} {\bibfnamefont {E.}~\bibnamefont {Garcia}}, \bibinfo {author}
  {\bibfnamefont {A.}~\bibnamefont {Reyes}}, \bibinfo {author} {\bibfnamefont
  {H.}~\bibnamefont {Lee}}, \bibinfo {author} {\bibfnamefont {I.}~\bibnamefont
  {Fisher}}, \ and\ \bibinfo {author} {\bibfnamefont {V.}~\bibnamefont
  {Mitrović}},\ }\href {\doibase https://doi.org/10.1016/j.physb.2017.08.062}
  {\bibfield  {journal} {\bibinfo  {journal} {Physica B: Condens. Matter}\
  }\textbf {\bibinfo {volume} {536}},\ \bibinfo {pages} {863} (\bibinfo {year}
  {2018})}\BibitemShut {NoStop}%
\bibitem [{\citenamefont {Ishikawa}\ \emph {et~al.}(2019)\citenamefont
  {Ishikawa}, \citenamefont {Takayama}, \citenamefont {Kremer}, \citenamefont
  {Nuss}, \citenamefont {Dinnebier}, \citenamefont {Kitagawa}, \citenamefont
  {Ishii},\ and\ \citenamefont {Takagi}}]{Ishikawa2019}%
  \BibitemOpen
  \bibfield  {author} {\bibinfo {author} {\bibfnamefont {H.}~\bibnamefont
  {Ishikawa}}, \bibinfo {author} {\bibfnamefont {T.}~\bibnamefont {Takayama}},
  \bibinfo {author} {\bibfnamefont {R.~K.}\ \bibnamefont {Kremer}}, \bibinfo
  {author} {\bibfnamefont {J.}~\bibnamefont {Nuss}}, \bibinfo {author}
  {\bibfnamefont {R.}~\bibnamefont {Dinnebier}}, \bibinfo {author}
  {\bibfnamefont {K.}~\bibnamefont {Kitagawa}}, \bibinfo {author}
  {\bibfnamefont {K.}~\bibnamefont {Ishii}}, \ and\ \bibinfo {author}
  {\bibfnamefont {H.}~\bibnamefont {Takagi}},\ }\href {\doibase
  10.1103/PhysRevB.100.045142} {\bibfield  {journal} {\bibinfo  {journal}
  {Phys. Rev. B}\ }\textbf {\bibinfo {volume} {100}},\ \bibinfo {pages}
  {045142} (\bibinfo {year} {2019})}\BibitemShut {NoStop}%
\bibitem [{\citenamefont {Mansouri~Tehrani}\ \emph {et~al.}(2023)\citenamefont
  {Mansouri~Tehrani}, \citenamefont {Soh}, \citenamefont {P\'asztorov\'a},
  \citenamefont {Merkel}, \citenamefont {\ifmmode \check{Z}\else
  \v{Z}\fi{}ivkovi\ifmmode~\acute{c}\else \'{c}\fi{}}, \citenamefont
  {R\o{}nnow},\ and\ \citenamefont {Spaldin}}]{Tehrani2023}%
  \BibitemOpen
  \bibfield  {author} {\bibinfo {author} {\bibfnamefont {A.}~\bibnamefont
  {Mansouri~Tehrani}}, \bibinfo {author} {\bibfnamefont {J.-R.}\ \bibnamefont
  {Soh}}, \bibinfo {author} {\bibfnamefont {J.}~\bibnamefont {P\'asztorov\'a}},
  \bibinfo {author} {\bibfnamefont {M.~E.}\ \bibnamefont {Merkel}}, \bibinfo
  {author} {\bibfnamefont {I.}~\bibnamefont {\ifmmode \check{Z}\else
  \v{Z}\fi{}ivkovi\ifmmode~\acute{c}\else \'{c}\fi{}}}, \bibinfo {author}
  {\bibfnamefont {H.~M.}\ \bibnamefont {R\o{}nnow}}, \ and\ \bibinfo {author}
  {\bibfnamefont {N.~A.}\ \bibnamefont {Spaldin}},\ }\href {\doibase
  10.1103/PhysRevResearch.5.L012010} {\bibfield  {journal} {\bibinfo  {journal}
  {Phys. Rev. Res.}\ }\textbf {\bibinfo {volume} {5}},\ \bibinfo {pages}
  {L012010} (\bibinfo {year} {2023})}\BibitemShut {NoStop}%
\bibitem [{\citenamefont {Hirai}\ \emph {et~al.}(2020)\citenamefont {Hirai},
  \citenamefont {Sagayama}, \citenamefont {Gao}, \citenamefont {Ohsumi},
  \citenamefont {Chen}, \citenamefont {Arima},\ and\ \citenamefont
  {Hiroi}}]{Hirai2020}%
  \BibitemOpen
  \bibfield  {author} {\bibinfo {author} {\bibfnamefont {D.}~\bibnamefont
  {Hirai}}, \bibinfo {author} {\bibfnamefont {H.}~\bibnamefont {Sagayama}},
  \bibinfo {author} {\bibfnamefont {S.}~\bibnamefont {Gao}}, \bibinfo {author}
  {\bibfnamefont {H.}~\bibnamefont {Ohsumi}}, \bibinfo {author} {\bibfnamefont
  {G.}~\bibnamefont {Chen}}, \bibinfo {author} {\bibfnamefont {T.-h.}\
  \bibnamefont {Arima}}, \ and\ \bibinfo {author} {\bibfnamefont
  {Z.}~\bibnamefont {Hiroi}},\ }\href {\doibase
  10.1103/PhysRevResearch.2.022063} {\bibfield  {journal} {\bibinfo  {journal}
  {Phys. Rev. Res.}\ }\textbf {\bibinfo {volume} {2}},\ \bibinfo {pages}
  {022063} (\bibinfo {year} {2020})}\BibitemShut {NoStop}%
\bibitem [{\citenamefont {Soh}\ \emph {et~al.}(2023)\citenamefont {Soh},
  \citenamefont {Merkel}, \citenamefont {Pourovskii}, \citenamefont
  {Živković}, \citenamefont {Malanyuk}, \citenamefont {Pásztorová},
  \citenamefont {Francoual}, \citenamefont {Hirai}, \citenamefont {Urru},
  \citenamefont {Tolj}, \citenamefont {Fiore-Mosca}, \citenamefont {Yazyev},
  \citenamefont {Spaldin}, \citenamefont {Ederer},\ and\ \citenamefont
  {Rønnow}}]{Soh2023}%
  \BibitemOpen
  \bibfield  {author} {\bibinfo {author} {\bibfnamefont {J.-R.}\ \bibnamefont
  {Soh}}, \bibinfo {author} {\bibfnamefont {M.~E.}\ \bibnamefont {Merkel}},
  \bibinfo {author} {\bibfnamefont {L.}~\bibnamefont {Pourovskii}}, \bibinfo
  {author} {\bibfnamefont {I.}~\bibnamefont {Živković}}, \bibinfo {author}
  {\bibfnamefont {O.}~\bibnamefont {Malanyuk}}, \bibinfo {author}
  {\bibfnamefont {J.}~\bibnamefont {Pásztorová}}, \bibinfo {author}
  {\bibfnamefont {S.}~\bibnamefont {Francoual}}, \bibinfo {author}
  {\bibfnamefont {D.}~\bibnamefont {Hirai}}, \bibinfo {author} {\bibfnamefont
  {A.}~\bibnamefont {Urru}}, \bibinfo {author} {\bibfnamefont {D.}~\bibnamefont
  {Tolj}}, \bibinfo {author} {\bibfnamefont {D.}~\bibnamefont {Fiore-Mosca}},
  \bibinfo {author} {\bibfnamefont {O.}~\bibnamefont {Yazyev}}, \bibinfo
  {author} {\bibfnamefont {N.~A.}\ \bibnamefont {Spaldin}}, \bibinfo {author}
  {\bibfnamefont {C.}~\bibnamefont {Ederer}}, \ and\ \bibinfo {author}
  {\bibfnamefont {H.~M.}\ \bibnamefont {Rønnow}},\ }\href@noop {} {} (\bibinfo
  {year} {2023}),\ \Eprint {http://arxiv.org/abs/2312.01767} {arXiv:2312.01767
  [cond-mat.str-el]} \BibitemShut {NoStop}%
\bibitem [{\citenamefont {da~Cruz Pinha~Barbosa}\ \emph
  {et~al.}(2022)\citenamefont {da~Cruz Pinha~Barbosa}, \citenamefont {Xiong},
  \citenamefont {Tran}, \citenamefont {McGuire}, \citenamefont {Yan},
  \citenamefont {Warren}, \citenamefont {Aguilar}, \citenamefont {Zhang},
  \citenamefont {Randeria}, \citenamefont {Trivedi}, \citenamefont {Haskel},\
  and\ \citenamefont {Woodward}}]{Barbosa2022}%
  \BibitemOpen
  \bibfield  {author} {\bibinfo {author} {\bibfnamefont {V.}~\bibnamefont
  {da~Cruz Pinha~Barbosa}}, \bibinfo {author} {\bibfnamefont {J.}~\bibnamefont
  {Xiong}}, \bibinfo {author} {\bibfnamefont {P.~M.}\ \bibnamefont {Tran}},
  \bibinfo {author} {\bibfnamefont {M.~A.}\ \bibnamefont {McGuire}}, \bibinfo
  {author} {\bibfnamefont {J.}~\bibnamefont {Yan}}, \bibinfo {author}
  {\bibfnamefont {M.~T.}\ \bibnamefont {Warren}}, \bibinfo {author}
  {\bibfnamefont {R.~V.}\ \bibnamefont {Aguilar}}, \bibinfo {author}
  {\bibfnamefont {W.}~\bibnamefont {Zhang}}, \bibinfo {author} {\bibfnamefont
  {M.}~\bibnamefont {Randeria}}, \bibinfo {author} {\bibfnamefont
  {N.}~\bibnamefont {Trivedi}}, \bibinfo {author} {\bibfnamefont
  {D.}~\bibnamefont {Haskel}}, \ and\ \bibinfo {author} {\bibfnamefont {P.~M.}\
  \bibnamefont {Woodward}},\ }\href {\doibase 10.1021/acs.chemmater.1c03456}
  {\bibfield  {journal} {\bibinfo  {journal} {Chem. Mater.}\ }\textbf {\bibinfo
  {volume} {34}},\ \bibinfo {pages} {1098} (\bibinfo {year}
  {2022})}\BibitemShut {NoStop}%
\bibitem [{\citenamefont {Moffitt}\ and\ \citenamefont
  {Thorson}(1957)}]{Moffitt1957a}%
  \BibitemOpen
  \bibfield  {author} {\bibinfo {author} {\bibfnamefont {W.}~\bibnamefont
  {Moffitt}}\ and\ \bibinfo {author} {\bibfnamefont {W.}~\bibnamefont
  {Thorson}},\ }\href {\doibase 10.1103/PhysRev.108.1251} {\bibfield  {journal}
  {\bibinfo  {journal} {Phys. Rev.}\ }\textbf {\bibinfo {volume} {108}},\
  \bibinfo {pages} {1251} (\bibinfo {year} {1957})}\BibitemShut {NoStop}%
\bibitem [{\citenamefont {Moffitt}\ and\ \citenamefont
  {Liehr}(1957)}]{Moffitt1957b}%
  \BibitemOpen
  \bibfield  {author} {\bibinfo {author} {\bibfnamefont {W.}~\bibnamefont
  {Moffitt}}\ and\ \bibinfo {author} {\bibfnamefont {A.~D.}\ \bibnamefont
  {Liehr}},\ }\href {\doibase 10.1103/PhysRev.106.1195} {\bibfield  {journal}
  {\bibinfo  {journal} {Phys. Rev.}\ }\textbf {\bibinfo {volume} {106}},\
  \bibinfo {pages} {1195} (\bibinfo {year} {1957})}\BibitemShut {NoStop}%
\bibitem [{\citenamefont {Longuet-Higgins}\ \emph {et~al.}(1958)\citenamefont
  {Longuet-Higgins}, \citenamefont {Öpik}, \citenamefont {Pryce},\ and\
  \citenamefont {Sack}}]{Longuet-Higgins1958}%
  \BibitemOpen
  \bibfield  {author} {\bibinfo {author} {\bibfnamefont {H.~C.}\ \bibnamefont
  {Longuet-Higgins}}, \bibinfo {author} {\bibfnamefont {U.}~\bibnamefont
  {Öpik}}, \bibinfo {author} {\bibfnamefont {M.~H.~L.}\ \bibnamefont {Pryce}},
  \ and\ \bibinfo {author} {\bibfnamefont {R.~A.}\ \bibnamefont {Sack}},\
  }\href {http://www.jstor.org/stable/100248} {\bibfield  {journal} {\bibinfo
  {journal} {Proceedings of the Royal Society of London. Series A, Mathematical
  and Physical Sciences}\ }\textbf {\bibinfo {volume} {244}},\ \bibinfo {pages}
  {1} (\bibinfo {year} {1958})}\BibitemShut {NoStop}%
\bibitem [{\citenamefont {Xu}\ \emph {et~al.}(2016)\citenamefont {Xu},
  \citenamefont {Bogdanov}, \citenamefont {Princep}, \citenamefont {Fulde},
  \citenamefont {van~den Brink},\ and\ \citenamefont {Hozoi}}]{Xu2016}%
  \BibitemOpen
  \bibfield  {author} {\bibinfo {author} {\bibfnamefont {L.}~\bibnamefont
  {Xu}}, \bibinfo {author} {\bibfnamefont {N.~A.}\ \bibnamefont {Bogdanov}},
  \bibinfo {author} {\bibfnamefont {A.}~\bibnamefont {Princep}}, \bibinfo
  {author} {\bibfnamefont {P.}~\bibnamefont {Fulde}}, \bibinfo {author}
  {\bibfnamefont {J.}~\bibnamefont {van~den Brink}}, \ and\ \bibinfo {author}
  {\bibfnamefont {L.}~\bibnamefont {Hozoi}},\ }\href {\doibase
  doi:10.1038/npjquantmats.2016.29} {\bibfield  {journal} {\bibinfo  {journal}
  {npj Quantum Mater.}\ }\textbf {\bibinfo {volume} {1}},\ \bibinfo {pages}
  {16029} (\bibinfo {year} {2016})}\BibitemShut {NoStop}%
\bibitem [{\citenamefont {Iwahara}\ \emph {et~al.}(2018)\citenamefont
  {Iwahara}, \citenamefont {Vieru},\ and\ \citenamefont
  {Chibotaru}}]{Iwahara2018}%
  \BibitemOpen
  \bibfield  {author} {\bibinfo {author} {\bibfnamefont {N.}~\bibnamefont
  {Iwahara}}, \bibinfo {author} {\bibfnamefont {V.}~\bibnamefont {Vieru}}, \
  and\ \bibinfo {author} {\bibfnamefont {L.~F.}\ \bibnamefont {Chibotaru}},\
  }\href {\doibase 10.1103/PhysRevB.98.075138} {\bibfield  {journal} {\bibinfo
  {journal} {Phys. Rev. B}\ }\textbf {\bibinfo {volume} {98}},\ \bibinfo
  {pages} {075138} (\bibinfo {year} {2018})}\BibitemShut {NoStop}%
\bibitem [{\citenamefont {Frontini}\ \emph {et~al.}(2024)\citenamefont
  {Frontini}, \citenamefont {Johnstone}, \citenamefont {Iwahara}, \citenamefont
  {Bhattacharyya}, \citenamefont {Bogdanov}, \citenamefont {Hozoi},
  \citenamefont {Upton}, \citenamefont {Casa}, \citenamefont {Hirai},\ and\
  \citenamefont {Kim}}]{Frontini2024}%
  \BibitemOpen
  \bibfield  {author} {\bibinfo {author} {\bibfnamefont {F.~I.}\ \bibnamefont
  {Frontini}}, \bibinfo {author} {\bibfnamefont {G.~H.~J.}\ \bibnamefont
  {Johnstone}}, \bibinfo {author} {\bibfnamefont {N.}~\bibnamefont {Iwahara}},
  \bibinfo {author} {\bibfnamefont {P.}~\bibnamefont {Bhattacharyya}}, \bibinfo
  {author} {\bibfnamefont {N.~A.}\ \bibnamefont {Bogdanov}}, \bibinfo {author}
  {\bibfnamefont {L.}~\bibnamefont {Hozoi}}, \bibinfo {author} {\bibfnamefont
  {M.~H.}\ \bibnamefont {Upton}}, \bibinfo {author} {\bibfnamefont {D.~M.}\
  \bibnamefont {Casa}}, \bibinfo {author} {\bibfnamefont {D.}~\bibnamefont
  {Hirai}}, \ and\ \bibinfo {author} {\bibfnamefont {Y.-J.}\ \bibnamefont
  {Kim}},\ }\href {\doibase 10.1103/PhysRevLett.133.036501} {\bibfield
  {journal} {\bibinfo  {journal} {Phys. Rev. Lett.}\ }\textbf {\bibinfo
  {volume} {133}},\ \bibinfo {pages} {036501} (\bibinfo {year}
  {2024})}\BibitemShut {NoStop}%
\bibitem [{\citenamefont {Iwahara}\ \emph {et~al.}(2024)\citenamefont
  {Iwahara}, \citenamefont {Soh}, \citenamefont {Hirai}, \citenamefont
  {\v{Z}ivkovi\'c}, \citenamefont {Wei}, \citenamefont {Zhang}, \citenamefont
  {Galdino}, \citenamefont {Yu}, \citenamefont {Ishii}, \citenamefont {Pisani},
  \citenamefont {Malanyuk}, \citenamefont {Schmitt},\ and\ \citenamefont
  {R{\o}nnow}}]{Iwahara2024}%
  \BibitemOpen
  \bibfield  {author} {\bibinfo {author} {\bibfnamefont {N.}~\bibnamefont
  {Iwahara}}, \bibinfo {author} {\bibfnamefont {J.-R.}\ \bibnamefont {Soh}},
  \bibinfo {author} {\bibfnamefont {D.}~\bibnamefont {Hirai}}, \bibinfo
  {author} {\bibfnamefont {I.}~\bibnamefont {\v{Z}ivkovi\'c}}, \bibinfo
  {author} {\bibfnamefont {Y.}~\bibnamefont {Wei}}, \bibinfo {author}
  {\bibfnamefont {W.}~\bibnamefont {Zhang}}, \bibinfo {author} {\bibfnamefont
  {C.}~\bibnamefont {Galdino}}, \bibinfo {author} {\bibfnamefont
  {T.}~\bibnamefont {Yu}}, \bibinfo {author} {\bibfnamefont {K.}~\bibnamefont
  {Ishii}}, \bibinfo {author} {\bibfnamefont {F.}~\bibnamefont {Pisani}},
  \bibinfo {author} {\bibfnamefont {O.}~\bibnamefont {Malanyuk}}, \bibinfo
  {author} {\bibfnamefont {T.}~\bibnamefont {Schmitt}}, \ and\ \bibinfo
  {author} {\bibfnamefont {H.~M.}\ \bibnamefont {R{\o}nnow}},\ }\href@noop {}
  {\enquote {\bibinfo {title} {Persistent quantum vibronic dynamics in a $5d^1$
  double perovskite oxide},}\ } (\bibinfo {year} {2024}),\ \bibinfo {note}
  {submitted}\BibitemShut {NoStop}%
\bibitem [{\citenamefont {Agrestini}\ \emph {et~al.}(2024)\citenamefont
  {Agrestini}, \citenamefont {Borgatti}, \citenamefont {Florio}, \citenamefont
  {Frassineti}, \citenamefont {Fiore~Mosca}, \citenamefont {Faure},
  \citenamefont {Detlefs}, \citenamefont {Sahle}, \citenamefont {Francoual},
  \citenamefont {Choi}, \citenamefont {Garcia-Fernandez}, \citenamefont {Zhou},
  \citenamefont {Mitrovi\ifmmode~\acute{c}\else \'{c}\fi{}}, \citenamefont
  {Woodward}, \citenamefont {Ghiringhelli}, \citenamefont {Franchini},
  \citenamefont {Boscherini}, \citenamefont {Sanna},\ and\ \citenamefont
  {Moretti~Sala}}]{Agrestini2024}%
  \BibitemOpen
  \bibfield  {author} {\bibinfo {author} {\bibfnamefont {S.}~\bibnamefont
  {Agrestini}}, \bibinfo {author} {\bibfnamefont {F.}~\bibnamefont {Borgatti}},
  \bibinfo {author} {\bibfnamefont {P.}~\bibnamefont {Florio}}, \bibinfo
  {author} {\bibfnamefont {J.}~\bibnamefont {Frassineti}}, \bibinfo {author}
  {\bibfnamefont {D.}~\bibnamefont {Fiore~Mosca}}, \bibinfo {author}
  {\bibfnamefont {Q.}~\bibnamefont {Faure}}, \bibinfo {author} {\bibfnamefont
  {B.}~\bibnamefont {Detlefs}}, \bibinfo {author} {\bibfnamefont {C.~J.}\
  \bibnamefont {Sahle}}, \bibinfo {author} {\bibfnamefont {S.}~\bibnamefont
  {Francoual}}, \bibinfo {author} {\bibfnamefont {J.}~\bibnamefont {Choi}},
  \bibinfo {author} {\bibfnamefont {M.}~\bibnamefont {Garcia-Fernandez}},
  \bibinfo {author} {\bibfnamefont {K.-J.}\ \bibnamefont {Zhou}}, \bibinfo
  {author} {\bibfnamefont {V.~F.}\ \bibnamefont {Mitrovi\ifmmode~\acute{c}\else
  \'{c}\fi{}}}, \bibinfo {author} {\bibfnamefont {P.~M.}\ \bibnamefont
  {Woodward}}, \bibinfo {author} {\bibfnamefont {G.}~\bibnamefont
  {Ghiringhelli}}, \bibinfo {author} {\bibfnamefont {C.}~\bibnamefont
  {Franchini}}, \bibinfo {author} {\bibfnamefont {F.}~\bibnamefont
  {Boscherini}}, \bibinfo {author} {\bibfnamefont {S.}~\bibnamefont {Sanna}}, \
  and\ \bibinfo {author} {\bibfnamefont {M.}~\bibnamefont {Moretti~Sala}},\
  }\href {\doibase 10.1103/PhysRevLett.133.066501} {\bibfield  {journal}
  {\bibinfo  {journal} {Phys. Rev. Lett.}\ }\textbf {\bibinfo {volume} {133}},\
  \bibinfo {pages} {066501} (\bibinfo {year} {2024})}\BibitemShut {NoStop}%
\bibitem [{\citenamefont {\v{Z}ivkovi\'c}\ \emph {et~al.}(2024)\citenamefont
  {\v{Z}ivkovi\'c}, \citenamefont {Soh}, \citenamefont {Malanyuk},
  \citenamefont {Yadav}, \citenamefont {Pisani}, \citenamefont {Tehrani},
  \citenamefont {Tolj}, \citenamefont {Pasztorova}, \citenamefont {Hirai},
  \citenamefont {Wei}, \citenamefont {Zhang}, \citenamefont {Galdino},
  \citenamefont {Yu}, \citenamefont {Ishii}, \citenamefont {Demuer},
  \citenamefont {Yazyev}, \citenamefont {Schmitt},\ and\ \citenamefont
  {R{\o}nnow}}]{Zivkovic2024}%
  \BibitemOpen
  \bibfield  {author} {\bibinfo {author} {\bibfnamefont {I.}~\bibnamefont
  {\v{Z}ivkovi\'c}}, \bibinfo {author} {\bibfnamefont {J.-R.}\ \bibnamefont
  {Soh}}, \bibinfo {author} {\bibfnamefont {O.}~\bibnamefont {Malanyuk}},
  \bibinfo {author} {\bibfnamefont {R.}~\bibnamefont {Yadav}}, \bibinfo
  {author} {\bibfnamefont {F.}~\bibnamefont {Pisani}}, \bibinfo {author}
  {\bibfnamefont {A.~M.}\ \bibnamefont {Tehrani}}, \bibinfo {author}
  {\bibfnamefont {D.}~\bibnamefont {Tolj}}, \bibinfo {author} {\bibfnamefont
  {J.}~\bibnamefont {Pasztorova}}, \bibinfo {author} {\bibfnamefont
  {D.}~\bibnamefont {Hirai}}, \bibinfo {author} {\bibfnamefont
  {Y.}~\bibnamefont {Wei}}, \bibinfo {author} {\bibfnamefont {W.}~\bibnamefont
  {Zhang}}, \bibinfo {author} {\bibfnamefont {C.}~\bibnamefont {Galdino}},
  \bibinfo {author} {\bibfnamefont {T.}~\bibnamefont {Yu}}, \bibinfo {author}
  {\bibfnamefont {K.}~\bibnamefont {Ishii}}, \bibinfo {author} {\bibfnamefont
  {A.}~\bibnamefont {Demuer}}, \bibinfo {author} {\bibfnamefont {O.~V.}\
  \bibnamefont {Yazyev}}, \bibinfo {author} {\bibfnamefont {T.}~\bibnamefont
  {Schmitt}}, \ and\ \bibinfo {author} {\bibfnamefont {H.~M.}\ \bibnamefont
  {R{\o}nnow}},\ }\href {\doibase 10.1038/s41467-024-52935-w} {\ \textbf
  {\bibinfo {volume} {15}},\ \bibinfo {pages} {8587} (\bibinfo {year}
  {2024})}\BibitemShut {NoStop}%
\bibitem [{\citenamefont {Takahashi}\ \emph {et~al.}(2021)\citenamefont
  {Takahashi}, \citenamefont {Suzuki}, \citenamefont {Bertinshaw},
  \citenamefont {Bette}, \citenamefont {M\"uhle}, \citenamefont {Nuss},
  \citenamefont {Dinnebier}, \citenamefont {Yaresko}, \citenamefont
  {Khaliullin}, \citenamefont {Gretarsson}, \citenamefont {Takayama},
  \citenamefont {Takagi},\ and\ \citenamefont {Keimer}}]{Takahashi2021}%
  \BibitemOpen
  \bibfield  {author} {\bibinfo {author} {\bibfnamefont {H.}~\bibnamefont
  {Takahashi}}, \bibinfo {author} {\bibfnamefont {H.}~\bibnamefont {Suzuki}},
  \bibinfo {author} {\bibfnamefont {J.}~\bibnamefont {Bertinshaw}}, \bibinfo
  {author} {\bibfnamefont {S.}~\bibnamefont {Bette}}, \bibinfo {author}
  {\bibfnamefont {C.}~\bibnamefont {M\"uhle}}, \bibinfo {author} {\bibfnamefont
  {J.}~\bibnamefont {Nuss}}, \bibinfo {author} {\bibfnamefont {R.}~\bibnamefont
  {Dinnebier}}, \bibinfo {author} {\bibfnamefont {A.}~\bibnamefont {Yaresko}},
  \bibinfo {author} {\bibfnamefont {G.}~\bibnamefont {Khaliullin}}, \bibinfo
  {author} {\bibfnamefont {H.}~\bibnamefont {Gretarsson}}, \bibinfo {author}
  {\bibfnamefont {T.}~\bibnamefont {Takayama}}, \bibinfo {author}
  {\bibfnamefont {H.}~\bibnamefont {Takagi}}, \ and\ \bibinfo {author}
  {\bibfnamefont {B.}~\bibnamefont {Keimer}},\ }\href {\doibase
  10.1103/PhysRevLett.127.227201} {\bibfield  {journal} {\bibinfo  {journal}
  {Phys. Rev. Lett.}\ }\textbf {\bibinfo {volume} {127}},\ \bibinfo {pages}
  {227201} (\bibinfo {year} {2021})}\BibitemShut {NoStop}%
\bibitem [{\citenamefont {Reig-i Plessis}\ \emph {et~al.}(2020)\citenamefont
  {Reig-i Plessis}, \citenamefont {Johnson}, \citenamefont {Lu}, \citenamefont
  {Chen}, \citenamefont {Ruff}, \citenamefont {Upton}, \citenamefont
  {Williams}, \citenamefont {Calder}, \citenamefont {Zhou}, \citenamefont
  {Clancy}, \citenamefont {Aczel},\ and\ \citenamefont
  {MacDougall}}]{Reigiplessis2020}%
  \BibitemOpen
  \bibfield  {author} {\bibinfo {author} {\bibfnamefont {D.}~\bibnamefont
  {Reig-i Plessis}}, \bibinfo {author} {\bibfnamefont {T.~A.}\ \bibnamefont
  {Johnson}}, \bibinfo {author} {\bibfnamefont {K.}~\bibnamefont {Lu}},
  \bibinfo {author} {\bibfnamefont {Q.}~\bibnamefont {Chen}}, \bibinfo {author}
  {\bibfnamefont {J.~P.~C.}\ \bibnamefont {Ruff}}, \bibinfo {author}
  {\bibfnamefont {M.~H.}\ \bibnamefont {Upton}}, \bibinfo {author}
  {\bibfnamefont {T.~J.}\ \bibnamefont {Williams}}, \bibinfo {author}
  {\bibfnamefont {S.}~\bibnamefont {Calder}}, \bibinfo {author} {\bibfnamefont
  {H.~D.}\ \bibnamefont {Zhou}}, \bibinfo {author} {\bibfnamefont {J.~P.}\
  \bibnamefont {Clancy}}, \bibinfo {author} {\bibfnamefont {A.~A.}\
  \bibnamefont {Aczel}}, \ and\ \bibinfo {author} {\bibfnamefont {G.~J.}\
  \bibnamefont {MacDougall}},\ }\href {\doibase
  10.1103/PhysRevMaterials.4.124407} {\bibfield  {journal} {\bibinfo  {journal}
  {Phys. Rev. Mater.}\ }\textbf {\bibinfo {volume} {4}},\ \bibinfo {pages}
  {124407} (\bibinfo {year} {2020})}\BibitemShut {NoStop}%
\bibitem [{\citenamefont {Khan}\ \emph {et~al.}(2021)\citenamefont {Khan},
  \citenamefont {Prishchenko}, \citenamefont {Upton}, \citenamefont
  {Mazurenko},\ and\ \citenamefont {Tsirlin}}]{Khan2021}%
  \BibitemOpen
  \bibfield  {author} {\bibinfo {author} {\bibfnamefont {N.}~\bibnamefont
  {Khan}}, \bibinfo {author} {\bibfnamefont {D.}~\bibnamefont {Prishchenko}},
  \bibinfo {author} {\bibfnamefont {M.~H.}\ \bibnamefont {Upton}}, \bibinfo
  {author} {\bibfnamefont {V.~G.}\ \bibnamefont {Mazurenko}}, \ and\ \bibinfo
  {author} {\bibfnamefont {A.~A.}\ \bibnamefont {Tsirlin}},\ }\href {\doibase
  10.1103/PhysRevB.103.125158} {\bibfield  {journal} {\bibinfo  {journal}
  {Phys. Rev. B}\ }\textbf {\bibinfo {volume} {103}},\ \bibinfo {pages}
  {125158} (\bibinfo {year} {2021})}\BibitemShut {NoStop}%
\bibitem [{\citenamefont {Revelli}\ \emph {et~al.}(2019)\citenamefont
  {Revelli}, \citenamefont {Loo}, \citenamefont {Kiese}, \citenamefont
  {Becker}, \citenamefont {Fr\"ohlich}, \citenamefont {Lorenz}, \citenamefont
  {Moretti~Sala}, \citenamefont {Monaco}, \citenamefont {Buessen},
  \citenamefont {Attig}, \citenamefont {Hermanns}, \citenamefont {Streltsov},
  \citenamefont {Khomskii}, \citenamefont {van~den Brink}, \citenamefont
  {Braden}, \citenamefont {van Loosdrecht}, \citenamefont {Trebst},
  \citenamefont {Paramekanti},\ and\ \citenamefont {Gr\"uninger}}]{Raveli2019}%
  \BibitemOpen
  \bibfield  {author} {\bibinfo {author} {\bibfnamefont {A.}~\bibnamefont
  {Revelli}}, \bibinfo {author} {\bibfnamefont {C.~C.}\ \bibnamefont {Loo}},
  \bibinfo {author} {\bibfnamefont {D.}~\bibnamefont {Kiese}}, \bibinfo
  {author} {\bibfnamefont {P.}~\bibnamefont {Becker}}, \bibinfo {author}
  {\bibfnamefont {T.}~\bibnamefont {Fr\"ohlich}}, \bibinfo {author}
  {\bibfnamefont {T.}~\bibnamefont {Lorenz}}, \bibinfo {author} {\bibfnamefont
  {M.}~\bibnamefont {Moretti~Sala}}, \bibinfo {author} {\bibfnamefont
  {G.}~\bibnamefont {Monaco}}, \bibinfo {author} {\bibfnamefont {F.~L.}\
  \bibnamefont {Buessen}}, \bibinfo {author} {\bibfnamefont {J.}~\bibnamefont
  {Attig}}, \bibinfo {author} {\bibfnamefont {M.}~\bibnamefont {Hermanns}},
  \bibinfo {author} {\bibfnamefont {S.~V.}\ \bibnamefont {Streltsov}}, \bibinfo
  {author} {\bibfnamefont {D.~I.}\ \bibnamefont {Khomskii}}, \bibinfo {author}
  {\bibfnamefont {J.}~\bibnamefont {van~den Brink}}, \bibinfo {author}
  {\bibfnamefont {M.}~\bibnamefont {Braden}}, \bibinfo {author} {\bibfnamefont
  {P.~H.~M.}\ \bibnamefont {van Loosdrecht}}, \bibinfo {author} {\bibfnamefont
  {S.}~\bibnamefont {Trebst}}, \bibinfo {author} {\bibfnamefont
  {A.}~\bibnamefont {Paramekanti}}, \ and\ \bibinfo {author} {\bibfnamefont
  {M.}~\bibnamefont {Gr\"uninger}},\ }\href {\doibase
  10.1103/PhysRevB.100.085139} {\bibfield  {journal} {\bibinfo  {journal}
  {Phys. Rev. B}\ }\textbf {\bibinfo {volume} {100}},\ \bibinfo {pages}
  {085139} (\bibinfo {year} {2019})}\BibitemShut {NoStop}%
\bibitem [{\citenamefont {Lee}\ \emph {et~al.}(2022)\citenamefont {Lee},
  \citenamefont {Kim}, \citenamefont {Seong},\ and\ \citenamefont
  {Choi}}]{Lee2022}%
  \BibitemOpen
  \bibfield  {author} {\bibinfo {author} {\bibfnamefont {S.}~\bibnamefont
  {Lee}}, \bibinfo {author} {\bibfnamefont {B.~H.}\ \bibnamefont {Kim}},
  \bibinfo {author} {\bibfnamefont {M.-J.}\ \bibnamefont {Seong}}, \ and\
  \bibinfo {author} {\bibfnamefont {K.-Y.}\ \bibnamefont {Choi}},\ }\href
  {\doibase 10.1103/PhysRevB.105.184433} {\bibfield  {journal} {\bibinfo
  {journal} {Phys. Rev. B}\ }\textbf {\bibinfo {volume} {105}},\ \bibinfo
  {pages} {184433} (\bibinfo {year} {2022})}\BibitemShut {NoStop}%
\bibitem [{\citenamefont {Warzanowski}\ \emph
  {et~al.}(2024{\natexlab{a}})\citenamefont {Warzanowski}, \citenamefont
  {Magnaterra}, \citenamefont {Sahle}, \citenamefont {Sala}, \citenamefont
  {Becker}, \citenamefont {Bohat\'y}, \citenamefont {C\'{i}sa\v{r}ov\'a},
  \citenamefont {Monaco}, \citenamefont {Lorenz}, \citenamefont {van
  Loosdrecht}, \citenamefont {van~den Brink},\ and\ \citenamefont
  {Gr\"{u}ninger}}]{Warzanowski2024}%
  \BibitemOpen
  \bibfield  {author} {\bibinfo {author} {\bibfnamefont {P.}~\bibnamefont
  {Warzanowski}}, \bibinfo {author} {\bibfnamefont {M.}~\bibnamefont
  {Magnaterra}}, \bibinfo {author} {\bibfnamefont {C.~J.}\ \bibnamefont
  {Sahle}}, \bibinfo {author} {\bibfnamefont {M.~M.}\ \bibnamefont {Sala}},
  \bibinfo {author} {\bibfnamefont {P.}~\bibnamefont {Becker}}, \bibinfo
  {author} {\bibfnamefont {L.}~\bibnamefont {Bohat\'y}}, \bibinfo {author}
  {\bibfnamefont {I.}~\bibnamefont {C\'{i}sa\v{r}ov\'a}}, \bibinfo {author}
  {\bibfnamefont {G.}~\bibnamefont {Monaco}}, \bibinfo {author} {\bibfnamefont
  {T.}~\bibnamefont {Lorenz}}, \bibinfo {author} {\bibfnamefont {P.~H.~M.}\
  \bibnamefont {van Loosdrecht}}, \bibinfo {author} {\bibfnamefont
  {J.}~\bibnamefont {van~den Brink}}, \ and\ \bibinfo {author} {\bibfnamefont
  {M.}~\bibnamefont {Gr\"{u}ninger}},\ }\href
  {https://arxiv.org/abs/2407.12133} {} (\bibinfo {year}
  {2024}{\natexlab{a}}),\ \Eprint {http://arxiv.org/abs/2407.12133}
  {arXiv:2407.12133 [cond-mat.str-el]} \BibitemShut {NoStop}%
\bibitem [{\citenamefont {\"{O}pik}\ and\ \citenamefont
  {Pryce}(1957)}]{Opik1957}%
  \BibitemOpen
  \bibfield  {author} {\bibinfo {author} {\bibfnamefont {U.}~\bibnamefont
  {\"{O}pik}}\ and\ \bibinfo {author} {\bibfnamefont {M.~H.~L.}\ \bibnamefont
  {Pryce}},\ }\href {http://www.jstor.org/stable/100101} {\bibfield  {journal}
  {\bibinfo  {journal} {Proceedings of the Royal Society of London. Series A,
  Mathematical and Physical Sciences}\ }\textbf {\bibinfo {volume} {238}},\
  \bibinfo {pages} {425} (\bibinfo {year} {1957})}\BibitemShut {NoStop}%
\bibitem [{\citenamefont {Kanamori}(1960)}]{Kanamori1960}%
  \BibitemOpen
  \bibfield  {author} {\bibinfo {author} {\bibfnamefont {J.}~\bibnamefont
  {Kanamori}},\ }\href {\doibase 10.1063/1.1984590} {\bibfield  {journal}
  {\bibinfo  {journal} {Journal of Applied Physics}\ }\textbf {\bibinfo
  {volume} {31}},\ \bibinfo {pages} {S14} (\bibinfo {year} {1960})}\BibitemShut
  {NoStop}%
\bibitem [{\citenamefont {Englman}(1972)}]{Englman1972}%
  \BibitemOpen
  \bibfield  {author} {\bibinfo {author} {\bibfnamefont {R.}~\bibnamefont
  {Englman}},\ }\href@noop {} {\emph {\bibinfo {title} {The Jahn-Teller Effect
  in Molecules and Crystals}}}\ (\bibinfo  {publisher} {John Wiley \& Sons
  Ltd},\ \bibinfo {address} {London},\ \bibinfo {year} {1972})\BibitemShut
  {NoStop}%
\bibitem [{\citenamefont {Gehring}\ and\ \citenamefont
  {Gehring}(1975)}]{Gehring1975}%
  \BibitemOpen
  \bibfield  {author} {\bibinfo {author} {\bibfnamefont {G.~A.}\ \bibnamefont
  {Gehring}}\ and\ \bibinfo {author} {\bibfnamefont {K.~A.}\ \bibnamefont
  {Gehring}},\ }\href {\doibase 10.1088/0034-4885/38/1/001} {\bibfield
  {journal} {\bibinfo  {journal} {Reports on Progress in Physics}\ }\textbf
  {\bibinfo {volume} {38}},\ \bibinfo {pages} {1} (\bibinfo {year}
  {1975})}\BibitemShut {NoStop}%
\bibitem [{\citenamefont {Kugel'}\ and\ \citenamefont
  {Khomski\v{i}}(1982)}]{Kugel1982}%
  \BibitemOpen
  \bibfield  {author} {\bibinfo {author} {\bibfnamefont {K.~I.}\ \bibnamefont
  {Kugel'}}\ and\ \bibinfo {author} {\bibfnamefont {D.~I.}\ \bibnamefont
  {Khomski\v{i}}},\ }\href {\doibase 10.1070/PU1982v025n04ABEH004537}
  {\bibfield  {journal} {\bibinfo  {journal} {Soviet Physics Uspekhi}\ }\textbf
  {\bibinfo {volume} {25}},\ \bibinfo {pages} {231} (\bibinfo {year}
  {1982})}\BibitemShut {NoStop}%
\bibitem [{\citenamefont {Kaplan}\ and\ \citenamefont
  {Vekhter}(1995)}]{Kaplan1995}%
  \BibitemOpen
  \bibfield  {author} {\bibinfo {author} {\bibfnamefont {M.~D.}\ \bibnamefont
  {Kaplan}}\ and\ \bibinfo {author} {\bibfnamefont {B.~G.}\ \bibnamefont
  {Vekhter}},\ }\href@noop {} {\emph {\bibinfo {title} {Cooperative Phenomena
  in Jahn-Teller Crystals}}}\ (\bibinfo  {publisher} {Plenum Press},\ \bibinfo
  {address} {New York and London},\ \bibinfo {year} {1995})\BibitemShut
  {NoStop}%
\bibitem [{\citenamefont {Tokura}\ and\ \citenamefont
  {Nagaosa}(2000)}]{Tokura2000}%
  \BibitemOpen
  \bibfield  {author} {\bibinfo {author} {\bibfnamefont {Y.}~\bibnamefont
  {Tokura}}\ and\ \bibinfo {author} {\bibfnamefont {N.}~\bibnamefont
  {Nagaosa}},\ }\href {\doibase 10.1126/science.288.5465.462} {\bibfield
  {journal} {\bibinfo  {journal} {Science}\ }\textbf {\bibinfo {volume}
  {288}},\ \bibinfo {pages} {462} (\bibinfo {year} {2000})}\BibitemShut
  {NoStop}%
\bibitem [{\citenamefont {Bersuker}(2006)}]{Bersuker2006}%
  \BibitemOpen
  \bibfield  {author} {\bibinfo {author} {\bibfnamefont {I.~B.}\ \bibnamefont
  {Bersuker}},\ }\href@noop {} {\emph {\bibinfo {title} {The Jahn-Teller
  Effect}}}\ (\bibinfo  {publisher} {Cambridge University Press},\ \bibinfo
  {address} {Cambridge},\ \bibinfo {year} {2006})\BibitemShut {NoStop}%
\bibitem [{\citenamefont {Khomskii}\ and\ \citenamefont
  {Streltsov}(2021)}]{Khomskii2021}%
  \BibitemOpen
  \bibfield  {author} {\bibinfo {author} {\bibfnamefont {D.~I.}\ \bibnamefont
  {Khomskii}}\ and\ \bibinfo {author} {\bibfnamefont {S.~V.}\ \bibnamefont
  {Streltsov}},\ }\href {\doibase 10.1021/acs.chemrev.0c00579} {\bibfield
  {journal} {\bibinfo  {journal} {Chemical Reviews}\ }\textbf {\bibinfo
  {volume} {121}},\ \bibinfo {pages} {2992} (\bibinfo {year}
  {2021})}\BibitemShut {NoStop}%
\bibitem [{\citenamefont {Bersuker}(1962)}]{Bersuker1962}%
  \BibitemOpen
  \bibfield  {author} {\bibinfo {author} {\bibfnamefont {I.~B.}\ \bibnamefont
  {Bersuker}},\ }\href {http://jetp.ras.ru/cgi-bin/dn/e_016_04_0933.pdf}
  {\bibfield  {journal} {\bibinfo  {journal} {J. Exp. Theor. Phys. (U.S.S.R.)}\
  }\textbf {\bibinfo {volume} {43}},\ \bibinfo {pages} {1315} (\bibinfo {year}
  {1962})},\ \bibinfo {note} {{English translation: Sov. Phys. JETP {\bf 16},
  933 (1963).}}\BibitemShut {Stop}%
\bibitem [{\citenamefont {O'Brien}(1964)}]{OBrien1964}%
  \BibitemOpen
  \bibfield  {author} {\bibinfo {author} {\bibfnamefont {M.~C.~M.}\
  \bibnamefont {O'Brien}},\ }\href {http://www.jstor.org/stable/2414883}
  {\bibfield  {journal} {\bibinfo  {journal} {Proceedings of the Royal Society
  of London. Series A, Mathematical and Physical Sciences}\ }\textbf {\bibinfo
  {volume} {281}},\ \bibinfo {pages} {323} (\bibinfo {year}
  {1964})}\BibitemShut {NoStop}%
\bibitem [{\citenamefont {Caner}\ and\ \citenamefont
  {Englman}(1966)}]{Caner1966}%
  \BibitemOpen
  \bibfield  {author} {\bibinfo {author} {\bibfnamefont {M.}~\bibnamefont
  {Caner}}\ and\ \bibinfo {author} {\bibfnamefont {R.}~\bibnamefont
  {Englman}},\ }\href {\doibase 10.1063/1.1726575} {\bibfield  {journal}
  {\bibinfo  {journal} {The Journal of Chemical Physics}\ }\textbf {\bibinfo
  {volume} {44}},\ \bibinfo {pages} {4054} (\bibinfo {year}
  {1966})}\BibitemShut {NoStop}%
\bibitem [{\citenamefont {Muramatsu}\ and\ \citenamefont
  {Sakamoto}(1978)}]{Muramatsu1978}%
  \BibitemOpen
  \bibfield  {author} {\bibinfo {author} {\bibfnamefont {S.}~\bibnamefont
  {Muramatsu}}\ and\ \bibinfo {author} {\bibfnamefont {N.}~\bibnamefont
  {Sakamoto}},\ }\href {\doibase 10.1143/JPSJ.44.1640} {\bibfield  {journal}
  {\bibinfo  {journal} {Journal of the Physical Society of Japan}\ }\textbf
  {\bibinfo {volume} {44}},\ \bibinfo {pages} {1640} (\bibinfo {year}
  {1978})}\BibitemShut {NoStop}%
\bibitem [{\citenamefont {Pooler}(1978)}]{Pooler1978}%
  \BibitemOpen
  \bibfield  {author} {\bibinfo {author} {\bibfnamefont {D.~R.}\ \bibnamefont
  {Pooler}},\ }\href {\doibase 10.1088/0305-4470/11/6/008} {\bibfield
  {journal} {\bibinfo  {journal} {Journal of Physics A: Mathematical and
  General}\ }\textbf {\bibinfo {volume} {11}},\ \bibinfo {pages} {1045}
  (\bibinfo {year} {1978})}\BibitemShut {NoStop}%
\bibitem [{\citenamefont {Pooler}(1980)}]{Pooler1980}%
  \BibitemOpen
  \bibfield  {author} {\bibinfo {author} {\bibfnamefont {D.~R.}\ \bibnamefont
  {Pooler}},\ }\href {\doibase 10.1088/0022-3719/13/6/013} {\bibfield
  {journal} {\bibinfo  {journal} {Journal of Physics C: Solid State Physics}\
  }\textbf {\bibinfo {volume} {13}},\ \bibinfo {pages} {1029} (\bibinfo {year}
  {1980})}\BibitemShut {NoStop}%
\bibitem [{\citenamefont {Auerbach}\ \emph {et~al.}(1994)\citenamefont
  {Auerbach}, \citenamefont {Manini},\ and\ \citenamefont
  {Tosatti}}]{Auerbach1994}%
  \BibitemOpen
  \bibfield  {author} {\bibinfo {author} {\bibfnamefont {A.}~\bibnamefont
  {Auerbach}}, \bibinfo {author} {\bibfnamefont {N.}~\bibnamefont {Manini}}, \
  and\ \bibinfo {author} {\bibfnamefont {E.}~\bibnamefont {Tosatti}},\ }\href
  {\doibase 10.1103/PhysRevB.49.12998} {\bibfield  {journal} {\bibinfo
  {journal} {Phys. Rev. B}\ }\textbf {\bibinfo {volume} {49}},\ \bibinfo
  {pages} {12998} (\bibinfo {year} {1994})}\BibitemShut {NoStop}%
\bibitem [{\citenamefont {Koizumi}\ and\ \citenamefont
  {Sugano}(1994)}]{Koizumi1994}%
  \BibitemOpen
  \bibfield  {author} {\bibinfo {author} {\bibfnamefont {H.}~\bibnamefont
  {Koizumi}}\ and\ \bibinfo {author} {\bibfnamefont {S.}~\bibnamefont
  {Sugano}},\ }\href {\doibase 10.1063/1.467412} {\bibfield  {journal}
  {\bibinfo  {journal} {The Journal of Chemical Physics}\ }\textbf {\bibinfo
  {volume} {101}},\ \bibinfo {pages} {4903} (\bibinfo {year}
  {1994})}\BibitemShut {NoStop}%
\bibitem [{\citenamefont {Koizumi}\ and\ \citenamefont
  {Sugano}(1995)}]{Koizumi1995}%
  \BibitemOpen
  \bibfield  {author} {\bibinfo {author} {\bibfnamefont {H.}~\bibnamefont
  {Koizumi}}\ and\ \bibinfo {author} {\bibfnamefont {S.}~\bibnamefont
  {Sugano}},\ }\href {\doibase 10.1063/1.469495} {\bibfield  {journal}
  {\bibinfo  {journal} {The Journal of Chemical Physics}\ }\textbf {\bibinfo
  {volume} {102}},\ \bibinfo {pages} {4472} (\bibinfo {year}
  {1995})}\BibitemShut {NoStop}%
\bibitem [{\citenamefont {O'Brien}(1996)}]{OBrien1996}%
  \BibitemOpen
  \bibfield  {author} {\bibinfo {author} {\bibfnamefont {M.~C.~M.}\
  \bibnamefont {O'Brien}},\ }\href {\doibase 10.1103/PhysRevB.53.3775}
  {\bibfield  {journal} {\bibinfo  {journal} {Phys. Rev. B}\ }\textbf {\bibinfo
  {volume} {53}},\ \bibinfo {pages} {3775} (\bibinfo {year}
  {1996})}\BibitemShut {NoStop}%
\bibitem [{\citenamefont {Sato}\ \emph {et~al.}(2005)\citenamefont {Sato},
  \citenamefont {Chibotaru},\ and\ \citenamefont {Ceulemans}}]{Sato2005}%
  \BibitemOpen
  \bibfield  {author} {\bibinfo {author} {\bibfnamefont {T.}~\bibnamefont
  {Sato}}, \bibinfo {author} {\bibfnamefont {L.~F.}\ \bibnamefont {Chibotaru}},
  \ and\ \bibinfo {author} {\bibfnamefont {A.}~\bibnamefont {Ceulemans}},\
  }\href {\doibase 10.1063/1.1836758} {\bibfield  {journal} {\bibinfo
  {journal} {The Journal of Chemical Physics}\ }\textbf {\bibinfo {volume}
  {122}},\ \bibinfo {pages} {054104} (\bibinfo {year} {2005})}\BibitemShut
  {NoStop}%
\bibitem [{\citenamefont {Iwahara}\ \emph {et~al.}(2012)\citenamefont
  {Iwahara}, \citenamefont {Sato}, \citenamefont {Tanaka},\ and\ \citenamefont
  {Chibotaru}}]{Iwahara2012}%
  \BibitemOpen
  \bibfield  {author} {\bibinfo {author} {\bibfnamefont {N.}~\bibnamefont
  {Iwahara}}, \bibinfo {author} {\bibfnamefont {T.}~\bibnamefont {Sato}},
  \bibinfo {author} {\bibfnamefont {K.}~\bibnamefont {Tanaka}}, \ and\ \bibinfo
  {author} {\bibfnamefont {L.~F.}\ \bibnamefont {Chibotaru}},\ }\href {\doibase
  10.1209/0295-5075/100/43001} {\bibfield  {journal} {\bibinfo  {journal}
  {Europhysics Letters}\ }\textbf {\bibinfo {volume} {100}},\ \bibinfo {pages}
  {43001} (\bibinfo {year} {2012})}\BibitemShut {NoStop}%
\bibitem [{\citenamefont {Requist}\ \emph {et~al.}(2016)\citenamefont
  {Requist}, \citenamefont {Tandetzky},\ and\ \citenamefont
  {Gross}}]{Requist2016}%
  \BibitemOpen
  \bibfield  {author} {\bibinfo {author} {\bibfnamefont {R.}~\bibnamefont
  {Requist}}, \bibinfo {author} {\bibfnamefont {F.}~\bibnamefont {Tandetzky}},
  \ and\ \bibinfo {author} {\bibfnamefont {E.~K.~U.}\ \bibnamefont {Gross}},\
  }\href {\doibase 10.1103/PhysRevA.93.042108} {\bibfield  {journal} {\bibinfo
  {journal} {Phys. Rev. A}\ }\textbf {\bibinfo {volume} {93}},\ \bibinfo
  {pages} {042108} (\bibinfo {year} {2016})}\BibitemShut {NoStop}%
\bibitem [{\citenamefont {Requist}\ \emph {et~al.}(2017)\citenamefont
  {Requist}, \citenamefont {Proetto},\ and\ \citenamefont
  {Gross}}]{Requist2017}%
  \BibitemOpen
  \bibfield  {author} {\bibinfo {author} {\bibfnamefont {R.}~\bibnamefont
  {Requist}}, \bibinfo {author} {\bibfnamefont {C.~R.}\ \bibnamefont
  {Proetto}}, \ and\ \bibinfo {author} {\bibfnamefont {E.~K.~U.}\ \bibnamefont
  {Gross}},\ }\href {\doibase 10.1103/PhysRevA.96.062503} {\bibfield  {journal}
  {\bibinfo  {journal} {Phys. Rev. A}\ }\textbf {\bibinfo {volume} {96}},\
  \bibinfo {pages} {062503} (\bibinfo {year} {2017})}\BibitemShut {NoStop}%
\bibitem [{\citenamefont {Child}\ and\ \citenamefont
  {Longuet-Higgins}(1961)}]{Child1961}%
  \BibitemOpen
  \bibfield  {author} {\bibinfo {author} {\bibfnamefont {M.~S.}\ \bibnamefont
  {Child}}\ and\ \bibinfo {author} {\bibfnamefont {H.~C.}\ \bibnamefont
  {Longuet-Higgins}},\ }\href {\doibase 10.1098/rsta.1961.0017} {\bibfield
  {journal} {\bibinfo  {journal} {Philosophical Transactions of the Royal
  Society of London. Series A, Mathematical and Physical Sciences}\ }\textbf
  {\bibinfo {volume} {254}},\ \bibinfo {pages} {259} (\bibinfo {year}
  {1961})}\BibitemShut {NoStop}%
\bibitem [{\citenamefont {Ham}(1965)}]{Ham1965}%
  \BibitemOpen
  \bibfield  {author} {\bibinfo {author} {\bibfnamefont {F.~S.}\ \bibnamefont
  {Ham}},\ }\href {\doibase 10.1103/PhysRev.138.A1727} {\bibfield  {journal}
  {\bibinfo  {journal} {Phys. Rev.}\ }\textbf {\bibinfo {volume} {138}},\
  \bibinfo {pages} {A1727} (\bibinfo {year} {1965})}\BibitemShut {NoStop}%
\bibitem [{\citenamefont {Toyozawa}\ and\ \citenamefont
  {Inoue}(1966)}]{Toyozawa1966}%
  \BibitemOpen
  \bibfield  {author} {\bibinfo {author} {\bibfnamefont {Y.}~\bibnamefont
  {Toyozawa}}\ and\ \bibinfo {author} {\bibfnamefont {M.}~\bibnamefont
  {Inoue}},\ }\href {\doibase 10.1143/JPSJ.21.1663} {\bibfield  {journal}
  {\bibinfo  {journal} {Journal of the Physical Society of Japan}\ }\textbf
  {\bibinfo {volume} {21}},\ \bibinfo {pages} {1663} (\bibinfo {year}
  {1966})}\BibitemShut {NoStop}%
\bibitem [{\citenamefont {Cho}(1968)}]{Cho1968}%
  \BibitemOpen
  \bibfield  {author} {\bibinfo {author} {\bibfnamefont {K.}~\bibnamefont
  {Cho}},\ }\href {\doibase 10.1143/JPSJ.25.1372} {\bibfield  {journal}
  {\bibinfo  {journal} {Journal of the Physical Society of Japan}\ }\textbf
  {\bibinfo {volume} {25}},\ \bibinfo {pages} {1372} (\bibinfo {year}
  {1968})}\BibitemShut {NoStop}%
\bibitem [{\citenamefont {Ham}(1968)}]{Ham1968}%
  \BibitemOpen
  \bibfield  {author} {\bibinfo {author} {\bibfnamefont {F.~S.}\ \bibnamefont
  {Ham}},\ }\href {\doibase 10.1103/PhysRev.166.307} {\bibfield  {journal}
  {\bibinfo  {journal} {Phys. Rev.}\ }\textbf {\bibinfo {volume} {166}},\
  \bibinfo {pages} {307} (\bibinfo {year} {1968})}\BibitemShut {NoStop}%
\bibitem [{\citenamefont {Fukuda}(1969)}]{Fukuda1969}%
  \BibitemOpen
  \bibfield  {author} {\bibinfo {author} {\bibfnamefont {A.}~\bibnamefont
  {Fukuda}},\ }\href {\doibase 10.1143/JPSJ.27.96} {\bibfield  {journal}
  {\bibinfo  {journal} {Journal of the Physical Society of Japan}\ }\textbf
  {\bibinfo {volume} {27}},\ \bibinfo {pages} {96} (\bibinfo {year}
  {1969})}\BibitemShut {NoStop}%
\bibitem [{\citenamefont {Kahn}\ and\ \citenamefont {Kettle}(1972)}]{Kahn1972}%
  \BibitemOpen
  \bibfield  {author} {\bibinfo {author} {\bibfnamefont {O.}~\bibnamefont
  {Kahn}}\ and\ \bibinfo {author} {\bibfnamefont {S.~F.~A.}\ \bibnamefont
  {Kettle}},\ }\href {\doibase 10.1007/BF01046364} {\bibfield  {journal}
  {\bibinfo  {journal} {Theor. chim. acta}\ }\textbf {\bibinfo {volume} {27}},\
  \bibinfo {pages} {187} (\bibinfo {year} {1972})}\BibitemShut {NoStop}%
\bibitem [{\citenamefont {Gunnarsson}\ \emph {et~al.}(1995)\citenamefont
  {Gunnarsson}, \citenamefont {Handschuh}, \citenamefont {Bechthold},
  \citenamefont {Kessler}, \citenamefont {Gantef\"{o}r},\ and\ \citenamefont
  {Eberhardt}}]{Gunnarsson1995}%
  \BibitemOpen
  \bibfield  {author} {\bibinfo {author} {\bibfnamefont {O.}~\bibnamefont
  {Gunnarsson}}, \bibinfo {author} {\bibfnamefont {H.}~\bibnamefont
  {Handschuh}}, \bibinfo {author} {\bibfnamefont {P.~S.}\ \bibnamefont
  {Bechthold}}, \bibinfo {author} {\bibfnamefont {B.}~\bibnamefont {Kessler}},
  \bibinfo {author} {\bibfnamefont {G.}~\bibnamefont {Gantef\"{o}r}}, \ and\
  \bibinfo {author} {\bibfnamefont {W.}~\bibnamefont {Eberhardt}},\ }\href
  {\doibase 10.1103/PhysRevLett.74.1875} {\bibfield  {journal} {\bibinfo
  {journal} {Phys. Rev. Lett.}\ }\textbf {\bibinfo {volume} {74}},\ \bibinfo
  {pages} {1875} (\bibinfo {year} {1995})}\BibitemShut {NoStop}%
\bibitem [{\citenamefont {W\"orner}\ \emph {et~al.}(2006)\citenamefont
  {W\"orner}, \citenamefont {van~der Veen},\ and\ \citenamefont
  {Merkt}}]{Worner2006}%
  \BibitemOpen
  \bibfield  {author} {\bibinfo {author} {\bibfnamefont {H.~J.}\ \bibnamefont
  {W\"orner}}, \bibinfo {author} {\bibfnamefont {R.}~\bibnamefont {van~der
  Veen}}, \ and\ \bibinfo {author} {\bibfnamefont {F.}~\bibnamefont {Merkt}},\
  }\href {\doibase 10.1103/PhysRevLett.97.173003} {\bibfield  {journal}
  {\bibinfo  {journal} {Phys. Rev. Lett.}\ }\textbf {\bibinfo {volume} {97}},\
  \bibinfo {pages} {173003} (\bibinfo {year} {2006})}\BibitemShut {NoStop}%
\bibitem [{\citenamefont {Fu}\ \emph {et~al.}(2009)\citenamefont {Fu},
  \citenamefont {Santori}, \citenamefont {Barclay}, \citenamefont {Rogers},
  \citenamefont {Manson},\ and\ \citenamefont {Beausoleil}}]{Fu2009}%
  \BibitemOpen
  \bibfield  {author} {\bibinfo {author} {\bibfnamefont {K.-M.~C.}\
  \bibnamefont {Fu}}, \bibinfo {author} {\bibfnamefont {C.}~\bibnamefont
  {Santori}}, \bibinfo {author} {\bibfnamefont {P.~E.}\ \bibnamefont
  {Barclay}}, \bibinfo {author} {\bibfnamefont {L.~J.}\ \bibnamefont {Rogers}},
  \bibinfo {author} {\bibfnamefont {N.~B.}\ \bibnamefont {Manson}}, \ and\
  \bibinfo {author} {\bibfnamefont {R.~G.}\ \bibnamefont {Beausoleil}},\ }\href
  {\doibase 10.1103/PhysRevLett.103.256404} {\bibfield  {journal} {\bibinfo
  {journal} {Phys. Rev. Lett.}\ }\textbf {\bibinfo {volume} {103}},\ \bibinfo
  {pages} {256404} (\bibinfo {year} {2009})}\BibitemShut {NoStop}%
\bibitem [{\citenamefont {Iwahara}\ \emph {et~al.}(2010)\citenamefont
  {Iwahara}, \citenamefont {Sato}, \citenamefont {Tanaka},\ and\ \citenamefont
  {Chibotaru}}]{Iwahara2010}%
  \BibitemOpen
  \bibfield  {author} {\bibinfo {author} {\bibfnamefont {N.}~\bibnamefont
  {Iwahara}}, \bibinfo {author} {\bibfnamefont {T.}~\bibnamefont {Sato}},
  \bibinfo {author} {\bibfnamefont {K.}~\bibnamefont {Tanaka}}, \ and\ \bibinfo
  {author} {\bibfnamefont {L.~F.}\ \bibnamefont {Chibotaru}},\ }\href {\doibase
  10.1103/PhysRevB.82.245409} {\bibfield  {journal} {\bibinfo  {journal} {Phys.
  Rev. B}\ }\textbf {\bibinfo {volume} {82}},\ \bibinfo {pages} {245409}
  (\bibinfo {year} {2010})}\BibitemShut {NoStop}%
\bibitem [{\citenamefont {Abtew}\ \emph {et~al.}(2011)\citenamefont {Abtew},
  \citenamefont {Sun}, \citenamefont {Shih}, \citenamefont {Dev}, \citenamefont
  {Zhang},\ and\ \citenamefont {Zhang}}]{Abtew2011}%
  \BibitemOpen
  \bibfield  {author} {\bibinfo {author} {\bibfnamefont {T.~A.}\ \bibnamefont
  {Abtew}}, \bibinfo {author} {\bibfnamefont {Y.~Y.}\ \bibnamefont {Sun}},
  \bibinfo {author} {\bibfnamefont {B.-C.}\ \bibnamefont {Shih}}, \bibinfo
  {author} {\bibfnamefont {P.}~\bibnamefont {Dev}}, \bibinfo {author}
  {\bibfnamefont {S.~B.}\ \bibnamefont {Zhang}}, \ and\ \bibinfo {author}
  {\bibfnamefont {P.}~\bibnamefont {Zhang}},\ }\href {\doibase
  10.1103/PhysRevLett.107.146403} {\bibfield  {journal} {\bibinfo  {journal}
  {Phys. Rev. Lett.}\ }\textbf {\bibinfo {volume} {107}},\ \bibinfo {pages}
  {146403} (\bibinfo {year} {2011})}\BibitemShut {NoStop}%
\bibitem [{\citenamefont {Saha}\ \emph {et~al.}(2015)\citenamefont {Saha},
  \citenamefont {Sarma}, \citenamefont {Yang}, \citenamefont {van~de
  Meerakker}, \citenamefont {Parker},\ and\ \citenamefont
  {Western}}]{Saha2015}%
  \BibitemOpen
  \bibfield  {author} {\bibinfo {author} {\bibfnamefont {A.~K.}\ \bibnamefont
  {Saha}}, \bibinfo {author} {\bibfnamefont {G.}~\bibnamefont {Sarma}},
  \bibinfo {author} {\bibfnamefont {C.-H.}\ \bibnamefont {Yang}}, \bibinfo
  {author} {\bibfnamefont {S.~Y.~T.}\ \bibnamefont {van~de Meerakker}},
  \bibinfo {author} {\bibfnamefont {D.~H.}\ \bibnamefont {Parker}}, \ and\
  \bibinfo {author} {\bibfnamefont {C.~M.}\ \bibnamefont {Western}},\ }\href
  {\doibase 10.1039/C5CP01299F} {\bibfield  {journal} {\bibinfo  {journal}
  {Phys. Chem. Chem. Phys.}\ }\textbf {\bibinfo {volume} {17}},\ \bibinfo
  {pages} {14145} (\bibinfo {year} {2015})}\BibitemShut {NoStop}%
\bibitem [{\citenamefont {Kayanuma}\ and\ \citenamefont
  {Nakamura}(2017)}]{Kayanuma2017}%
  \BibitemOpen
  \bibfield  {author} {\bibinfo {author} {\bibfnamefont {Y.}~\bibnamefont
  {Kayanuma}}\ and\ \bibinfo {author} {\bibfnamefont {K.~G.}\ \bibnamefont
  {Nakamura}},\ }\href {\doibase 10.1103/PhysRevB.95.104302} {\bibfield
  {journal} {\bibinfo  {journal} {Phys. Rev. B}\ }\textbf {\bibinfo {volume}
  {95}},\ \bibinfo {pages} {104302} (\bibinfo {year} {2017})}\BibitemShut
  {NoStop}%
\bibitem [{\citenamefont {Zinchenko}\ \emph {et~al.}(2023)\citenamefont
  {Zinchenko}, \citenamefont {Ardana-Lamas}, \citenamefont {Lanfaloni},
  \citenamefont {Monahan}, \citenamefont {Seidu}, \citenamefont {Schuurman},
  \citenamefont {Neville},\ and\ \citenamefont {Wörner}}]{Zinchenko2023}%
  \BibitemOpen
  \bibfield  {author} {\bibinfo {author} {\bibfnamefont {K.~S.}\ \bibnamefont
  {Zinchenko}}, \bibinfo {author} {\bibfnamefont {F.}~\bibnamefont
  {Ardana-Lamas}}, \bibinfo {author} {\bibfnamefont {V.~U.}\ \bibnamefont
  {Lanfaloni}}, \bibinfo {author} {\bibfnamefont {N.}~\bibnamefont {Monahan}},
  \bibinfo {author} {\bibfnamefont {I.}~\bibnamefont {Seidu}}, \bibinfo
  {author} {\bibfnamefont {M.~S.}\ \bibnamefont {Schuurman}}, \bibinfo {author}
  {\bibfnamefont {S.~P.}\ \bibnamefont {Neville}}, \ and\ \bibinfo {author}
  {\bibfnamefont {H.~J.}\ \bibnamefont {Wörner}},\ }\href {\doibase
  10.1063/4.0000217} {\bibfield  {journal} {\bibinfo  {journal} {Structural
  Dynamics}\ }\textbf {\bibinfo {volume} {10}},\ \bibinfo {pages} {064303}
  (\bibinfo {year} {2023})}\BibitemShut {NoStop}%
\bibitem [{\citenamefont {Nandipati}\ and\ \citenamefont
  {Vendrell}(2023)}]{Nandipati2023}%
  \BibitemOpen
  \bibfield  {author} {\bibinfo {author} {\bibfnamefont {K.~R.}\ \bibnamefont
  {Nandipati}}\ and\ \bibinfo {author} {\bibfnamefont {O.}~\bibnamefont
  {Vendrell}},\ }\href {\doibase 10.1103/PhysRevA.107.L061101} {\bibfield
  {journal} {\bibinfo  {journal} {Phys. Rev. A}\ }\textbf {\bibinfo {volume}
  {107}},\ \bibinfo {pages} {L061101} (\bibinfo {year} {2023})}\BibitemShut
  {NoStop}%
\bibitem [{\citenamefont {Lyakhov}\ \emph {et~al.}(2023)\citenamefont
  {Lyakhov}, \citenamefont {Ovchinnikov}, \citenamefont {Bostrem},\ and\
  \citenamefont {Kishine}}]{Lyakhov2023}%
  \BibitemOpen
  \bibfield  {author} {\bibinfo {author} {\bibfnamefont {A.~D.}\ \bibnamefont
  {Lyakhov}}, \bibinfo {author} {\bibfnamefont {A.~S.}\ \bibnamefont
  {Ovchinnikov}}, \bibinfo {author} {\bibfnamefont {I.~G.}\ \bibnamefont
  {Bostrem}}, \ and\ \bibinfo {author} {\bibfnamefont {J.}~\bibnamefont
  {Kishine}},\ }\href {\doibase 10.1103/PhysRevB.108.115429} {\bibfield
  {journal} {\bibinfo  {journal} {Phys. Rev. B}\ }\textbf {\bibinfo {volume}
  {108}},\ \bibinfo {pages} {115429} (\bibinfo {year} {2023})}\BibitemShut
  {NoStop}%
\bibitem [{\citenamefont {Magoni}\ \emph {et~al.}(2023)\citenamefont {Magoni},
  \citenamefont {Joshi},\ and\ \citenamefont {Lesanovsky}}]{Magoni2023}%
  \BibitemOpen
  \bibfield  {author} {\bibinfo {author} {\bibfnamefont {M.}~\bibnamefont
  {Magoni}}, \bibinfo {author} {\bibfnamefont {R.}~\bibnamefont {Joshi}}, \
  and\ \bibinfo {author} {\bibfnamefont {I.}~\bibnamefont {Lesanovsky}},\
  }\href {\doibase 10.1103/PhysRevLett.131.093002} {\bibfield  {journal}
  {\bibinfo  {journal} {Phys. Rev. Lett.}\ }\textbf {\bibinfo {volume} {131}},\
  \bibinfo {pages} {093002} (\bibinfo {year} {2023})}\BibitemShut {NoStop}%
\bibitem [{\citenamefont {Sarychev}\ \emph {et~al.}(2024)\citenamefont
  {Sarychev}, \citenamefont {Zhevstovskikh}, \citenamefont {Ofitserova},
  \citenamefont {Korostelin}, \citenamefont {Ulanov}, \citenamefont {Egranov},
  \citenamefont {Surikov}, \citenamefont {Averkiev},\ and\ \citenamefont
  {Gudkov}}]{Sarychev2024}%
  \BibitemOpen
  \bibfield  {author} {\bibinfo {author} {\bibfnamefont {M.~N.}\ \bibnamefont
  {Sarychev}}, \bibinfo {author} {\bibfnamefont {I.~V.}\ \bibnamefont
  {Zhevstovskikh}}, \bibinfo {author} {\bibfnamefont {N.~Y.}\ \bibnamefont
  {Ofitserova}}, \bibinfo {author} {\bibfnamefont {Y.~V.}\ \bibnamefont
  {Korostelin}}, \bibinfo {author} {\bibfnamefont {V.~A.}\ \bibnamefont
  {Ulanov}}, \bibinfo {author} {\bibfnamefont {A.~V.}\ \bibnamefont {Egranov}},
  \bibinfo {author} {\bibfnamefont {V.~T.}\ \bibnamefont {Surikov}}, \bibinfo
  {author} {\bibfnamefont {N.~S.}\ \bibnamefont {Averkiev}}, \ and\ \bibinfo
  {author} {\bibfnamefont {V.~V.}\ \bibnamefont {Gudkov}},\ }\href {\doibase
  10.1103/PhysRevB.109.214104} {\bibfield  {journal} {\bibinfo  {journal}
  {Phys. Rev. B}\ }\textbf {\bibinfo {volume} {109}},\ \bibinfo {pages}
  {214104} (\bibinfo {year} {2024})}\BibitemShut {NoStop}%
\bibitem [{\citenamefont {Silkinis}\ \emph {et~al.}(2024)\citenamefont
  {Silkinis}, \citenamefont {\ifmmode~\check{Z}\else \v{Z}\fi{}alandauskas},
  \citenamefont {Thiering}, \citenamefont {Gali}, \citenamefont {Van~de Walle},
  \citenamefont {Alkauskas},\ and\ \citenamefont {Razinkovas}}]{Silkinis2024}%
  \BibitemOpen
  \bibfield  {author} {\bibinfo {author} {\bibfnamefont {R.}~\bibnamefont
  {Silkinis}}, \bibinfo {author} {\bibfnamefont {V.}~\bibnamefont
  {\ifmmode~\check{Z}\else \v{Z}\fi{}alandauskas}}, \bibinfo {author}
  {\bibfnamefont {G.~m.~H.}\ \bibnamefont {Thiering}}, \bibinfo {author}
  {\bibfnamefont {A.}~\bibnamefont {Gali}}, \bibinfo {author} {\bibfnamefont
  {C.~G.}\ \bibnamefont {Van~de Walle}}, \bibinfo {author} {\bibfnamefont
  {A.}~\bibnamefont {Alkauskas}}, \ and\ \bibinfo {author} {\bibfnamefont
  {L.}~\bibnamefont {Razinkovas}},\ }\href {\doibase
  10.1103/PhysRevB.110.075303} {\bibfield  {journal} {\bibinfo  {journal}
  {Phys. Rev. B}\ }\textbf {\bibinfo {volume} {110}},\ \bibinfo {pages}
  {075303} (\bibinfo {year} {2024})}\BibitemShut {NoStop}%
\bibitem [{\citenamefont {Yanagisawa}\ \emph {et~al.}(2024)\citenamefont
  {Yanagisawa}, \citenamefont {Hibino}, \citenamefont {Hidaka}, \citenamefont
  {Amitsuka}, \citenamefont {Tashima}, \citenamefont {Akatsu}, \citenamefont
  {Nemoto}, \citenamefont {Zherlitsyn},\ and\ \citenamefont
  {Wosnitza}}]{Yanagisawa2024}%
  \BibitemOpen
  \bibfield  {author} {\bibinfo {author} {\bibfnamefont {T.}~\bibnamefont
  {Yanagisawa}}, \bibinfo {author} {\bibfnamefont {R.}~\bibnamefont {Hibino}},
  \bibinfo {author} {\bibfnamefont {H.}~\bibnamefont {Hidaka}}, \bibinfo
  {author} {\bibfnamefont {H.}~\bibnamefont {Amitsuka}}, \bibinfo {author}
  {\bibfnamefont {T.}~\bibnamefont {Tashima}}, \bibinfo {author} {\bibfnamefont
  {M.}~\bibnamefont {Akatsu}}, \bibinfo {author} {\bibfnamefont
  {Y.}~\bibnamefont {Nemoto}}, \bibinfo {author} {\bibfnamefont
  {S.}~\bibnamefont {Zherlitsyn}}, \ and\ \bibinfo {author} {\bibfnamefont
  {J.}~\bibnamefont {Wosnitza}},\ }\href {https://arxiv.org/abs/2401.03377} {}
  (\bibinfo {year} {2024}),\ \Eprint {http://arxiv.org/abs/2401.03377}
  {arXiv:2401.03377 [cond-mat.mtrl-sci]} \BibitemShut {NoStop}%
\bibitem [{\citenamefont {Abragam}\ and\ \citenamefont
  {Bleaney}(1970)}]{Abragam1970}%
  \BibitemOpen
  \bibfield  {author} {\bibinfo {author} {\bibfnamefont {A.}~\bibnamefont
  {Abragam}}\ and\ \bibinfo {author} {\bibfnamefont {B.}~\bibnamefont
  {Bleaney}},\ }\href@noop {} {\emph {\bibinfo {title} {{Electron Paramagnetic
  Resonance of Transition Ions}}}}\ (\bibinfo  {publisher} {Clarendon Press},\
  \bibinfo {address} {Oxford},\ \bibinfo {year} {1970})\BibitemShut {NoStop}%
\bibitem [{\citenamefont {Bersuker}\ and\ \citenamefont
  {Polinger}(1989)}]{Bersuker1989}%
  \BibitemOpen
  \bibfield  {author} {\bibinfo {author} {\bibfnamefont {I.~B.}\ \bibnamefont
  {Bersuker}}\ and\ \bibinfo {author} {\bibfnamefont {V.~Z.}\ \bibnamefont
  {Polinger}},\ }\href {\doibase 10.1007/978-3-642-83479-0} {\emph {\bibinfo
  {title} {Vibronic Interactions in Molecules and Crystals}}}\ (\bibinfo
  {publisher} {Springer-Verlag},\ \bibinfo {address} {Berlin and Heidelberg},\
  \bibinfo {year} {1989})\BibitemShut {NoStop}%
\bibitem [{\citenamefont {Kim}\ \emph {et~al.}(2008)\citenamefont {Kim},
  \citenamefont {Jin}, \citenamefont {Moon}, \citenamefont {Kim}, \citenamefont
  {Park}, \citenamefont {Leem}, \citenamefont {Yu}, \citenamefont {Noh},
  \citenamefont {Kim}, \citenamefont {Oh}, \citenamefont {Park}, \citenamefont
  {Durairaj}, \citenamefont {Cao},\ and\ \citenamefont {Rotenberg}}]{Kim2008}%
  \BibitemOpen
  \bibfield  {author} {\bibinfo {author} {\bibfnamefont {B.~J.}\ \bibnamefont
  {Kim}}, \bibinfo {author} {\bibfnamefont {H.}~\bibnamefont {Jin}}, \bibinfo
  {author} {\bibfnamefont {S.~J.}\ \bibnamefont {Moon}}, \bibinfo {author}
  {\bibfnamefont {J.-Y.}\ \bibnamefont {Kim}}, \bibinfo {author} {\bibfnamefont
  {B.-G.}\ \bibnamefont {Park}}, \bibinfo {author} {\bibfnamefont {C.~S.}\
  \bibnamefont {Leem}}, \bibinfo {author} {\bibfnamefont {J.}~\bibnamefont
  {Yu}}, \bibinfo {author} {\bibfnamefont {T.~W.}\ \bibnamefont {Noh}},
  \bibinfo {author} {\bibfnamefont {C.}~\bibnamefont {Kim}}, \bibinfo {author}
  {\bibfnamefont {S.-J.}\ \bibnamefont {Oh}}, \bibinfo {author} {\bibfnamefont
  {J.-H.}\ \bibnamefont {Park}}, \bibinfo {author} {\bibfnamefont
  {V.}~\bibnamefont {Durairaj}}, \bibinfo {author} {\bibfnamefont
  {G.}~\bibnamefont {Cao}}, \ and\ \bibinfo {author} {\bibfnamefont
  {E.}~\bibnamefont {Rotenberg}},\ }\href {\doibase
  10.1103/PhysRevLett.101.076402} {\bibfield  {journal} {\bibinfo  {journal}
  {Phys. Rev. Lett.}\ }\textbf {\bibinfo {volume} {101}},\ \bibinfo {pages}
  {076402} (\bibinfo {year} {2008})}\BibitemShut {NoStop}%
\bibitem [{\citenamefont {Kim}\ \emph {et~al.}(2009)\citenamefont {Kim},
  \citenamefont {Ohsumi}, \citenamefont {Komesu}, \citenamefont {Sakai},
  \citenamefont {Morita}, \citenamefont {Takagi},\ and\ \citenamefont
  {Arima}}]{Kim2009}%
  \BibitemOpen
  \bibfield  {author} {\bibinfo {author} {\bibfnamefont {B.~J.}\ \bibnamefont
  {Kim}}, \bibinfo {author} {\bibfnamefont {H.}~\bibnamefont {Ohsumi}},
  \bibinfo {author} {\bibfnamefont {T.}~\bibnamefont {Komesu}}, \bibinfo
  {author} {\bibfnamefont {S.}~\bibnamefont {Sakai}}, \bibinfo {author}
  {\bibfnamefont {T.}~\bibnamefont {Morita}}, \bibinfo {author} {\bibfnamefont
  {H.}~\bibnamefont {Takagi}}, \ and\ \bibinfo {author} {\bibfnamefont
  {T.}~\bibnamefont {Arima}},\ }\href {\doibase 10.1126/science.1167106}
  {\bibfield  {journal} {\bibinfo  {journal} {Science}\ }\textbf {\bibinfo
  {volume} {323}},\ \bibinfo {pages} {1329} (\bibinfo {year}
  {2009})}\BibitemShut {NoStop}%
\bibitem [{\citenamefont {Gangopadhyay}\ and\ \citenamefont
  {Pickett}(2015)}]{Gangopadhyay2015}%
  \BibitemOpen
  \bibfield  {author} {\bibinfo {author} {\bibfnamefont {S.}~\bibnamefont
  {Gangopadhyay}}\ and\ \bibinfo {author} {\bibfnamefont {W.~E.}\ \bibnamefont
  {Pickett}},\ }\href {\doibase 10.1103/PhysRevB.91.045133} {\bibfield
  {journal} {\bibinfo  {journal} {Phys. Rev. B}\ }\textbf {\bibinfo {volume}
  {91}},\ \bibinfo {pages} {045133} (\bibinfo {year} {2015})}\BibitemShut
  {NoStop}%
\bibitem [{\citenamefont {Gangopadhyay}\ and\ \citenamefont
  {Pickett}(2016)}]{Gangopadhyay2016}%
  \BibitemOpen
  \bibfield  {author} {\bibinfo {author} {\bibfnamefont {S.}~\bibnamefont
  {Gangopadhyay}}\ and\ \bibinfo {author} {\bibfnamefont {W.~E.}\ \bibnamefont
  {Pickett}},\ }\href {\doibase 10.1103/PhysRevB.93.155126} {\bibfield
  {journal} {\bibinfo  {journal} {Phys. Rev. B}\ }\textbf {\bibinfo {volume}
  {93}},\ \bibinfo {pages} {155126} (\bibinfo {year} {2016})}\BibitemShut
  {NoStop}%
\bibitem [{\citenamefont {Grosso}\ and\ \citenamefont
  {Parravicini}(2014)}]{Grosso2014}%
  \BibitemOpen
  \bibfield  {author} {\bibinfo {author} {\bibfnamefont {G.}~\bibnamefont
  {Grosso}}\ and\ \bibinfo {author} {\bibfnamefont {G.}~\bibnamefont
  {Parravicini}},\ }\href@noop {} {\emph {\bibinfo {title} {Solid State
  Physics, 2nd ed.}}}\ (\bibinfo  {publisher} {Academic Press},\ \bibinfo
  {address} {Amsterdam},\ \bibinfo {year} {2014})\BibitemShut {NoStop}%
\bibitem [{\citenamefont {Sugano}\ \emph {et~al.}(1970)\citenamefont {Sugano},
  \citenamefont {Tanabe},\ and\ \citenamefont {Kamimura}}]{Sugano1970}%
  \BibitemOpen
  \bibfield  {author} {\bibinfo {author} {\bibfnamefont {S.}~\bibnamefont
  {Sugano}}, \bibinfo {author} {\bibfnamefont {Y.}~\bibnamefont {Tanabe}}, \
  and\ \bibinfo {author} {\bibfnamefont {H.}~\bibnamefont {Kamimura}},\
  }\href@noop {} {\emph {\bibinfo {title} {{Multiplets of Transition-Metal Ions
  in Crystals}}}}\ (\bibinfo  {publisher} {Academic Press},\ \bibinfo {address}
  {New York},\ \bibinfo {year} {1970})\BibitemShut {NoStop}%
\bibitem [{\citenamefont {Tanabe}\ and\ \citenamefont
  {Sugano}(1954{\natexlab{a}})}]{Tanabe1954I}%
  \BibitemOpen
  \bibfield  {author} {\bibinfo {author} {\bibfnamefont {Y.}~\bibnamefont
  {Tanabe}}\ and\ \bibinfo {author} {\bibfnamefont {S.}~\bibnamefont
  {Sugano}},\ }\href {\doibase 10.1143/JPSJ.9.753} {\bibfield  {journal}
  {\bibinfo  {journal} {Journal of the Physical Society of Japan}\ }\textbf
  {\bibinfo {volume} {9}},\ \bibinfo {pages} {753} (\bibinfo {year}
  {1954}{\natexlab{a}})}\BibitemShut {NoStop}%
\bibitem [{\citenamefont {Koster}\ \emph {et~al.}(1963)\citenamefont {Koster},
  \citenamefont {Dimmock}, \citenamefont {Wheeler},\ and\ \citenamefont
  {Statz}}]{Koster1963}%
  \BibitemOpen
  \bibfield  {author} {\bibinfo {author} {\bibfnamefont {G.~F.}\ \bibnamefont
  {Koster}}, \bibinfo {author} {\bibfnamefont {J.~O.}\ \bibnamefont {Dimmock}},
  \bibinfo {author} {\bibfnamefont {R.~G.}\ \bibnamefont {Wheeler}}, \ and\
  \bibinfo {author} {\bibfnamefont {H.}~\bibnamefont {Statz}},\ }\href@noop {}
  {\emph {\bibinfo {title} {Properties of the thirty-two point groups}}}\
  (\bibinfo  {publisher} {MIT press},\ \bibinfo {address} {Massachusetts},\
  \bibinfo {year} {1963})\BibitemShut {NoStop}%
\bibitem [{\citenamefont {Inui}\ \emph {et~al.}(1990)\citenamefont {Inui},
  \citenamefont {Tanabe},\ and\ \citenamefont {Onodera}}]{Inui1990}%
  \BibitemOpen
  \bibfield  {author} {\bibinfo {author} {\bibfnamefont {T.}~\bibnamefont
  {Inui}}, \bibinfo {author} {\bibfnamefont {Y.}~\bibnamefont {Tanabe}}, \ and\
  \bibinfo {author} {\bibfnamefont {Y.}~\bibnamefont {Onodera}},\ }\href@noop
  {} {\emph {\bibinfo {title} {Group Theory and Its Applications in Physics}}}\
  (\bibinfo  {publisher} {Springer-Verlag},\ \bibinfo {address} {Berlin and
  Heidelberg},\ \bibinfo {year} {1990})\BibitemShut {NoStop}%
\bibitem [{\citenamefont {Sugano}\ and\ \citenamefont
  {Shulman}(1963)}]{Sugano1963}%
  \BibitemOpen
  \bibfield  {author} {\bibinfo {author} {\bibfnamefont {S.}~\bibnamefont
  {Sugano}}\ and\ \bibinfo {author} {\bibfnamefont {R.~G.}\ \bibnamefont
  {Shulman}},\ }\href {\doibase 10.1103/PhysRev.130.517} {\bibfield  {journal}
  {\bibinfo  {journal} {Phys. Rev.}\ }\textbf {\bibinfo {volume} {130}},\
  \bibinfo {pages} {517} (\bibinfo {year} {1963})}\BibitemShut {NoStop}%
\bibitem [{Note1()}]{Note1}%
  \BibitemOpen
  \bibinfo {note} {In the octahedral environment, the degeneracy of the
  $t_{2g}$ orbitals ($yz$, $zx$, and $xy$) is intuitively clear because their
  physical equivalence in the octahedral environment is apparent. In contrast,
  the degeneracy of the $e_g$ orbitals is unclear. We can understand the
  degeneracy of the $2z^2-x^2-y^2$ ($z^2$ for simplicity) and $x^2-y^2$
  orbitals from the fact that their linear combinations generate physically
  equivalent pairs, $x^2$ and $y^2-z^2$ orbitals and $y^2$ and $z^2-x^2$
  orbitals.}\BibitemShut {Stop}%
\bibitem [{\citenamefont {Tanabe}\ and\ \citenamefont
  {Sugano}(1954{\natexlab{b}})}]{Tanabe1954II}%
  \BibitemOpen
  \bibfield  {author} {\bibinfo {author} {\bibfnamefont {Y.}~\bibnamefont
  {Tanabe}}\ and\ \bibinfo {author} {\bibfnamefont {S.}~\bibnamefont
  {Sugano}},\ }\href {\doibase 10.1143/JPSJ.9.766} {\bibfield  {journal}
  {\bibinfo  {journal} {Journal of the Physical Society of Japan}\ }\textbf
  {\bibinfo {volume} {9}},\ \bibinfo {pages} {766} (\bibinfo {year}
  {1954}{\natexlab{b}})}\BibitemShut {NoStop}%
\bibitem [{Note2()}]{Note2}%
  \BibitemOpen
  \bibinfo {note} {In the orbitally nondegenerate high-spin states, a weak JT
  effect can arise via the hybridization between the nondegenerate states and
  excited orbitally degenerate states \cite {Warren1984, Halliday1988,
  Iwahara2017, Streltsov2020, Streltsov2022, Warzanowski2024b}.}\BibitemShut
  {Stop}%
\bibitem [{\citenamefont {Landau}\ and\ \citenamefont
  {Lifshitz}(1977)}]{LandauQM}%
  \BibitemOpen
  \bibfield  {author} {\bibinfo {author} {\bibfnamefont {L.~D.}\ \bibnamefont
  {Landau}}\ and\ \bibinfo {author} {\bibfnamefont {E.~M.}\ \bibnamefont
  {Lifshitz}},\ }\href@noop {} {\emph {\bibinfo {title} {Quantum Mechanics
  (Non-Relativistic Theory), Third Edition}}}\ (\bibinfo  {publisher}
  {Butterworth-Heinemann},\ \bibinfo {address} {Oxford},\ \bibinfo {year}
  {1977})\BibitemShut {NoStop}%
\bibitem [{\citenamefont {Voleti}\ \emph {et~al.}(2020)\citenamefont {Voleti},
  \citenamefont {Maharaj}, \citenamefont {Gaulin}, \citenamefont {Luke},\ and\
  \citenamefont {Paramekanti}}]{Voleti2020}%
  \BibitemOpen
  \bibfield  {author} {\bibinfo {author} {\bibfnamefont {S.}~\bibnamefont
  {Voleti}}, \bibinfo {author} {\bibfnamefont {D.~D.}\ \bibnamefont {Maharaj}},
  \bibinfo {author} {\bibfnamefont {B.~D.}\ \bibnamefont {Gaulin}}, \bibinfo
  {author} {\bibfnamefont {G.}~\bibnamefont {Luke}}, \ and\ \bibinfo {author}
  {\bibfnamefont {A.}~\bibnamefont {Paramekanti}},\ }\href {\doibase
  10.1103/PhysRevB.101.155118} {\bibfield  {journal} {\bibinfo  {journal}
  {Phys. Rev. B}\ }\textbf {\bibinfo {volume} {101}},\ \bibinfo {pages}
  {155118} (\bibinfo {year} {2020})}\BibitemShut {NoStop}%
\bibitem [{\citenamefont {Maharaj}\ \emph {et~al.}(2020)\citenamefont
  {Maharaj}, \citenamefont {Sala}, \citenamefont {Stone}, \citenamefont
  {Kermarrec}, \citenamefont {Ritter}, \citenamefont {Fauth}, \citenamefont
  {Marjerrison}, \citenamefont {Greedan}, \citenamefont {Paramekanti},\ and\
  \citenamefont {Gaulin}}]{Maharaj2020}%
  \BibitemOpen
  \bibfield  {author} {\bibinfo {author} {\bibfnamefont {D.~D.}\ \bibnamefont
  {Maharaj}}, \bibinfo {author} {\bibfnamefont {G.}~\bibnamefont {Sala}},
  \bibinfo {author} {\bibfnamefont {M.~B.}\ \bibnamefont {Stone}}, \bibinfo
  {author} {\bibfnamefont {E.}~\bibnamefont {Kermarrec}}, \bibinfo {author}
  {\bibfnamefont {C.}~\bibnamefont {Ritter}}, \bibinfo {author} {\bibfnamefont
  {F.}~\bibnamefont {Fauth}}, \bibinfo {author} {\bibfnamefont {C.~A.}\
  \bibnamefont {Marjerrison}}, \bibinfo {author} {\bibfnamefont {J.~E.}\
  \bibnamefont {Greedan}}, \bibinfo {author} {\bibfnamefont {A.}~\bibnamefont
  {Paramekanti}}, \ and\ \bibinfo {author} {\bibfnamefont {B.~D.}\ \bibnamefont
  {Gaulin}},\ }\href {\doibase 10.1103/PhysRevLett.124.087206} {\bibfield
  {journal} {\bibinfo  {journal} {Phys. Rev. Lett.}\ }\textbf {\bibinfo
  {volume} {124}},\ \bibinfo {pages} {087206} (\bibinfo {year}
  {2020})}\BibitemShut {NoStop}%
\bibitem [{\citenamefont {Pradhan}\ \emph {et~al.}(2024)\citenamefont
  {Pradhan}, \citenamefont {Paramekanti},\ and\ \citenamefont
  {Saha-Dasgupta}}]{Pradhan2024}%
  \BibitemOpen
  \bibfield  {author} {\bibinfo {author} {\bibfnamefont {K.}~\bibnamefont
  {Pradhan}}, \bibinfo {author} {\bibfnamefont {A.}~\bibnamefont
  {Paramekanti}}, \ and\ \bibinfo {author} {\bibfnamefont {T.}~\bibnamefont
  {Saha-Dasgupta}},\ }\href {\doibase 10.1103/PhysRevB.109.184416} {\bibfield
  {journal} {\bibinfo  {journal} {Phys. Rev. B}\ }\textbf {\bibinfo {volume}
  {109}},\ \bibinfo {pages} {184416} (\bibinfo {year} {2024})}\BibitemShut
  {NoStop}%
\bibitem [{\citenamefont {Thompson}\ \emph {et~al.}(2014)\citenamefont
  {Thompson}, \citenamefont {Carlo}, \citenamefont {Flacau}, \citenamefont
  {Aharen}, \citenamefont {Leahy}, \citenamefont {Pollichemi}, \citenamefont
  {Munsie}, \citenamefont {Medina}, \citenamefont {Luke}, \citenamefont
  {Munevar}, \citenamefont {Cheung}, \citenamefont {Goko}, \citenamefont
  {Uemura},\ and\ \citenamefont {Greedan}}]{Thompson2014}%
  \BibitemOpen
  \bibfield  {author} {\bibinfo {author} {\bibfnamefont {C.~M.}\ \bibnamefont
  {Thompson}}, \bibinfo {author} {\bibfnamefont {J.~P.}\ \bibnamefont {Carlo}},
  \bibinfo {author} {\bibfnamefont {R.}~\bibnamefont {Flacau}}, \bibinfo
  {author} {\bibfnamefont {T.}~\bibnamefont {Aharen}}, \bibinfo {author}
  {\bibfnamefont {I.~A.}\ \bibnamefont {Leahy}}, \bibinfo {author}
  {\bibfnamefont {J.~R.}\ \bibnamefont {Pollichemi}}, \bibinfo {author}
  {\bibfnamefont {T.~J.~S.}\ \bibnamefont {Munsie}}, \bibinfo {author}
  {\bibfnamefont {T.}~\bibnamefont {Medina}}, \bibinfo {author} {\bibfnamefont
  {G.~M.}\ \bibnamefont {Luke}}, \bibinfo {author} {\bibfnamefont
  {J.}~\bibnamefont {Munevar}}, \bibinfo {author} {\bibfnamefont
  {S.}~\bibnamefont {Cheung}}, \bibinfo {author} {\bibfnamefont
  {T.}~\bibnamefont {Goko}}, \bibinfo {author} {\bibfnamefont {Y.~J.}\
  \bibnamefont {Uemura}}, \ and\ \bibinfo {author} {\bibfnamefont {J.~E.}\
  \bibnamefont {Greedan}},\ }\href {\doibase 10.1088/0953-8984/26/30/306003}
  {\bibfield  {journal} {\bibinfo  {journal} {Journal of Physics: Condensed
  Matter}\ }\textbf {\bibinfo {volume} {26}},\ \bibinfo {pages} {306003}
  (\bibinfo {year} {2014})}\BibitemShut {NoStop}%
\bibitem [{\citenamefont {Marjerrison}\ \emph
  {et~al.}(2016{\natexlab{b}})\citenamefont {Marjerrison}, \citenamefont
  {Thompson}, \citenamefont {Sharma}, \citenamefont {Hallas}, \citenamefont
  {Wilson}, \citenamefont {Munsie}, \citenamefont {Flacau}, \citenamefont
  {Wiebe}, \citenamefont {Gaulin}, \citenamefont {Luke},\ and\ \citenamefont
  {Greedan}}]{Marjerrison2016b}%
  \BibitemOpen
  \bibfield  {author} {\bibinfo {author} {\bibfnamefont {C.~A.}\ \bibnamefont
  {Marjerrison}}, \bibinfo {author} {\bibfnamefont {C.~M.}\ \bibnamefont
  {Thompson}}, \bibinfo {author} {\bibfnamefont {A.~Z.}\ \bibnamefont
  {Sharma}}, \bibinfo {author} {\bibfnamefont {A.~M.}\ \bibnamefont {Hallas}},
  \bibinfo {author} {\bibfnamefont {M.~N.}\ \bibnamefont {Wilson}}, \bibinfo
  {author} {\bibfnamefont {T.~J.~S.}\ \bibnamefont {Munsie}}, \bibinfo {author}
  {\bibfnamefont {R.}~\bibnamefont {Flacau}}, \bibinfo {author} {\bibfnamefont
  {C.~R.}\ \bibnamefont {Wiebe}}, \bibinfo {author} {\bibfnamefont {B.~D.}\
  \bibnamefont {Gaulin}}, \bibinfo {author} {\bibfnamefont {G.~M.}\
  \bibnamefont {Luke}}, \ and\ \bibinfo {author} {\bibfnamefont {J.~E.}\
  \bibnamefont {Greedan}},\ }\href {\doibase 10.1103/PhysRevB.94.134429}
  {\bibfield  {journal} {\bibinfo  {journal} {Phys. Rev. B}\ }\textbf {\bibinfo
  {volume} {94}},\ \bibinfo {pages} {134429} (\bibinfo {year}
  {2016}{\natexlab{b}})}\BibitemShut {NoStop}%
\bibitem [{\citenamefont {Nilsen}\ \emph {et~al.}(2021)\citenamefont {Nilsen},
  \citenamefont {Thompson}, \citenamefont {Marjerisson}, \citenamefont
  {Badrtdinov}, \citenamefont {Tsirlin},\ and\ \citenamefont
  {Greedan}}]{Nilsen2021}%
  \BibitemOpen
  \bibfield  {author} {\bibinfo {author} {\bibfnamefont {G.~J.}\ \bibnamefont
  {Nilsen}}, \bibinfo {author} {\bibfnamefont {C.~M.}\ \bibnamefont
  {Thompson}}, \bibinfo {author} {\bibfnamefont {C.}~\bibnamefont
  {Marjerisson}}, \bibinfo {author} {\bibfnamefont {D.~I.}\ \bibnamefont
  {Badrtdinov}}, \bibinfo {author} {\bibfnamefont {A.~A.}\ \bibnamefont
  {Tsirlin}}, \ and\ \bibinfo {author} {\bibfnamefont {J.~E.}\ \bibnamefont
  {Greedan}},\ }\href {\doibase 10.1103/PhysRevB.103.104430} {\bibfield
  {journal} {\bibinfo  {journal} {Phys. Rev. B}\ }\textbf {\bibinfo {volume}
  {103}},\ \bibinfo {pages} {104430} (\bibinfo {year} {2021})}\BibitemShut
  {NoStop}%
\bibitem [{\citenamefont {Jahn}\ and\ \citenamefont {Teller}(1937)}]{Jahn1937}%
  \BibitemOpen
  \bibfield  {author} {\bibinfo {author} {\bibfnamefont {H.~A.}\ \bibnamefont
  {Jahn}}\ and\ \bibinfo {author} {\bibfnamefont {E.}~\bibnamefont {Teller}},\
  }\href {\doibase 10.1098/rspa.1937.0142} {\bibfield  {journal} {\bibinfo
  {journal} {Proceedings of the Royal Society of London. Series A -
  Mathematical and Physical Sciences}\ }\textbf {\bibinfo {volume} {161}},\
  \bibinfo {pages} {220} (\bibinfo {year} {1937})}\BibitemShut {NoStop}%
\bibitem [{\citenamefont {Jahn}(1938)}]{Jahn1938}%
  \BibitemOpen
  \bibfield  {author} {\bibinfo {author} {\bibfnamefont {H.~A.}\ \bibnamefont
  {Jahn}},\ }\href {\doibase 10.1098/rspa.1938.0008} {\bibfield  {journal}
  {\bibinfo  {journal} {Proceedings of the Royal Society of London. Series A -
  Mathematical and Physical Sciences}\ }\textbf {\bibinfo {volume} {164}},\
  \bibinfo {pages} {117} (\bibinfo {year} {1938})}\BibitemShut {NoStop}%
\bibitem [{\citenamefont {{E. B. Wilson, Jr.}}\ \emph
  {et~al.}(1980)\citenamefont {{E. B. Wilson, Jr.}}, \citenamefont {Decius},\
  and\ \citenamefont {Cross}}]{Wilson1980}%
  \BibitemOpen
  \bibfield  {author} {\bibinfo {author} {\bibnamefont {{E. B. Wilson, Jr.}}},
  \bibinfo {author} {\bibfnamefont {J.~C.}\ \bibnamefont {Decius}}, \ and\
  \bibinfo {author} {\bibfnamefont {P.~C.}\ \bibnamefont {Cross}},\ }\href@noop
  {} {\emph {\bibinfo {title} {Molecular Vibrations: The Theory of Infrared and
  Raman Vibrational Spectra}}}\ (\bibinfo  {publisher} {Dover},\ \bibinfo
  {address} {New York},\ \bibinfo {year} {1980})\BibitemShut {NoStop}%
\bibitem [{\citenamefont {Iwahara}\ and\ \citenamefont
  {Shikano}(2023)}]{Iwahara2023Ru}%
  \BibitemOpen
  \bibfield  {author} {\bibinfo {author} {\bibfnamefont {N.}~\bibnamefont
  {Iwahara}}\ and\ \bibinfo {author} {\bibfnamefont {S.}~\bibnamefont
  {Shikano}},\ }\href {\doibase 10.1103/PhysRevResearch.5.023051} {\bibfield
  {journal} {\bibinfo  {journal} {Phys. Rev. Res.}\ }\textbf {\bibinfo {volume}
  {5}},\ \bibinfo {pages} {023051} (\bibinfo {year} {2023})}\BibitemShut
  {NoStop}%
\bibitem [{\citenamefont {Liehr}(1963)}]{Liehr1963}%
  \BibitemOpen
  \bibfield  {author} {\bibinfo {author} {\bibfnamefont {A.~D.}\ \bibnamefont
  {Liehr}},\ }\href {\doibase 10.1021/j100796a043} {\bibfield  {journal}
  {\bibinfo  {journal} {The Journal of Physical Chemistry}\ }\textbf {\bibinfo
  {volume} {67}},\ \bibinfo {pages} {389} (\bibinfo {year} {1963})}\BibitemShut
  {NoStop}%
\bibitem [{Note3()}]{Note3}%
  \BibitemOpen
  \bibinfo {note} {Note that this rotation is not the real rotation of the
  octahedra.}\BibitemShut {Stop}%
\bibitem [{\citenamefont {Berry}(1984)}]{Berry1984}%
  \BibitemOpen
  \bibfield  {author} {\bibinfo {author} {\bibfnamefont {M.~V.}\ \bibnamefont
  {Berry}},\ }\href {http://www.jstor.org/stable/2397741} {\bibfield  {journal}
  {\bibinfo  {journal} {Proceedings of the Royal Society of London. Series A,
  Mathematical and Physical Sciences}\ }\textbf {\bibinfo {volume} {392}},\
  \bibinfo {pages} {45} (\bibinfo {year} {1984})}\BibitemShut {NoStop}%
\bibitem [{\citenamefont {Warren}(1984)}]{Warren1984}%
  \BibitemOpen
  \bibfield  {author} {\bibinfo {author} {\bibfnamefont {K.~D.}\ \bibnamefont
  {Warren}},\ }in\ \href {\doibase https://doi.org/10.1007/BFb0111455} {\emph
  {\bibinfo {booktitle} {Complex Chemistry}}}\ (\bibinfo {organization}
  {Springer},\ \bibinfo {year} {1984})\ p.\ \bibinfo {pages} {119}\BibitemShut
  {NoStop}%
\bibitem [{\citenamefont {Streltsov}\ and\ \citenamefont
  {Khomskii}(2020)}]{Streltsov2020}%
  \BibitemOpen
  \bibfield  {author} {\bibinfo {author} {\bibfnamefont {S.~V.}\ \bibnamefont
  {Streltsov}}\ and\ \bibinfo {author} {\bibfnamefont {D.~I.}\ \bibnamefont
  {Khomskii}},\ }\href {\doibase 10.1103/PhysRevX.10.031043} {\bibfield
  {journal} {\bibinfo  {journal} {Phys. Rev. X}\ }\textbf {\bibinfo {volume}
  {10}},\ \bibinfo {pages} {031043} (\bibinfo {year} {2020})}\BibitemShut
  {NoStop}%
\bibitem [{\citenamefont {Streltsov}\ \emph {et~al.}(2022)\citenamefont
  {Streltsov}, \citenamefont {Temnikov}, \citenamefont {Kugel},\ and\
  \citenamefont {Khomskii}}]{Streltsov2022}%
  \BibitemOpen
  \bibfield  {author} {\bibinfo {author} {\bibfnamefont {S.~V.}\ \bibnamefont
  {Streltsov}}, \bibinfo {author} {\bibfnamefont {F.~V.}\ \bibnamefont
  {Temnikov}}, \bibinfo {author} {\bibfnamefont {K.~I.}\ \bibnamefont {Kugel}},
  \ and\ \bibinfo {author} {\bibfnamefont {D.~I.}\ \bibnamefont {Khomskii}},\
  }\href {\doibase 10.1103/PhysRevB.105.205142} {\bibfield  {journal} {\bibinfo
   {journal} {Phys. Rev. B}\ }\textbf {\bibinfo {volume} {105}},\ \bibinfo
  {pages} {205142} (\bibinfo {year} {2022})}\BibitemShut {NoStop}%
\bibitem [{\citenamefont {P\'{a}sztorov\'{a}}\ \emph
  {et~al.}(2023{\natexlab{a}})\citenamefont {P\'{a}sztorov\'{a}}, \citenamefont
  {Bi}, \citenamefont {Gaal}, \citenamefont {Kr\"{a}mer}, \citenamefont
  {\v{Z}ivkovi\'{a}},\ and\ \citenamefont {R{\o}nnow}}]{Pasztorova2023}%
  \BibitemOpen
  \bibfield  {author} {\bibinfo {author} {\bibfnamefont {J.}~\bibnamefont
  {P\'{a}sztorov\'{a}}}, \bibinfo {author} {\bibfnamefont {W.~H.}\ \bibnamefont
  {Bi}}, \bibinfo {author} {\bibfnamefont {R.}~\bibnamefont {Gaal}}, \bibinfo
  {author} {\bibfnamefont {K.}~\bibnamefont {Kr\"{a}mer}}, \bibinfo {author}
  {\bibfnamefont {I.}~\bibnamefont {\v{Z}ivkovi\'{a}}}, \ and\ \bibinfo
  {author} {\bibfnamefont {H.~M.}\ \bibnamefont {R{\o}nnow}},\ }\href {\doibase
  https://doi.org/10.1016/j.jssc.2023.124184} {\bibfield  {journal} {\bibinfo
  {journal} {J. Solid State Chem.}\ }\textbf {\bibinfo {volume} {326}},\
  \bibinfo {pages} {124184} (\bibinfo {year} {2023}{\natexlab{a}})}\BibitemShut
  {NoStop}%
\bibitem [{\citenamefont {Yuan}\ \emph {et~al.}(2017)\citenamefont {Yuan},
  \citenamefont {Clancy}, \citenamefont {Cook}, \citenamefont {Thompson},
  \citenamefont {Greedan}, \citenamefont {Cao}, \citenamefont {Jeon},
  \citenamefont {Noh}, \citenamefont {Upton}, \citenamefont {Casa},
  \citenamefont {Gog}, \citenamefont {Paramekanti},\ and\ \citenamefont
  {Kim}}]{Yuan2017}%
  \BibitemOpen
  \bibfield  {author} {\bibinfo {author} {\bibfnamefont {B.}~\bibnamefont
  {Yuan}}, \bibinfo {author} {\bibfnamefont {J.~P.}\ \bibnamefont {Clancy}},
  \bibinfo {author} {\bibfnamefont {A.~M.}\ \bibnamefont {Cook}}, \bibinfo
  {author} {\bibfnamefont {C.~M.}\ \bibnamefont {Thompson}}, \bibinfo {author}
  {\bibfnamefont {J.}~\bibnamefont {Greedan}}, \bibinfo {author} {\bibfnamefont
  {G.}~\bibnamefont {Cao}}, \bibinfo {author} {\bibfnamefont {B.~C.}\
  \bibnamefont {Jeon}}, \bibinfo {author} {\bibfnamefont {T.~W.}\ \bibnamefont
  {Noh}}, \bibinfo {author} {\bibfnamefont {M.~H.}\ \bibnamefont {Upton}},
  \bibinfo {author} {\bibfnamefont {D.}~\bibnamefont {Casa}}, \bibinfo {author}
  {\bibfnamefont {T.}~\bibnamefont {Gog}}, \bibinfo {author} {\bibfnamefont
  {A.}~\bibnamefont {Paramekanti}}, \ and\ \bibinfo {author} {\bibfnamefont
  {Y.-J.}\ \bibnamefont {Kim}},\ }\href {\doibase 10.1103/PhysRevB.95.235114}
  {\bibfield  {journal} {\bibinfo  {journal} {Phys. Rev. B}\ }\textbf {\bibinfo
  {volume} {95}},\ \bibinfo {pages} {235114} (\bibinfo {year}
  {2017})}\BibitemShut {NoStop}%
\bibitem [{Note4()}]{Note4}%
  \BibitemOpen
  \bibinfo {note} {We do not consider the pseudo-spin sector because it is
  irrelevant to the vibronic coupling.}\BibitemShut {Stop}%
\bibitem [{\citenamefont {Ham}(1987)}]{Ham1987}%
  \BibitemOpen
  \bibfield  {author} {\bibinfo {author} {\bibfnamefont {F.~S.}\ \bibnamefont
  {Ham}},\ }\href {\doibase 10.1103/PhysRevLett.58.725} {\bibfield  {journal}
  {\bibinfo  {journal} {Phys. Rev. Lett.}\ }\textbf {\bibinfo {volume} {58}},\
  \bibinfo {pages} {725} (\bibinfo {year} {1987})}\BibitemShut {NoStop}%
\bibitem [{Note5()}]{Note5}%
  \BibitemOpen
  \bibinfo {note} {The warping is much smaller than the JT stabilization
  energy.}\BibitemShut {Stop}%
\bibitem [{\citenamefont {Koizumi}\ and\ \citenamefont
  {Bersuker}(1999)}]{Koizumi1999}%
  \BibitemOpen
  \bibfield  {author} {\bibinfo {author} {\bibfnamefont {H.}~\bibnamefont
  {Koizumi}}\ and\ \bibinfo {author} {\bibfnamefont {I.~B.}\ \bibnamefont
  {Bersuker}},\ }\href {\doibase 10.1103/PhysRevLett.83.3009} {\bibfield
  {journal} {\bibinfo  {journal} {Phys. Rev. Lett.}\ }\textbf {\bibinfo
  {volume} {83}},\ \bibinfo {pages} {3009} (\bibinfo {year}
  {1999})}\BibitemShut {NoStop}%
\bibitem [{\citenamefont {O’Brien}\ and\ \citenamefont
  {Chancey}(1993)}]{OBrien1993}%
  \BibitemOpen
  \bibfield  {author} {\bibinfo {author} {\bibfnamefont {M.~C.~M.}\
  \bibnamefont {O’Brien}}\ and\ \bibinfo {author} {\bibfnamefont {C.~C.}\
  \bibnamefont {Chancey}},\ }\href {\doibase 10.1119/1.17197} {\bibfield
  {journal} {\bibinfo  {journal} {American Journal of Physics}\ }\textbf
  {\bibinfo {volume} {61}},\ \bibinfo {pages} {688} (\bibinfo {year}
  {1993})}\BibitemShut {NoStop}%
\bibitem [{\citenamefont {Sakurai}(1994)}]{Sakurai1994}%
  \BibitemOpen
  \bibfield  {author} {\bibinfo {author} {\bibfnamefont {J.~J.}\ \bibnamefont
  {Sakurai}},\ }\href@noop {} {\emph {\bibinfo {title} {Modern Quantum
  Mechanics, Revised Ed.}}}\ (\bibinfo  {publisher} {Addison-Wesley Publishing
  Company},\ \bibinfo {address} {Massachusettsu},\ \bibinfo {year}
  {1994})\BibitemShut {NoStop}%
\bibitem [{\citenamefont {Bohm}\ \emph {et~al.}(2003)\citenamefont {Bohm},
  \citenamefont {Mostafazadeh}, \citenamefont {Koizumi}, \citenamefont {Niu},\
  and\ \citenamefont {Zwanziger}}]{Bohm2003}%
  \BibitemOpen
  \bibfield  {author} {\bibinfo {author} {\bibfnamefont {A.}~\bibnamefont
  {Bohm}}, \bibinfo {author} {\bibfnamefont {A.}~\bibnamefont {Mostafazadeh}},
  \bibinfo {author} {\bibfnamefont {H.}~\bibnamefont {Koizumi}}, \bibinfo
  {author} {\bibfnamefont {Q.}~\bibnamefont {Niu}}, \ and\ \bibinfo {author}
  {\bibfnamefont {J.}~\bibnamefont {Zwanziger}},\ }\href {\doibase
  https://doi.org/10.1007/978-3-662-10333-3} {\emph {\bibinfo {title} {The
  Geometric Phase in Quantum Systems}}}\ (\bibinfo  {publisher} {Springer
  Berlin},\ \bibinfo {address} {Heidelberg},\ \bibinfo {year}
  {2003})\BibitemShut {NoStop}%
\bibitem [{\citenamefont {Iwahara}\ and\ \citenamefont
  {Chibotaru}(2023)}]{Iwahara2023}%
  \BibitemOpen
  \bibfield  {author} {\bibinfo {author} {\bibfnamefont {N.}~\bibnamefont
  {Iwahara}}\ and\ \bibinfo {author} {\bibfnamefont {L.~F.}\ \bibnamefont
  {Chibotaru}},\ }\href {\doibase 10.1103/PhysRevB.107.L220404} {\bibfield
  {journal} {\bibinfo  {journal} {Phys. Rev. B}\ }\textbf {\bibinfo {volume}
  {107}},\ \bibinfo {pages} {L220404} (\bibinfo {year} {2023})}\BibitemShut
  {NoStop}%
\bibitem [{\citenamefont {Ament}\ \emph {et~al.}(2011)\citenamefont {Ament},
  \citenamefont {van Veenendaal}, \citenamefont {Devereaux}, \citenamefont
  {Hill},\ and\ \citenamefont {van~den Brink}}]{RIXS}%
  \BibitemOpen
  \bibfield  {author} {\bibinfo {author} {\bibfnamefont {L.~J.~P.}\
  \bibnamefont {Ament}}, \bibinfo {author} {\bibfnamefont {M.}~\bibnamefont
  {van Veenendaal}}, \bibinfo {author} {\bibfnamefont {T.~P.}\ \bibnamefont
  {Devereaux}}, \bibinfo {author} {\bibfnamefont {J.~P.}\ \bibnamefont {Hill}},
  \ and\ \bibinfo {author} {\bibfnamefont {J.}~\bibnamefont {van~den Brink}},\
  }\href {\doibase 10.1103/RevModPhys.83.705} {\bibfield  {journal} {\bibinfo
  {journal} {Rev. Mod. Phys.}\ }\textbf {\bibinfo {volume} {83}},\ \bibinfo
  {pages} {705} (\bibinfo {year} {2011})}\BibitemShut {NoStop}%
\bibitem [{\citenamefont {de~Groot}\ \emph {et~al.}(2024)\citenamefont
  {de~Groot}, \citenamefont {Haverkort}, \citenamefont {Elnaggar},
  \citenamefont {Juhin}, \citenamefont {Zhou},\ and\ \citenamefont
  {Glatzel}}]{RIXS2}%
  \BibitemOpen
  \bibfield  {author} {\bibinfo {author} {\bibfnamefont {F.~M.~F.}\
  \bibnamefont {de~Groot}}, \bibinfo {author} {\bibfnamefont {M.~W.}\
  \bibnamefont {Haverkort}}, \bibinfo {author} {\bibfnamefont {H.}~\bibnamefont
  {Elnaggar}}, \bibinfo {author} {\bibfnamefont {A.}~\bibnamefont {Juhin}},
  \bibinfo {author} {\bibfnamefont {K.-J.}\ \bibnamefont {Zhou}}, \ and\
  \bibinfo {author} {\bibfnamefont {P.}~\bibnamefont {Glatzel}},\ }\href
  {\doibase 10.1038/s43586-024-00322-6} {\bibfield  {journal} {\bibinfo
  {journal} {Nature Reviews Methods Primers}\ }\textbf {\bibinfo {volume}
  {4}},\ \bibinfo {pages} {45} (\bibinfo {year} {2024})}\BibitemShut {NoStop}%
\bibitem [{\citenamefont {Khan}\ \emph {et~al.}(2019)\citenamefont {Khan},
  \citenamefont {Prishchenko}, \citenamefont {Skourski}, \citenamefont
  {Mazurenko},\ and\ \citenamefont {Tsirlin}}]{Khan2019}%
  \BibitemOpen
  \bibfield  {author} {\bibinfo {author} {\bibfnamefont {N.}~\bibnamefont
  {Khan}}, \bibinfo {author} {\bibfnamefont {D.}~\bibnamefont {Prishchenko}},
  \bibinfo {author} {\bibfnamefont {Y.}~\bibnamefont {Skourski}}, \bibinfo
  {author} {\bibfnamefont {V.~G.}\ \bibnamefont {Mazurenko}}, \ and\ \bibinfo
  {author} {\bibfnamefont {A.~A.}\ \bibnamefont {Tsirlin}},\ }\href {\doibase
  10.1103/PhysRevB.99.144425} {\bibfield  {journal} {\bibinfo  {journal} {Phys.
  Rev. B}\ }\textbf {\bibinfo {volume} {99}},\ \bibinfo {pages} {144425}
  (\bibinfo {year} {2019})}\BibitemShut {NoStop}%
\bibitem [{Note6()}]{Note6}%
  \BibitemOpen
  \bibinfo {note} {Recent high-resolution x-ray diffraction data suggest the
  development of tiny deformation below $T_N \approx 3.2$ \cite
  {Wang2024}.}\BibitemShut {Stop}%
\bibitem [{\citenamefont {Bhaskaran}\ \emph {et~al.}(2021)\citenamefont
  {Bhaskaran}, \citenamefont {Ponomaryov}, \citenamefont {Wosnitza},
  \citenamefont {Khan}, \citenamefont {Tsirlin}, \citenamefont {Zhitomirsky},\
  and\ \citenamefont {Zvyagin}}]{Bhaskaran2021}%
  \BibitemOpen
  \bibfield  {author} {\bibinfo {author} {\bibfnamefont {L.}~\bibnamefont
  {Bhaskaran}}, \bibinfo {author} {\bibfnamefont {A.~N.}\ \bibnamefont
  {Ponomaryov}}, \bibinfo {author} {\bibfnamefont {J.}~\bibnamefont
  {Wosnitza}}, \bibinfo {author} {\bibfnamefont {N.}~\bibnamefont {Khan}},
  \bibinfo {author} {\bibfnamefont {A.~A.}\ \bibnamefont {Tsirlin}}, \bibinfo
  {author} {\bibfnamefont {M.~E.}\ \bibnamefont {Zhitomirsky}}, \ and\ \bibinfo
  {author} {\bibfnamefont {S.~A.}\ \bibnamefont {Zvyagin}},\ }\href {\doibase
  10.1103/PhysRevB.104.184404} {\bibfield  {journal} {\bibinfo  {journal}
  {Phys. Rev. B}\ }\textbf {\bibinfo {volume} {104}},\ \bibinfo {pages}
  {184404} (\bibinfo {year} {2021})}\BibitemShut {NoStop}%
\bibitem [{\citenamefont {Iwahara}\ and\ \citenamefont
  {Furukawa}(2023)}]{Iwahara2023Ir}%
  \BibitemOpen
  \bibfield  {author} {\bibinfo {author} {\bibfnamefont {N.}~\bibnamefont
  {Iwahara}}\ and\ \bibinfo {author} {\bibfnamefont {W.}~\bibnamefont
  {Furukawa}},\ }\href {\doibase 10.1103/PhysRevB.108.075136} {\bibfield
  {journal} {\bibinfo  {journal} {Phys. Rev. B}\ }\textbf {\bibinfo {volume}
  {108}},\ \bibinfo {pages} {075136} (\bibinfo {year} {2023})}\BibitemShut
  {NoStop}%
\bibitem [{\citenamefont {Sakurai}(1967)}]{Sakurai1967}%
  \BibitemOpen
  \bibfield  {author} {\bibinfo {author} {\bibfnamefont {J.~J.}\ \bibnamefont
  {Sakurai}},\ }\href@noop {} {\emph {\bibinfo {title} {Advanced Quantum
  Mechanics}}}\ (\bibinfo  {publisher} {Addison-Wesley},\ \bibinfo {address}
  {Massachusetts},\ \bibinfo {year} {1967})\BibitemShut {NoStop}%
\bibitem [{\citenamefont {Blume}(1985)}]{Blume1985}%
  \BibitemOpen
  \bibfield  {author} {\bibinfo {author} {\bibfnamefont {M.}~\bibnamefont
  {Blume}},\ }\href {\doibase 10.1063/1.335023} {\bibfield  {journal} {\bibinfo
   {journal} {Journal of Applied Physics}\ }\textbf {\bibinfo {volume} {57}},\
  \bibinfo {pages} {3615} (\bibinfo {year} {1985})}\BibitemShut {NoStop}%
\bibitem [{\citenamefont {Luo}\ \emph {et~al.}(1993)\citenamefont {Luo},
  \citenamefont {Trammell},\ and\ \citenamefont {Hannon}}]{Luo1993}%
  \BibitemOpen
  \bibfield  {author} {\bibinfo {author} {\bibfnamefont {J.}~\bibnamefont
  {Luo}}, \bibinfo {author} {\bibfnamefont {G.~T.}\ \bibnamefont {Trammell}}, \
  and\ \bibinfo {author} {\bibfnamefont {J.~P.}\ \bibnamefont {Hannon}},\
  }\href {\doibase 10.1103/PhysRevLett.71.287} {\bibfield  {journal} {\bibinfo
  {journal} {Phys. Rev. Lett.}\ }\textbf {\bibinfo {volume} {71}},\ \bibinfo
  {pages} {287} (\bibinfo {year} {1993})}\BibitemShut {NoStop}%
\bibitem [{\citenamefont {van Veenendaal}(2006)}]{vanVeenendaal2006}%
  \BibitemOpen
  \bibfield  {author} {\bibinfo {author} {\bibfnamefont {M.}~\bibnamefont {van
  Veenendaal}},\ }\href {\doibase 10.1103/PhysRevLett.96.117404} {\bibfield
  {journal} {\bibinfo  {journal} {Phys. Rev. Lett.}\ }\textbf {\bibinfo
  {volume} {96}},\ \bibinfo {pages} {117404} (\bibinfo {year}
  {2006})}\BibitemShut {NoStop}%
\bibitem [{\citenamefont {Clancy}\ \emph {et~al.}(2012)\citenamefont {Clancy},
  \citenamefont {Chen}, \citenamefont {Kim}, \citenamefont {Chen},
  \citenamefont {Plumb}, \citenamefont {Jeon}, \citenamefont {Noh},\ and\
  \citenamefont {Kim}}]{Clancy2012}%
  \BibitemOpen
  \bibfield  {author} {\bibinfo {author} {\bibfnamefont {J.~P.}\ \bibnamefont
  {Clancy}}, \bibinfo {author} {\bibfnamefont {N.}~\bibnamefont {Chen}},
  \bibinfo {author} {\bibfnamefont {C.~Y.}\ \bibnamefont {Kim}}, \bibinfo
  {author} {\bibfnamefont {W.~F.}\ \bibnamefont {Chen}}, \bibinfo {author}
  {\bibfnamefont {K.~W.}\ \bibnamefont {Plumb}}, \bibinfo {author}
  {\bibfnamefont {B.~C.}\ \bibnamefont {Jeon}}, \bibinfo {author}
  {\bibfnamefont {T.~W.}\ \bibnamefont {Noh}}, \ and\ \bibinfo {author}
  {\bibfnamefont {Y.-J.}\ \bibnamefont {Kim}},\ }\href {\doibase
  10.1103/PhysRevB.86.195131} {\bibfield  {journal} {\bibinfo  {journal} {Phys.
  Rev. B}\ }\textbf {\bibinfo {volume} {86}},\ \bibinfo {pages} {195131}
  (\bibinfo {year} {2012})}\BibitemShut {NoStop}%
\bibitem [{\citenamefont {Toyozawa}(2003)}]{Toyozawa2003}%
  \BibitemOpen
  \bibfield  {author} {\bibinfo {author} {\bibfnamefont {Y.}~\bibnamefont
  {Toyozawa}},\ }\href@noop {} {\emph {\bibinfo {title} {Optical Processes in
  Solids}}}\ (\bibinfo  {publisher} {Cambridge University Press},\ \bibinfo
  {address} {Cambridge},\ \bibinfo {year} {2003})\BibitemShut {NoStop}%
\bibitem [{\citenamefont {Suzuki}\ \emph {et~al.}(2021)\citenamefont {Suzuki},
  \citenamefont {Liu}, \citenamefont {Bertinshaw}, \citenamefont {Ueda},
  \citenamefont {Kim}, \citenamefont {Laha}, \citenamefont {Weber},
  \citenamefont {Yang}, \citenamefont {Wang}, \citenamefont {Takahashi},
  \citenamefont {F\"{u}rsich}, \citenamefont {Minola}, \citenamefont {Lotsch},
  \citenamefont {Kim}, \citenamefont {Yava\c{s}}, \citenamefont {Daghofer},
  \citenamefont {Chaloupka}, \citenamefont {Khaliullin}, \citenamefont
  {Gretarsson},\ and\ \citenamefont {Keimer}}]{Suzuki2021}%
  \BibitemOpen
  \bibfield  {author} {\bibinfo {author} {\bibfnamefont {H.}~\bibnamefont
  {Suzuki}}, \bibinfo {author} {\bibfnamefont {H.}~\bibnamefont {Liu}},
  \bibinfo {author} {\bibfnamefont {J.}~\bibnamefont {Bertinshaw}}, \bibinfo
  {author} {\bibfnamefont {K.}~\bibnamefont {Ueda}}, \bibinfo {author}
  {\bibfnamefont {H.}~\bibnamefont {Kim}}, \bibinfo {author} {\bibfnamefont
  {S.}~\bibnamefont {Laha}}, \bibinfo {author} {\bibfnamefont {D.}~\bibnamefont
  {Weber}}, \bibinfo {author} {\bibfnamefont {Z.}~\bibnamefont {Yang}},
  \bibinfo {author} {\bibfnamefont {L.}~\bibnamefont {Wang}}, \bibinfo {author}
  {\bibfnamefont {H.}~\bibnamefont {Takahashi}}, \bibinfo {author}
  {\bibfnamefont {K.}~\bibnamefont {F\"{u}rsich}}, \bibinfo {author}
  {\bibfnamefont {M.}~\bibnamefont {Minola}}, \bibinfo {author} {\bibfnamefont
  {B.~V.}\ \bibnamefont {Lotsch}}, \bibinfo {author} {\bibfnamefont {B.~J.}\
  \bibnamefont {Kim}}, \bibinfo {author} {\bibfnamefont {H.}~\bibnamefont
  {Yava\c{s}}}, \bibinfo {author} {\bibfnamefont {M.}~\bibnamefont {Daghofer}},
  \bibinfo {author} {\bibfnamefont {J.}~\bibnamefont {Chaloupka}}, \bibinfo
  {author} {\bibfnamefont {G.}~\bibnamefont {Khaliullin}}, \bibinfo {author}
  {\bibfnamefont {H.}~\bibnamefont {Gretarsson}}, \ and\ \bibinfo {author}
  {\bibfnamefont {B.}~\bibnamefont {Keimer}},\ }\href {\doibase
  10.1038/s41467-021-24722-4} {\bibfield  {journal} {\bibinfo  {journal} {Nat.
  Commun.}\ }\textbf {\bibinfo {volume} {12}},\ \bibinfo {pages} {4512}
  (\bibinfo {year} {2021})}\BibitemShut {NoStop}%
\bibitem [{\citenamefont {Paramekanti}\ \emph {et~al.}(2018)\citenamefont
  {Paramekanti}, \citenamefont {Singh}, \citenamefont {Yuan}, \citenamefont
  {Casa}, \citenamefont {Said}, \citenamefont {Kim},\ and\ \citenamefont
  {Christianson}}]{Paramekanti2018}%
  \BibitemOpen
  \bibfield  {author} {\bibinfo {author} {\bibfnamefont {A.}~\bibnamefont
  {Paramekanti}}, \bibinfo {author} {\bibfnamefont {D.~J.}\ \bibnamefont
  {Singh}}, \bibinfo {author} {\bibfnamefont {B.}~\bibnamefont {Yuan}},
  \bibinfo {author} {\bibfnamefont {D.}~\bibnamefont {Casa}}, \bibinfo {author}
  {\bibfnamefont {A.}~\bibnamefont {Said}}, \bibinfo {author} {\bibfnamefont
  {Y.-J.}\ \bibnamefont {Kim}}, \ and\ \bibinfo {author} {\bibfnamefont
  {A.~D.}\ \bibnamefont {Christianson}},\ }\href {\doibase
  10.1103/PhysRevB.97.235119} {\bibfield  {journal} {\bibinfo  {journal} {Phys.
  Rev. B}\ }\textbf {\bibinfo {volume} {97}},\ \bibinfo {pages} {235119}
  (\bibinfo {year} {2018})}\BibitemShut {NoStop}%
\bibitem [{\citenamefont {Fujimori}(2024)}]{Fujimori}%
  \BibitemOpen
  \bibfield  {author} {\bibinfo {author} {\bibfnamefont {A.}~\bibnamefont
  {Fujimori}},\ }\href@noop {} {\enquote {\bibinfo {title} {Ligand field and
  charge transfer in transition-metal compounds},}\ } (\bibinfo {year}
  {2024}),\ \bibinfo {note} {70 years of the Tanabe-Sugano
  diagrams}\BibitemShut {NoStop}%
\bibitem [{\citenamefont {Moriya}(1960)}]{Moriya1960}%
  \BibitemOpen
  \bibfield  {author} {\bibinfo {author} {\bibfnamefont {T.}~\bibnamefont
  {Moriya}},\ }\href {\doibase 10.1103/PhysRev.120.91} {\bibfield  {journal}
  {\bibinfo  {journal} {Phys. Rev.}\ }\textbf {\bibinfo {volume} {120}},\
  \bibinfo {pages} {91} (\bibinfo {year} {1960})}\BibitemShut {NoStop}%
\bibitem [{\citenamefont {Khaliullin}\ \emph {et~al.}(2021)\citenamefont
  {Khaliullin}, \citenamefont {Churchill}, \citenamefont {Stavropoulos},\ and\
  \citenamefont {Kee}}]{Khaliullin2021}%
  \BibitemOpen
  \bibfield  {author} {\bibinfo {author} {\bibfnamefont {G.}~\bibnamefont
  {Khaliullin}}, \bibinfo {author} {\bibfnamefont {D.}~\bibnamefont
  {Churchill}}, \bibinfo {author} {\bibfnamefont {P.~P.}\ \bibnamefont
  {Stavropoulos}}, \ and\ \bibinfo {author} {\bibfnamefont {H.-Y.}\
  \bibnamefont {Kee}},\ }\href {\doibase 10.1103/PhysRevResearch.3.033163}
  {\bibfield  {journal} {\bibinfo  {journal} {Phys. Rev. Res.}\ }\textbf
  {\bibinfo {volume} {3}},\ \bibinfo {pages} {033163} (\bibinfo {year}
  {2021})}\BibitemShut {NoStop}%
\bibitem [{\citenamefont {Chibotaru}(2005)}]{Chibotaru2005}%
  \BibitemOpen
  \bibfield  {author} {\bibinfo {author} {\bibfnamefont {L.~F.}\ \bibnamefont
  {Chibotaru}},\ }\href {\doibase 10.1103/PhysRevLett.94.186405} {\bibfield
  {journal} {\bibinfo  {journal} {Phys. Rev. Lett.}\ }\textbf {\bibinfo
  {volume} {94}},\ \bibinfo {pages} {186405} (\bibinfo {year}
  {2005})}\BibitemShut {NoStop}%
\bibitem [{\citenamefont {Iwahara}\ and\ \citenamefont
  {Chibotaru}(2013)}]{Iwahara2013}%
  \BibitemOpen
  \bibfield  {author} {\bibinfo {author} {\bibfnamefont {N.}~\bibnamefont
  {Iwahara}}\ and\ \bibinfo {author} {\bibfnamefont {L.~F.}\ \bibnamefont
  {Chibotaru}},\ }\href {\doibase 10.1103/PhysRevLett.111.056401} {\bibfield
  {journal} {\bibinfo  {journal} {Phys. Rev. Lett.}\ }\textbf {\bibinfo
  {volume} {111}},\ \bibinfo {pages} {056401} (\bibinfo {year}
  {2013})}\BibitemShut {NoStop}%
\bibitem [{\citenamefont {Nasu}\ and\ \citenamefont
  {Ishihara}(2013)}]{Nasu2013}%
  \BibitemOpen
  \bibfield  {author} {\bibinfo {author} {\bibfnamefont {J.}~\bibnamefont
  {Nasu}}\ and\ \bibinfo {author} {\bibfnamefont {S.}~\bibnamefont
  {Ishihara}},\ }\href {\doibase 10.1103/PhysRevB.88.094408} {\bibfield
  {journal} {\bibinfo  {journal} {Phys. Rev. B}\ }\textbf {\bibinfo {volume}
  {88}},\ \bibinfo {pages} {094408} (\bibinfo {year} {2013})}\BibitemShut
  {NoStop}%
\bibitem [{\citenamefont {Nasu}\ and\ \citenamefont
  {Ishihara}(2014)}]{Nasu2014e}%
  \BibitemOpen
  \bibfield  {author} {\bibinfo {author} {\bibfnamefont {J.}~\bibnamefont
  {Nasu}}\ and\ \bibinfo {author} {\bibfnamefont {S.}~\bibnamefont
  {Ishihara}},\ }\href {\doibase 10.1103/PhysRevB.90.179903} {\bibfield
  {journal} {\bibinfo  {journal} {Phys. Rev. B}\ }\textbf {\bibinfo {volume}
  {90}},\ \bibinfo {pages} {179903} (\bibinfo {year} {2014})}\BibitemShut
  {NoStop}%
\bibitem [{\citenamefont {Anderson}(1959)}]{Anderson1959}%
  \BibitemOpen
  \bibfield  {author} {\bibinfo {author} {\bibfnamefont {P.~W.}\ \bibnamefont
  {Anderson}},\ }\href {\doibase 10.1103/PhysRev.115.2} {\bibfield  {journal}
  {\bibinfo  {journal} {Phys. Rev.}\ }\textbf {\bibinfo {volume} {115}},\
  \bibinfo {pages} {2} (\bibinfo {year} {1959})}\BibitemShut {NoStop}%
\bibitem [{\citenamefont {Slater}\ and\ \citenamefont
  {Koster}(1954)}]{Slater1954}%
  \BibitemOpen
  \bibfield  {author} {\bibinfo {author} {\bibfnamefont {J.~C.}\ \bibnamefont
  {Slater}}\ and\ \bibinfo {author} {\bibfnamefont {G.~F.}\ \bibnamefont
  {Koster}},\ }\href {\doibase 10.1103/PhysRev.94.1498} {\bibfield  {journal}
  {\bibinfo  {journal} {Phys. Rev.}\ }\textbf {\bibinfo {volume} {94}},\
  \bibinfo {pages} {1498} (\bibinfo {year} {1954})}\BibitemShut {NoStop}%
\bibitem [{\citenamefont {Tsunetsugu}\ \emph {et~al.}(2021)\citenamefont
  {Tsunetsugu}, \citenamefont {Ishitobi},\ and\ \citenamefont
  {Hattori}}]{Tsunetsugu2021}%
  \BibitemOpen
  \bibfield  {author} {\bibinfo {author} {\bibfnamefont {H.}~\bibnamefont
  {Tsunetsugu}}, \bibinfo {author} {\bibfnamefont {T.}~\bibnamefont
  {Ishitobi}}, \ and\ \bibinfo {author} {\bibfnamefont {K.}~\bibnamefont
  {Hattori}},\ }\href {\doibase 10.7566/JPSJ.90.043701} {\bibfield  {journal}
  {\bibinfo  {journal} {J. Phys. Soc. Jpn.}\ }\textbf {\bibinfo {volume}
  {90}},\ \bibinfo {pages} {043701} (\bibinfo {year} {2021})}\BibitemShut
  {NoStop}%
\bibitem [{\citenamefont {Hattori}\ \emph {et~al.}(2023)\citenamefont
  {Hattori}, \citenamefont {Ishitobi},\ and\ \citenamefont
  {Tsunetsugu}}]{Hattori2023}%
  \BibitemOpen
  \bibfield  {author} {\bibinfo {author} {\bibfnamefont {K.}~\bibnamefont
  {Hattori}}, \bibinfo {author} {\bibfnamefont {T.}~\bibnamefont {Ishitobi}}, \
  and\ \bibinfo {author} {\bibfnamefont {H.}~\bibnamefont {Tsunetsugu}},\
  }\href {\doibase 10.1103/PhysRevB.107.205126} {\bibfield  {journal} {\bibinfo
   {journal} {Phys. Rev. B}\ }\textbf {\bibinfo {volume} {107}},\ \bibinfo
  {pages} {205126} (\bibinfo {year} {2023})}\BibitemShut {NoStop}%
\bibitem [{\citenamefont {Sugihara}(1959)}]{Sugihara1959}%
  \BibitemOpen
  \bibfield  {author} {\bibinfo {author} {\bibfnamefont {K.}~\bibnamefont
  {Sugihara}},\ }\href {\doibase 10.1143/JPSJ.14.1231} {\bibfield  {journal}
  {\bibinfo  {journal} {Journal of the Physical Society of Japan}\ }\textbf
  {\bibinfo {volume} {14}},\ \bibinfo {pages} {1231} (\bibinfo {year}
  {1959})}\BibitemShut {NoStop}%
\bibitem [{\citenamefont {Erickson}\ \emph {et~al.}(2007)\citenamefont
  {Erickson}, \citenamefont {Misra}, \citenamefont {Miller}, \citenamefont
  {Gupta}, \citenamefont {Schlesinger}, \citenamefont {Harrison}, \citenamefont
  {Kim},\ and\ \citenamefont {Fisher}}]{Erickson2007}%
  \BibitemOpen
  \bibfield  {author} {\bibinfo {author} {\bibfnamefont {A.~S.}\ \bibnamefont
  {Erickson}}, \bibinfo {author} {\bibfnamefont {S.}~\bibnamefont {Misra}},
  \bibinfo {author} {\bibfnamefont {G.~J.}\ \bibnamefont {Miller}}, \bibinfo
  {author} {\bibfnamefont {R.~R.}\ \bibnamefont {Gupta}}, \bibinfo {author}
  {\bibfnamefont {Z.}~\bibnamefont {Schlesinger}}, \bibinfo {author}
  {\bibfnamefont {W.~A.}\ \bibnamefont {Harrison}}, \bibinfo {author}
  {\bibfnamefont {J.~M.}\ \bibnamefont {Kim}}, \ and\ \bibinfo {author}
  {\bibfnamefont {I.~R.}\ \bibnamefont {Fisher}},\ }\href {\doibase
  10.1103/PhysRevLett.99.016404} {\bibfield  {journal} {\bibinfo  {journal}
  {Phys. Rev. Lett.}\ }\textbf {\bibinfo {volume} {99}},\ \bibinfo {pages}
  {016404} (\bibinfo {year} {2007})}\BibitemShut {NoStop}%
\bibitem [{\citenamefont {Willa}\ \emph {et~al.}(2019)\citenamefont {Willa},
  \citenamefont {Willa}, \citenamefont {Welp}, \citenamefont {Fisher},
  \citenamefont {Rydh}, \citenamefont {Kwok},\ and\ \citenamefont
  {Islam}}]{Willa2019}%
  \BibitemOpen
  \bibfield  {author} {\bibinfo {author} {\bibfnamefont {K.}~\bibnamefont
  {Willa}}, \bibinfo {author} {\bibfnamefont {R.}~\bibnamefont {Willa}},
  \bibinfo {author} {\bibfnamefont {U.}~\bibnamefont {Welp}}, \bibinfo {author}
  {\bibfnamefont {I.~R.}\ \bibnamefont {Fisher}}, \bibinfo {author}
  {\bibfnamefont {A.}~\bibnamefont {Rydh}}, \bibinfo {author} {\bibfnamefont
  {W.-K.}\ \bibnamefont {Kwok}}, \ and\ \bibinfo {author} {\bibfnamefont
  {Z.}~\bibnamefont {Islam}},\ }\href {\doibase 10.1103/PhysRevB.100.041108}
  {\bibfield  {journal} {\bibinfo  {journal} {Phys. Rev. B}\ }\textbf {\bibinfo
  {volume} {100}},\ \bibinfo {pages} {041108(R)} (\bibinfo {year}
  {2019})}\BibitemShut {NoStop}%
\bibitem [{\citenamefont {Hirai}\ and\ \citenamefont
  {Hiroi}(2019)}]{Hirai2019}%
  \BibitemOpen
  \bibfield  {author} {\bibinfo {author} {\bibfnamefont {D.}~\bibnamefont
  {Hirai}}\ and\ \bibinfo {author} {\bibfnamefont {Z.}~\bibnamefont {Hiroi}},\
  }\href {\doibase 10.7566/JPSJ.88.064712} {\bibfield  {journal} {\bibinfo
  {journal} {Journal of the Physical Society of Japan}\ }\textbf {\bibinfo
  {volume} {88}},\ \bibinfo {pages} {064712} (\bibinfo {year}
  {2019})}\BibitemShut {NoStop}%
\bibitem [{\citenamefont {Huang}\ \emph {et~al.}(2024)\citenamefont {Huang},
  \citenamefont {Iwahara},\ and\ \citenamefont {Chibotaru}}]{Huang2024}%
  \BibitemOpen
  \bibfield  {author} {\bibinfo {author} {\bibfnamefont {Z.}~\bibnamefont
  {Huang}}, \bibinfo {author} {\bibfnamefont {N.}~\bibnamefont {Iwahara}}, \
  and\ \bibinfo {author} {\bibfnamefont {L.~F.}\ \bibnamefont {Chibotaru}},\
  }\href@noop {} {} (\bibinfo {year} {2024}),\ \Eprint
  {http://arxiv.org/abs/2401.08204} {arXiv:2401.08204 [cond-mat.mtrl-sci]}
  \BibitemShut {NoStop}%
\bibitem [{\citenamefont {Stitzer}\ \emph {et~al.}(2002)\citenamefont
  {Stitzer}, \citenamefont {Smith},\ and\ \citenamefont {{zur
  Loye}}}]{Stitzer2002}%
  \BibitemOpen
  \bibfield  {author} {\bibinfo {author} {\bibfnamefont {K.~E.}\ \bibnamefont
  {Stitzer}}, \bibinfo {author} {\bibfnamefont {M.~D.}\ \bibnamefont {Smith}},
  \ and\ \bibinfo {author} {\bibfnamefont {H.-C.}\ \bibnamefont {{zur Loye}}},\
  }\href {\doibase https://doi.org/10.1016/S1293-2558(01)01257-2} {\bibfield
  {journal} {\bibinfo  {journal} {Solid State Sci.}\ }\textbf {\bibinfo
  {volume} {4}},\ \bibinfo {pages} {311} (\bibinfo {year} {2002})}\BibitemShut
  {NoStop}%
\bibitem [{\citenamefont {Steele}\ \emph {et~al.}(2011)\citenamefont {Steele},
  \citenamefont {Baker}, \citenamefont {Lancaster}, \citenamefont {Pratt},
  \citenamefont {Franke}, \citenamefont {Ghannadzadeh}, \citenamefont
  {Goddard}, \citenamefont {Hayes}, \citenamefont {Prabhakaran},\ and\
  \citenamefont {Blundell}}]{Steele2011}%
  \BibitemOpen
  \bibfield  {author} {\bibinfo {author} {\bibfnamefont {A.~J.}\ \bibnamefont
  {Steele}}, \bibinfo {author} {\bibfnamefont {P.~J.}\ \bibnamefont {Baker}},
  \bibinfo {author} {\bibfnamefont {T.}~\bibnamefont {Lancaster}}, \bibinfo
  {author} {\bibfnamefont {F.~L.}\ \bibnamefont {Pratt}}, \bibinfo {author}
  {\bibfnamefont {I.}~\bibnamefont {Franke}}, \bibinfo {author} {\bibfnamefont
  {S.}~\bibnamefont {Ghannadzadeh}}, \bibinfo {author} {\bibfnamefont {P.~A.}\
  \bibnamefont {Goddard}}, \bibinfo {author} {\bibfnamefont {W.}~\bibnamefont
  {Hayes}}, \bibinfo {author} {\bibfnamefont {D.}~\bibnamefont {Prabhakaran}},
  \ and\ \bibinfo {author} {\bibfnamefont {S.~J.}\ \bibnamefont {Blundell}},\
  }\href {\doibase 10.1103/PhysRevB.84.144416} {\bibfield  {journal} {\bibinfo
  {journal} {Phys. Rev. B}\ }\textbf {\bibinfo {volume} {84}},\ \bibinfo
  {pages} {144416} (\bibinfo {year} {2011})}\BibitemShut {NoStop}%
\bibitem [{\citenamefont {Cong}(2022)}]{Cong2022}%
  \BibitemOpen
  \bibfield  {author} {\bibinfo {author} {\bibfnamefont {R.}~\bibnamefont
  {Cong}},\ }\emph {\bibinfo {title} {Magnetic and Structural Properties of 5d
  Osmate Double Perovskites Probed by Nuclear Magnetic Resonance}},\ \href
  {\doibase 10.26300/7tx8-gq83} {Ph.D. thesis},\ \bibinfo  {school} {Brown
  University} (\bibinfo {year} {2022})\BibitemShut {NoStop}%
\bibitem [{\citenamefont {P\'{a}sztorov\'{a}}\ \emph
  {et~al.}(2023{\natexlab{b}})\citenamefont {P\'{a}sztorov\'{a}}, \citenamefont
  {Tehrani}, \citenamefont {\v{Z}ivkovi\'c}, \citenamefont {Spaldin},\ and\
  \citenamefont {R{\o}nnow}}]{Pasztorova2023b}%
  \BibitemOpen
  \bibfield  {author} {\bibinfo {author} {\bibfnamefont {J.}~\bibnamefont
  {P\'{a}sztorov\'{a}}}, \bibinfo {author} {\bibfnamefont {A.~M.}\ \bibnamefont
  {Tehrani}}, \bibinfo {author} {\bibfnamefont {I.}~\bibnamefont
  {\v{Z}ivkovi\'c}}, \bibinfo {author} {\bibfnamefont {N.~A.}\ \bibnamefont
  {Spaldin}}, \ and\ \bibinfo {author} {\bibfnamefont {H.~M.}\ \bibnamefont
  {R{\o}nnow}},\ }\href {\doibase 10.1088/1361-648X/acc62a} {\bibfield
  {journal} {\bibinfo  {journal} {Journal of Physics: Condensed Matter}\
  }\textbf {\bibinfo {volume} {35}},\ \bibinfo {pages} {245603} (\bibinfo
  {year} {2023}{\natexlab{b}})}\BibitemShut {NoStop}%
\bibitem [{\citenamefont {Yamamura}\ \emph {et~al.}(2006)\citenamefont
  {Yamamura}, \citenamefont {Wakeshima},\ and\ \citenamefont
  {Hinatsu}}]{Yamaura2006}%
  \BibitemOpen
  \bibfield  {author} {\bibinfo {author} {\bibfnamefont {K.}~\bibnamefont
  {Yamamura}}, \bibinfo {author} {\bibfnamefont {M.}~\bibnamefont {Wakeshima}},
  \ and\ \bibinfo {author} {\bibfnamefont {Y.}~\bibnamefont {Hinatsu}},\ }\href
  {\doibase https://doi.org/10.1016/j.jssc.2005.10.003} {\bibfield  {journal}
  {\bibinfo  {journal} {J. Solid. State. Chem.}\ }\textbf {\bibinfo {volume}
  {179}},\ \bibinfo {pages} {605} (\bibinfo {year} {2006})}\BibitemShut
  {NoStop}%
\bibitem [{\citenamefont {Ishikawa}\ \emph
  {et~al.}(2021{\natexlab{a}})\citenamefont {Ishikawa}, \citenamefont {Hirai},
  \citenamefont {Ikeda}, \citenamefont {Gen}, \citenamefont {Yajima},
  \citenamefont {Matsuo}, \citenamefont {Matsuda}, \citenamefont {Hiroi},\ and\
  \citenamefont {Kindo}}]{Ishikawa2021b}%
  \BibitemOpen
  \bibfield  {author} {\bibinfo {author} {\bibfnamefont {H.}~\bibnamefont
  {Ishikawa}}, \bibinfo {author} {\bibfnamefont {D.}~\bibnamefont {Hirai}},
  \bibinfo {author} {\bibfnamefont {A.}~\bibnamefont {Ikeda}}, \bibinfo
  {author} {\bibfnamefont {M.}~\bibnamefont {Gen}}, \bibinfo {author}
  {\bibfnamefont {T.}~\bibnamefont {Yajima}}, \bibinfo {author} {\bibfnamefont
  {A.}~\bibnamefont {Matsuo}}, \bibinfo {author} {\bibfnamefont {Y.~H.}\
  \bibnamefont {Matsuda}}, \bibinfo {author} {\bibfnamefont {Z.}~\bibnamefont
  {Hiroi}}, \ and\ \bibinfo {author} {\bibfnamefont {K.}~\bibnamefont
  {Kindo}},\ }\href {\doibase 10.1103/PhysRevB.104.174422} {\bibfield
  {journal} {\bibinfo  {journal} {Phys. Rev. B}\ }\textbf {\bibinfo {volume}
  {104}},\ \bibinfo {pages} {174422} (\bibinfo {year}
  {2021}{\natexlab{a}})}\BibitemShut {NoStop}%
\bibitem [{\citenamefont {Hirai}\ and\ \citenamefont
  {Hiroi}(2021)}]{Hirai2021}%
  \BibitemOpen
  \bibfield  {author} {\bibinfo {author} {\bibfnamefont {D.}~\bibnamefont
  {Hirai}}\ and\ \bibinfo {author} {\bibfnamefont {Z.}~\bibnamefont {Hiroi}},\
  }\href {\doibase 10.1088/1361-648x/abda79} {\bibfield  {journal} {\bibinfo
  {journal} {J. Phys.: Condens. Matter}\ }\textbf {\bibinfo {volume} {33}},\
  \bibinfo {pages} {135603} (\bibinfo {year} {2021})}\BibitemShut {NoStop}%
\bibitem [{\citenamefont {Bramnik}\ \emph {et~al.}(2003)\citenamefont
  {Bramnik}, \citenamefont {Ehrenberg}, \citenamefont {Dehn},\ and\
  \citenamefont {Fuess}}]{Bramnik2003}%
  \BibitemOpen
  \bibfield  {author} {\bibinfo {author} {\bibfnamefont {K.~G.}\ \bibnamefont
  {Bramnik}}, \bibinfo {author} {\bibfnamefont {H.}~\bibnamefont {Ehrenberg}},
  \bibinfo {author} {\bibfnamefont {J.~K.}\ \bibnamefont {Dehn}}, \ and\
  \bibinfo {author} {\bibfnamefont {H.}~\bibnamefont {Fuess}},\ }\href
  {\doibase https://doi.org/10.1016/S1293-2558(02)00097-3} {\bibfield
  {journal} {\bibinfo  {journal} {Solid State Sciences}\ }\textbf {\bibinfo
  {volume} {5}},\ \bibinfo {pages} {235} (\bibinfo {year} {2003})}\BibitemShut
  {NoStop}%
\bibitem [{\citenamefont {Gao}\ \emph {et~al.}(2020)\citenamefont {Gao},
  \citenamefont {Hirai}, \citenamefont {Sagayama}, \citenamefont {Ohsumi},
  \citenamefont {Hiroi},\ and\ \citenamefont {Arima}}]{Gao2020}%
  \BibitemOpen
  \bibfield  {author} {\bibinfo {author} {\bibfnamefont {S.}~\bibnamefont
  {Gao}}, \bibinfo {author} {\bibfnamefont {D.}~\bibnamefont {Hirai}}, \bibinfo
  {author} {\bibfnamefont {H.}~\bibnamefont {Sagayama}}, \bibinfo {author}
  {\bibfnamefont {H.}~\bibnamefont {Ohsumi}}, \bibinfo {author} {\bibfnamefont
  {Z.}~\bibnamefont {Hiroi}}, \ and\ \bibinfo {author} {\bibfnamefont {T.-h.}\
  \bibnamefont {Arima}},\ }\href {\doibase 10.1103/PhysRevB.101.220412}
  {\bibfield  {journal} {\bibinfo  {journal} {Phys. Rev. B}\ }\textbf {\bibinfo
  {volume} {101}},\ \bibinfo {pages} {220412} (\bibinfo {year}
  {2020})}\BibitemShut {NoStop}%
\bibitem [{\citenamefont {Ishikawa}\ \emph
  {et~al.}(2021{\natexlab{b}})\citenamefont {Ishikawa}, \citenamefont {Yajima},
  \citenamefont {Matsuo},\ and\ \citenamefont {Kindo}}]{Ishikawa2021}%
  \BibitemOpen
  \bibfield  {author} {\bibinfo {author} {\bibfnamefont {H.}~\bibnamefont
  {Ishikawa}}, \bibinfo {author} {\bibfnamefont {T.}~\bibnamefont {Yajima}},
  \bibinfo {author} {\bibfnamefont {A.}~\bibnamefont {Matsuo}}, \ and\ \bibinfo
  {author} {\bibfnamefont {K.}~\bibnamefont {Kindo}},\ }\href {\doibase
  10.1088/1361-648x/abd7b5} {\bibfield  {journal} {\bibinfo  {journal} {J.
  Phys.: Condens. Matter}\ }\textbf {\bibinfo {volume} {33}},\ \bibinfo {pages}
  {125802} (\bibinfo {year} {2021}{\natexlab{b}})}\BibitemShut {NoStop}%
\bibitem [{\citenamefont {Cussen}\ \emph {et~al.}(2006)\citenamefont {Cussen},
  \citenamefont {Lynham},\ and\ \citenamefont {Rogers}}]{Cussen2006}%
  \BibitemOpen
  \bibfield  {author} {\bibinfo {author} {\bibfnamefont {E.~J.}\ \bibnamefont
  {Cussen}}, \bibinfo {author} {\bibfnamefont {D.~R.}\ \bibnamefont {Lynham}},
  \ and\ \bibinfo {author} {\bibfnamefont {J.}~\bibnamefont {Rogers}},\ }\href
  {\doibase 10.1021/cm0602388} {\bibfield  {journal} {\bibinfo  {journal}
  {Chem. Mater.}\ }\textbf {\bibinfo {volume} {18}},\ \bibinfo {pages} {2855}
  (\bibinfo {year} {2006})}\BibitemShut {NoStop}%
\bibitem [{\citenamefont {de~Vries}\ \emph {et~al.}(2010)\citenamefont
  {de~Vries}, \citenamefont {Mclaughlin},\ and\ \citenamefont
  {Bos}}]{deVries2010}%
  \BibitemOpen
  \bibfield  {author} {\bibinfo {author} {\bibfnamefont {M.~A.}\ \bibnamefont
  {de~Vries}}, \bibinfo {author} {\bibfnamefont {A.~C.}\ \bibnamefont
  {Mclaughlin}}, \ and\ \bibinfo {author} {\bibfnamefont {J.~W.~G.}\
  \bibnamefont {Bos}},\ }\href {\doibase 10.1103/PhysRevLett.104.177202}
  {\bibfield  {journal} {\bibinfo  {journal} {Phys. Rev. Lett.}\ }\textbf
  {\bibinfo {volume} {104}},\ \bibinfo {pages} {177202} (\bibinfo {year}
  {2010})}\BibitemShut {NoStop}%
\bibitem [{\citenamefont {Carlo}\ \emph {et~al.}(2011)\citenamefont {Carlo},
  \citenamefont {Clancy}, \citenamefont {Aharen}, \citenamefont {Yamani},
  \citenamefont {Ruff}, \citenamefont {Wagman}, \citenamefont {Van~Gastel},
  \citenamefont {Noad}, \citenamefont {Granroth}, \citenamefont {Greedan},
  \citenamefont {Dabkowska},\ and\ \citenamefont {Gaulin}}]{Carlo2011}%
  \BibitemOpen
  \bibfield  {author} {\bibinfo {author} {\bibfnamefont {J.~P.}\ \bibnamefont
  {Carlo}}, \bibinfo {author} {\bibfnamefont {J.~P.}\ \bibnamefont {Clancy}},
  \bibinfo {author} {\bibfnamefont {T.}~\bibnamefont {Aharen}}, \bibinfo
  {author} {\bibfnamefont {Z.}~\bibnamefont {Yamani}}, \bibinfo {author}
  {\bibfnamefont {J.~P.~C.}\ \bibnamefont {Ruff}}, \bibinfo {author}
  {\bibfnamefont {J.~J.}\ \bibnamefont {Wagman}}, \bibinfo {author}
  {\bibfnamefont {G.~J.}\ \bibnamefont {Van~Gastel}}, \bibinfo {author}
  {\bibfnamefont {H.~M.~L.}\ \bibnamefont {Noad}}, \bibinfo {author}
  {\bibfnamefont {G.~E.}\ \bibnamefont {Granroth}}, \bibinfo {author}
  {\bibfnamefont {J.~E.}\ \bibnamefont {Greedan}}, \bibinfo {author}
  {\bibfnamefont {H.~A.}\ \bibnamefont {Dabkowska}}, \ and\ \bibinfo {author}
  {\bibfnamefont {B.~D.}\ \bibnamefont {Gaulin}},\ }\href {\doibase
  10.1103/PhysRevB.84.100404} {\bibfield  {journal} {\bibinfo  {journal} {Phys.
  Rev. B}\ }\textbf {\bibinfo {volume} {84}},\ \bibinfo {pages} {100404}
  (\bibinfo {year} {2011})}\BibitemShut {NoStop}%
\bibitem [{\citenamefont {de~Vries}\ \emph {et~al.}(2013)\citenamefont
  {de~Vries}, \citenamefont {Piatek}, \citenamefont {Misek}, \citenamefont
  {Lord}, \citenamefont {R{\o}nnow},\ and\ \citenamefont {Bos}}]{deVries2013}%
  \BibitemOpen
  \bibfield  {author} {\bibinfo {author} {\bibfnamefont {M.~A.}\ \bibnamefont
  {de~Vries}}, \bibinfo {author} {\bibfnamefont {J.~O.}\ \bibnamefont
  {Piatek}}, \bibinfo {author} {\bibfnamefont {M.}~\bibnamefont {Misek}},
  \bibinfo {author} {\bibfnamefont {J.~S.}\ \bibnamefont {Lord}}, \bibinfo
  {author} {\bibfnamefont {H.~M.}\ \bibnamefont {R{\o}nnow}}, \ and\ \bibinfo
  {author} {\bibfnamefont {J.~W.~G.}\ \bibnamefont {Bos}},\ }\href {\doibase
  10.1088/1367-2630/15/4/043024} {\bibfield  {journal} {\bibinfo  {journal}
  {New J. Phys.}\ }\textbf {\bibinfo {volume} {15}},\ \bibinfo {pages} {043024}
  (\bibinfo {year} {2013})}\BibitemShut {NoStop}%
\bibitem [{\citenamefont {Coomer}\ and\ \citenamefont
  {Cussen}(2013)}]{Coomer2013}%
  \BibitemOpen
  \bibfield  {author} {\bibinfo {author} {\bibfnamefont {F.~C.}\ \bibnamefont
  {Coomer}}\ and\ \bibinfo {author} {\bibfnamefont {E.~J.}\ \bibnamefont
  {Cussen}},\ }\href {\doibase 10.1088/0953-8984/25/8/082202} {\bibfield
  {journal} {\bibinfo  {journal} {J. Phys.: Condens. Matter}\ }\textbf
  {\bibinfo {volume} {25}},\ \bibinfo {pages} {082202} (\bibinfo {year}
  {2013})}\BibitemShut {NoStop}%
\bibitem [{\citenamefont {Mustonen}\ \emph {et~al.}(2022)\citenamefont
  {Mustonen}, \citenamefont {Mutch}, \citenamefont {Walker}, \citenamefont
  {Baker}, \citenamefont {Coomer}, \citenamefont {Perry}, \citenamefont
  {Pughe}, \citenamefont {Stenning}, \citenamefont {Liu}, \citenamefont
  {Dutton},\ and\ \citenamefont {Cussen}}]{Mustonen2022}%
  \BibitemOpen
  \bibfield  {author} {\bibinfo {author} {\bibfnamefont {O.~H.}\ \bibnamefont
  {Mustonen}}, \bibinfo {author} {\bibfnamefont {H.~N.}\ \bibnamefont {Mutch}},
  \bibinfo {author} {\bibfnamefont {H.~C.}\ \bibnamefont {Walker}}, \bibinfo
  {author} {\bibfnamefont {P.~J.}\ \bibnamefont {Baker}}, \bibinfo {author}
  {\bibfnamefont {F.~C.}\ \bibnamefont {Coomer}}, \bibinfo {author}
  {\bibfnamefont {R.~S.}\ \bibnamefont {Perry}}, \bibinfo {author}
  {\bibfnamefont {C.}~\bibnamefont {Pughe}}, \bibinfo {author} {\bibfnamefont
  {G.~B.~G.}\ \bibnamefont {Stenning}}, \bibinfo {author} {\bibfnamefont
  {C.}~\bibnamefont {Liu}}, \bibinfo {author} {\bibfnamefont {S.~E.}\
  \bibnamefont {Dutton}}, \ and\ \bibinfo {author} {\bibfnamefont {E.~J.}\
  \bibnamefont {Cussen}},\ }\href {\doibase 10.1038/s41535-022-00480-4}
  {\bibfield  {journal} {\bibinfo  {journal} {npj Quantum Mater.}\ }\textbf
  {\bibinfo {volume} {7}},\ \bibinfo {pages} {74} (\bibinfo {year}
  {2022})}\BibitemShut {NoStop}%
\bibitem [{\citenamefont {Lee}\ \emph {et~al.}(2021)\citenamefont {Lee},
  \citenamefont {Lee}, \citenamefont {Guohua}, \citenamefont {Ma},
  \citenamefont {Zhou}, \citenamefont {Lee}, \citenamefont {Choi},\ and\
  \citenamefont {Choi}}]{Lee2021}%
  \BibitemOpen
  \bibfield  {author} {\bibinfo {author} {\bibfnamefont {S.}~\bibnamefont
  {Lee}}, \bibinfo {author} {\bibfnamefont {W.}~\bibnamefont {Lee}}, \bibinfo
  {author} {\bibfnamefont {W.}~\bibnamefont {Guohua}}, \bibinfo {author}
  {\bibfnamefont {J.}~\bibnamefont {Ma}}, \bibinfo {author} {\bibfnamefont
  {H.}~\bibnamefont {Zhou}}, \bibinfo {author} {\bibfnamefont {M.}~\bibnamefont
  {Lee}}, \bibinfo {author} {\bibfnamefont {E.~S.}\ \bibnamefont {Choi}}, \
  and\ \bibinfo {author} {\bibfnamefont {K.-Y.}\ \bibnamefont {Choi}},\ }\href
  {\doibase 10.1103/PhysRevB.103.224430} {\bibfield  {journal} {\bibinfo
  {journal} {Phys. Rev. B}\ }\textbf {\bibinfo {volume} {103}},\ \bibinfo
  {pages} {224430} (\bibinfo {year} {2021})}\BibitemShut {NoStop}%
\bibitem [{\citenamefont {Arima}\ \emph {et~al.}(2022)\citenamefont {Arima},
  \citenamefont {Oshita}, \citenamefont {Hirai}, \citenamefont {Hiroi},\ and\
  \citenamefont {Matsubayashi}}]{Arima2022}%
  \BibitemOpen
  \bibfield  {author} {\bibinfo {author} {\bibfnamefont {H.}~\bibnamefont
  {Arima}}, \bibinfo {author} {\bibfnamefont {Y.}~\bibnamefont {Oshita}},
  \bibinfo {author} {\bibfnamefont {D.}~\bibnamefont {Hirai}}, \bibinfo
  {author} {\bibfnamefont {Z.}~\bibnamefont {Hiroi}}, \ and\ \bibinfo {author}
  {\bibfnamefont {K.}~\bibnamefont {Matsubayashi}},\ }\href {\doibase
  10.7566/JPSJ.91.013702} {\bibfield  {journal} {\bibinfo  {journal} {J. Phys.
  Soc. Jpn.}\ }\textbf {\bibinfo {volume} {91}},\ \bibinfo {pages} {013702}
  (\bibinfo {year} {2022})}\BibitemShut {NoStop}%
\bibitem [{\citenamefont {Mosca}\ \emph {et~al.}(2024)\citenamefont {Mosca},
  \citenamefont {Franchini},\ and\ \citenamefont {Pourovskii}}]{Mosca2024}%
  \BibitemOpen
  \bibfield  {author} {\bibinfo {author} {\bibfnamefont {D.~F.}\ \bibnamefont
  {Mosca}}, \bibinfo {author} {\bibfnamefont {C.}~\bibnamefont {Franchini}}, \
  and\ \bibinfo {author} {\bibfnamefont {L.~V.}\ \bibnamefont {Pourovskii}},\
  }\href@noop {} {} (\bibinfo {year} {2024}),\ \Eprint
  {http://arxiv.org/abs/2402.15564} {arXiv:2402.15564 [cond-mat.str-el]}
  \BibitemShut {NoStop}%
\bibitem [{\citenamefont {Kesavan}\ \emph {et~al.}(2020)\citenamefont
  {Kesavan}, \citenamefont {Fiore~Mosca}, \citenamefont {Sanna}, \citenamefont
  {Borgatti}, \citenamefont {Schuck}, \citenamefont {Tran}, \citenamefont
  {Woodward}, \citenamefont {Mitrovi\'{c}}, \citenamefont {Franchini},\ and\
  \citenamefont {Boscherini}}]{Kesavan2020}%
  \BibitemOpen
  \bibfield  {author} {\bibinfo {author} {\bibfnamefont {J.~K.}\ \bibnamefont
  {Kesavan}}, \bibinfo {author} {\bibfnamefont {D.}~\bibnamefont
  {Fiore~Mosca}}, \bibinfo {author} {\bibfnamefont {S.}~\bibnamefont {Sanna}},
  \bibinfo {author} {\bibfnamefont {F.}~\bibnamefont {Borgatti}}, \bibinfo
  {author} {\bibfnamefont {G.}~\bibnamefont {Schuck}}, \bibinfo {author}
  {\bibfnamefont {P.~M.}\ \bibnamefont {Tran}}, \bibinfo {author}
  {\bibfnamefont {P.~M.}\ \bibnamefont {Woodward}}, \bibinfo {author}
  {\bibfnamefont {V.~F.}\ \bibnamefont {Mitrovi\'{c}}}, \bibinfo {author}
  {\bibfnamefont {C.}~\bibnamefont {Franchini}}, \ and\ \bibinfo {author}
  {\bibfnamefont {F.}~\bibnamefont {Boscherini}},\ }\href {\doibase
  10.1021/acs.jpcc.0c04807} {\bibfield  {journal} {\bibinfo  {journal} {The
  Journal of Physical Chemistry C}\ }\textbf {\bibinfo {volume} {124}},\
  \bibinfo {pages} {16577} (\bibinfo {year} {2020})}\BibitemShut {NoStop}%
\bibitem [{\citenamefont {Cong}\ \emph {et~al.}(2023)\citenamefont {Cong},
  \citenamefont {Garcia}, \citenamefont {Forino}, \citenamefont {Tassetti},
  \citenamefont {Allodi}, \citenamefont {Reyes}, \citenamefont {Tran},
  \citenamefont {Woodward}, \citenamefont {Franchini}, \citenamefont {Sanna},\
  and\ \citenamefont {Mitrovi\ifmmode~\acute{c}\else \'{c}\fi{}}}]{Cong2023}%
  \BibitemOpen
  \bibfield  {author} {\bibinfo {author} {\bibfnamefont {R.}~\bibnamefont
  {Cong}}, \bibinfo {author} {\bibfnamefont {E.}~\bibnamefont {Garcia}},
  \bibinfo {author} {\bibfnamefont {P.~C.}\ \bibnamefont {Forino}}, \bibinfo
  {author} {\bibfnamefont {A.}~\bibnamefont {Tassetti}}, \bibinfo {author}
  {\bibfnamefont {G.}~\bibnamefont {Allodi}}, \bibinfo {author} {\bibfnamefont
  {A.~P.}\ \bibnamefont {Reyes}}, \bibinfo {author} {\bibfnamefont {P.~M.}\
  \bibnamefont {Tran}}, \bibinfo {author} {\bibfnamefont {P.~M.}\ \bibnamefont
  {Woodward}}, \bibinfo {author} {\bibfnamefont {C.}~\bibnamefont {Franchini}},
  \bibinfo {author} {\bibfnamefont {S.}~\bibnamefont {Sanna}}, \ and\ \bibinfo
  {author} {\bibfnamefont {V.~F.}\ \bibnamefont {Mitrovi\ifmmode~\acute{c}\else
  \'{c}\fi{}}},\ }\href {\doibase 10.1103/PhysRevMaterials.7.084409} {\bibfield
   {journal} {\bibinfo  {journal} {Phys. Rev. Mater.}\ }\textbf {\bibinfo
  {volume} {7}},\ \bibinfo {pages} {084409} (\bibinfo {year}
  {2023})}\BibitemShut {NoStop}%
\bibitem [{\citenamefont {Voleti}\ \emph {et~al.}(2023)\citenamefont {Voleti},
  \citenamefont {Pradhan}, \citenamefont {Bhattacharjee}, \citenamefont
  {Saha-Dasgupta},\ and\ \citenamefont {Paramekanti}}]{Voleti2023}%
  \BibitemOpen
  \bibfield  {author} {\bibinfo {author} {\bibfnamefont {S.}~\bibnamefont
  {Voleti}}, \bibinfo {author} {\bibfnamefont {K.}~\bibnamefont {Pradhan}},
  \bibinfo {author} {\bibfnamefont {S.}~\bibnamefont {Bhattacharjee}}, \bibinfo
  {author} {\bibfnamefont {T.}~\bibnamefont {Saha-Dasgupta}}, \ and\ \bibinfo
  {author} {\bibfnamefont {A.}~\bibnamefont {Paramekanti}},\ }\href {\doibase
  10.1038/s41535-023-00575-6} {\bibfield  {journal} {\bibinfo  {journal} {npj
  Quantum materials}\ }\textbf {\bibinfo {volume} {8}},\ \bibinfo {pages} {42}
  (\bibinfo {year} {2023})}\BibitemShut {NoStop}%
\bibitem [{\citenamefont {Celiberti}\ \emph {et~al.}(2024)\citenamefont
  {Celiberti}, \citenamefont {Mosca}, \citenamefont {Allodi}, \citenamefont
  {Pourovskii}, \citenamefont {Tassetti}, \citenamefont {Forino}, \citenamefont
  {Cong}, \citenamefont {Garcia}, \citenamefont {Tran}, \citenamefont {Renzi},
  \citenamefont {Woodward}, \citenamefont {Mitrovi\'{c}}, \citenamefont
  {Sanna},\ and\ \citenamefont {Franchini}}]{Celiberti2024}%
  \BibitemOpen
  \bibfield  {author} {\bibinfo {author} {\bibfnamefont {L.}~\bibnamefont
  {Celiberti}}, \bibinfo {author} {\bibfnamefont {D.~F.}\ \bibnamefont
  {Mosca}}, \bibinfo {author} {\bibfnamefont {G.}~\bibnamefont {Allodi}},
  \bibinfo {author} {\bibfnamefont {L.~V.}\ \bibnamefont {Pourovskii}},
  \bibinfo {author} {\bibfnamefont {A.}~\bibnamefont {Tassetti}}, \bibinfo
  {author} {\bibfnamefont {P.~C.}\ \bibnamefont {Forino}}, \bibinfo {author}
  {\bibfnamefont {R.}~\bibnamefont {Cong}}, \bibinfo {author} {\bibfnamefont
  {E.}~\bibnamefont {Garcia}}, \bibinfo {author} {\bibfnamefont {P.~M.}\
  \bibnamefont {Tran}}, \bibinfo {author} {\bibfnamefont {R.~D.}\ \bibnamefont
  {Renzi}}, \bibinfo {author} {\bibfnamefont {P.~M.}\ \bibnamefont {Woodward}},
  \bibinfo {author} {\bibfnamefont {V.~F.}\ \bibnamefont {Mitrovi\'{c}}},
  \bibinfo {author} {\bibfnamefont {S.}~\bibnamefont {Sanna}}, \ and\ \bibinfo
  {author} {\bibfnamefont {C.}~\bibnamefont {Franchini}},\ }\href {\doibase
  10.1038/s41467-024-46621-0} {\bibfield  {journal} {\bibinfo  {journal} {Nat.
  Commun.}\ }\textbf {\bibinfo {volume} {15}},\ \bibinfo {pages} {2429}
  (\bibinfo {year} {2024})}\BibitemShut {NoStop}%
\bibitem [{\citenamefont {Cong}\ \emph {et~al.}(2020)\citenamefont {Cong},
  \citenamefont {Nanguneri}, \citenamefont {Rubenstein},\ and\ \citenamefont
  {Mitrović}}]{Cong2020}%
  \BibitemOpen
  \bibfield  {author} {\bibinfo {author} {\bibfnamefont {R.}~\bibnamefont
  {Cong}}, \bibinfo {author} {\bibfnamefont {R.}~\bibnamefont {Nanguneri}},
  \bibinfo {author} {\bibfnamefont {B.}~\bibnamefont {Rubenstein}}, \ and\
  \bibinfo {author} {\bibfnamefont {V.~F.}\ \bibnamefont {Mitrović}},\ }\href
  {\doibase 10.1088/1361-648X/ab9056} {\bibfield  {journal} {\bibinfo
  {journal} {Journal of Physics: Condensed Matter}\ }\textbf {\bibinfo {volume}
  {32}},\ \bibinfo {pages} {405802} (\bibinfo {year} {2020})}\BibitemShut
  {NoStop}%
\bibitem [{\citenamefont {Fiore~Mosca}\ \emph {et~al.}(2021)\citenamefont
  {Fiore~Mosca}, \citenamefont {Pourovskii}, \citenamefont {Kim}, \citenamefont
  {Liu}, \citenamefont {Sanna}, \citenamefont {Boscherini}, \citenamefont
  {Khmelevskyi},\ and\ \citenamefont {Franchini}}]{Mosca2021}%
  \BibitemOpen
  \bibfield  {author} {\bibinfo {author} {\bibfnamefont {D.}~\bibnamefont
  {Fiore~Mosca}}, \bibinfo {author} {\bibfnamefont {L.~V.}\ \bibnamefont
  {Pourovskii}}, \bibinfo {author} {\bibfnamefont {B.~H.}\ \bibnamefont {Kim}},
  \bibinfo {author} {\bibfnamefont {P.}~\bibnamefont {Liu}}, \bibinfo {author}
  {\bibfnamefont {S.}~\bibnamefont {Sanna}}, \bibinfo {author} {\bibfnamefont
  {F.}~\bibnamefont {Boscherini}}, \bibinfo {author} {\bibfnamefont
  {S.}~\bibnamefont {Khmelevskyi}}, \ and\ \bibinfo {author} {\bibfnamefont
  {C.}~\bibnamefont {Franchini}},\ }\href {\doibase
  10.1103/PhysRevB.103.104401} {\bibfield  {journal} {\bibinfo  {journal}
  {Phys. Rev. B}\ }\textbf {\bibinfo {volume} {103}},\ \bibinfo {pages}
  {104401} (\bibinfo {year} {2021})}\BibitemShut {NoStop}%
\bibitem [{\citenamefont {Weng}\ and\ \citenamefont {Dong}(2021)}]{Weng2021}%
  \BibitemOpen
  \bibfield  {author} {\bibinfo {author} {\bibfnamefont {Y.}~\bibnamefont
  {Weng}}\ and\ \bibinfo {author} {\bibfnamefont {S.}~\bibnamefont {Dong}},\
  }\href {\doibase 10.1103/PhysRevB.104.165150} {\bibfield  {journal} {\bibinfo
   {journal} {Phys. Rev. B}\ }\textbf {\bibinfo {volume} {104}},\ \bibinfo
  {pages} {165150} (\bibinfo {year} {2021})}\BibitemShut {NoStop}%
\bibitem [{\citenamefont {Mansouri~Tehrani}\ and\ \citenamefont
  {Spaldin}(2021)}]{Tehrani2021}%
  \BibitemOpen
  \bibfield  {author} {\bibinfo {author} {\bibfnamefont {A.}~\bibnamefont
  {Mansouri~Tehrani}}\ and\ \bibinfo {author} {\bibfnamefont {N.~A.}\
  \bibnamefont {Spaldin}},\ }\href {\doibase 10.1103/PhysRevMaterials.5.104410}
  {\bibfield  {journal} {\bibinfo  {journal} {Phys. Rev. Mater.}\ }\textbf
  {\bibinfo {volume} {5}},\ \bibinfo {pages} {104410} (\bibinfo {year}
  {2021})}\BibitemShut {NoStop}%
\bibitem [{\citenamefont {Merkel}\ \emph {et~al.}(2024)\citenamefont {Merkel},
  \citenamefont {Tehrani},\ and\ \citenamefont {Ederer}}]{Merkel2023}%
  \BibitemOpen
  \bibfield  {author} {\bibinfo {author} {\bibfnamefont {M.~E.}\ \bibnamefont
  {Merkel}}, \bibinfo {author} {\bibfnamefont {A.~M.}\ \bibnamefont {Tehrani}},
  \ and\ \bibinfo {author} {\bibfnamefont {C.}~\bibnamefont {Ederer}},\ }\href
  {\doibase 10.1103/PhysRevResearch.6.023233} {\bibfield  {journal} {\bibinfo
  {journal} {Phys. Rev. Res.}\ }\textbf {\bibinfo {volume} {6}},\ \bibinfo
  {pages} {023233} (\bibinfo {year} {2024})}\BibitemShut {NoStop}%
\bibitem [{\citenamefont {Morgan}\ \emph {et~al.}(2023)\citenamefont {Morgan},
  \citenamefont {Kent}, \citenamefont {Zohar}, \citenamefont {O’Dea},
  \citenamefont {Wu}, \citenamefont {Cheetham},\ and\ \citenamefont
  {Seshadri}}]{Morgan2023}%
  \BibitemOpen
  \bibfield  {author} {\bibinfo {author} {\bibfnamefont {E.~E.}\ \bibnamefont
  {Morgan}}, \bibinfo {author} {\bibfnamefont {G.~T.}\ \bibnamefont {Kent}},
  \bibinfo {author} {\bibfnamefont {A.}~\bibnamefont {Zohar}}, \bibinfo
  {author} {\bibfnamefont {A.}~\bibnamefont {O’Dea}}, \bibinfo {author}
  {\bibfnamefont {G.}~\bibnamefont {Wu}}, \bibinfo {author} {\bibfnamefont
  {A.~K.}\ \bibnamefont {Cheetham}}, \ and\ \bibinfo {author} {\bibfnamefont
  {R.}~\bibnamefont {Seshadri}},\ }\href {\doibase
  10.1021/acs.chemmater.3c01300} {\bibfield  {journal} {\bibinfo  {journal}
  {Chemistry of Materials}\ }\textbf {\bibinfo {volume} {35}},\ \bibinfo
  {pages} {7032} (\bibinfo {year} {2023})}\BibitemShut {NoStop}%
\bibitem [{\citenamefont {Paramekanti}\ \emph {et~al.}(2020)\citenamefont
  {Paramekanti}, \citenamefont {Maharaj},\ and\ \citenamefont
  {Gaulin}}]{Paramekanti2020}%
  \BibitemOpen
  \bibfield  {author} {\bibinfo {author} {\bibfnamefont {A.}~\bibnamefont
  {Paramekanti}}, \bibinfo {author} {\bibfnamefont {D.~D.}\ \bibnamefont
  {Maharaj}}, \ and\ \bibinfo {author} {\bibfnamefont {B.~D.}\ \bibnamefont
  {Gaulin}},\ }\href {\doibase 10.1103/PhysRevB.101.054439} {\bibfield
  {journal} {\bibinfo  {journal} {Phys. Rev. B}\ }\textbf {\bibinfo {volume}
  {101}},\ \bibinfo {pages} {054439} (\bibinfo {year} {2020})}\BibitemShut
  {NoStop}%
\bibitem [{\citenamefont {Pourovskii}\ \emph {et~al.}(2021)\citenamefont
  {Pourovskii}, \citenamefont {Mosca},\ and\ \citenamefont
  {Franchini}}]{Pourovskii2021}%
  \BibitemOpen
  \bibfield  {author} {\bibinfo {author} {\bibfnamefont {L.~V.}\ \bibnamefont
  {Pourovskii}}, \bibinfo {author} {\bibfnamefont {D.~F.}\ \bibnamefont
  {Mosca}}, \ and\ \bibinfo {author} {\bibfnamefont {C.}~\bibnamefont
  {Franchini}},\ }\href {\doibase 10.1103/PhysRevLett.127.237201} {\bibfield
  {journal} {\bibinfo  {journal} {Phys. Rev. Lett.}\ }\textbf {\bibinfo
  {volume} {127}},\ \bibinfo {pages} {237201} (\bibinfo {year}
  {2021})}\BibitemShut {NoStop}%
\bibitem [{\citenamefont {Azumi}\ and\ \citenamefont
  {Matsuzaki}(1977)}]{Azumi1977}%
  \BibitemOpen
  \bibfield  {author} {\bibinfo {author} {\bibfnamefont {T.}~\bibnamefont
  {Azumi}}\ and\ \bibinfo {author} {\bibfnamefont {K.}~\bibnamefont
  {Matsuzaki}},\ }\href {\doibase
  https://doi.org/10.1111/j.1751-1097.1977.tb06918.x} {\bibfield  {journal}
  {\bibinfo  {journal} {Photochemistry and Photobiology}\ }\textbf {\bibinfo
  {volume} {25}},\ \bibinfo {pages} {315} (\bibinfo {year} {1977})}\BibitemShut
  {NoStop}%
\bibitem [{\citenamefont {Sato}\ \emph {et~al.}(2009)\citenamefont {Sato},
  \citenamefont {Tokunaga}, \citenamefont {Iwahara}, \citenamefont {Shizu},\
  and\ \citenamefont {Tanaka}}]{Sato2009}%
  \BibitemOpen
  \bibfield  {author} {\bibinfo {author} {\bibfnamefont {T.}~\bibnamefont
  {Sato}}, \bibinfo {author} {\bibfnamefont {K.}~\bibnamefont {Tokunaga}},
  \bibinfo {author} {\bibfnamefont {N.}~\bibnamefont {Iwahara}}, \bibinfo
  {author} {\bibfnamefont {K.}~\bibnamefont {Shizu}}, \ and\ \bibinfo {author}
  {\bibfnamefont {K.}~\bibnamefont {Tanaka}},\ }\enquote {\bibinfo {title}
  {Vibronic coupling constant and vibronic coupling density},}\ in\ \href
  {\doibase 10.1007/978-3-642-03432-9_5} {\emph {\bibinfo {booktitle} {The
  Jahn-Teller Effect: Fundamentals and Implications for Physics and
  Chemistry}}},\ \bibinfo {editor} {edited by\ \bibinfo {editor} {\bibfnamefont
  {H.}~\bibnamefont {K{\"o}ppel}}, \bibinfo {editor} {\bibfnamefont {D.~R.}\
  \bibnamefont {Yarkony}}, \ and\ \bibinfo {editor} {\bibfnamefont
  {H.}~\bibnamefont {Barentzen}}}\ (\bibinfo  {publisher} {Springer Berlin
  Heidelberg},\ \bibinfo {address} {Berlin, Heidelberg},\ \bibinfo {year}
  {2009})\ pp.\ \bibinfo {pages} {99--129}\BibitemShut {NoStop}%
\bibitem [{\citenamefont {Halliday}\ and\ \citenamefont
  {Tucker}(1988)}]{Halliday1988}%
  \BibitemOpen
  \bibfield  {author} {\bibinfo {author} {\bibfnamefont {I.}~\bibnamefont
  {Halliday}}\ and\ \bibinfo {author} {\bibfnamefont {J.~W.}\ \bibnamefont
  {Tucker}},\ }\href@noop {} {\bibfield  {journal} {\bibinfo  {journal} {J.
  Phys. C: Solid State Phys.}\ }\textbf {\bibinfo {volume} {21}},\ \bibinfo
  {pages} {5403} (\bibinfo {year} {1988})}\BibitemShut {NoStop}%
\bibitem [{\citenamefont {Iwahara}\ \emph {et~al.}(2017)\citenamefont
  {Iwahara}, \citenamefont {Vieru}, \citenamefont {Ungur},\ and\ \citenamefont
  {Chibotaru}}]{Iwahara2017}%
  \BibitemOpen
  \bibfield  {author} {\bibinfo {author} {\bibfnamefont {N.}~\bibnamefont
  {Iwahara}}, \bibinfo {author} {\bibfnamefont {V.}~\bibnamefont {Vieru}},
  \bibinfo {author} {\bibfnamefont {L.}~\bibnamefont {Ungur}}, \ and\ \bibinfo
  {author} {\bibfnamefont {L.~F.}\ \bibnamefont {Chibotaru}},\ }\href {\doibase
  10.1103/PhysRevB.96.064416} {\bibfield  {journal} {\bibinfo  {journal} {Phys.
  Rev. B}\ }\textbf {\bibinfo {volume} {96}},\ \bibinfo {pages} {064416}
  (\bibinfo {year} {2017})}\BibitemShut {NoStop}%
\bibitem [{\citenamefont {Warzanowski}\ \emph
  {et~al.}(2024{\natexlab{b}})\citenamefont {Warzanowski}, \citenamefont
  {Magnaterra}, \citenamefont {Schlicht}, \citenamefont {Faure}, \citenamefont
  {Sahle}, \citenamefont {Becker}, \citenamefont {Bohat\'y}, \citenamefont
  {Sala}, \citenamefont {Monaco}, \citenamefont {Hermanns}, \citenamefont {van
  Loosdrecht},\ and\ \citenamefont {Gr\"uninger}}]{Warzanowski2024b}%
  \BibitemOpen
  \bibfield  {author} {\bibinfo {author} {\bibfnamefont {P.}~\bibnamefont
  {Warzanowski}}, \bibinfo {author} {\bibfnamefont {M.}~\bibnamefont
  {Magnaterra}}, \bibinfo {author} {\bibfnamefont {G.}~\bibnamefont
  {Schlicht}}, \bibinfo {author} {\bibfnamefont {Q.}~\bibnamefont {Faure}},
  \bibinfo {author} {\bibfnamefont {C.~J.}\ \bibnamefont {Sahle}}, \bibinfo
  {author} {\bibfnamefont {P.}~\bibnamefont {Becker}}, \bibinfo {author}
  {\bibfnamefont {L.}~\bibnamefont {Bohat\'y}}, \bibinfo {author}
  {\bibfnamefont {M.~M.}\ \bibnamefont {Sala}}, \bibinfo {author}
  {\bibfnamefont {G.}~\bibnamefont {Monaco}}, \bibinfo {author} {\bibfnamefont
  {M.}~\bibnamefont {Hermanns}}, \bibinfo {author} {\bibfnamefont {P.~H.~M.}\
  \bibnamefont {van Loosdrecht}}, \ and\ \bibinfo {author} {\bibfnamefont
  {M.}~\bibnamefont {Gr\"uninger}},\ }\href {\doibase
  10.1103/PhysRevB.109.155149} {\bibfield  {journal} {\bibinfo  {journal}
  {Phys. Rev. B}\ }\textbf {\bibinfo {volume} {109}},\ \bibinfo {pages}
  {155149} (\bibinfo {year} {2024}{\natexlab{b}})}\BibitemShut {NoStop}%
\bibitem [{\citenamefont {Wang}\ \emph {et~al.}(2024)\citenamefont {Wang},
  \citenamefont {de~la Torre}, \citenamefont {Rodriguez-Rivera}, \citenamefont
  {Podlesnyak}, \citenamefont {Tian}, \citenamefont {Aczel}, \citenamefont
  {Matsuda}, \citenamefont {Ryan}, \citenamefont {Kim}, \citenamefont {Rau},\
  and\ \citenamefont {Plumb}}]{Wang2024}%
  \BibitemOpen
  \bibfield  {author} {\bibinfo {author} {\bibfnamefont {Q.}~\bibnamefont
  {Wang}}, \bibinfo {author} {\bibfnamefont {A.}~\bibnamefont {de~la Torre}},
  \bibinfo {author} {\bibfnamefont {J.~A.}\ \bibnamefont {Rodriguez-Rivera}},
  \bibinfo {author} {\bibfnamefont {A.~A.}\ \bibnamefont {Podlesnyak}},
  \bibinfo {author} {\bibfnamefont {W.}~\bibnamefont {Tian}}, \bibinfo {author}
  {\bibfnamefont {A.~A.}\ \bibnamefont {Aczel}}, \bibinfo {author}
  {\bibfnamefont {M.}~\bibnamefont {Matsuda}}, \bibinfo {author} {\bibfnamefont
  {P.~J.}\ \bibnamefont {Ryan}}, \bibinfo {author} {\bibfnamefont {J.-W.}\
  \bibnamefont {Kim}}, \bibinfo {author} {\bibfnamefont {J.~G.}\ \bibnamefont
  {Rau}}, \ and\ \bibinfo {author} {\bibfnamefont {K.~W.}\ \bibnamefont
  {Plumb}},\ }\href {https://arxiv.org/abs/2407.17559} {\enquote {\bibinfo
  {title} {Pulling order back from the brink of disorder: Observation of a
  nodal line spin liquid and fluctuation stabilized order in k$_2$ircl$_6$},}\
  } (\bibinfo {year} {2024}),\ \Eprint {http://arxiv.org/abs/2407.17559}
  {arXiv:2407.17559 [cond-mat.str-el]} \BibitemShut {NoStop}%
\end{thebibliography}%
\end{document}